\pdfoutput=1
\documentclass{aa}
\makeatletter
\renewcommand*\aa@pageof{, page \thepage{} of \pageref*{LastPage}}
\makeatother

\usepackage{tikz}
\usepackage{graphicx}
\usepackage[varg]{txfonts}
\usepackage[T1]{fontenc}
\usepackage{ae,aecompl}
\usepackage{amsmath}	
\usepackage{amssymb}	
\usepackage{times} 
\usepackage{mathptmx}
\usepackage{listings}
\usepackage{gensymb}
\usepackage{amsmath,xspace}
\usepackage{pdflscape}
\usepackage{rotating,booktabs}
 \usepackage{tabularx,tablefootnote} 

\usepackage[breaklinks=true]{hyperref} 
\usepackage{natbib}
\bibpunct{(}{)}{;}{a}{}{,} 
\usepackage{caption}

\usepackage[normalem]{ulem}
\definecolor{green}{rgb}{0.3,0.7,0.}

\let\oldAA\AA
\renewcommand{\AA}{\text{\normalfont\oldAA}}
\newcommand{\rmca}{RMC~127\xspace}

\newcommand{\rmc}{RMC$\,$143\xspace}

\begin{document}

   \title{The contribution by luminous blue variable stars\\ to the dust content of the Magellanic Clouds
   \thanks{{Based on observations made with ESO Telescopes at the La Silla Paranal Observatory under programme IDs: 097.D-0612(A,B) and 0100.D-0469(A,B). 
   } 
}\fnmsep
}

   \author{C. Agliozzo 
          \inst{1},
         N.  Phillips\inst{1},
         A. Mehner\inst{2},
         D. Baade\inst{1}, 
         P. Scicluna\inst{2},    
         F. Kemper\inst{1,3},
         D. Asmus\inst{2,4,5},
        W.-J. de Wit\inst{2},
         G. Pignata\inst{6,7}
          }
   \institute{European Southern Observatory, Karl-Schwarzschild-Strasse 2, 85748 Garching bei M\"unchen, Germany\\
   \email{claudia.agliozzo@eso.org}
         \and
             European Southern Observatory, Alonso de Cordova 3107, Vitacura,  Santiago de Chile, Chile
            \and 
              Institute of Astronomy and Astrophysics, Academia Sinica, No. 1, Sec. 4, Roosevelt Rd., Taipei 10617, Taiwan 
        \and     
         School of Physics \& Astronomy, University of Southampton, Southampton SO17 1BJ, UK
        \and
           Gymnasium Schwarzenbek, Buschkoppel 7, 21493 Schwarzenbek, Germany  
             \and
             Departamento de Ciencias Fisicas, Universidad Andres Bello,  
            Avda. Republica 252, Santiago, 8320000, Chile
            \and
            Millennium Institute of Astrophysics (MAS), Nuncio Monse{\~{n}}or S{\'{o}}tero Sanz 100, Providencia, Santiago, Chile  
             }
   \date{}

 
  \abstract
   {Previous studies have concluded that low- and intermediate-mass stars cannot account for the interstellar dust yield in the Magellanic Clouds inferred from far-infrared and sub-millimetre observations.}
   {Luminous blue variable stars (LBVs) form dust as a result of episodic, violent mass loss.  
To investigate their contribution as dust producers in the Magellanic Clouds, we analyse 31 confirmed and candidate LBVs from a recent census.}
   {We built a maximally complete multi-wavelength dataset of these sources from archival space telescope images and catalogues from near-infrared to millimetre wavelengths. We also present new Very Large Telescope VISIR observations of three sources  in the Large Magellanic Cloud (LMC). 
    We review the LBV classification on the basis of the infrared spectral energy distribution. 
   To derive characteristic dust parameters, we fitted the photometry resulting from a stacking analysis, which consists of co-adding images of the same wavelength band of several targets to improve the signal-to-noise.  
   For comparison we also stacked the images of low- and intermediate-mass evolved stars in the LMC.}
   {We find four classes of sources: 1) LBVs showing mid-infrared dust emission plus near-infrared free-free emission from an ionised stellar wind (Class 1a) or only mid-infrared dust emission (Class 1b); 2) LBVs with a near-infrared excess due to free-free emission only (Class 2); 3) objects with an sgB[e] classification in the literature, displaying a distinctive hot dust component; and 4) objects with no detected stellar winds and no circumstellar matter in their SEDs.  
From the stacking analysis of the 18 Class 1 and 2 objects in the LMC, we derived an integrated dust mass of $0.11^{+0.06}_{-0.03}\,M_{\odot}$. 
   This is two orders of magnitude larger than the value inferred from stacking 1342 extreme-asymptotic giant branch (AGB) stars. The dust mass of individual LBVs  does not correlate with the stellar parameters, possibly suggesting that the dust production mechanism is independent of the initial stellar mass or that the stars have different evolutionary histories. The total dust yield from LBVs over the age of the LMC is $\sim 10^{4}-10^{5} \, M_{\odot}$. The one order of magnitude uncertainty is mainly due to uncertainties of the LBV population, star formation history, and initial mass function. }
  {LBVs are potentially the second most important source of dust in normal galaxies. The role of dust destruction in LBV nebulae by a possible subsequent supernova (SN) blast wave has yet to be determined. Recent theoretical developments in the field of dust processing by SN shocks highlight the potential survival of dust grains from the pre-existing circumstellar nebula.
}

\titlerunning{LBVs as dust factories at sub-solar metallicities}
\authorrunning{Agliozzo et al.}
\maketitle 

%

\section{Introduction}

The origin of dust on cosmic scales is of great astrophysical interest. The cosmic importance of dust is illustrated by the fact that about $50\%$ of starlight is reprocessed by dust \citep{2006Dole}. Thus, dust dominates how we see galaxies. 
Dust is very important for the thermal balance of the interstellar medium (ISM) of galaxies and accordingly for star formation \citep[][and references therein]{2016McKinnon}.
Large amounts of dust have been observed in high redshift galaxies \citep[$z\gtrapprox6$, e.g. ][]{2017Riechers, 2017Laporte}, 
where low- and intermediate-mass stars
did not have time to evolve and thus to inject dust into the ISM. It is widely accepted now that dust must arise from rapidly evolving (1--10 Myr)
massive stars, particularly in the ejecta of core-collapse supernovae (CC-SNe) \citep{2006Sugerman}. In fact, large amounts of dust ($0.1-1\,M_{\odot}$) were postulated by \citet{2001TF} and \citet{2003Nozawa}, under the assumption of favourable conditions in the SN ejecta. Nevertheless, for nearby CC-SNe, the dust masses inferred from mid-infrared observations are in general of the order of $10^{-4}-10^{-2}\,M_{\odot}$ \citep[e.g.][]{2009Kotak,2011Fox}. Dust masses of the order of $10^{-1}\,M_{\odot}$ were reported from sub-millimetre observations of SN1987A  \citep[e.g.][]{2014Indebetouw, 2015Matsuura}, suggesting that perhaps cold dust emission is not captured by mid-infrared observations. However, such estimates are difficult to compare because they are often based on different observational constraints and assumptions.  It is also expected that
only a fraction of the newly formed dust reaches the ISM, as it will be heavily processed and destroyed by the reverse shock \citep[see][]{2018Micelotta, 2018Gall}.  
Dust can also grow in dense ISM from existing dust grains \citep[e.g.][]{2009Draine},
although this is still not sufficient, together with the contribution from SNe, to explain dust yields in galaxies \citep[e.g.][]{2019Tamura}. The topic is  also controversial with respect to the nearby  Magellanic Clouds (MCs).

The MCs are two irregular dwarf galaxies with metallicities 0.5 and 0.25 times Solar for the Large Magellanic Cloud (LMC) and the Small Magellanic Cloud (SMC), respectively.
Their relative proximity to the Milky Way (MW), accurate distance measurements and low line-of-sight extinction make them an ideal laboratory, resembling the physical conditions of galaxies at the peak of star formation. 
In the past decades, several efforts were made to estimate dust yields in these galaxies and determine the primary dust sources at low metallicities. With the advent of \emph{Herschel} this became possible. Particularly the HERITAGE \emph{Herschel} key project \citep{2013Meixner}, observing in five bands from 100 to 500$\,\rm\mu m$, offered a plethora of data that advanced our understanding of the dust budget problem. Recent estimates report integrated dust masses of $(7.3\pm1.7)\times10^{5}$ and $(8.3\pm2.1)\times10^{4}\,M_{\odot}$ for the LMC and SMC, respectively \citep{2014Gordon}, although later \citet{2017Chastenet} derived dust masses about a factor of two smaller. Several teams tried to identify the main sources of dust in such galaxies. \citet{2009Srinivasan}, \citet{2009Matsuura}, \citet{2012Riebel}, \citet{2012Boyer} and \citet{2016Srinivasan} studied the dust production rate from low- and intermediate-mass stars in the MCs using mid-infrared observations. 
\citet{2015Jones} identified some tens of new evolved stellar objects from HERITAGE. 
These works generally agree on the result that the global dust production from the numerous population of asymptotic giant branch (AGB) and red supergiant (RSG) stars cannot account for the ISM dust reported from far-infrared and sub-millimetre observations. Another result is that the majority of dust input from low- and intermediate-mass stellar sources comes from ``extreme'' AGB   \citep{2009Srinivasan, 2012Riebel, 2012Boyer}.

\emph{Spitzer} observations from the SAGE legacy survey \citep{2006Meixner} were used by \citet{2009Bonanos,2010Bonanos} to investigate the infrared properties of massive stars. They found an excess of emission at 24$\,\rm \mu m$ due to dust from a handful of luminous blue variable (LBV) stars. Some of these, and other LBVs with known extended nebulae detected with {\it{HST\/}} \citep{2003Weis}, were then followed up and large masses of dust were measured \citep{2014Niyogi,2017AgliozzoA, 2017AgliozzoB, 2019Agliozzo}. 

In the classical scenario, LBVs are transitional objects from the main sequence of O, B stars to the Wolf-Rayet (WR) phase \citep{1984Conti}. This represents a short phase of evolution of the most massive ($M\geq25\,M_{\odot}$) stars, ranging in duration from $10^{4}$ to $10^{5}\,\rm yr$ \citep[e.g.][]{1989Maeder, 2007Massey, 2014Smith}. Observationally, LBVs are blue supergiants (BSGs) that experience instabilities, observed in the form of spectroscopic and photometric variabilities, usually in the optical and near-infrared. 
These instabilities are poorly understood but, on the basis of their observational properties, are phenomenologically divided into three main groups: giant eruptions, S~Doradus-type variability, and micro-variabilities \citep{1994HD,vanGenderen2001}. Although the physical mechanisms underlying these instabilities are possibly different \citep[e.g. the vicinity to the Eddington limit, sub-photospheric instabilities, bistability jump, envelope inflation,  wind-envelope interaction, fast rotation, binarity, stellar merger;][]{1994HD,1989Gallagher,1999Vink,2009Groh,2012A&A...538A..40G,2016Portegies,2017Owocki, 2021Grassitelli}, observations suggest increases in the mass-loss rates by about a factor of three in S~Doradus-type variability (e.g. \citealt{2002Vink}) or by several orders of magnitude in giant eruptions, like that of $\eta$~Carinae (e.g. \citealt{1994HD}). During the quiescent state, LBV mass-loss rates range between   $10^{-7}$ and $10^{-5}\,M_{\odot}\,\rm yr^{-1}$, typical of BSGs as explained by the radiatively-driven stellar-wind theory \citep{1975CAK, 2019A&A...632A.126S}. 

Particularly enigmatic are the rare giant eruptions, initially often mistaken for CC-SNe, because of their brightness,   
and subsequently named SN impostors \citep[e.g.][]{2012ApJ...746..179V, 2013Mauerhan, 2016ApJ...823L..23T, 2017Pastorello, 2018NElias, 2019Reguitti}. Famous examples in our Galaxy are $\eta$~Car and P~Cygni.  
Observationally, mass-loss rates exceeding $10^{-3}-10^{-2}\,M_{\odot}\,\rm yr^{-1}$ are estimated \citep[e.g.][]{2014Smith}. In the case of $\eta$ Car, more than $40 \,M_{\odot}$ \citep{2017Morris} were released during the Great Eruption in the XIX~century, which lasted about 20 years, forming the Homunculus nebula. In eruptions of SN impostors, nebular masses of the order of $0.1\,M_{\odot}$ are observed \citep[e.g.][]{2014Margutti}. Several  observational and theoretical works suggest that some LBVs could be the immediate progenitors of SNe \citep[e.g. ][]{2006Kotak,2008Trundle,2009Natur.458..865G,2007ApJ...666.1116S,2008ApJ...686..467S,2013A&A...550L...7G,groh14b,boian18}. 

Luminous blue variable stars are often surrounded by dusty circumstellar nebulae. In the Milky Way typical dust masses ranging between $10^{-3}$ and $10^{-1}\,M_{\odot}$ are reported \citep{1997Hutsemekers,2012Umana,2013Vamvatira,2015Vamvatira,2014Agliozzo,2014Lau,2018Arneson}. Furthermore, several tens of candidate LBVs were identified on the basis of the detection of infrared circumstellar nebulae \citep{2002Egan, 2003Clark, 2010Gvaramadze} and of an infrared excess usually peaking between 24 and $70~\mu$m. The central star, which is a BSG during quiescence, is very hot, and one does not expect dust condensation in its wind. However, the physical conditions in the optically thick and cool ``pseudo-photospheres'' formed during eruptions are favourable for dust condensation and growth \citep{2011Kochanek}. 
\citeauthor{2011Kochanek} showed analytically that the required temperature and particle density for dust formation are met for mass-loss rates larger than $10^{-2.5}\,M_{\odot}\,\rm yr^{-1}$, that are typical of giant eruptions. 
Grain growth and dust evolution may be observed in S~Doradus variables (which have mass-loss rates that are a couple of orders of magnitudes lower), as suggested by the disappearance of the silicate bump around $10-13\,\rm \mu m$ during the most recent S~Dor outburst of the Magellanic LBV RMC~71 \citep{2017Mehner}. 
Another way to form dust around a massive star is in the colliding winds of close binaries, as in some Wolf Rayet stars. In these systems dust seems to form episodically, near the periastron passage, or persistently, creating  pinwheel-like nebulae \citep[e.g.][]{2007Crowther,2020Lau}. 
This mechanism of dust formation is presumably the case of  the Galactic LBV binary HR~Car \citep{2016Boffin}, a model of which to explain the clumpy and dusty inner nebula is an Archimedean spiral centred on the binary system \citep[alternative to the expanding bipolar lobe model, see][ and references therein]{2017Buemi}. Episodic dust formation also  potentially occurs in the shocked regions of $\eta$~Car's binary colliding winds \citep{2010SmithN}.
\begin{figure*}
\centering
   \includegraphics[width=1\linewidth]{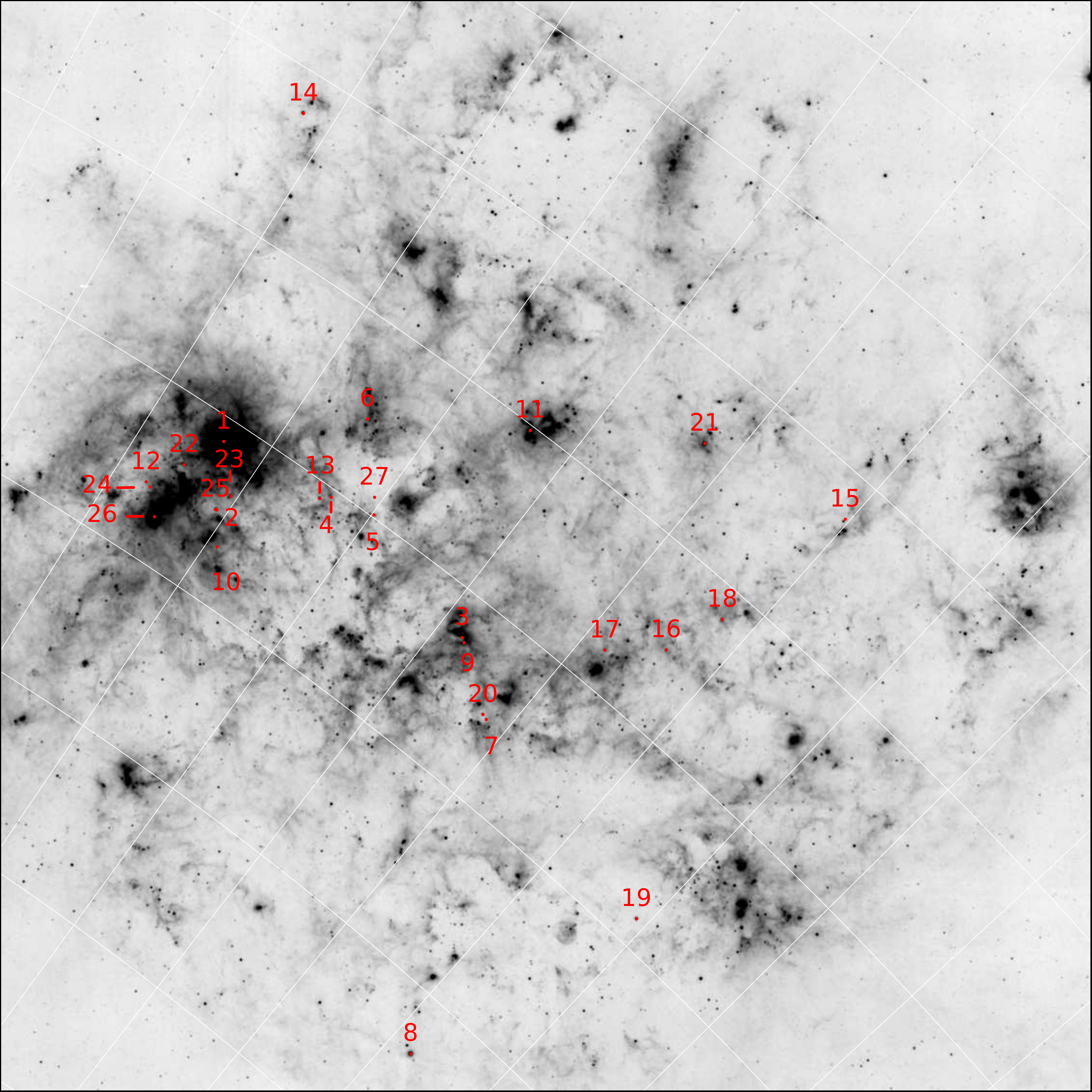}
     \caption{Cutout of the SAGE MIPS24 image \citep{2006Meixner} of the LMC and positions of the candidate and confirmed LBVs analysed in this work. The numbers are the IDs in Table~\ref{tab:sample}, and the red circles have a radius of 30\arcsec. North is towards the top right. }
     \label{fig:wholeLMC}
\end{figure*}
\begin{figure}
\centering
   \includegraphics[trim={1cm 1cm 1cm 1cm},clip,width=1\linewidth]{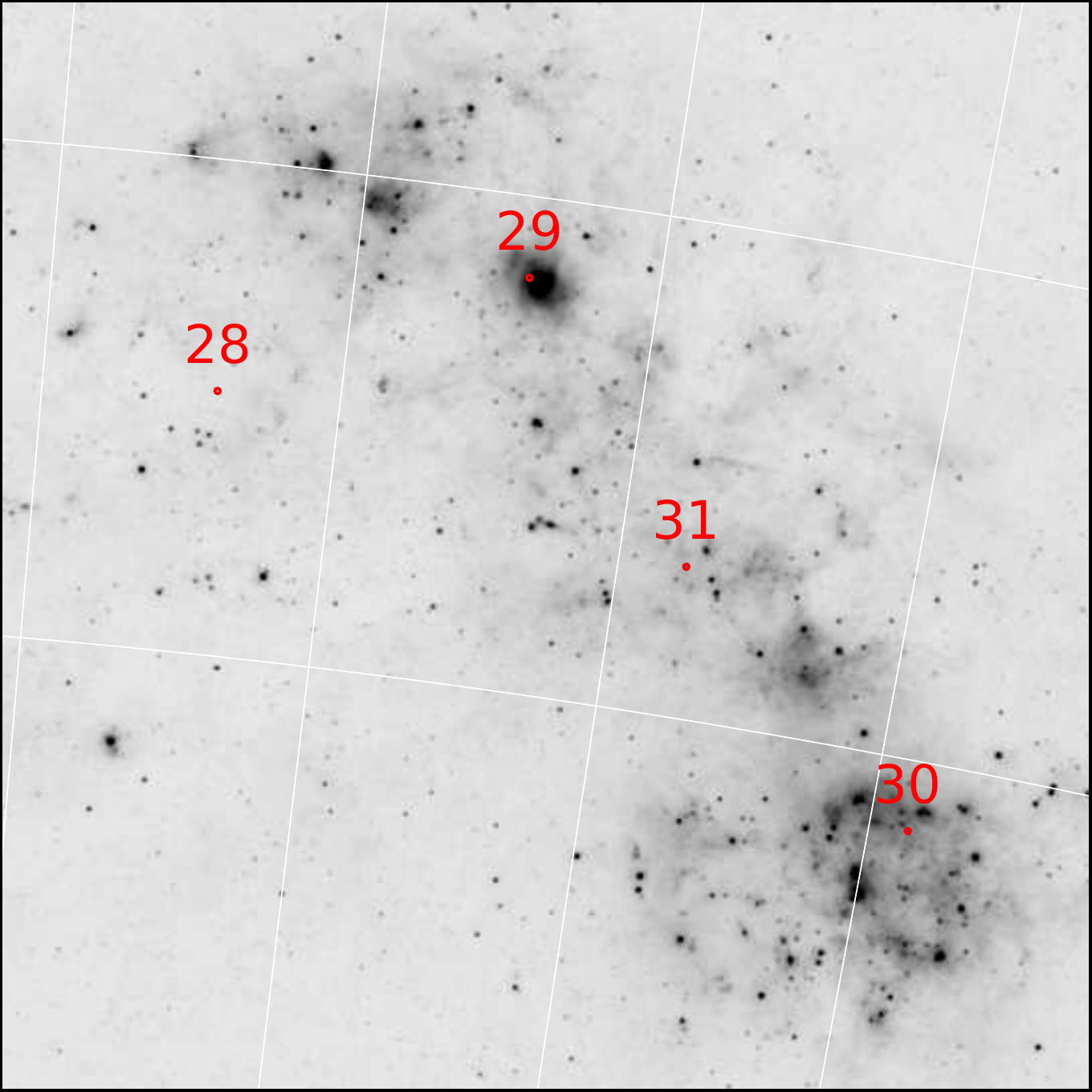}
     \caption{Cutout of the SAGE MIPS24 image \citep{2006Meixner} of the SMC and positions of the candidate and confirmed LBVs analysed in this work. The numbers are the IDs in Table~\ref{tab:sample}, and the red circles have a radius of 30\arcsec.  North is up and east is left. }
     \label{fig:wholeSMC}
\end{figure}

In the MCs only up to one third of the LBV sample appears in studies of the primary dust sources (see above), and the dust masses from these objects remain unconstrained. Here we aim to expand these studies to the full population of LBVs in the MCs. Figs.~\ref{fig:wholeLMC} and \ref{fig:wholeSMC} show the locations of the candidate and confirmed LBVs in the LMC and SMC that are analysed in this work. We have built a maximally complete multi-wavelength dataset of 31 LBVs mostly from archival space telescope images and catalogues, but also have included ground-based mid-infrared observations of three sources. 
In addition, we derived new photometric values for sources missing in catalogues or with low-accuracy photometry (Sects.~\ref{sec:dataset} and \ref{sec:photometry}).
The goals of this work are: 
\begin{enumerate}
    \item Review the status of the MC LBV sample by means of their infrared properties, combined with literature information (Sect.\  \ref{sec:review}). 
    \item  Estimate the dust masses of LBVs in the MCs by modelling individual sources 
    (Sect.~\ref{sec:modelling:grey}) and modelling photometry from stacked images of the LBVs in the LMC (Sect.~\ref{sec:stacking}).
    \item Evaluate the importance of LBVs in producing dust by comparing the derived dust mass with that inferred from the stacking of lower-mass stars and with literature values of ISM dust yields (Sects.\ \ref{sec:stackingAGBs} and \ref{sec:discussion}). 
\end{enumerate}

\section{The multiwavelength dataset}
\label{sec:dataset}
\begin{table*}
\caption{Sources, celestial coordinates \citep{2018yCat.1345....0G}, LBV status (previous: \citealt{2018RNAAS...2c.121R}; new: this work), and SED-based classification proposed in this work. 
}
\label{tab:sample}
\centering
\begin{tabular}{rllccccc}
\hline\hline
ID & Name 1   & Name 2 	& RA (ICRS, J2000) &  Dec (ICRS, J2000) &  \multicolumn{2}{c}{LBV 	status}	& SED	\\
   &          &         &                  &                    &  Previous&New	& Class\tablefootmark{a}	\\
\hline
 1 & RMC$\,$143          & CPD$\,-69\, 463$   & 05 38 51.617 & $-69$ $08$ $07.31$ & LBV & LBV        &   1 \\
 2 & RMC$\,$127          & HD$\,$269858       & 05 36 43.694 & $-69$ $29$ $47.45$ & LBV & LBV        &   1 \\
 3 & S$\,$Doradus        & HD$\,$5343         & 05 18 14.357 & $-69$ $15$ $01.15$ & LBV & LBV        &   1 \\
 4 & RMC$\,$110          & HD$\,$269662       & 05 30 51.476 & $-69$ $02$ $58.59$ & LBV & LBV        & 1 \\
 5 & HD$\,$269582        & LHA$\,$120-S$\,$83 & 05 27 52.662 & $-68$ $59$ $08.49$ & LBV & LBV        &   2 \\
 6 & RMC$\,$116          & HD$\,$269700       & 05 31 52.282 & $-68$ $32$ $38.86$ & LBV & LBV        &   2 \\
 7 & HD$\,$269216        & LHA$\,$120-S$\,$88 & 05 13 30.781 & $-69$ $32$ $23.65$ & LBV & LBV        &   1 \\
 8 & RMC$\,$71           & HD$\,$269006       & 05 02 07.394 & $-71$ $20$ $13.12$ & LBV & LBV        &   1 \\
 9 & RMC$\,$85	         & HD$\,$269321       & 05 17 56.074 & $-69$ $16$ $03.81$ & LBV	& LBV        & --- \\ 		
10 & RMC$\,$123	         & HD$\,$37836        & 05 35 16.633 & $-69$ $40$ $38.44$ & cLBV& cLBV	    &   2 \\
11 & RMC$\,$99	         & HD$\,$269445       & 05 22 59.785 & $-68$ $01$ $46.63$ & cLBV& cLBV	    &   2 \\
12 & Sk$\,-69\, 279$	 & ---                & 05 41 44.656 & $-69$ $35$ $14.90$ & cLBV& cLBV	    &   1 \\
13 & LHA$\,$120-S$\,$119 & HD$\,$269687       & 05 31 25.525 & $-69$ $05$ $38.56$ & cLBV& cLBV        &   1 \\
14 & LHA$\,$120-S$\,$61  & AL$\,$418          & 05 45 51.939 & $-67$ $14$ $25.94$ & cLBV& cLBV        &   1 \\ 
15 & RMC$\,$74	         & HD$\,$268939       & 05 04 14.909 & $-67$ $15$ $05.25$ & cLBV& cLBV        &   2 \\
16 & RMC$\,$78           & HD$\,$269050       & 05 07 20.422 & $-68$ $32$ $08.57$ & cLBV& cLBV        &   1 \\
17 & RMC$\,$81           & HD$\,$269128       & 05 10 22.789 & $-68$ $46$ $23.82$ & cLBV& cLBV        &   1 \\
18 & LHA$\,$120-S$\,$18  & Sk$\,-68\,42$      & 05 05 53.981 & $-68$ $10$ $50.54$ & cLBV& cLBV        &   1 \\
19 & RMC$\,$66           & HD$\,$268835       & 04 56 47.080 & $-69$ $50$ $24.77$ & cLBV&sgB[e]\tablefootmark{b}      &   3 \\
20 & RMC$\,$84           & HD$\,$269227       & 05 13 54.280 & $-69$ $31$ $46.66$ & cLBV&sgB[e]\tablefootmark{c}      &   3 \\
21 & HD$\,$34664         & Sk$\,-67\,64$      & 05 13 52.994 & $-67$ $26$ $54.82$ & cLBV&sgB[e]\tablefootmark{b}      &   3 \\
22 & HD$\,$38489         & Sk$\,-69\,259$     & 05 40 13.321 & $-69$ $22$ $46.49$ & cLBV&sgB[e]\tablefootmark{b}      &   3 \\
23 & RMC$\,$126          & HD$\,$37974        & 05 36 25.854 & $-69$ $22$ $55.79$ & cLBV&sgB[e]\tablefootmark{b}      &   3 \\
24 & Sk$\,-69\,271$      & CPD$\,-69\,500$    & 05 41 20.126 & $-69$ $36$ $22.89$ & cLBV&YSG\tablefootmark{d}        &   4 \\
25 & RMC$\,$128          & HD$\,$269859       & 05 36 47.188 & $-69$ $29$ $52.09$ & cLBV&B2Ia\tablefootmark{e}        &   4 \\
26 & RMC$\,$149          & Sk$\,-69\,257$     & 05 39 58.745 & $-69$ $44$ $04.07$ & cLBV&O8.5 II((f))\tablefootmark{f}        &   4 \\
27 & HD$\,$269604        & Sk$\,-68\,93$      & 05 28 31.367 & $-68$ $53$ $55.75$ & cLBV&A1 Ia-0\tablefootmark{g}        &   4 \\
\hline
28 & RMC$\,$14           & HD$\,$5980	      & 00 59 26.585 & $-72$ $09$ $53.93$ & LBV & LBV        &   2 \\
29 & RMC$\,$40           & HD$\,$6884	      & 01 07 18.218 & $-72$ $28$ $03.66$ & LBV& LBV	        &   1 \\
30 & RMC$\,$4            & LHA$\,$115-S$\,$6  & 00 46 55.030 & $-73$ $08$ $34.14$ & cLBV& cLBV        &   3 \\
31 & LHA$\,$115-S$\,$18  & ---                & 00 54 09.542 & $-72$ $41$ $43.29$ & cLBV& cLBV        &   3 \\
\hline
\end{tabular}
\tablefoot{The sample includes confirmed LBV  and candidate LBV (cLBV) stars in the MCs. 
\tablefoottext{a}{Class~1: sources with a warm/cool dusty nebula peaking in the mid- to far-infrared. Class~2: sources  displaying from 1 to $\sim24\,\rm \mu m$ only free-free emission from the stellar wind in addition to the stellar photosphere. Class~3: sources with an SED dominated at all wavelengths $\gtrsim 2\,\rm \mu m$} by a hot dusty component. Class~4: featureless sources only displaying the stellar photosphere. \tablefoottext{b}{\citet{1986Zickgraf}.}\tablefoottext{c}{\citet{1984Stahl}.}\tablefoottext{d}{\citet{2012Neugent}.}
\tablefoottext{e}{\citet{1991Fitzpatrick}.}
\tablefoottext{f}{\citet{2009Farigna}.}
\tablefoottext{g}{\citet{1973Osmer}.}
}
\end{table*}

The sample was extracted from the most recent census of LBVs in the Local Group \citep{2018RNAAS...2c.121R}, which contains 31 objects in the MCs. Table~\ref{tab:sample} includes the names and celestial coordinates of all the stars included in this work. More information on individual sources 
is provided in Sect.~\ref{sec:review}. 

\subsection{VISIR observations}
 \begin{figure*}
\centering
   \includegraphics[width=16cm]{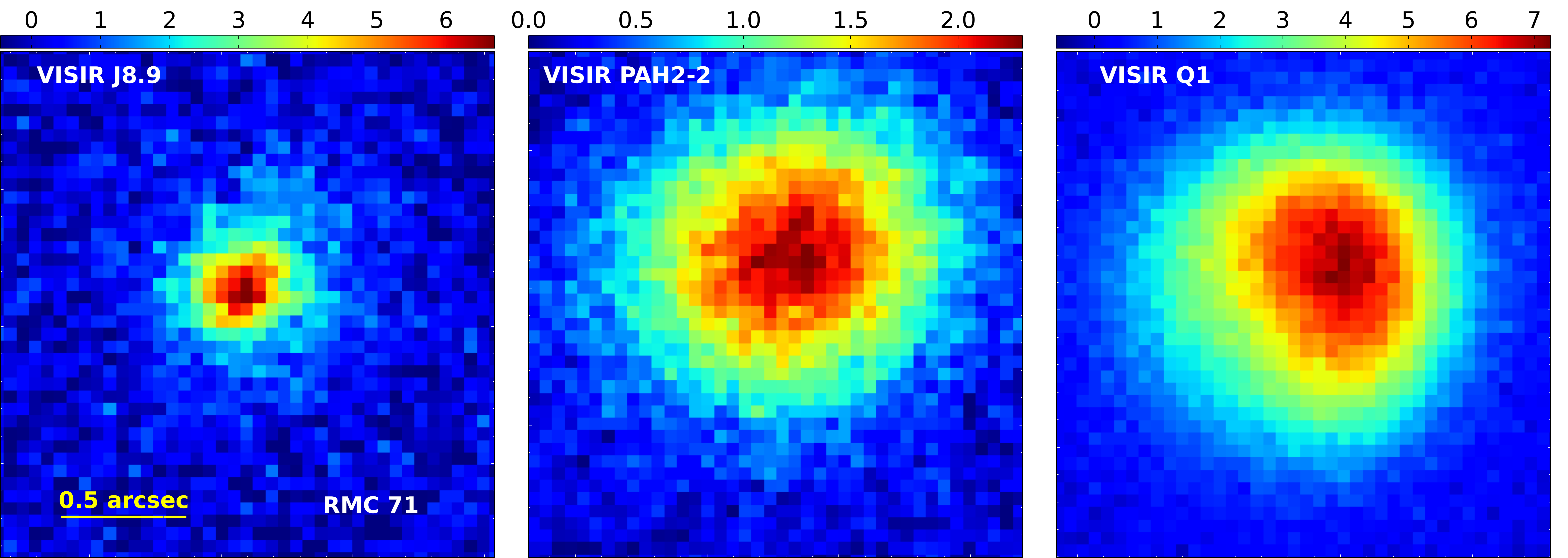}
     \medskip
     \includegraphics[width=16cm]{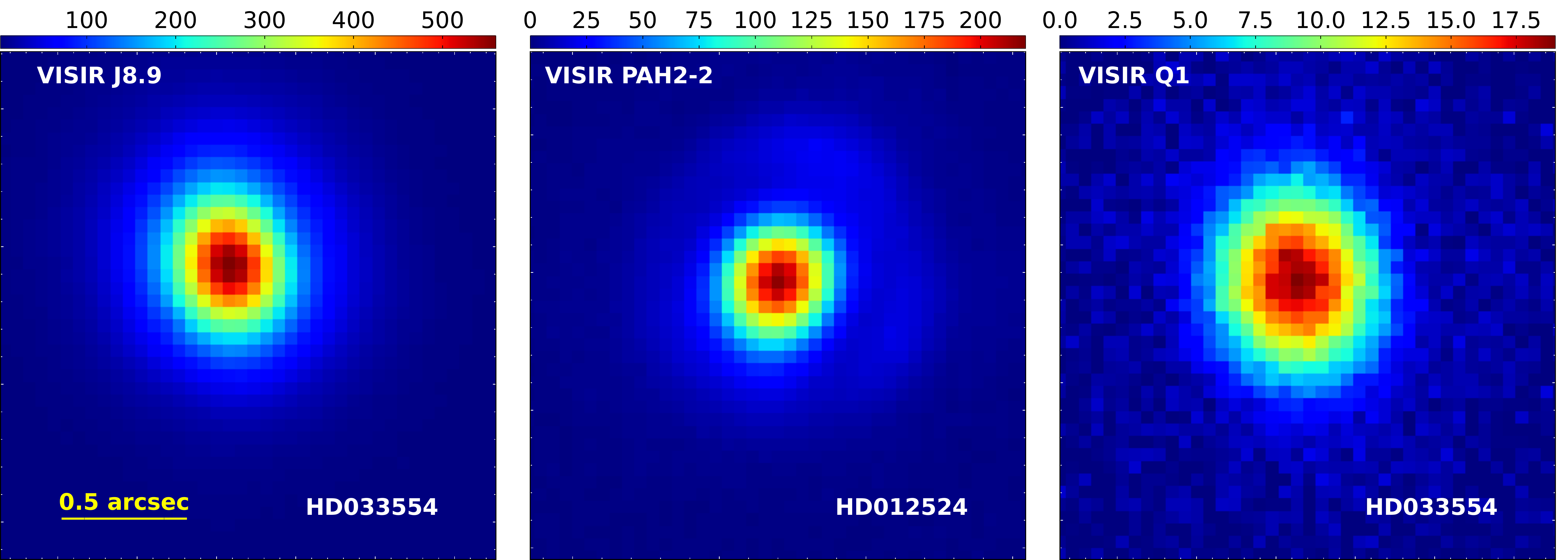}
     \caption{VISIR images of RMC$\,$71 (top panels), and the standard calibrators observed on the same nights (bottom panels, with identifications given in the lower right corners). North is up and east is left. The colour-scale is in arbitrary units.  
     }
     \label{visirR71}
\end{figure*}
\begin{table*}
\caption{Journal of VISIR observations. }
\label{tab:obsVISIR}
\centering
\begin{tabular}{llccccc}
\hline\hline    
Source       & Filter    & UT start time     & Airmass  &  IQ at $\lambda$   & Exposure & {$F_\nu$} \\
             &           &                   &          & (arcsec) & (s)      & (mJy)    \\
    \hline\\
    RMC~71   & J8.9      & 2017-10-07 07:47  & 1.47     &  N/A        & 90   & $136\pm19$\\
             & J8.9      & 2017-10-11 07:30  & 1.47     &    0.42      & 90   & $122\pm12$\\
    & PAH2\_2 & 2016-08-14 08:57 & 1.63 & 0.35 & 1892 & $408\pm 41$ \\ 
    & PAH2\_2 & 2016-08-18 08:31 & 1.65 & 0.36 & 1892 & $482\pm 48$ \\ 
    & PAH2\_2 & 2016-09-05 08:11  & 1.57 & 0.35 & 1892 & $408\pm 41$ \\ 
             & Q1        & 2016-08-17 08:58 & 1.64 & 0.46 & 588 & $3261\pm326$ \\
             & Q1        & 2017-10-07 07:56  & 1.46     &   0.46    & 1980 &  $3207\pm321$ \\
             & Q1        & 2017-10-11 07:35  & 1.46     &   0.47    & 1980 & $3490\pm349$\\
    \\
    HD~34664 & J8.9      & 2017-10-08 07:49  & 1.38     &   0.29       & 90   & $1051\pm105$\\
             & PAH2$\_$2 & 2016-09-05 08:54  & 1.43     &    0.38      & 1260 & $944\pm94$\\
             & PAH2$\_$2 & 2016-09-08 09:02  & 1.40     &    0.38      & 1260 & $903\pm90$\\
             & Q1        & 2017-10-08 07:55  & 1.37     &    0.46      & 1980 & $580\pm58$\\
    \\
    HD~38489 & J8.9      & 2017-10-27 07:33 &  1.41     &    N/A      & 90 &  $905\pm129$ \\
             & PAH2$\_$2 & 2016-09-22 09:05 &  1.42     &    0.32      & 990  & $781\pm78$\\
             & Q1        & 2017-10-27 07:54 &  1.41     &     0.46     & 1980 & $561\pm56$\\
\hline
\end{tabular}
\tablefoot{
IQ at $\lambda$ gives the approximate image quality at the observing wavelength at the zenith as measured from the 
flux standard stars observed before or after the science targets. 
Filter centre wavelengths are J8.9: $8.7\,\mu{\rm m}$, PAH2$\_$2: $11.68\,\mu{\rm m}$, Q1: $17.65\,\mu{\rm m}$.
}
\end{table*}

RMC$\,$71, HD$\,$38489, and HD$\,$34664 were observed in 2016 and 2017 (ESO program IDs: 097.D-0612(A,B) and 0100.D-0469(A,B)) with the upgraded Vlt Imager and Spectrometer for the mid-InfraRed (VISIR; \citealt{2004Lagage,2015Kaufl,2016Kerber}) on the Very Large Telescope \citep[VLT, ][]{1991Enard}.
Some of the data of RMC$\,$71 from program 097.D-0612 were previously published in \citet{2017Mehner}. The images were taken in standard imaging mode through filters J8.9 ($8.7 \pm0.37\,\mu$m), PAH2\_2 ($11.88 \pm0.19\,\mu$m) and Q1 ($17.65 \pm0.42\,\mu$m) with perpendicular nodding and a chop/nod throw of $20\arcsec$ in 097.D-0612 and $10\arcsec$ in 0100.D-0469, respectively.
The individual observations are listed in Table~\ref{tab:obsVISIR}.  J8.9 was used for target acquisition only (hence the short exposure times).
For flux calibration and point spread function (PSF) reference, the science observations were preceded or followed by the observation of mid-infrared standard stars \citep{1999Cohen}, which provide a systematic flux uncertainty of $\le10\%$.
The data were reduced with the custom Python tool VISIR and ISAAC Pipeline Environment\footnote{\url{https://github.com/danielasmus/vipe}} (VIPE).
Fluxes and source sizes were measured with Gaussian-fitting photometry performed as part of the VIPE reduction.

Both epochs of the PAH2\_2 observations of HD$\,$34664 were affected by a chopping-induced PSF instability at the VLT, leading to an artificial elongation of the source. 
However, in the J8.9 and Q1 filters, the object is consistent with being unresolved (FWHM $\le 0.4\arcsec$ in J8.9 and $\le0.5\arcsec$ in Q1). 

The PAH2\_2 observation of HD$\,$38489 suffers from the same PSF-instability problem while, during the Q1 observation, the source was offset such that it was on the detector in only half of the chopping positions. 
The latter fact has been taken into account during the data reduction.
Like HD$\,$34664, HD$\,$38489 is not obviously resolved, with upper limits in FWHM of $\le 0.37\arcsec$ in J8.9 and $\le0.5\arcsec$ in Q1, respectively.

RMC$\,$71 appears resolved in all images, with the following FWHM dimensions:
$\sim 0.46\arcsec \times 0.42\arcsec$ in J8.9 (PA$\sim 126$\,deg),
$\sim 0.96\arcsec \times 0.83\arcsec$ in PAH2\_2 (PA$\sim110$\,deg), and 
$\sim 0.87\arcsec$ round in Q1. 
These are to be regarded as upper limits because none of the observations was made under diffraction-limited conditions (also because of the low declination of the object). Thus the differences between the two axes in each filter are considered insignificant. 
However, the extended nature of this source is robust, because of the consistent appearance at both epochs, including the position angle, and the significantly larger FWHM than found for the calibrator stars. 
The circumstellar environment of RMC$\,$71 appears in agreement with a broad circular shell and an asymmetric core, which is offset from the centre by $\sim0.2\arcsec$ toward the north-west (Fig.~\ref{visirR71}). 
\subsection{Archival data}
\label{sec:archive}
We consulted the infrared catalogues with the CDS VizieR \citep{2000Ochsenbein} and IRSA Gator\footnote{ \url{https://irsa.ipac.caltech.edu/applications/Gator/}} tools and retrieved photometry from the 2MASS Point Source Catalogue  \citep{2003Cutri}, the \textsl{Spitzer} SAGE legacy survey \citep[Data Release 3/final]{2006Meixner}, the ``Optically bright post-AGB population of LMC''  \citep{2011vanAarle} and ``\textsl{Spitzer} Atlas of Stellar Spectra'' \citep{2010Ardila} catalogues, the \textsl{AKARI} IRC all-sky survey and LMC point source catalogues \citep[][]{2010aIshihara,2012Kato}, \textsl{WISE} surveys \citep{2012Cutri},  the MSX Point Source Catalogue \citep{2003Egan}, and \textsl{Herschel} HERITAGE survey \citep{2013Meixner}. We evaluated all stars for possible contamination by neighbouring sources by visually inspecting the infrared images when available (for each source, these are shown in Figs.~\ref{fig:group1}, \ref{fig:group1b}, \ref{fig:group2}, \ref{fig:group3}, \ref{fig:group4} of Appendix \ref{appendix:imagesSEDs}). 

We checked for ALMA observations of the sample. ALMA Band~7 (343 GHz) observations of RMC\,127, RMC\,143 and LHA$\,$120-S$\,$61 were previously reported in \citet{2017AgliozzoA,2017AgliozzoB,2019Agliozzo}. Unpublished ALMA/ACA data of a mosaic region containing RMC\,14 are available in the archive, in Bands 3 ($\sim\!93\,\rm GHz$; project ID: 2018.A.00049.S) and 6 ($\sim\!220\,\rm GHz$; project ID: 2017.A.00054.S). RMC\,14 is not detected, with $3~\sigma$ upper limits of $1.41\,\rm mJy$ at $93\,\rm GHz$ and $6.0\,\rm mJy$ at $220\,\rm GHz$. 
We also made use of the SPT+\emph{Planck} maps of the MCs at 1.4~mm \citep{2016PlanckMCs} for the stacking analysis in  Sect.~\ref{sec:stacking}. None of the sources is detected in these maps obtained with the large beam of 1.5\arcmin. 

Finally, we used the available Infrared Spectrograph \citep[IRS;][]{2004Houck} spectra from the Combined Atlas of Sources with \emph{Spitzer} IRS Spectra \citep[CASSIS;][]{2011CASSIS-LR,2015CASSIS-HR} or from the IRSA data archive (as SSC Enhanced Products). 
RMC$\,$127, LHA$\,$120-S$\,$61, and RMC~84 were observed under Program Name/Id: STELLARATLAS/485 and published by \citet{2010Ardila}. The RMC$\,$66, HD$\,$38489, HD$\,$34665, and LHA$\,$115-S$\,$18 spectra were published in \citet{2010Kastner}, although we found that the spectrum of LHA$\,$115-S$\,$18 is not on source. The spectra of RMC$\,$14 and RMC$\,$40 were discussed in \citet{2015Ruffle}, and those of RMC$\,$110 and S~Dor in \citet{2017Jones}. Data of RMC$\,$99 and Sk$\,-69\,279$ are available in the archive: we are not aware of any previous publications. 
The IRS spectra are shown with grey continuous lines in the flux density distribution plots in Appendix \ref{appendix:imagesSEDs}.  
RMC$\,$71 and RMC$\,$66 also have \emph{Infrared Space Observatory (ISO)} data\footnote{\url{https://irsa.ipac.caltech.edu/data/SWS/} or \url{https://www.cosmos.esa.int/web/iso/access-the-archive}} and low resolution Multiband Imaging Photometer for \emph{Spitzer} \citep[MIPS;][]{2004Rieke} SED-mode spectra \citep[presented in][which also include RMC~126]{2010Kemper,2010vanLoon} but they are not considered in this work. The IRS spectra are mainly used to guide the interpretation of the infrared photometry when no measurement at 24~$\rm \mu$m is possible due to confusion.

\section{Data analysis}
\label{sec:photometry}
\subsection{Photometry from {\it Spitzer} MIPS24/MIPS70}
\begin{table*}
\caption{MIPS-24 aperture photometry.}

\label{tab:photMIPS24}
\centering
\begin{tabular}{llcccccl}
\hline\hline    
ID &Source & Ap. radius & Sky annulus & Ap. corr. & Measured $F_\nu$ & Catalogue $F_\nu$
& Comment \\
 &      & (\arcsec)        & (\arcsec)    &                     & (mJy)            & (mJy)      &         \\
\hline
1&RMC$\,$143      &   --- & --- & --- & $\mathbf{482.2\pm 96.4}$ & --- & from    \citet{2019Agliozzo} \\
2&RMC$\,$127      &   20 & 20--32 & 1.15 & $1255\pm10$ & $1180\pm60$ & \\
3&S$\,$Doradus    &   20 & 20--32 & 1.15 &  $848\pm90$ & $898\pm22$  &  \\
4&RMC$\,$110      &   7 & 7--13 & 2.05 & $\mathbf{10\pm2}$ & $7.39\pm0.14$ & small aperture due to  \\
& &   &  & & && confusion\\
5&HD$\,$269582    &   7 & 7--13 & 2.05 & $7.7\pm0.4$ & $7.783\pm0.225$ &  \\
6&RMC$\,$116      &   7 & 7--13 & 2.05 & $1.4\pm0.8$ & $1.293\pm0.155$ & marginally detected \\
7&HD$\,$269216    &   20 & 20--32 & 1.15 & $\mathbf{21\pm5}$ & $6.952\pm0.134$ & marginally resolved \\
8&RMC$\,$71       &   --- & --- & --- & --- & $>4100\pm410$ & saturated \\
9&RMC$\,$85       &   --- & --- & --- & --- & --- & not detected/confusion \\
10&RMC$\,$123     &   7 & 7--13 &2.05  & $15.0\pm1.2$ & $16.13\pm0.28$ &  small aperture due to  \\ 
& &   &  & & && confusion\\
11&RMC$\,$99      &    --- & --- & --- & --- & $49.4\pm0.5$ &  not detected/confusion  \\   
12&Sk$\,-69\,279$ &   20 & 20--32 & 1.15 & $\mathbf{22\pm5}$ & --- & bubble \\  
13&LHA$\,$120-S$\,119$ & 35 & 40--50 & 1.08 & $\mathbf{703\pm23}$ & $537.1\pm5.3$ & extended \\
14&LHA$\,$120-S$\,$61&   20 & 20--32 & 1.15 & $1474\pm10$ & $1470\pm74$ & extended \\
15&RMC$\,$74      &    7 & 7--13 & 2.05 & $3.5\pm0.3$ & $3.074\pm0.145$ &  \\   
16&RMC$\,$78      &   20 & 20--32 & 1.15 & $\mathbf{59\pm19}$ & --- & double source \\  
17&RMC$\,$81      &   35 & 40--50 & 1.08 & $\mathbf{108\pm29}$ & --- & extended \\  
18&LHA$\,$120-S$\,$18 & 7 & 7--13 & 2.05 & $\mathbf{11.5\pm1.5}$ & $7.882\pm0.101$ & smaller aperture \\
& &   &  & & && due to confusion,\\
& &   &  & & && lower limit?\\
19&RMC$\,$66          & 20 & 20--32 & 1.15 & $856\pm8$ & $813\pm6$ &  \\
20&RMC$\,$84          & 20 & 20--32 & 1.15 & $104\pm4$ & $99.3\pm0.9$ &  \\
21&HD$\,$34664   &   7 & 7--13 & 2.05 & $389\pm19$ & $411\pm3$ &  point source in the bright\\  
& &   &  & & && ring of another star \\
22&HD$\,$38489   &   20 & 20--32 & 1.15 & $490\pm42$ & $498\pm4$ &  \\       
23 & RMC$\,$126          & 20 &20--32 & 1.15 & $1098\pm15$& $1120\pm10$ & \\
24 & Sk$\,-69\,271$     & 20 &20--32 & 1.15 & < 23& --- & \\
25&RMC$\,$128    &   --- & --- & --- & --- & --- & not detected \\   
&    &   &  &  &  &  & in the vicinity of LBV [2] \\
26&RMC$\,$149    &   --- & --- & --- & --- & --- & confusion \\       
27&HD$\,$269604  &  7 & 7--13 & 2.05 & $\mathbf{0.6\pm0.2}$ & $0.737$ &  \\    
\hline
28&RMC$\,$14     &  --- & --- & --- & --- & $1.634\pm0.131$ &  \\     
29&RMC$\,$40     &  20 & 20--32 & 1.15 & $29.4\pm1$ & $27.0\pm0.4$ &  \\     
30&RMC$\,$4      &  20 & 20--32 & 1.15 & $\mathbf{67.3\pm8.8}$ & $45.4\pm0.5$ & \\      
31&LHA$\,$115-S$\,$18 &  20 & 20--32 & 1.15 & $82.1\pm2.6$ & $87.0\pm0.6$ &  \\ 
\hline
\end{tabular}
\tablefoot{
The numbers highlighted in bold are new measurements with no counterpart in the catalogue or with revised photometry, which are adopted in the analysis.
}
\end{table*}

\begin{table*}
\caption{MIPS-70 aperture photometry and 3$\sigma$ upper limits.}
\label{tab:photMIPS70}
\centering
\begin{tabular}{llcccccl}
\hline\hline    
ID &Source & Aperture radius & Sky annulus & Ap. corr. & Measured $F_\nu$ & Catalogue $F_\nu$ & Comment \\
 &      & (\arcsec)        & (\arcsec)    &                     & (mJy)            & (mJy)      &         \\
\hline
1&RMC$\,$143      & 16 & 18--39 & 2.07 & $<14551$ 	& --- 			& confusion \\ 
2&RMC$\,$127      & 16 & 18--39 & 2.07 & $1021.3\pm54.4$ 	& $987\pm27$ 	& \\
3&S$\,$Doradus    & 16 & 18--39 & 2.07 & $<2090$ 	& --- 			& confusion \\
4&RMC$\,$110      & 16 & 18--39 & 2.07 & $<1243$ 	& --- 			& confusion \\ 
5&HD$\,$269582    & 16 & 18--39 & 2.07 & $<155$ 	& --- 			& confusion\\ 
6&RMC$\,$116      & 16 & 18--39 & 2.07 & $<663$ 	& --- 			& confusion \\  
7&HD$\,$269216    & 16 & 18--39 & 2.07 & $\mathbf{128\pm71}$ & --- 		& detected \\ 
& &   &  & & && but confused\\
8&RMC$\,$71       & 16 & 18--39 & 2.07 & $2029.4\pm49.6$ 	& $1858\pm24$ 	& \\
9&RMC$\,$85       & 16 & 18--39 & 2.07 & $<3864$ 	& --- 			& confusion \\  
10&RMC$\,$123     & 16 & 18--39 & 2.07 & $<477$ 	& --- 			& confusion \\  
11&RMC$\,$99      & 16 & 18--39 & 2.07 & $<867$ 	& --- 			& confusion \\ 
12&Sk$\,-69\,279$ & 16 & 18--39 & 2.07 & $<214$  	& --- 			& confusion \\
13&LHA$\,$120-S$\,119$ & 16 & 18--39 & 2.07 & $447.2\pm51.4$ 	& $428.3\pm15.1$ 	& marginally \\
& &   &  & & && resolved\\
14&LHA$\,$120-S$\,$61   & 16 & 18--39 & 2.07 & $635.2\pm22.1$ 	& $579.3\pm8.8$ 	& \\
15&RMC$\,$74      & 16 & 18--39 & 2.07 & $<58$ 	& --- 			& noise \\ 
16&RMC$\,$78      & 16 & 18--39 & 2.07 & $<350$	     & ---		     &  confusion\\  
17&RMC$\,$81      & 16 & 18--39 & 2.07 & $\mathbf{150\pm56}$ 	& --- 		    & possible detection \\  
& &   &  & & && but confused\\
18&LHA$\,$120-S$\,$18 & 16 & 18--39 & 2.07 & $<330$		& --- 			& confusion  \\
19&RMC$\,$66      & 35 & 39--65 & 1.24 & $531.5\pm56.2$ 	& $528\pm9$ 	& \\
20&RMC$\,$84      & 16 & 18--39 & 2.07 & $\mathbf{104\pm29}$ 	& --- 			& possible detection \\  
& &   &  & & && but confused\\
21&HD$\,$34664    & 16 & 18--39 & 2.07 & $<4484$ 	& --- 			& contaminating \\ 
& &   &  & & &&nearby source \\
22&HD$\,$38489    & 16 & 18--39 & 2.07 & $<769$ 	& --- 			& confusion \\ 
23 & RMC$\,$126          & 16 & 18--39 & 2.07 & $299\pm126$& $380\pm24$ & inhomougeneous \\
& &   &  & & &&background \\
24 & Sk$\,-69\,271$    & 16 & 18--39 & 2.07 & <298 & --- & confusion \\
25&RMC$\,$128     & 16 & 18--39 & 2.07 & $<218$ 	& --- 			& contaminating \\ 
& &   &  & & &&nearby source \\
26&RMC$\,$149     & 16 & 18--39 & 2.07 & $<16700$ 	& --- 			& confusion \\ 
27&HD$\,$269604   & 16 & 18--39 & 2.07 & $<53$ 	& --- 			& marginal detection \\ 
\hline
28&RMC$\,$14    & 16 & 18--39 & 2.07 & $<1612$	& --- 	& confusion \\
29&RMC$\,$40     &  16 & 18--39 & 2.07 & $\mathbf{58.7\pm8.0}$ & --- &  \\     
30&RMC$\,$4       & 16 & 18--39 & 2.07 & $<338$ 	& --- 			& confusion \\ 
31&LHA$\,$115-S$\,$18 & 16 & 18--39 & 2.07 & $<120$ 	& --- 			& confusion \\  
\hline
\end{tabular}
\tablefoot{
 The numbers highlighted in bold are new measurements with no counterpart in the catalogue or with revised photometry, which are adopted in the analysis.
}
\end{table*}

\begin{table*}
\caption{PACS-100 point source extraction and 3$\sigma$ upper limits. }
\label{tab:photPACS100}
\centering
\begin{tabular}{llcccl}
\hline\hline    
ID &Source & Measured $F_\nu$ & Catalogue $F_\nu$ & Comment \\
 &       & (mJy)            & (mJy)      &         \\
\hline
1&RMC$\,$143     & 	---  &  $636\pm254$		  &  \\ 
2&RMC$\,$127     & $535\pm65$      & $493\pm43$	          & & \\
3&S$\,$Doradus   & $<1100$	  & --- 		  & not detected \\
 &&&& due to confusion \\
4&RMC$\,$110     & $<350$	  & --- 		  &  \\ 
5&HD$\,$269582   & $<270$	  & --- 		  &  \\ 
6&RMC$\,$116     & $<560$	  & --- 		  &  \\  
7&HD$\,$269216       & $<283$	  & --- 	  	  &  \\ 
8&RMC$\,$71      & $669.0\pm68.5$& $706\pm48$	  & & \\ \\ 
9&RMC$\,$85      & $<1850$	  & --- 		  &  \\  
10&RMC$\,$123    & $<206$	  & --- 		  &  \\  
11&RMC$\,$99     & $<699$	  & --- 		  &  \\ 
12&SK$\,$-69279  & $<137$	  & --- 		  &  \\
13&LHA$\,$120-S$\,119$ & $\mathbf{204.7 \pm 33.6}$& ---         &  marginally resolved\\
14&LHA$\,$120-S$\,$61   & $292.8\pm37.6$         & $305\pm31$ &  & \\
15&RMC$\,$74      & $<38$  	  & ---		          &  \\ 
16&RMC$\,$78      & $<195$	  & ---			  &  \\  
17&RMC$\,$81      & $<105$    & ---	                  &  \\  
18&LHA$\,$120-S$\,$18 & $<132$    & ---		          &  \\
19&RMC$\,$66      & $  318\pm33   $& $320\pm26$       &   & \\
20&RMC$\,$84      & $<59$    & ---		          &  \\ 
21&HD$\,$34664    & $<1805$	  & ---		          &  \\ 
22&HD$\,$38489    & $\mathbf{123.3\pm 90.8}$&---		  	  &  marginal detection\\ 
 &&&& due to confusion \\
23 & RMC$\,$126          &  $\mathbf{199\pm52}$& $136\pm29$ & \\
24 & Sk$\,-69\,271$    & <151 & --- &  \\ 
25&RMC$\,$128     & $<94$	  & ---		          &  \\ 
26&RMC$\,$149     & $<34800$	  & ---		          &  \\ 
27&HD$\,$269604   & $<37$         & ---		          &  \\ 
28&RMC$\,$14      & $<440$              & --- 	  &  \\
29&RMC$\,$40      & $<40$  & ---                   &  \\	
30&RMC$\,$4       &   $<434$	  & ---		          &  \\ 
31&LHA$\,$115-S$\,$18 & $<62$    & ---		          &  \\  
\hline
\end{tabular}
\tablefoot{
The numbers highlighted in bold are new measurements with no counterpart in the catalogue or with revised photometry, which are adopted in the analysis.
}
\end{table*}

A small number of sources are not included in the {\it Spitzer} SAGE catalogues, or they have poor catalogue photometry, most likely because of the strict requirements adopted to identify point-sources \citep{2006Meixner}. Indeed we find that these sources are usually extended or embedded in a confused region, requiring a manual extraction of the flux measurement. 
The most remarkable case is Sk$\,-69\,279$, which appears as a well detected extended bubble in the MIPS24 map, but has no catalogue  photometry.  
We performed aperture photometry for all sources that were either lacking measurements or are extended in the images.
As a cross-check, we also measured all targets with catalogue values. Following this verification, we chose to use the existing catalogue photometry for the non-extended sources, unless indicated otherwise.

Our aperture photometry was performed with the astropy package ``photoutils''. 
The aperture radii, sky annuli, and aperture corrections, as recommended in the MIPS instrument handbook\footnote{\url{https://irsa.ipac.caltech.edu/data/SPITZER/docs/mips/mipsinstrumenthandbook/50/}}, are listed in Tables \ref{tab:photMIPS24} and \ref{tab:photMIPS70}, together with the measured flux densities, comparison with the catalogue value when available, and relevant comments on the sources and their environments. 
The reported flux density errors only account for the uncertainty in determining the background (defined as the standard deviation in the background annulus integrated over the aperture), which generally dominates over the 
absolute flux calibration error and statistical noise in the source aperture. 
For non-detected sources, we provide $3\sigma$~upper limits. 

We checked the quality of our photometry against the catalogues (see Tables \ref{tab:photMIPS24} and \ref{tab:photMIPS70}):  
the results are consistent with the published values. In bold we highlight our measurements that are used in the later analysis instead of the catalogue values.

\subsection{Photometry and upper limits from {\it Herschel}}

Only a few sources were well detected with the Photodetector Array Camera and Spectrometer \citep[PACS, ][]{2010PACS} at 100 and 160\,$\mu$m, and their flux densities were taken from the HERITAGE catalogue \citep{2013Meixner}. Additionally, we report in Table \ref{tab:photPACS100} new PACS100 photometry for LHA$\,$120-S$\,$119 and HD$\,$38489 (although the latter is only a marginal detection). To maximise the signal-to-noise ratio, we adopted a procedure equivalent to Gaussian fitting. Local peaks were searched for in boxes with size equivalent to the beam. The regions used to evaluate the sky contribution were separated by one or two beams. 
The $1\,\sigma$ total uncertainty is determined by summing in quadrature the standard deviation of the background and the flux calibration uncertainty of 10\%. Because of contamination of S~Dor from a strong nearby nebular region, we reduced the background spacing to one beam. Nevertheless, the source was not unambiguously detected. 
For the sources with no detection (either in this work or in the catalogue), we report the $3\,\sigma$ upper limit at $100\mu\rm m$ determined as described above. 

Finally, in PACS160  we only find a single new detection, LHA$\,$120-S$\,$61, which has a flux density of $60\pm16\,\rm mJy$. 
We do not report any new detection obtained at longer wavelengths with the Spectral and Photometric Imaging Receiver (SPIRE) instrument \citep{2010Griffin}. 
The upper limits at $160\,\rm \mu m$ and longer wavelengths are not provided as they do not add useful information owing to the larger beam sizes and high interstellar dust brightness at such wavelengths.

\section{The LBV SED  classification scheme and colour-magnitude diagram}
\label{sec:review}
\label{sec:sedreview}
\begin{figure*}
    \begin{minipage}{.45\textwidth}
        \centering
        \includegraphics[width=1\linewidth]{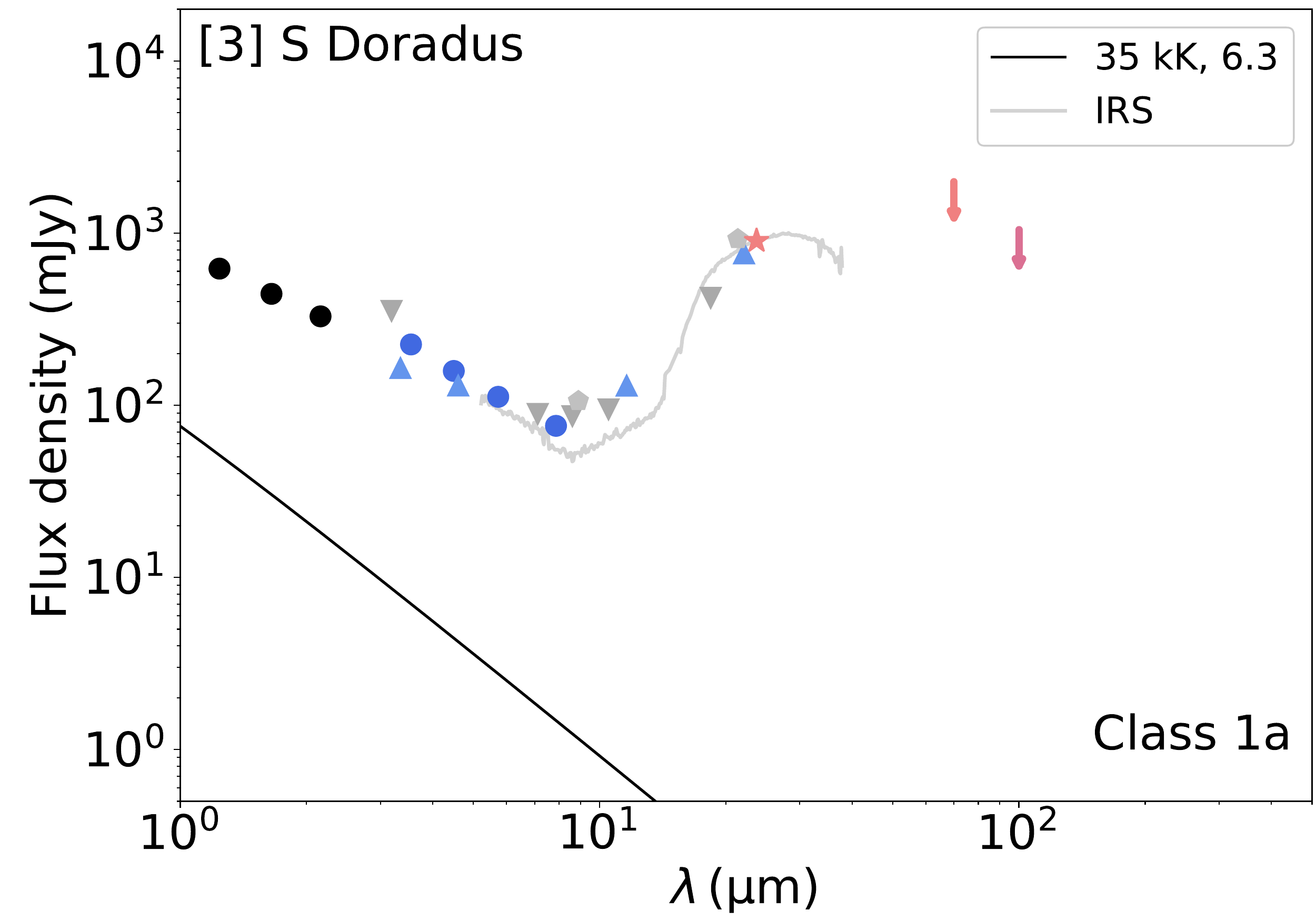} 
       \includegraphics[width=1\linewidth]{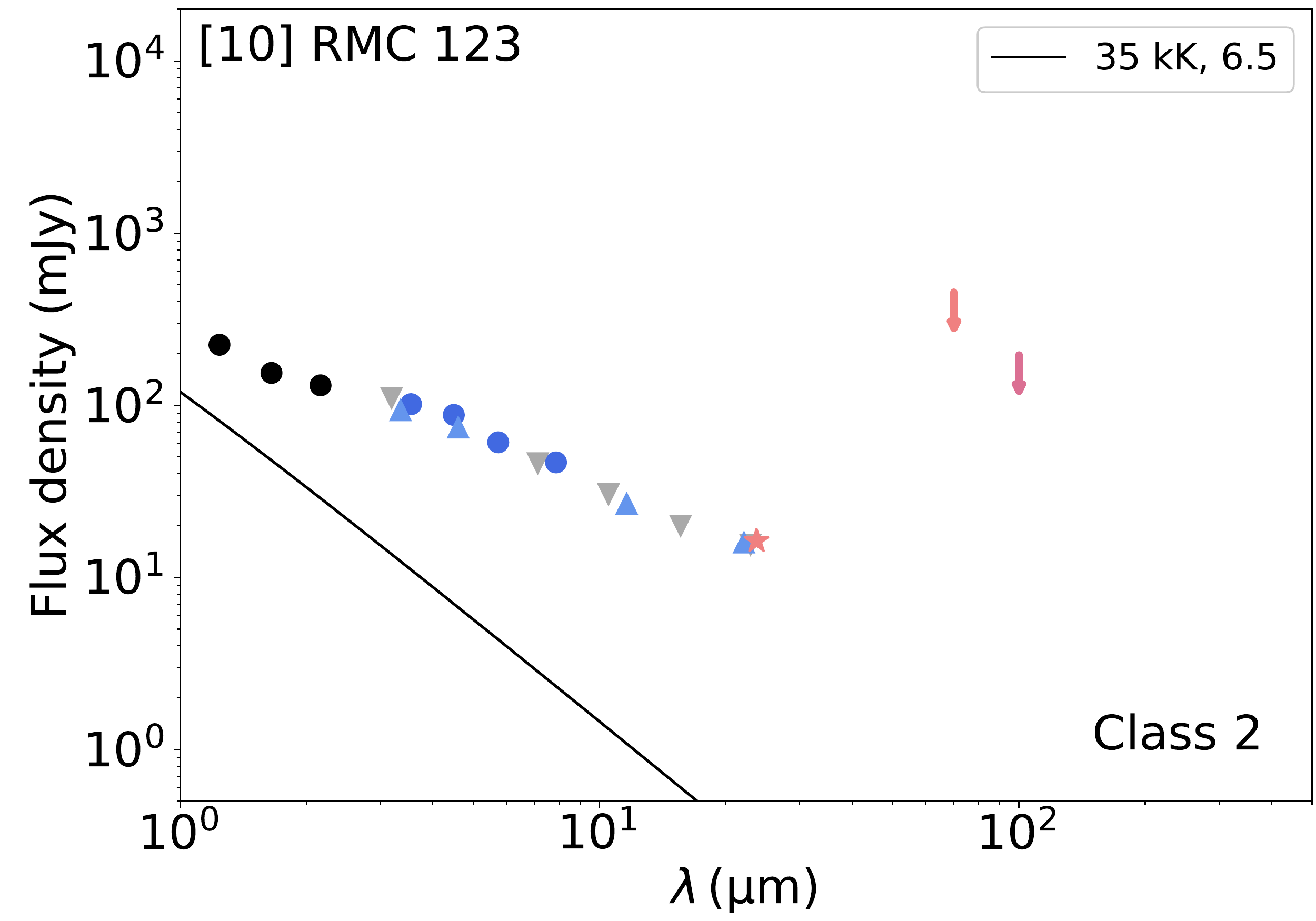} 

    \end{minipage}  
    \hfill
    \begin{minipage}{.45\textwidth}
    \centering
       \includegraphics[width=1\linewidth]{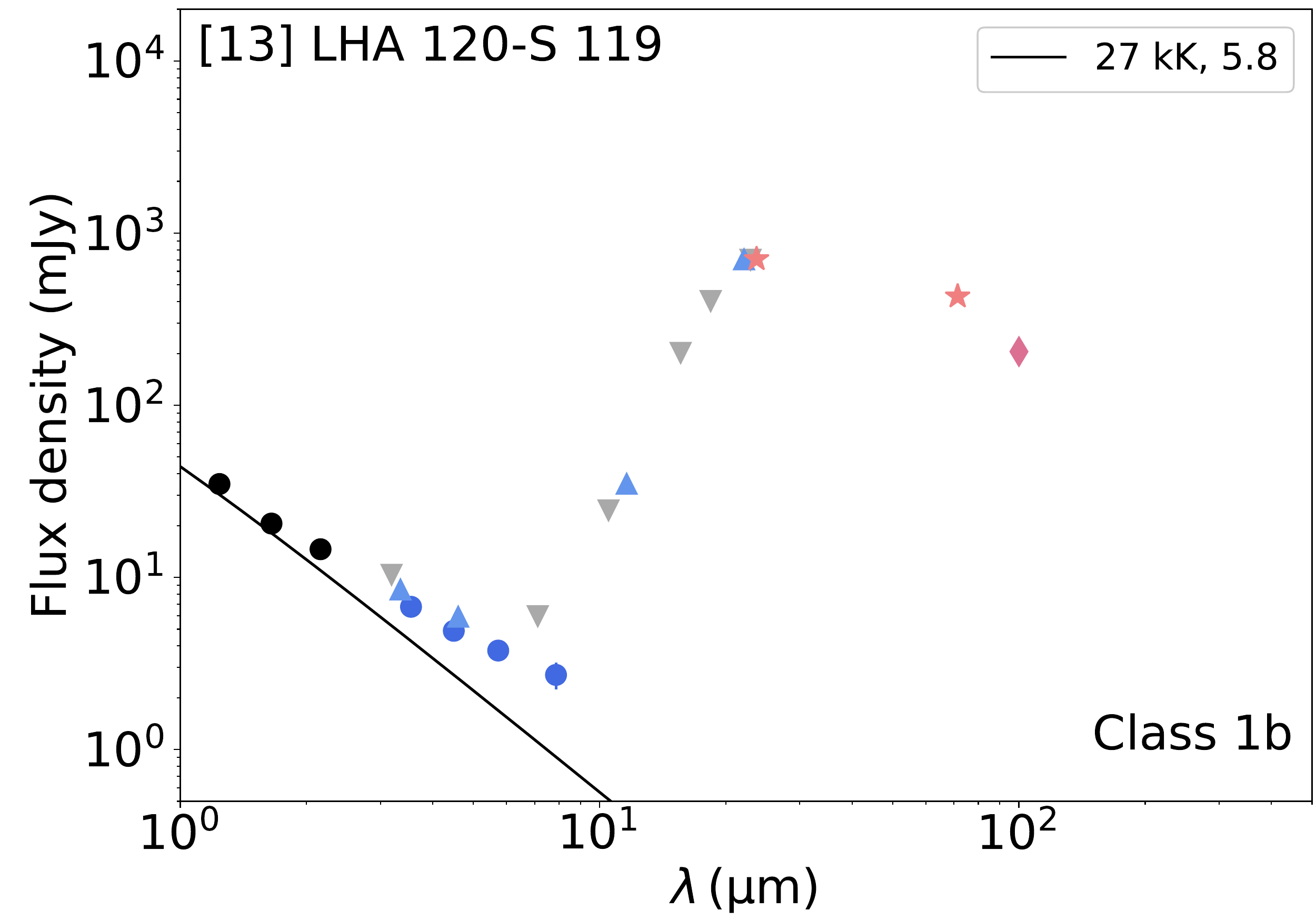}
       \includegraphics[width=1\linewidth]{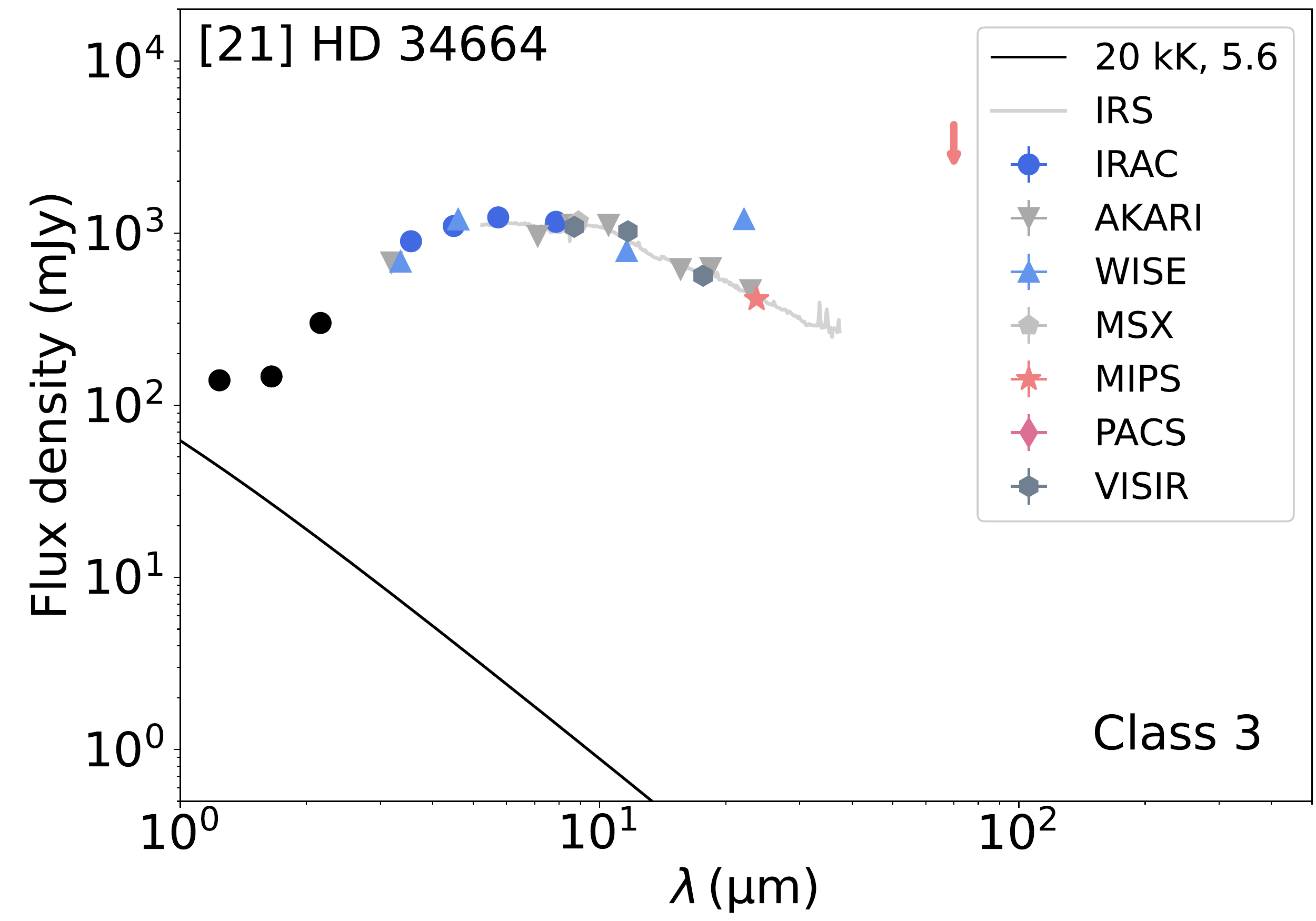}        
    \end{minipage} 
    
\begin{minipage}{1\textwidth}
     \centering    
 \includegraphics[width=.45\linewidth]{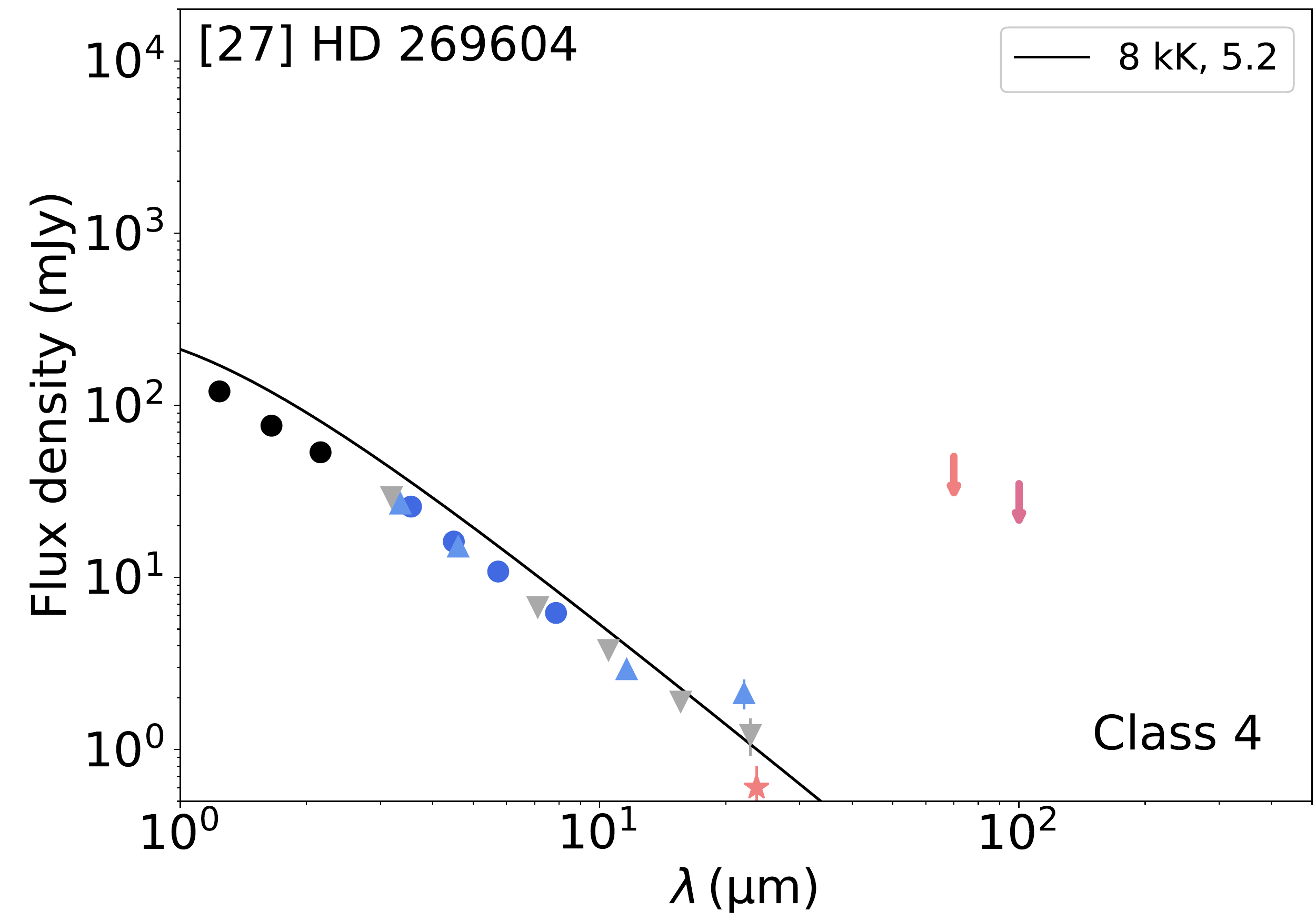}
\end{minipage}  
\caption{Examples of each of the four LBV SED classes defined in Sect.~\ref{sec:review}.
}
\label{fig:example_sed}
\end{figure*}

\begin{figure*}
\centering
   \includegraphics[scale=0.5]{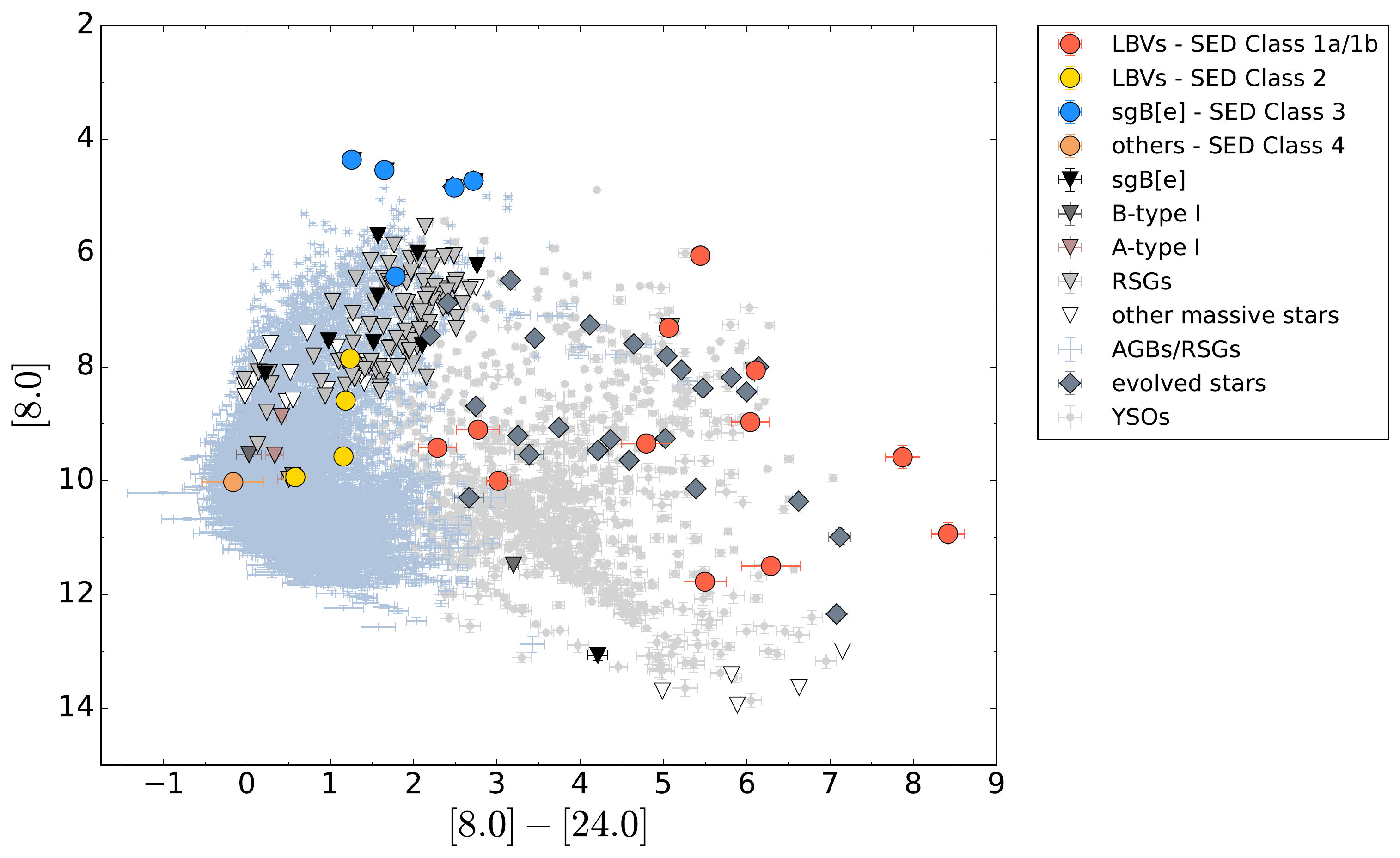}
   \caption{Mid-infrared colour-magnitude diagram for stars in the LMC. The LBVs are shown with filled circles in different colours, depending on the SED class defined in this work. Several types of supergiant stars (RSGs, sgB[e], supergiants A and B, six LBVs and other massive stars) as in \citet{2009Bonanos} are displayed with triangles of different colours as identified in the legend. Dusty evolved stars from 
   \citet{2015Jones} are shown with diamond symbols (these include post-AGBs, planetary nebulae, sgB[e] stars, one LBV, one WR and two SN remnants). Pixel symbols indicate the AGBs and RSGs from \citet{2012Riebel} and point symbols are confirmed and candidate YSOs from \citet{2008Whitney}.}
   \label{fig:cmd-LMC}
\end{figure*}
We first need to ensure to limit our sample to LBV stars and 
find a classification
criterion that does not rely on the spectroscopic and photometric variations only,
but also on properties of the stellar ejecta. 
Table \ref{tab:sample} summarises the relevant information on our sample, including the stellar classification from the literature.  Stellar parameters, extinction, and nebular size when known are summarised in Table \ref{tab:stellparam} of Appendix \ref{appendix:stellarparam}.  
Unfortunately, we cannot provide information on the binary status, as to our knowledge this is known only for RMC~14 in the SMC and RMC~81 in the LMC. The complete set of photometry from this work and from the catalogues are in Tables~\ref{tab:PhotometryNearIR}, \ref{tab:PhotometryMidIR} and \ref{tab:PhotometryFarIR} of Appendix \ref{appendix:allphoto}. 
For all sources, the flux density distribution  starting from $\rm 1~\mu$m are shown in  Figs.~\ref{fig:group1}, \ref{fig:group1b}, \ref{fig:group2},
\ref{fig:ungrouped}, \ref{fig:group3}, \ref{fig:group4}  of Appendix~\ref{appendix:imagesSEDs}, where in addition to the photometric points the stellar continuum (on the basis of stellar parameters as in Table~\ref{tab:stellparam}) and the IRS spectra are also included when available. 

In their sample of 1750 massive stars in the LMC, \citet{2009Bonanos} included six LBVs (S~Dor, RMC~127, RMC~71, RMC~110, RMC~85, HD~269582) and concluded that these stars do not represent a homogeneous class of objects on the basis of their mid-infrared colours.  We identify four classes based on the individual infrared spectral energy distributions (SED):  

\begin{enumerate}
    \item Sources with a dusty nebula peaking in the mid- to far-infrared (Class~1). These can be further divided into two groups: with or without a significant free-free excess above the stellar photosphere (Classes 1a and 1b, respectively). An example can be found in the top-right panel in Fig.~\ref{fig:example_sed}, where the object shows a moderate free-free continuum up to $\rm 10~\mu$m, and at longer wavelengths a dominant infrared excess due to dust in the extended nebula, previously imaged by \citet{2003Weis-s119} and  \citet{2012Agliozzo}. By contrast, the top left panel shows an example of strong free-free emission. 
    \item Sources with only a free-free emission excess (Class~2).  An example is given in the middle-left panel in Fig.~\ref{fig:example_sed}. For these sources the MIPS24 flux density is consistent with a power law describing the free-free continuum. 
    \item Sources dominated at all wavelengths $\gtrsim2\mu\rm m$ by a hot dusty component, suggestive of a disc close to the star (Class~3). In the literature, these are all classified as B[e] supergiants (see  Table~\ref{tab:stellparam}). An example of a Class~3 SED is HD~34664, shown in the second-right panel of Fig.~\ref{fig:example_sed}. 
    \item Featureless sources with a single component consistent with the stellar photosphere at all wavelengths (Class~4).  An example is presented in the bottom panel of Fig.~\ref{fig:example_sed}. 
\end{enumerate}

Table~\ref{tab:sample} summarises the classification of the sources in the scheme proposed here.  
Among them, only one object cannot be classified,  RMC~85. This is an active LBV \citep{1998vG}. The star is embedded in a very crowded region.  Because of the absence of reliable photometry longward of $\sim20\mu{\rm m}$ \citep[confirming][]{2009Bonanos}, we are not able to distinguish whether it is SED Class~1 or Class~2 (see Fig.~\ref{fig:ungrouped}).

The dust composition of sources in Class~1 typically comprises amorphous silicates, although a mixed chemistry is sometimes reported, like in the case of RMC~71 in the LMC \citep{2014Niyogi} or HD~168625 in the Milky Way \citep{2010Umana}.  Crystalline silicates are also found in LBVs, albeit rarely \citep[e.g. RMC~71 and  HR~Car, ][respectively]{1999Voors,2009Umana}, 
while they seem to be a common feature of sgB[e] stars \citep[e.g.][]{2006Kastner, 2010Kastner,2014deWit}. 

In Fig. \ref{fig:cmd-LMC} we present the [8.0] versus [8.0]-[24] colour-magnitude diagram (CMD) of stars in the LMC, 
depicting the position of the LBV sample analysed in this work (filled circles of various colours representing the different LBV SED classes as described in the legend); early and late supergiants as in \citet{2009Bonanos} (downward pointing triangle symbols); grey diamonds are low- and intermediate-mass, but also some high mass stars,  planetary nebulae and SN remnants from \citet{2015Jones}; AGBs and RSGs from \citet{2012Riebel} and Young Stellar Objects (YSOs) from \citet{2008Whitney}. Stars with SED Classes 2, 3 and 4 occupy the same region as RSGs and other types of massive stars in the diagram, which Bonanos et al. found indistinguishable from AGBs and young stellar objects, highlighting the importance of using wide-baseline photometry. However, we notice some differences: Class 2 objects tend to be fainter than RSGs at 8$\,\rm \mu m$, due to lacking a hot 
dusty envelope; Class 3 are usually brighter than RSGs,  due to their hot dusty disc; the only Class 4 object with MIPS24 photometry is the faintest in this region of the diagram, consistent with no excess above the photosphere.  For stars in the SED Class~1, the [8.0]-[24] colours coincide with the sample of stars in Jones et al. However, the nature of 7 out of their 35 objects was not spectroscopically confirmed. These could all be sources with a dusty gaseous nebula (e.g. SN remnants, planetary nebulae, LBV nebulae). 
For stars in the SMC we refer to the colour-diagram by \citet{2010Bonanos}, who included all the LBVs in the census \citep{2018RNAAS...2c.121R}. Bonanos et al.\ found that LBVs are among the most luminous stars at $24\,\rm \mu m$, but they are not clearly separated from other types of star.  

\subsection{Class~1: sources with dusty nebulae peaking in the mid- to far-infrared}

This class contains well-known extended MCs LBV nebulae (LBVNe), extensively studied in the literature. Some of them are associated with very active LBVs \citep{2017Walborn}. 
Some stars are classified as candidate LBVs: these are stars that do not satisfy all LBV classification criteria (both observed spectroscopic and photometric variability; \citealt{1994HD}), however the presence of a circumstellar nebula around some of them suggests that they are ex/dormant LBVs.

Depending on whether or not they have an excess of near-infrared emission due to the ionised stellar wind, stars in this class can be further divided into: Class 1a, which are objects that present free-free emission from the wind; Class 1b, which are objects with no detected free-free  emission from the wind. For two objects, Sk$\,-69\,279$ and RMC\,78, we report the first identification of a dusty circumstellar nebula. The dust component in these sources peaks between 24 and 100~$\rm \mu$m, similarly to several Galactic objects with extended nebulae \citep[typically with sizes of the order of $0.1-1$~pc, e.g. G79.29+0.46, Pistol Star, AG Car, AFGL2298, HR Car; ][and references therein]{2014Agliozzo,2014Lau, 2015Vamvatira,2010Buemi, 2017Buemi}.
Among the sources in this class, five (RMC$\,$127, RMC$\,$71, S~Dor, LHA$\,$120-S$\,$61, and RMC$\,$40) have an IRS spectrum and all of them show, together with the thermal continuum, the silicate bump around $10-12\,\rm \mu m$. 

The objects classified as belonging to Class~1 are briefly introduced (see Fig.~ \ref{fig:group1} and \ref{fig:group1b} in the Appendix):
\\
\textbf{S~Dor} is the prototype of the most common LBV variability \citep[a summary of the pioneering works can be found in][]{1989Wolf}. The infrared excess from a dusty nebula was  reported in \citet{2009Bonanos} from mid-infrared photometry and in \citet{2017Jones} from the IRS spectrum. The power-law behaviour in the IRS data below $10~\rm \mu m$ is more likely consistent with strong free-free emission in the ionised stellar wind rather than the stellar photosphere. 
The bright surrounding environment does not allow for the  disentanglement of the nebula at $70~\rm \mu m$ and beyond. The SED Class of this object is 1a.\\
\textbf{RMC\,127}: this star belongs to the group of active LBVs \citep{2008Walborn,2017Walborn}. This is another object with  SED Class 1a. See \citet{2017AgliozzoB} for a multiwavelength study of its  nebula.\\
\textbf{LHA\,120-S\,61}: candidate LBV \citep{1994HD} with SED Class 1a, its dusty nebula was modelled by  \citet{2017AgliozzoA}.\\
\textbf{Sk$\mathbf{\,-69\,279}$}: candidate LBV with SED Class 1a, the nebula was previously imaged by {\it{HST\/}} \citep{2002Weis}. 
We find that at $24\,\rm \mu m$ it appears as a bubble, resembling the morphology revealed in the optical. 
To our knowledge, this is the first report of dust in this object. \\
\textbf{LHA\,120-S\,119}: candidate LBV \citep{vanGenderen2001} with a known nebula (SED Class 1a) that was previously imaged at different wavelengths \citep[][and references therein]{2012Agliozzo}. In the infrared it is possible to notice a thermal component of dust, detected up to $100\,\rm \mu m$. The nebula is marginally resolved at $24\,\rm \mu m$, revealing a marked asymmetry similar to the optical and radio images. This asymmetry was explained as a bow-shock, given the run-away nature of its associated star \citep{2001D&C}. \\
\textbf{RMC\,81}: infrared emission from cool dust around this candidate LBV \citep{vanGenderen2001} was first reportedly detected by {\it{IRAS}} at $60\,\mu\rm m$ \citep{1988iras....1.....B,2002Tubb}. However, inspection of the MIPS70 image, comparison with the MIPS photometry, the lack of $25$ or $12\,\mu\rm m$ \emph{IRAS} detections, and considering the $1\sigma$ position offset, it appears that this flux was greatly elevated by ISM confusion. The object is  detected in both MIPS bands, even at $70\,\rm \mu m$ where the surrounding environment\,is bright. The star is an eclipsing binary \citep{1987StahlR81}. The SED Class of this object is 1a. \\
\textbf{LHA$\,$120-S$\,$18}: at $24\,\rm \mu m$ and beyond, the environment around this candidate LBV \citep{vanGenderen2001} is bright and the object is not distinguishable from the background. The SED Class of this object is 1a.\\
\textbf{RMC$\,$143}: the  dusty and massive nebula (SED Class 1b) was extensively studied by \citet{2019Agliozzo}, who also confirmed the LBV status of this star.  \\
\textbf{RMC$\,$71}: this star also belongs to the group of very active LBVs \citep{2013Mehner, 2017Walborn}, with SED Class 1b. It is possible to observe variability in the infrared SED (compare the mid-infrared photometry with the IRS spectrum), as the data were acquired at different epochs during the most recent stellar outburst \citep{2017Mehner}. Here we show a similar dataset, but extended to longer wavelengths. The entire SED was previously modelled by \citet{2014Niyogi}. 
\\
\textbf{RMC$\,$78}: at $24\,\rm \mu m$ the object \citep[a candidate LBV,][]{1999vanG} appears as a double source (SED Class  1b). \\
\textbf{RMC$\,$40}: this star belongs to the group of confirmed LBVs \citep{1993Szeifert} in the SMC. The first detection at $24\,\rm \mu m$ was reported by \citet{2010Bonanos}. Here we add a new clear detection at $70\,\rm \mu m$. The SED Class of this object is 1b. An IRS spectrum is available and suggests that the dust emission peaks between the two MIPS bands.\\
\textbf{HD$\,$269216}: a confirmed LBV \citep{1991Prinja}, this object is well detected and marginally resolved in MIPS~24. We add a new measurement as the catalogue PSF fitting may have missed a large fraction of its flux density. The SED Class of this object is 1b.
\\
\textbf{RMC$\,$110}: this is a confirmed LBV manifesting the eruptive phenomenon a few times in the past decades \citep{1990Stahl,2018Campagnolo}. Because of confusion at wavelengths beyond  $\sim20\,\rm \mu m$ it is not possible to obtain reliable photometry of dust in its ejecta (SED Class  1b).

\subsection{Class~2: sources with free-free excess only}

This class consists of the following objects (see Fig.~\ref{fig:group2})
\\
\textbf{RMC\,116} is a strongly active S~Dor variable  \citep{vanGenderen2001}. Its SED shows an ionised stellar wind in the form of a free-free excess above the stellar photosphere detected up to $24\rm \mu$m. \\
\textbf{RMC\,123} and \textbf{RMC\,99} belong to a sub-group proposed by \citet{1984Stahl} and \citet{1987S&W}, characterised by an excess of free-free emission from a gaseous disc, rich in forbidden lines, but with no dust features. The free-free emission of RMC\,123 is detected up to $24\rm \mu$m. At longer wavelengths, the SED of RMC\,99 is dominated by contamination from neighbouring sources so that the presence of an infrared excess is questionable. Both stars are catalogued as weakly active S~Dor variables in \citet{vanGenderen2001} and are considered candidate LBVs. RMC~99 was observed with IRS, but the spectrum is likely dominated by interstellar gas and dust.\\
\textbf{HD\,269582} is a very active LBV \citep{2017Walborn} with a strong ionised stellar wind detected up to $24\rm \mu$m. Differences between 2MASS and IRAC data suggest that the star was active (highly variable) in the years of the infrared observations. \\ 
\textbf{RMC\,74} is a candidate LBV \citep{vanGenderen2001} with a SED similar to that of RMC~116. \\
\textbf{RMC\,14}, a well-known multiple-system containing an active LBV in the SMC \citep[e.g.][]{1996Barba}, shows an SED similar to HD~269582, with a strong free-free excess above the continuum \citep[a similar SED can also be found in][]{2010Bonanos}. This is also seen in two other active LBVs \citep[RMC\,127,][and S~Dor]{2017AgliozzoA}. The system was observed with IRS, but, because of the large slit width, the spectrum is likely dominated by interstellar gas and dust.\\

\subsection{Class~3: sources dominated by a hot dusty disc}

\textbf{ RMC\,66, RMC\,84, HD\,38489, HD\,34664, RMC\,126, RMC\,4, and LHA\,115-S\,18}: 
These stars are all known in the literature as supergiant B[e] stars \citep[e.g.][]{1996Zickgraf,1999vanG,2012Aret, 2009Bonanos, 2010Bonanos,2006Kastner,  2010Kastner}, but sometimes they are also listed in the sample of candidate LBVs because of other observational commonalities \citep[e.g. variability of emission lines and P~Cyg profiles,][]{vanGenderen2001}. Their SEDs are suggestive of a different dust formation mechanism. Dust in these stars forms in the equatorial plane (perhaps as a consequence of fast rotation or binarity), and resides in a circumstellar disc-like configuration, resembling the mid-infrared spectra of T~Tauri and disc-enshrouded Herbig Ae/Be stars \citep{2006Kastner, 2010Kastner}. 
Furthermore, observations suggest the silicates to be crystalline, which requires long-lived dust particles permanently close to the star, as opposed to outflowing
material like in the case of LBV eruptions 
\citep[for a review, see][]{2014deWit}. The IRS spectrum, and IRAC and MIPS images of HD~38489, RMC~66, LHA~115-S~18, RMC~126 and HD~34664 first appeared in \citet{2006Kastner,2010Kastner}.

\subsection{Class~4: sources with no evident free-free or dust excess}

This class comprises the following objects:
\\
\textbf{HD\,269604} was classified as A1~Ia-0 by \citet{1973Osmer}. It appears in several works as a candidate LBV, but \citet{vanGenderen2001} flagged it as non-candidate. We do not find evidence of free-free and/or dust infrared excess.\\
\textbf{RMC\,128} \citep[B2~Ia with N weak, ][]{1991Fitzpatrick} was identified as non-candidate by van~Genderen, but it is mentioned as a candidate LBV in several other works. RMC~128 is only $20\arcsec$ away from RMC~127, which is much more luminous and the star is hard to discern in the mid-infrared. In the near-infrared, there is no evidence of free-free and dust excess. We do not find any observational argument to keep the star in the list of candidate LBVs, although its spatial proximity to the very active RMC~127 certainly makes for an interesting target (e.g. if the two stars are co-eval). \\
\textbf{RMC\,149} was characterised by \citet{vanGenderen2001} as a weakly active S~Dor variable (with low-amplitude variations over decades). However, \citet{2009Farigna} re-classified it as O8.5 II((f)). We do not find typical signatures of LBV nebular features (ionised gas and/or dust excess). The infrared images reveal that the source is embedded in a crowded environment. \\
\textbf{Sk}$\mathbf{\,-69\,271}$ is a yellow supergiant (YSG) \citep{2012Neugent}. It appeared in the list of candidate LBVs because of nearby nebular material resembling a half shell, most likely of interstellar origin \citep{1997Weis}. This source is particularly faint already at $8\,\rm \mu m$ and does not show any features at longer wavelengths.  
For lack of other evidence to keep these stars in the LBV sample, we discard them from further analysis and also suggest their removal from the LBV census \citep[the most recent study to mention is ][]{2018RNAAS...2c.121R}. 

\section{Modelling the infrared spectral energy distributions with grey-body fitting}
\subsection{Individual stars}
\label{sec:modelling:grey}

\begin{table*}
\caption{Summary of estimated dust properties.}
\label{tab:dustproperties}
\centering

\begin{tabular}{lcccr}
\hline\hline 
Source & Dust Mass & Dust $T$ & $\beta$& Ref. \\ 
& ($10^{-2}\,\rm M_\odot$)& (K) & &  \\ 
\hline
LHA~120-S~119& $1.2^{+2.2}_{-0.7}$& $108^{+9}_{-10}$&$1.03\pm0.045$&this work\\
\rmca& $0.2-2$& 71$-$90&1.5$-$2.0&[1]\\
LHA~120-S~61 &$0.5-3$& 105$-$145&0.55$-$1.5&[2]\\
\rmc &$5.5\pm1.8$ & $62-80$ & $1.0-1.6$ & [3] \\
RMC$\,$71 & 1& $107^{+10}_{-7}$& N/A&[4], [5]\\
& $0.9^{+0.3}_{-0.2}$& $100\pm10$&$1.8\pm0.2$&this work\\
\hline
HD~269216 & $0.8\pm0.7$& $62\pm6$& 1.3 (fixed) &this work\\
RMC$\,$81 & $0.4\pm0.2$& $81\pm8$& 1.3 (fixed)&this work\\
RMC$\,$40 & $0.31\pm0.07$& $75\pm2$& 1.3 (fixed)&this work\\
\hline
\end{tabular}

\tablebib{
 [1] \citet{2017AgliozzoB}; [2] \citet{2017AgliozzoA}; [3] \citet{2019Agliozzo};
 [4] \citet{2014Niyogi}; [5] \citet{2010vanLoon}.}

\end{table*}

\begin{figure}
    \centering
    \includegraphics[trim={1cm 0.5cm 2cm 2cm},clip,width =1.0\columnwidth]{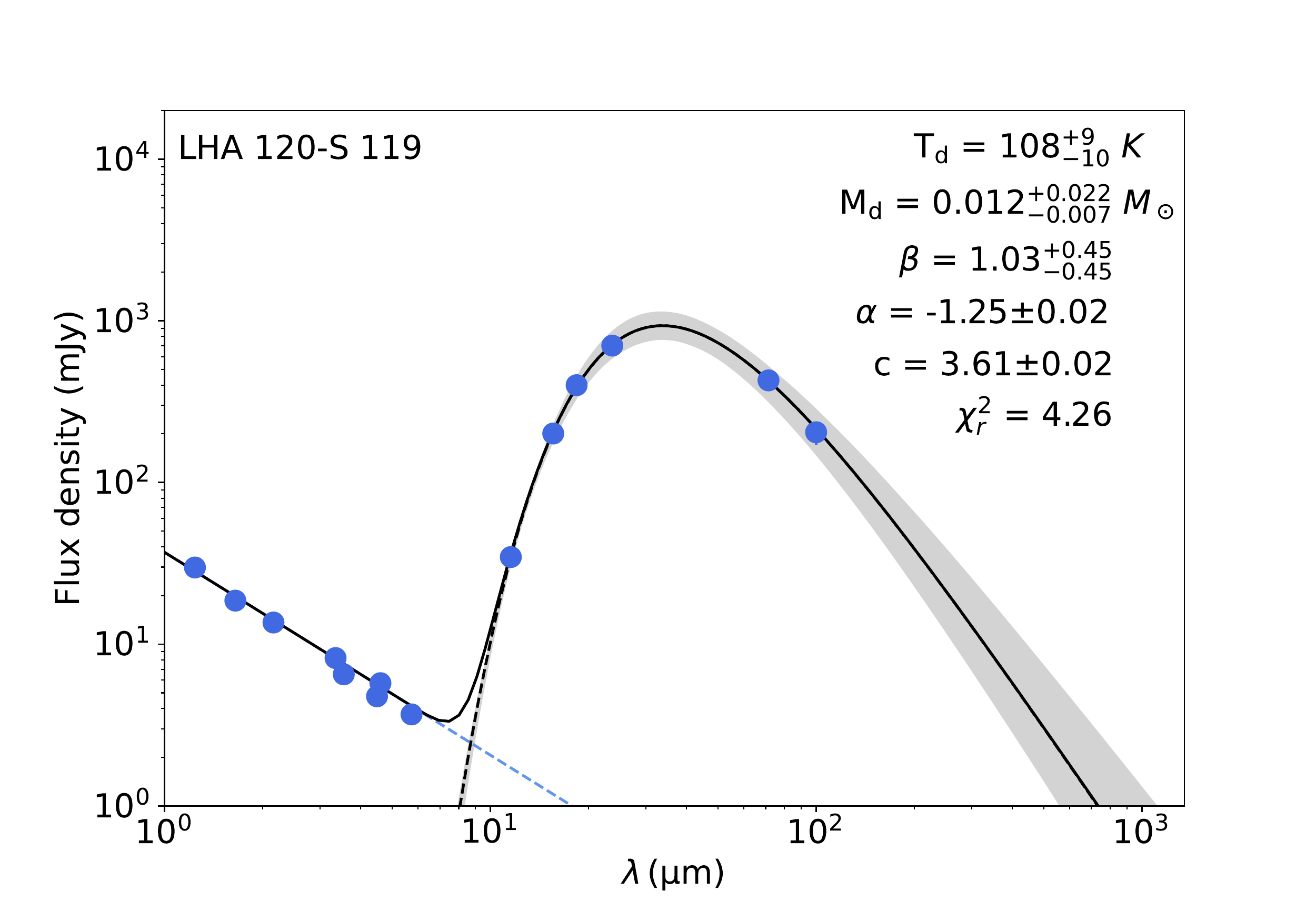}
    \caption{Example of single grey-body fit of the {\it{2MASS}, \it{WISE},  \it{Spitzer}}, and {\it{Herschel}} data, similarly to \citet{2017AgliozzoA,2017AgliozzoB,2019Agliozzo}.}
    \label{fig:s119fit}
\end{figure}

Several dusty LBV nebulae are  extended. Optical instruments detected H$_{\alpha}$ emission, almost unextinguished by the dust grains \citep{2012Agliozzo}. Indeed the extinction A$_{\rm H_{\alpha}}$ was mapped in three nebulae, revealing relatively low values and thus suggesting an optically thin configuration for the dust \citep{2017AgliozzoA,2017AgliozzoB,2019Agliozzo}. For such sources we can assume that radiative transfer effects are negligible. Among the stars classified as Class~1, five have a detection in at least one PACS filter, namely RMC~71, RMC~127, RMC~143, LHA~120-S~61, and LHA~120-S~119. With the exception of LHA~120-S~119, the dust mass for these sources has been previously determined in the literature (see Table~\ref{tab:dustproperties}). For LHA~120-S~119, we thus fit the infrared SED with a  greybody to derive the characteristic temperature, dust mass and $\beta$ parameter (Fig.~\ref{fig:s119fit}), using an absorption coefficient value $\kappa_{\rm 850\mu m}$ of  $\rm 1.7\, cm^{2}\, g^{-1}$ \citep{1987Sodroski}, the same as \citet{2017AgliozzoA,2017AgliozzoB,2019Agliozzo}.

We assume a distance of $49.97\,\rm kpc$ for the LMC \citep{2013Pietrzynski} for all sources. The distances to individual sources will vary around this nominal value due to the inclined geometry and finite thickness of the LMC, but the errors due to this single distance assumption are expected to be less than 2~kpc or 4\% \citep[][and considering a maximum separation of any source from the LMC centre of $2^\circ$]{2001AJ....122.1807V}. This uncertainty ($\sigma_{D}\sim1\,\rm kpc\sim2\%$) is an almost negligible contribution to the dust mass uncertainty compared to the systematic uncertainty in $\kappa$, and the typical measurement uncertainties of the far-infrared photometry.

Radiative transfer modelling was adopted to fit the data of RMC~71 by \citet{2014Niyogi}. For comparison, we also fit the data of RMC~71 with a grey-body and find a dust mass which agrees with the literature value. 
Three other sources (RMC~40, RMC~81, HD~269216) are detected in both the $24$ and $70\,\rm \mu m$ MIPS bands but not in the PACS bands. Their measurements suggest that the  thermal emission peaks between 24 and 70$\rm \mu m$.
We fit these data by fixing the parameter $\beta$=1.3 (the median of the $\beta$ values determined for the five LMC LBVs above, see also Table~\ref{tab:dustproperties}). The distance assumed for the SMC LBV RMC~40 is $62.44\, \rm kpc$ \citep{2020Graczyk}. We note that the uncertainty of HD~269216 is nearly 100$\%$ due to the marginal detection in MIPS70. For the other Class~1 sources the lack of detection longward of $24\,\rm \mu m$ does not allow to constrain the dust parameters. 
The results of this analysis are summarised in Table~\ref{tab:dustproperties}.

\subsection{Stack analysis of LBVs in the LMC}
\label{sec:stacking}

\begin{table*}
\caption{Flux densities from the stacked data.}
\label{tab:stackedphoto}
\centering
\begin{tabular}{lcccc}
\hline\hline
Filter &$\lambda_{\rm eff}$  &  LBVs	& LBVs &  x-AGBs \\
 &  &  	&  (w/o RMC143)&  \\
&($\rm \mu m$)&(Jy) &(Jy) & (Jy)\\
\hline
$J$\tablefootmark{a}	 & 1.235  & 3.430$\pm$0.068  & 3.330$\pm$0.066 & 5.951  $\pm$ 0.595 \\
$H$\tablefootmark{a}	 & 1.662  & 2.761$\pm$0.070  & 2.680$\pm$0.068 & 16.128 $\pm$ 1.613 \\
$K_{s}$\tablefootmark{a}	 & 2.159  & 2.430$\pm$0.050  & 2.37$0\pm$0.050 & 32.604 $\pm$ 3.260 \\
IRAC3.6\tablefootmark{a}	 & 3.550  & 0.776$\pm$0.024  & 0.735$\pm$0.023 & 83.965 $\pm$ 8.397 \\
IRAC4.5\tablefootmark{a}	 & 4.493  & 0.604$\pm$0.017  & 0.574$\pm$0.016 &  93.59 $\pm$ 9.36  \\
IRAC5.8\tablefootmark{a}	 & 5.731  & 0.470$\pm$0.013  & 0.447$\pm$0.012 &  104.07$\pm$10.42  \\
IRAC8.0\tablefootmark{a}	 & 7.872  & 0.541$\pm$0.018  & 0.500$\pm$0.020 &  104.18$\pm$10.42  \\
WISE3\tablefootmark{a}	 & 11.56  & 0.996$\pm$ 0.082 & 0.795$\pm$0.080 &  --  \\
MIPS24\tablefootmark{b}	 & 23.675 & 12.4$\pm$1.7  & 11.7$\pm$1.5 &  44.95 $\pm$4.64   \\
MIPS70\tablefootmark{b}	 & 71.42  & 4.3$\pm$2.6  & 4.1$\pm$0.8 &  7.2   $\pm$1.6    \\
                 
PACS100\tablefootmark{b}	 & 100.0  & 2.34$\pm$0.31  & 1.66$\pm$0.220 &  4.1   $\pm$1.4    \\
PACS160\tablefootmark{b}	 & 160.0  & 0.82$\pm$ 0.40 & $>0.41\pm0.16$\tablefootmark{c} &  $<4.26$\tablefootmark{d}    \\
SPIRE250\tablefootmark{b}	 & 250.0  & $>0.122\pm0.037$\tablefootmark{c}& --- & ---  \\

\hline
\end{tabular}
\tablefoot{\centering
\tablefoottext{a}{Summed photometry from catalogues.}\tablefoottext{b}{Photometry of point-source at the centre of stacked images.}\tablefoottext{c}{Lower limit fluxes are measured fluxes from stacks of subsets of the least confused or brightest sources (see text).}\tablefoottext{d}{$3\sigma$ statistical upper limit from the noise in the stacked image.}}
\end{table*}

 \begin{figure*}[!htb]
    \begin{minipage}{.5\linewidth}
        \centerline{\sf \large Stack of 18 LBVs in the LMC\hspace{3em}}
    \end{minipage}%
    \begin{minipage}{.5\linewidth}
        \centerline{\sf \large  Stack of 17 LBVs in the LMC (w/o RMC143)\hspace{1.5em}}
    \end{minipage}
    \vspace{5pt}
   \includegraphics[scale=0.31]{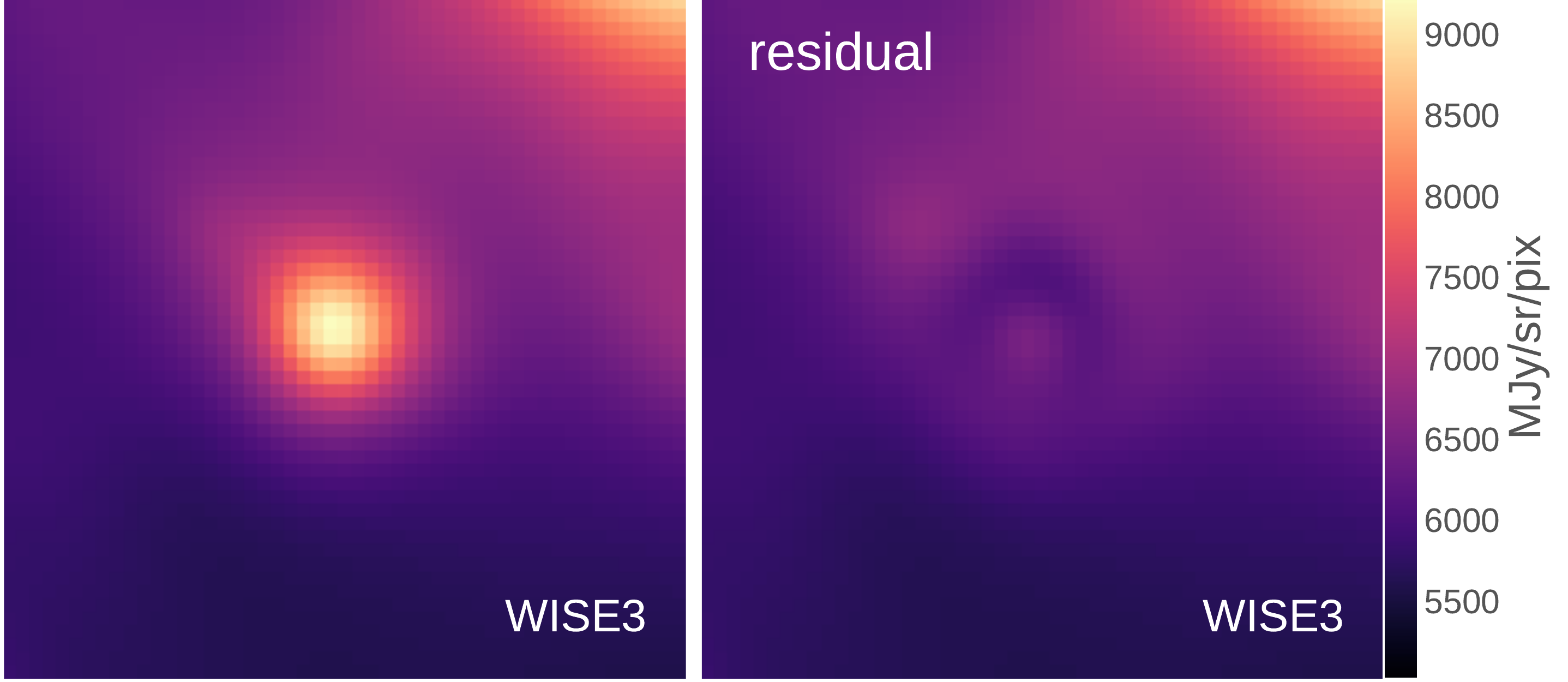}
      \vspace{1pt}
      \includegraphics[scale=0.31]{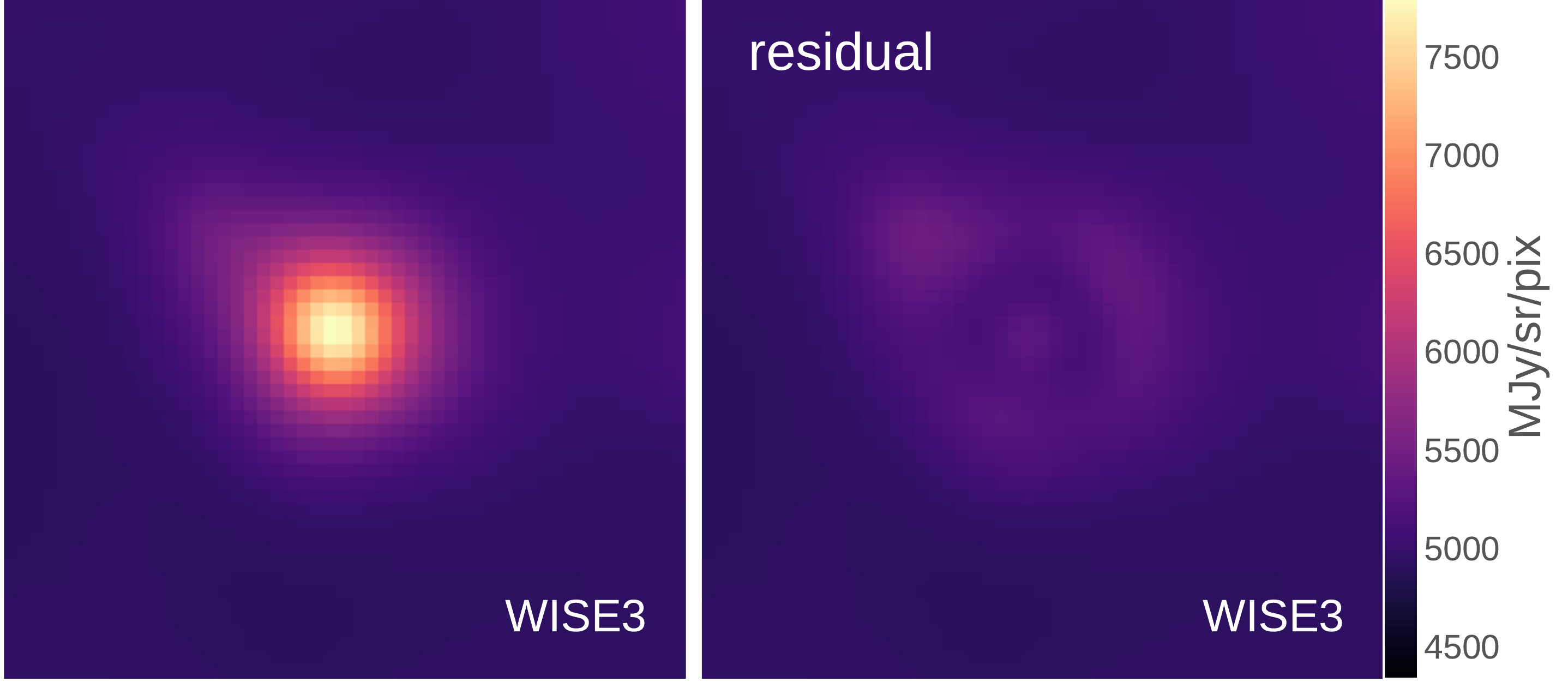}
            \vspace{1pt}
   \includegraphics[scale=0.31]{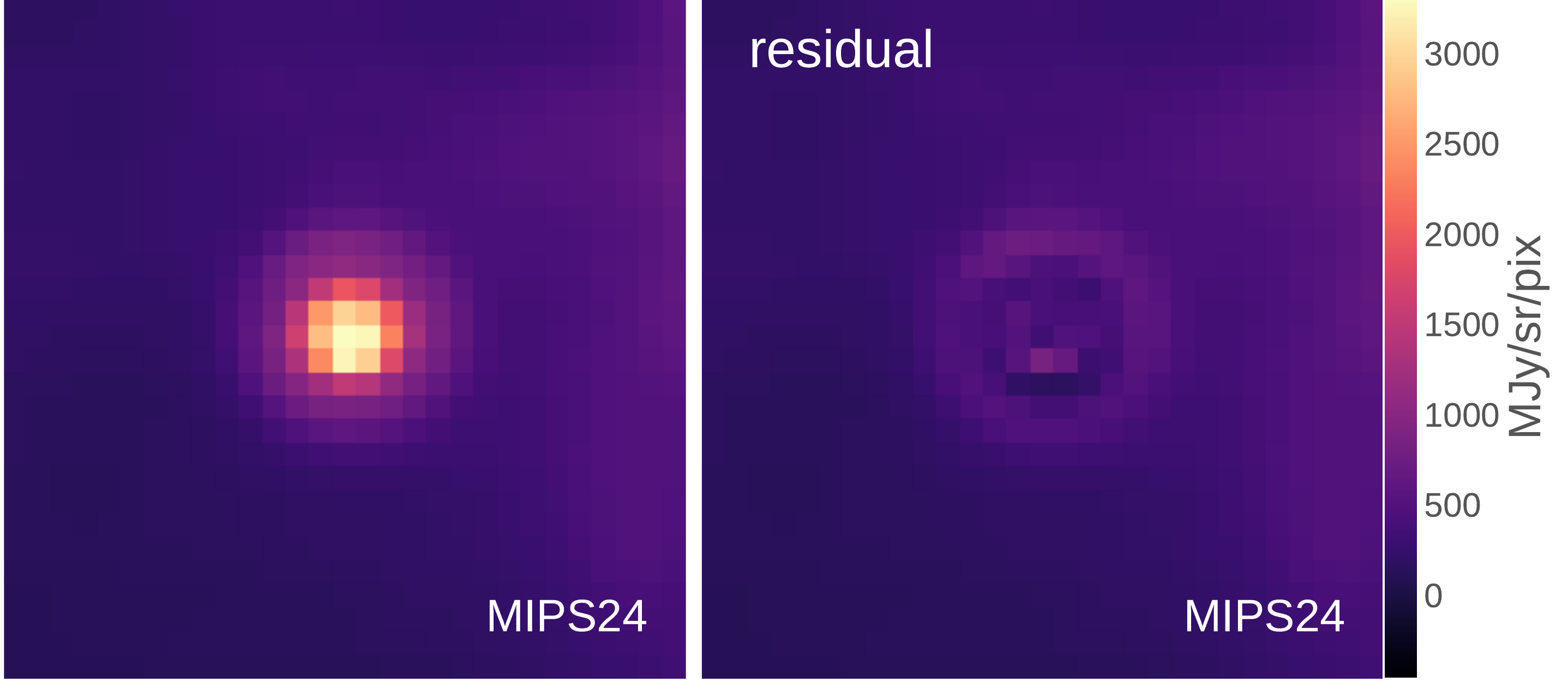}
         \vspace{1pt}
      \includegraphics[scale=0.31]{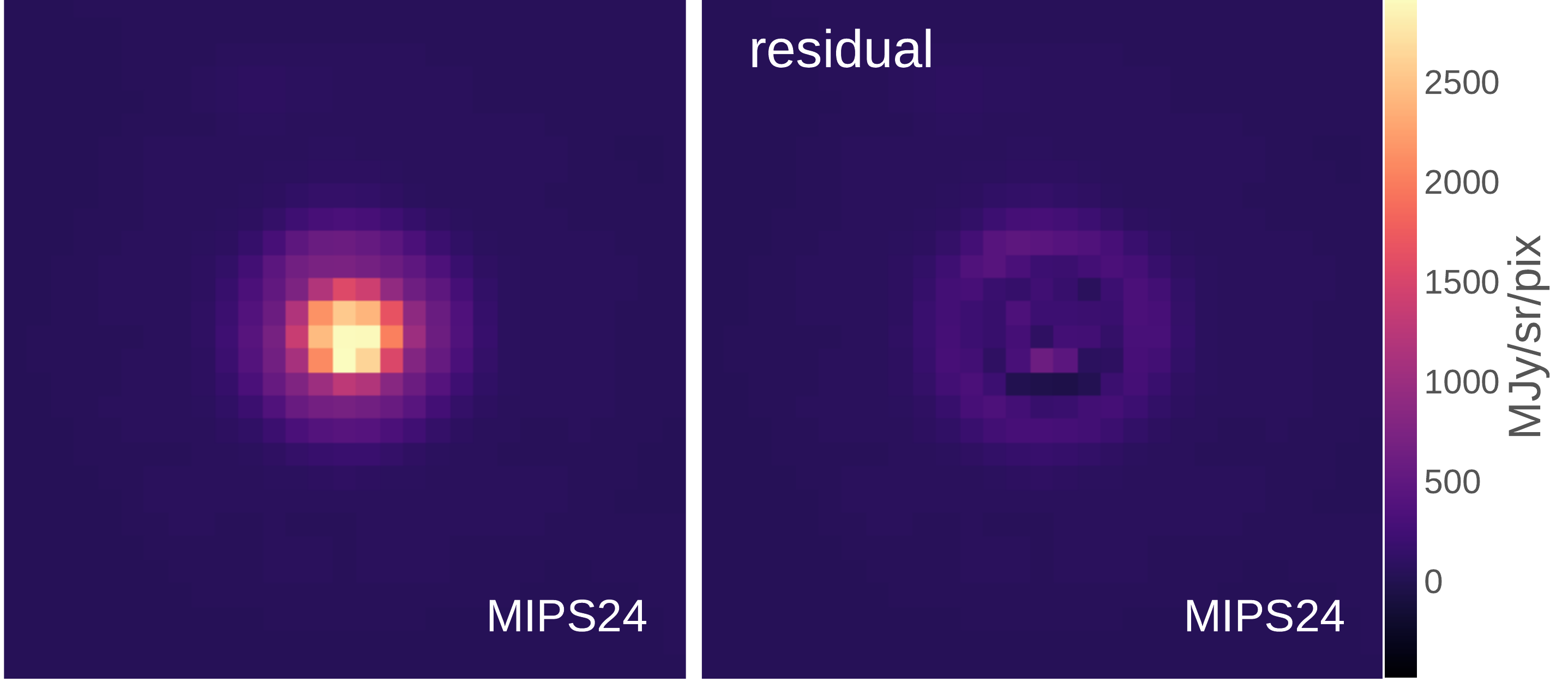}
            \vspace{1pt}
   \includegraphics[scale=0.31]{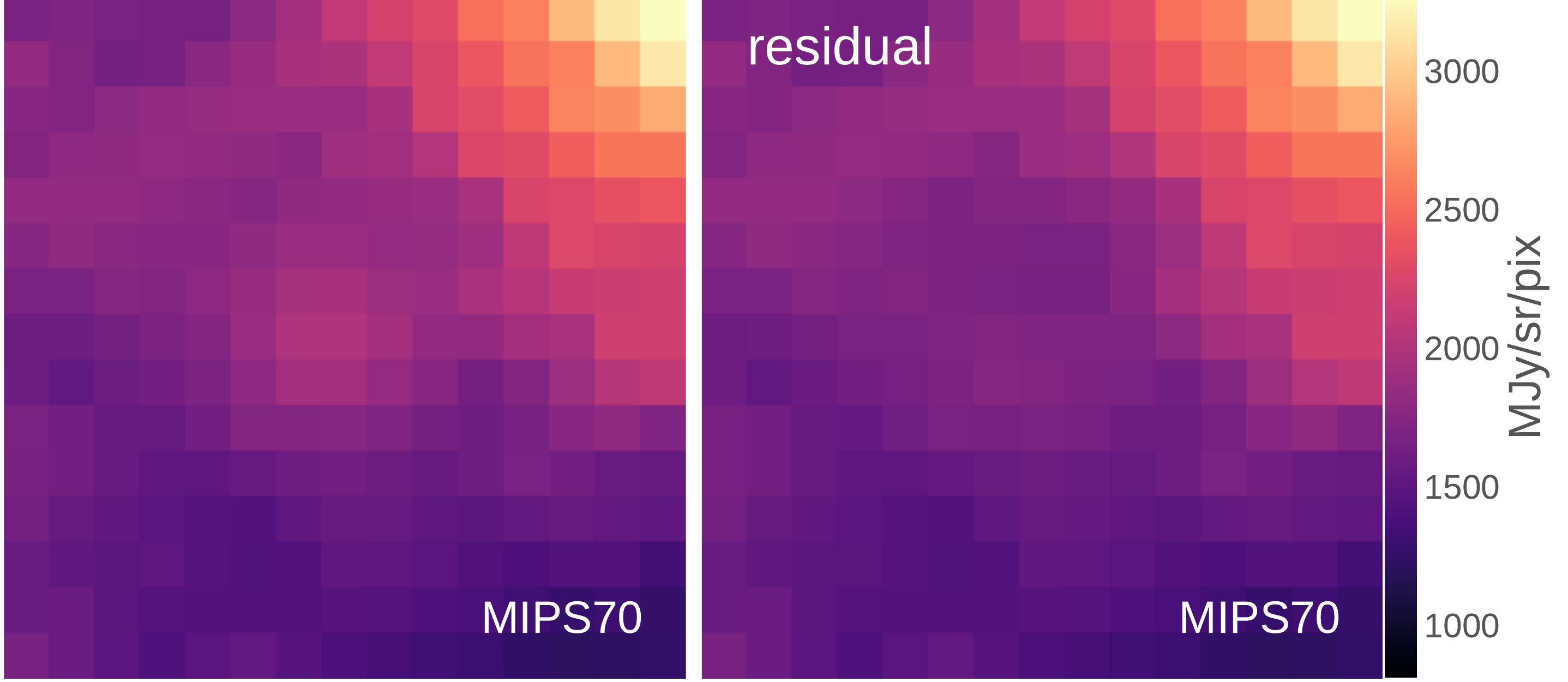}
         \vspace{1pt}
      \includegraphics[scale=0.31]{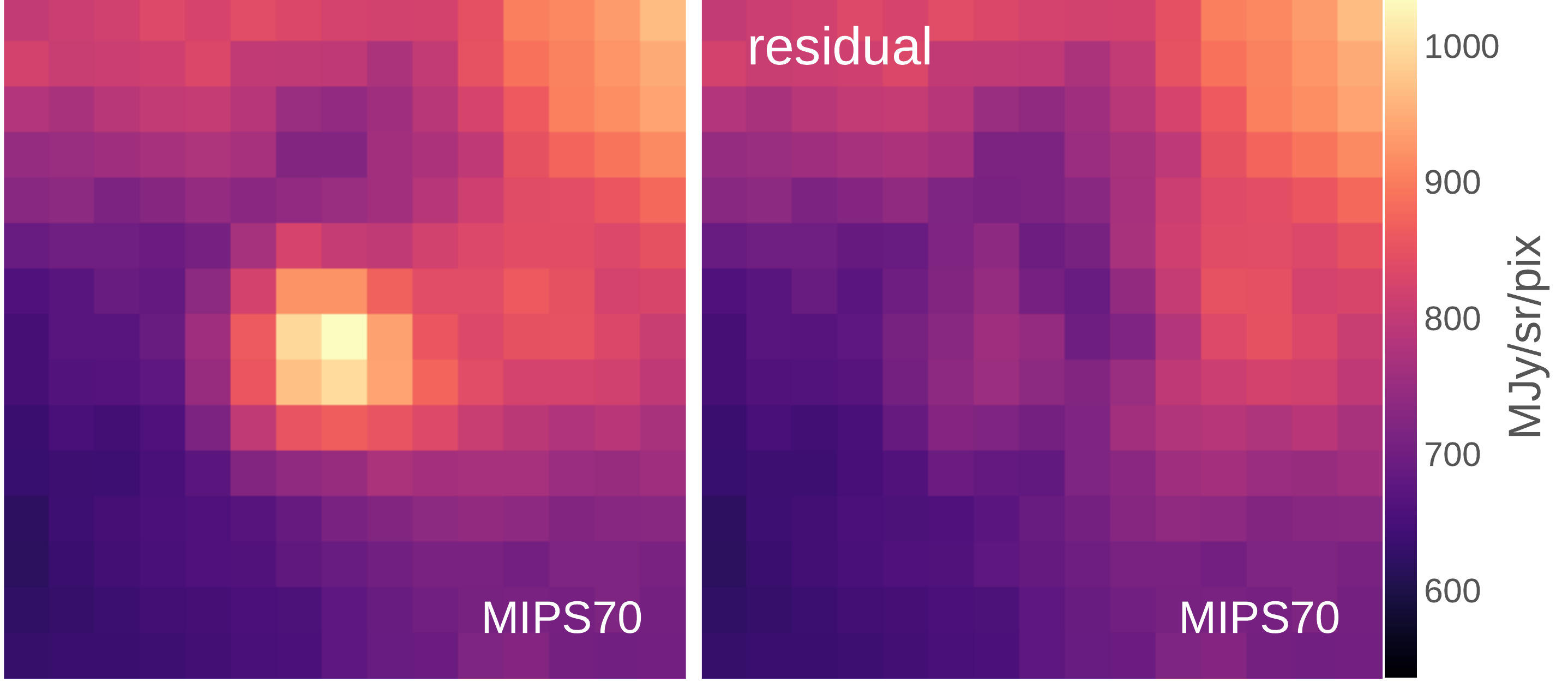}
            \vspace{1pt}
   \includegraphics[scale=0.31]{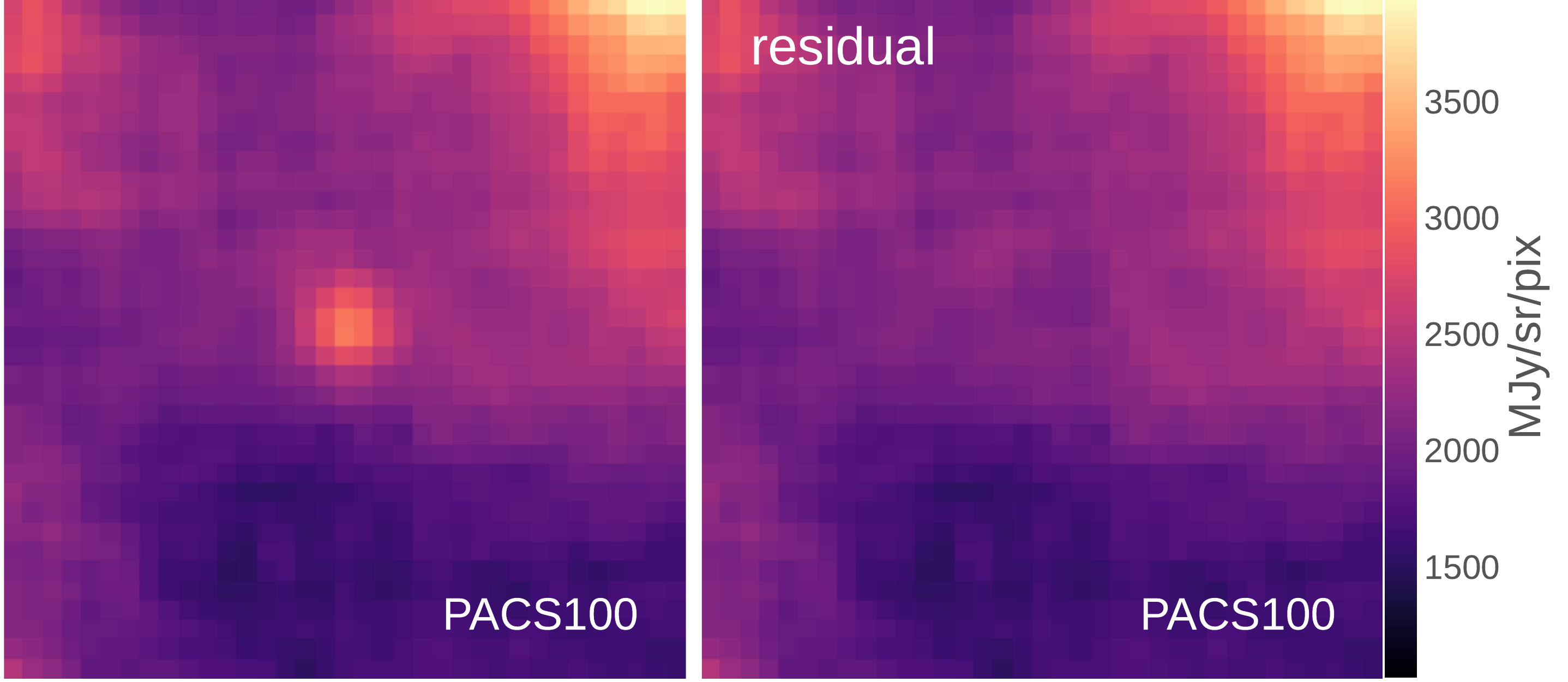}
         \vspace{1pt}
      \includegraphics[scale=0.31]{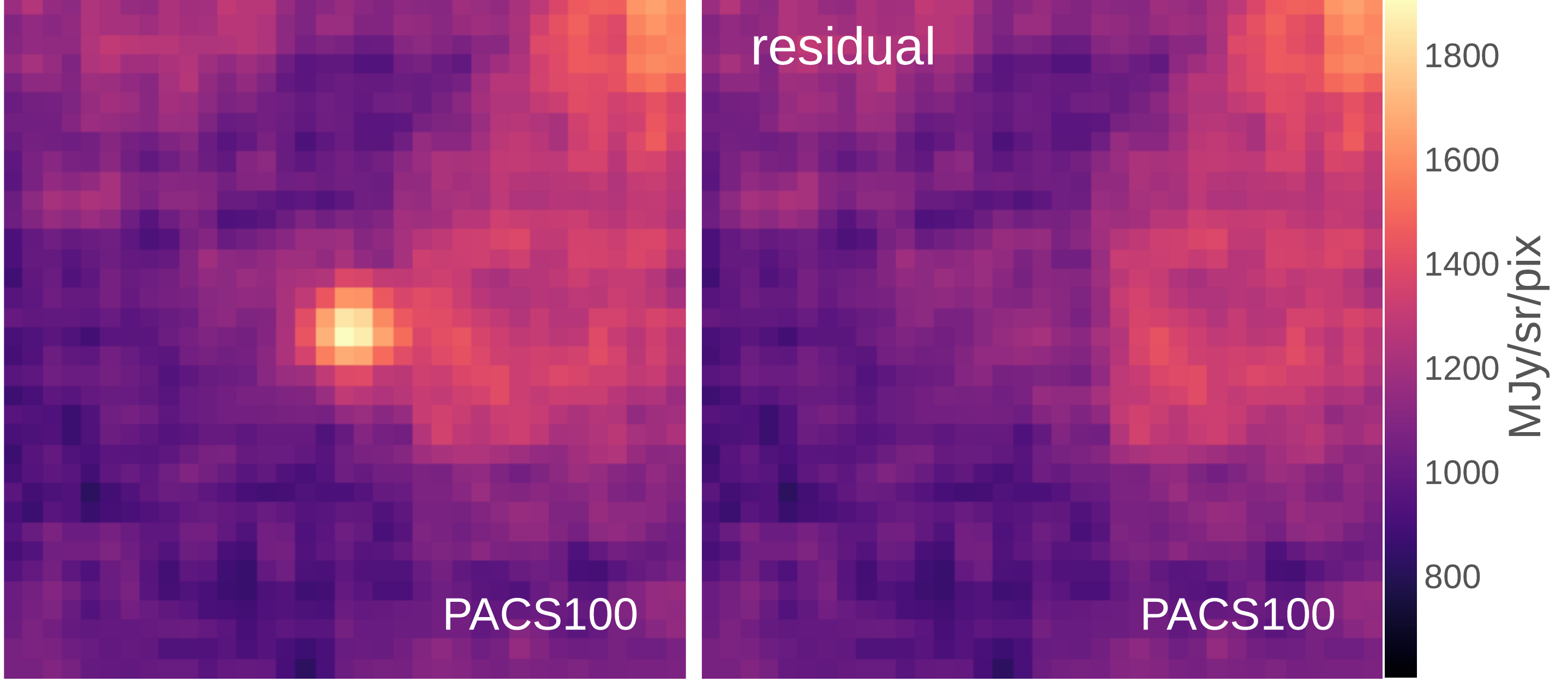}
            \vspace{1pt}
   \includegraphics[scale=0.31]{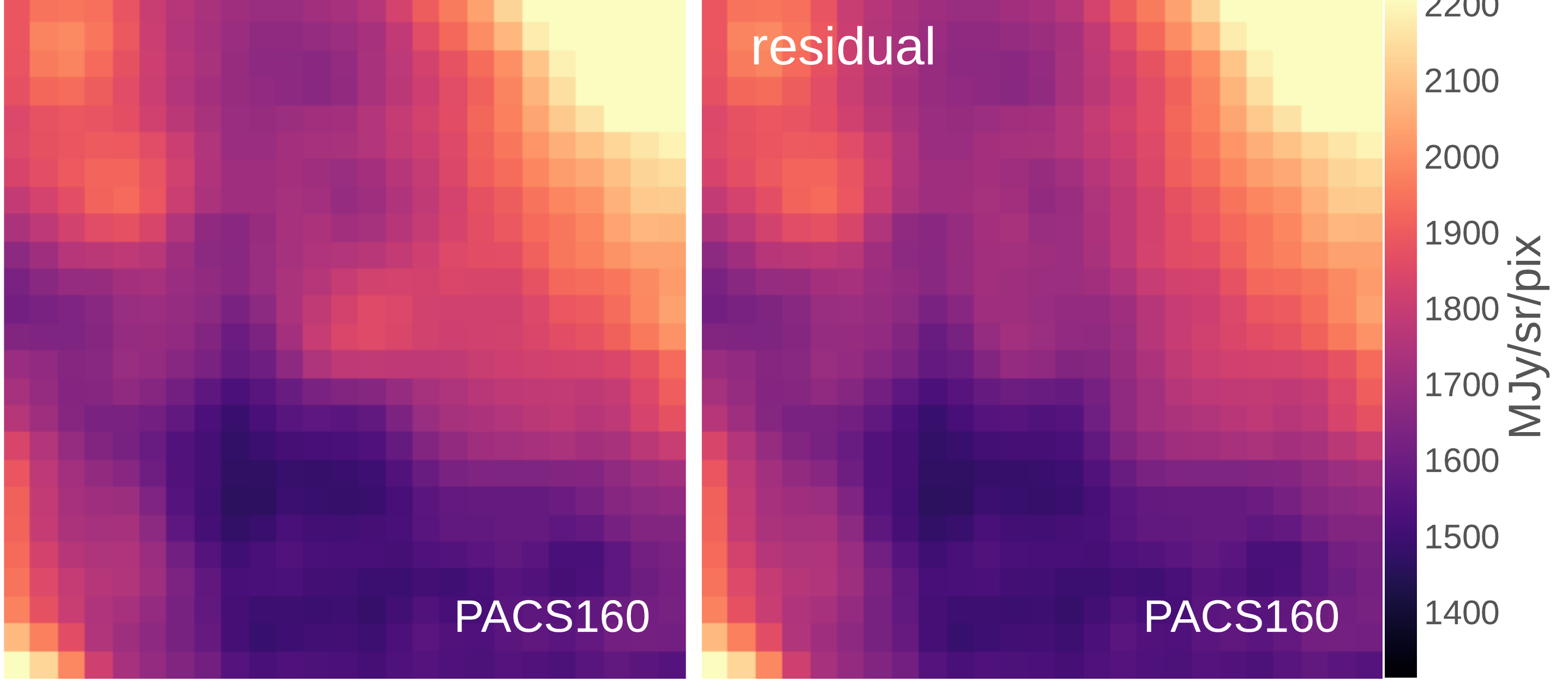}
   \hfill
      \includegraphics[scale=0.31]{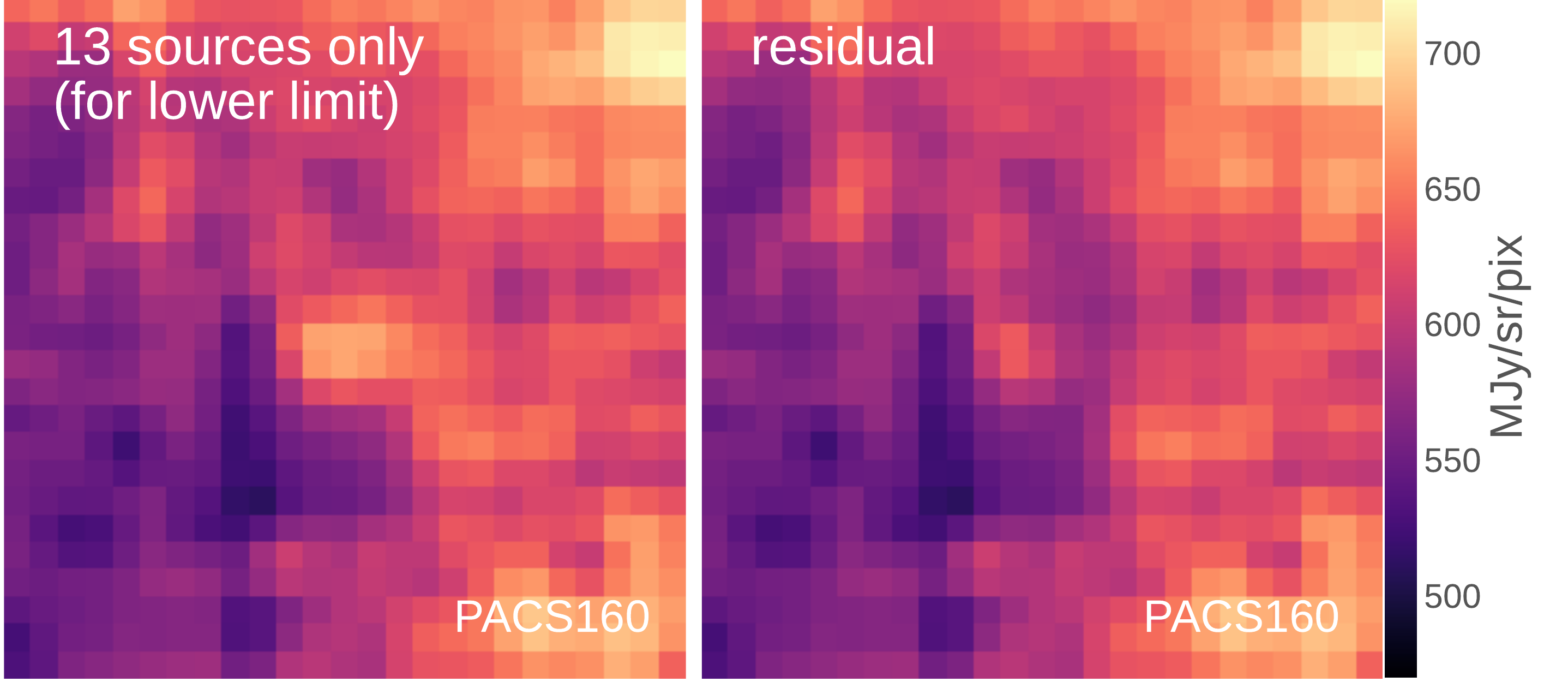}

     \caption{Left column: stacked images of LMC LBVs (SED Classes 1 and 2 in Table~\ref{tab:sample}) and residuals after Gaussian fitting. Right column: same as before but without RMC143.    
     The field of view corresponds approximately to $70\arcsec \times 70\arcsec$. The stack image at 160$\rm \mu m$ on the bottom right was derived with a smaller number of sources (IDs: 2, 5, 6, 7, 8, 10, 12, 13, 14, 15, 16, 17, 18) and used as a lower limit.
     }
     \label{fig:residuals-LMC}
\end{figure*}

 \begin{figure}
\centering
   \includegraphics[scale=0.38]{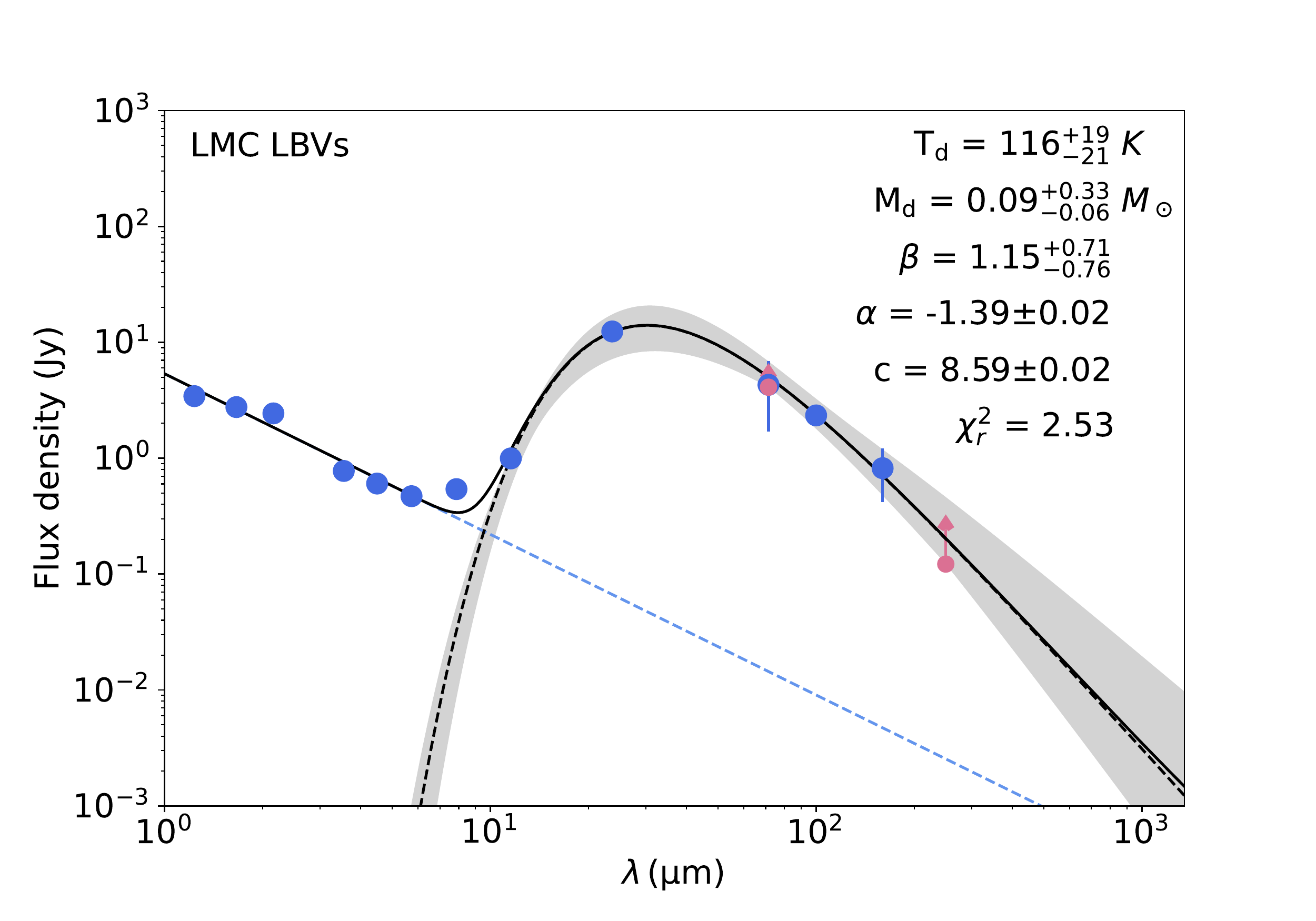}
   \includegraphics[scale=0.38]{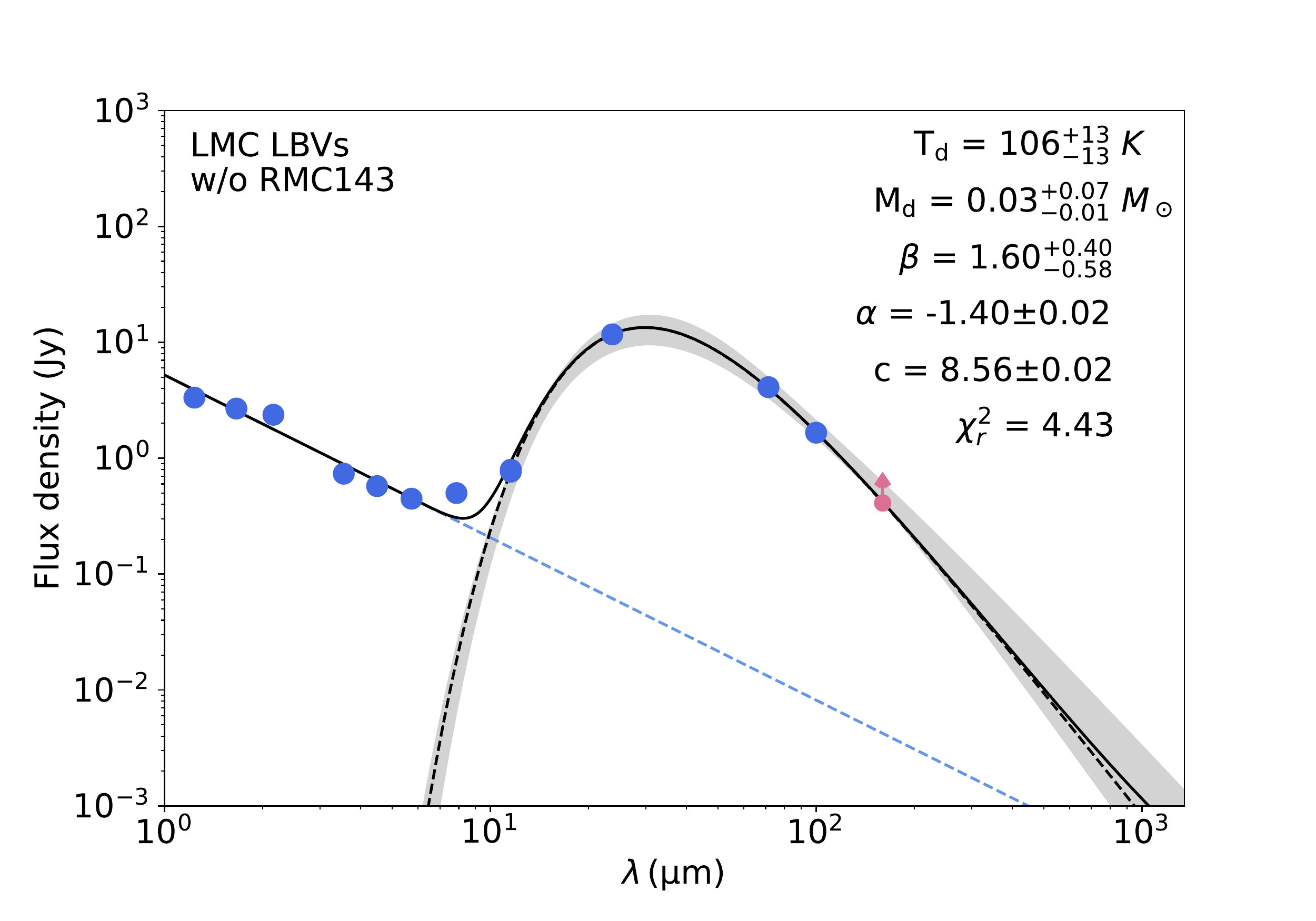}
     \caption{Flux density distribution of stacks of the LBVs in the LMC (SED Classes 1 and 2 in Table~\ref{tab:sample}), with (top) and without (bottom) RMC~143. 2MASS and IRAC values are the sum of the individual source fluxes, while WISE-W3, MIPS and PACS values are from photometry performed on the stacked images (Fig.~\ref{fig:residuals-LMC}). Parameters from the fitting procedure are shown ($\alpha$ and $c$ are the spectral index and offset of the best-fit  power-law describing the near-infrared data; $T_{\rm d}$, $M_{\rm d}$ and $\beta$ are the dust temperature, mass and opacity power law index of the best-fit greybody describing the thermal emission from dust). 
     The lower limits (red points and arrow symbols) for MIPS70 and SPIRE250 in the upper plot and PACS160 in the lower plot are from stacks of subsets of sources (see text for the subsets in each case), as there is not a reliable detection in the stacked images of the whole sample.
     }
     \label{fig:sedstot}

\end{figure}

 \begin{figure}
\centering
   \includegraphics[scale=0.38]{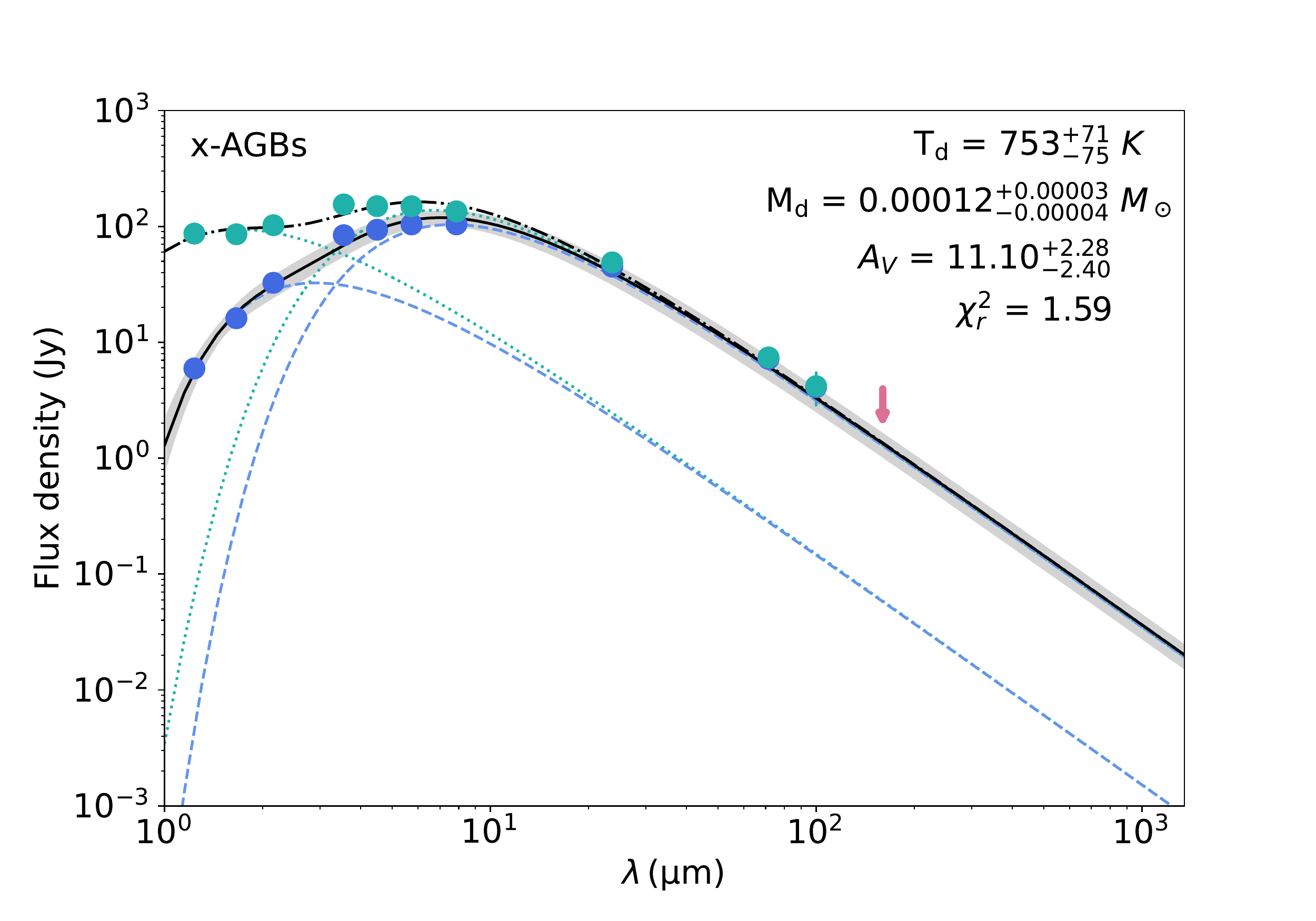}

     \caption{Flux density distribution of the 1342 extreme x-AGBs in the LMC from the catalogue by \citet{2012Riebel}, 
     constructed from the Gaussian fitting of the central detection in the stacked images (at wavelengths shorter than $\sim10~\rm \mu$m the value is just the sum of the catalogue measurements). Blue points: observed phtometry; green points: de-reddened photometry; red arrow: $3\sigma$ PACS160 upper limit.
     }
     \label{fig:sedstot-xAGBs}
\end{figure}

 \begin{figure}
        \centerline{\sf \large Stack of 1342  extreme AGBs in the LMC\hspace{2em}}
        \vspace{5pt}
   \includegraphics[scale=0.31]{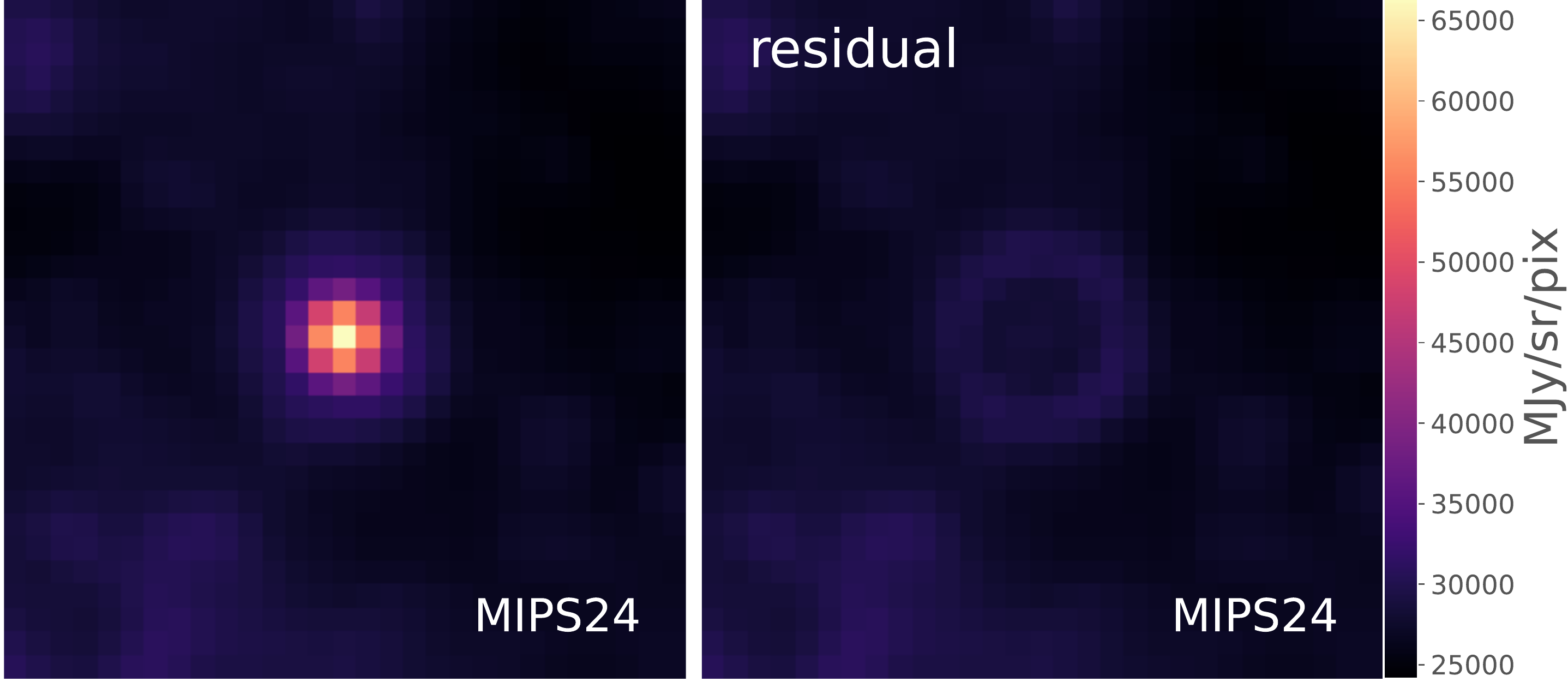}
   \vspace{1pt}
   \includegraphics[scale=0.31]{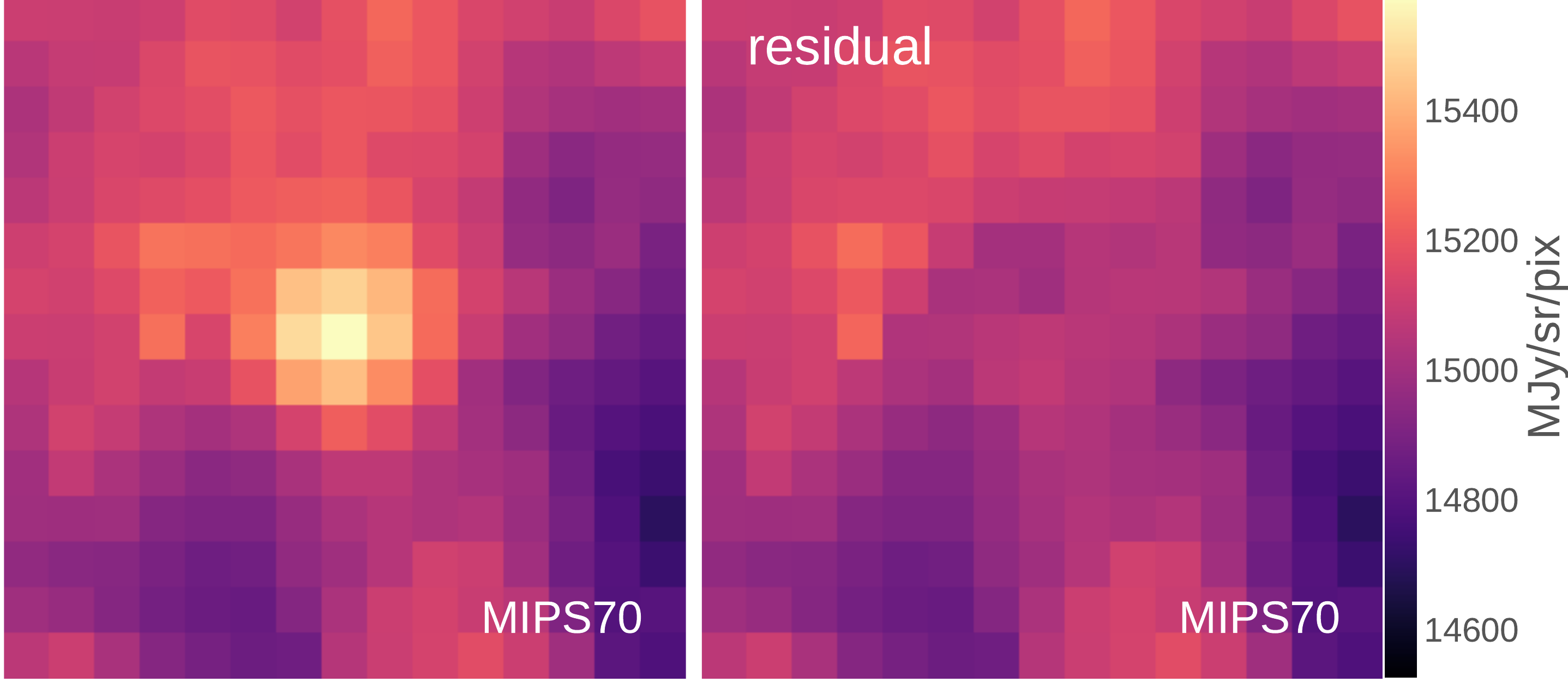}
   \vspace{1pt}
   \includegraphics[scale=0.31]{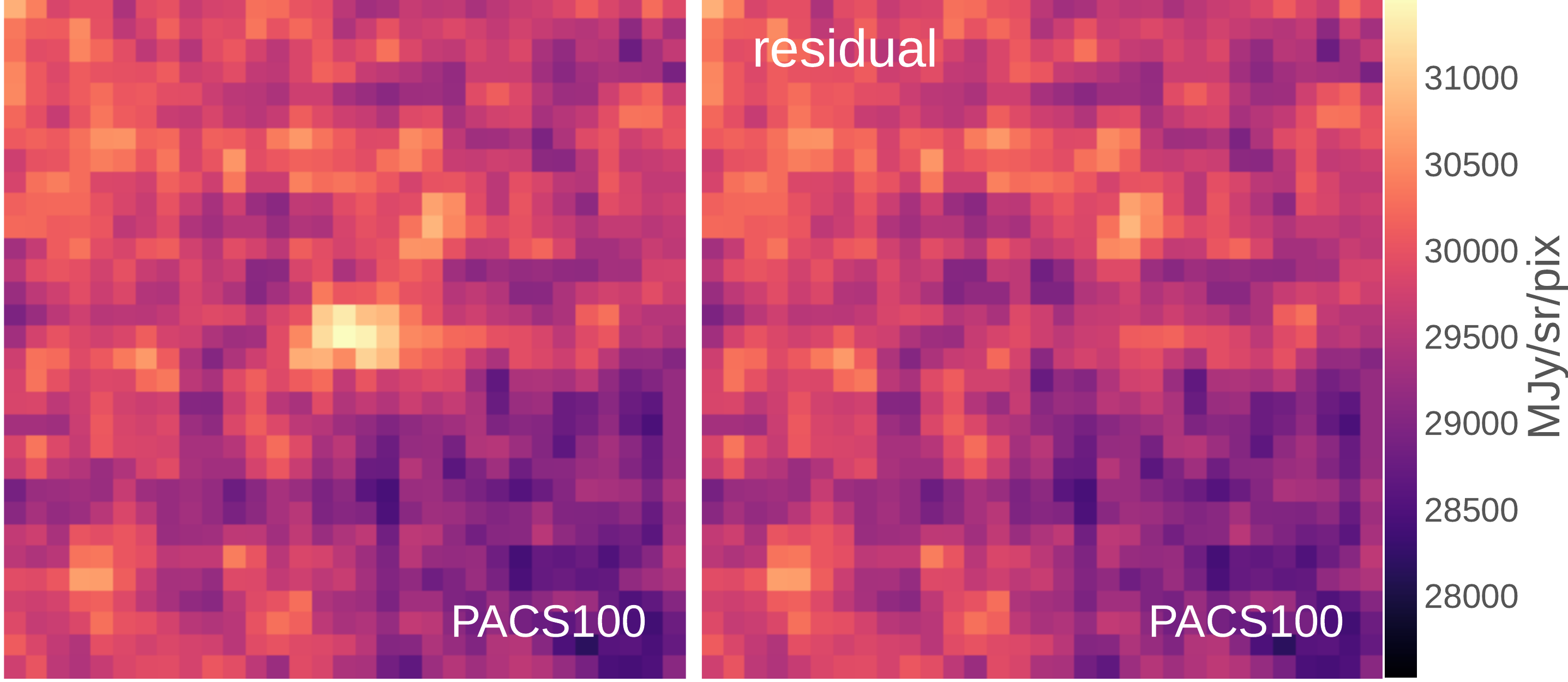}
   \vspace{1pt}
   \includegraphics[scale=0.31]{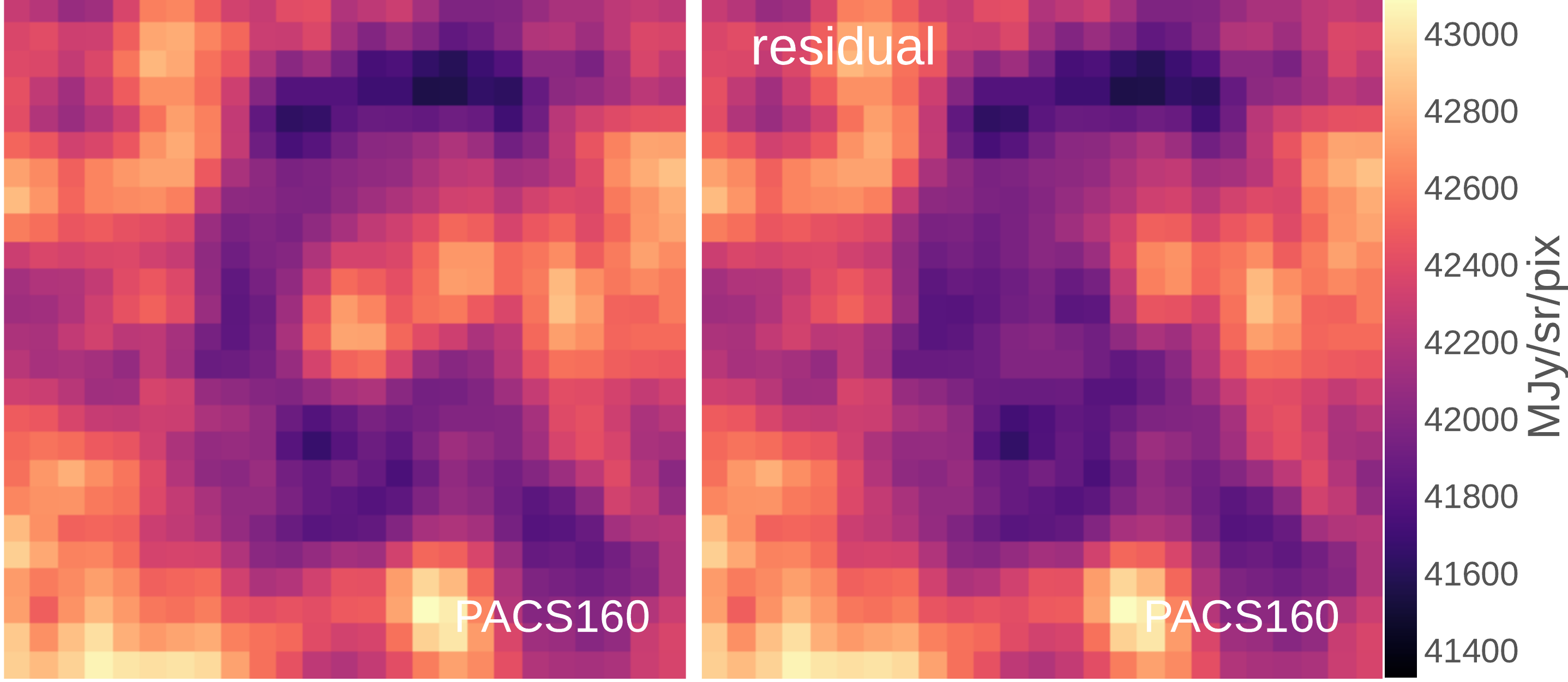}   
   \vfill
     \caption{Stacked images of the extreme-AGBs from \citet{2012Riebel} and residuals after Gaussian source extraction. The field of view corresponds approximately to $70\arcsec \times 70\arcsec$. Ten out of 1352 sources were excluded (see text), because the emission in their field is dominated by bright large-scale structures or because they are another type of stellar objects (e.g. YSOs).
     }
     \label{fig:residuals-xAGBs}
\end{figure}

Given the low detection rates of individual sources at $70\,\mu\rm m$ and longer, largely due to confusion, 
we have also carried out a stacking analysis to provide a statistical measure of total dust content. The stacking analysis adopted here consists in co-adding the images, at a given wavelength, of individual sources with known coordinates, in order to reduce the noise by averaging, resulting in an aggregate flux density measurement. This technique is particularly useful when the target sources are not detected. The stacking procedure is  described in \citet{2010Kurczynski}, and references therein. Here, we do not adopt any deblending method described by the authors, as the confusion in our case is dominated by diffuse ISM emission, of which a-priori positional information is difficult to obtain. 
For the stacking analysis we considered all confirmed and candidate LBVs (SED Classes 1 and 2), thus excluding sgB[e] stars and the sources discarded in Sect.~\ref{sec:review}. The LMC stack sample thus comprises the first 18 entries in Table~\ref{tab:sample}. In the SMC there are only two sources to consider (ID number 28 and 29 in the same table), of which only one clearly shows a dust component (number 29). For this reason, we perform the stacking analysis on the LBVs (Class 1 and 2 in Table~\ref{tab:sample}) of the LMC only. 

To obtain stacked images, for each source we created cutouts of the SAGE and HERITAGE mosaic  of the LMC  centred, to the nearest pixel, at the sky coordinates as in Table~\ref{tab:sample}, after cross-checking the coordinates with the actual detections in the map. As a  \emph{Wide-Field Infrared Survey Explorer (WISE)} image of the whole LMC galaxy is not available, in this case we queried the WISE image service in the IRSA website and downloaded images centred at the sky coordinates of each source. For none of the instruments it was  necessary to re-sample the images as they originate from the same grid. 
Images of the following bands were stacked:  WISE~W3, MIPS24, MIPS70, PACS100, PACS160, and SPIRE250. 
For each band we summed the cutouts of all sources. The resulting stacked images are shown in Fig.~\ref{fig:residuals-LMC}, while the associated photometry is summarised in Table~\ref{tab:stackedphoto}. In the same figure, we also show the same stack images without RMC$\,$143, which although the most massive LBV nebula recorded so far in the LMC \citep{2019Agliozzo}, is embedded in a very bright and crowded region, just south of the 30~Doradus star forming complex. The comparison of the resulting photometry allows us to assess the relative contribution of the other stars to the infrared emission.

Following the same method, we also stacked cutout images extracted from the LMC $1.4\,{\rm mm}$ SPT+\emph{Planck} map of \citet{2016PlanckMCs}. However, we did not obtain any detection above the noise ($\sim\!250\,{\rm mJy/beam}$) and so do not show the resulting image here.

We then determined the aggregate flux density of the combined source by performing a 2-D Gaussian fitting in each image. Peak height, $x$ and $y$ centre positions and background level 
are fitted parameters; the width parameters $\sigma_a$ and $\sigma_b$ were fixed to the instrumental beam $\sigma$ for MIPS70, PACS100, and PACS160 images, while they were left as free parameters (along with a position angle) for the WISE W3 and MIPS24 fitting because at these wavelengths sources were slightly resolved. We used the non-linear least-squares method of the \texttt{Scipy Optimize} package \citep{2020Scipy} 
for minimising the fitting function residuals. To account for systematic errors in this procedure (e.g. missing flux in sidelobes), we corrected the values by a factor determined in each filter by comparing our own estimates of individual test sources with catalogue measurements. 
Figure~\ref{fig:residuals-LMC} shows the stacked images and their residuals after subtraction of the 2-D Gaussian. We do not include SPIRE~250 because there is no source detected, owing to source confusion over the large beam.
For the stack analysis at the IRAC bands, we only summed the photometry from the catalogue, as all the  individual sources were well detected. 
We expect the amount of flux density missed by the small catalogue aperture to be negligible since the IRAC images do not show hints of resolved nebular emission at these wavelengths. 

For each passband, the resulting flux densities from the stack analysis are plotted in the SEDs in Fig.~\ref{fig:sedstot} as blue circles. The top figure visualises the sum of all 18 candidate and confirmed LBVs in the LMC, while the bottom shows the same without RMC~143. 
In the top figure the pink symbols represent lower limits.  The $70~\mu$m limit is given by the measurement in the stacked image without RMC~143, where the central source is better detected without the contamination by the 30~Dor region (as is evident from the comparison of the $70~\mu$m error bars in the top and bottom panels).  The $250~\mu$m limit is due to a source detected at the central position of a stack image built from the only four well detected sources in this filter (namely RMC~127, LHA~120-S~119, LHA~120-S~61, and RMC~71). In the lower plot, the $160~\mu$m lower limit is set based on the detection in the stack image obtained from fewer 13 sources (IDs: 2, 5, 6--8, 10, 12--18) that are less affected by crowding. 

The total SED from about 1 to $250~\mu$m resembles that of LBV stars with dense ionised stellar winds, surrounded by extended dusty nebulae (Class~1a). 
We initially fit the 2MASS $J$, $H$, $K_{\rm s}$ and IRAC~$3.6$, $4.5$ and $5.8\,\mu{\rm m}$ data with a power law representing the sum of stellar winds and photospheres
\citep[power-law free-free emission from ionised stellar winds, ][and $F_\nu\propto\lambda^{-2}$ Rayleigh-Jeans photosphere emission]{1975PF}, 
and then we added the resulting best fit and uncertainties to the grey-body function built to fit the WISE~W3, MIPS24, MIPS70, PACS100, and PACS160 measurements. In the figure, the best-fit total model is given by a black continuous line. The power-law fit of the wind component is the blue dashed line. The grey shaded region represents all the grey-body components that fit the data producing a $\chi^2$ which is twice the minimum value of the best fit. Because of the large uncertainty of the $70~\mu$m produced by the crowded region around RMC~143, we set as lower limit the photometric measurement at $70~\mu$m without RMC~143 to constrain the range of possible models. We also introduce another constraint at longer wavelengths by using the lower limit at $250~\mu$m. The resulting dust temperature, mass, and parameter $\beta$ are listed in the figures. The dust mass of the aggregate source is computed by assuming the flux density extracted from the best fit at $\rm 850\,\mu$m. The adopted absorption coefficient $\kappa_{\rm 850\mu m}$ is $\rm 1.7\, cm^{2}\, g^{-1}$ \citep{1987Sodroski}.

\subsection{Stack analysis of other evolved stars in the LMC}
\label{sec:stackingAGBs}
The same stacking analysis was performed for the sample of RSGs and AGBs in the catalogue of \citet{2012Riebel}, which contains $33\,501$ sources ($6\,709$ carbon-rich AGB, $19\,566$ oxygen-rich AGB, $5\,874$ RSG and $1\,352$ extreme ``x-AGB'' stars). Previously \citet{dharmawardena} stacked the PACS100 images of the sources in Riebel et al., split up by mass-loss rate bin and chemical class, and found some detections from carbon-rich and oxygen-rich AGB stars with the highest mass-loss rates. The wavelengths covered here are those of the four IRAC filters at 3.6, 4.5, 5.8 and 8~$\rm \mu$m, the two MIPS filters at 24 and 70~$\rm \mu$m, and the PACS filters at 100 and 160~$\rm \mu$m. For completeness, we also considered the SPIRE image at 250~$\rm \mu$m, although with low expectation of obtaining a detection because of source confusion at those wavelengths. 
We cross-matched the list of LBVs and the catalogues of Riebel et al. and found two common sources (RMC~66 and RMC~84), which we removed from their sample (they were catalogued as RSGs). 
Various attempts were made to obtain a detection at wavelengths longer than $24\,\rm \mu$m. Initially, the images of individual categories of stars were stacked separately, with and without applying a confusion threshold on the basis of the standard deviation evaluated from all the pixels in each field (measuring about 50$\times$50 pixels). Finally, all the $33\,501$ sources were summed together.  In the $70\,\mu{\rm m}$ and the stack images at longer wavelength, we did not obtain any statistically meaningful detection for the RSGs, the C- and O-rich AGBs, or the whole sample.

In the case of the ``extreme'' AGBs, which are considered the main dust producers among all AGBs and RSGs in the MCs  \citep{2012Boyer,2012Riebel}, we evaluated possible detections at 24, 100, 160, 250, 350 and $500\,\rm \mu$m. We noticed that at far-infrared wavelengths the stacked images are dominated by confusion when the following fields are included: J053929.34-694614.8, J053911.41-690824.7, J054515.83-694648.1, J051913.89-693818.3, J045506.54-691708.6. We also found that five other sources have been identified as YSOs or other types of stars \citep[J053840.77-690603.3, J053856.58-690417.3, J053839.68-690537.9, J045400.16-691155.4, J053238.60-682522.1, ][]{2011Evans,2013WalbornBarba,2020vanGelder,2019Oliveira}. We thus excluded these ten sources from further analysis. The final sample of extreme AGB stars from \citet{2012Riebel} that we stacked comprises 1342 objects. We obtained significant detections up to $100\,\mu\rm m$, and a marginal detection at $160\,\mu\rm m$, as shown in Fig.~\ref{fig:residuals-xAGBs}. To produce the composite continuum spectrum in the figure, no weighting based on the stellar luminosity or normalisation was applied. The spectrum simply comprises the aggregated flux densities measured from the stacked images at each wavelength.

We extracted the photometry as described in the preceding subsection. 
At $24\,\rm \mu$m we compared our photometry extracted from the stack image with the stacked photometry  \citep[obtained by summing the photometry in][]{2012Riebel} and found a consistent value. 
We fitted the data with two components: a black body of fixed temperature (3000~K) describing the stellar photosphere,  and a grey body with dust spectral index $\beta=0$. Because the near-infrared data of AGBs are extremely reddened by the optically thick dust, we also fitted the extinction by using the extinction curve of \citet{1999PASP..111...63F}, with selective parameter $R_{V} = 3.1$. 
Fig.~\ref{fig:sedstot-xAGBs} shows the photometry and resulting fit with and without extinction correction. 
For comparison, the integrated flux density at 100~$\mu$m from 18 LBVs is of the same order of magnitude of that from 1342 extreme-AGB stars (see Table~\ref{tab:stackedphoto}).
The dust mass derived from this fitting assumes that the dust is optically thin (at the wavelength at which $\kappa$ is defined), and thus must be considered a lower limit.

\section{Discussion}
\label{sec:discussion}
\subsection{LBV dust masses versus stellar luminosity and nebular size}
\begin{figure}
    \centering
    \includegraphics[width=.9\linewidth]{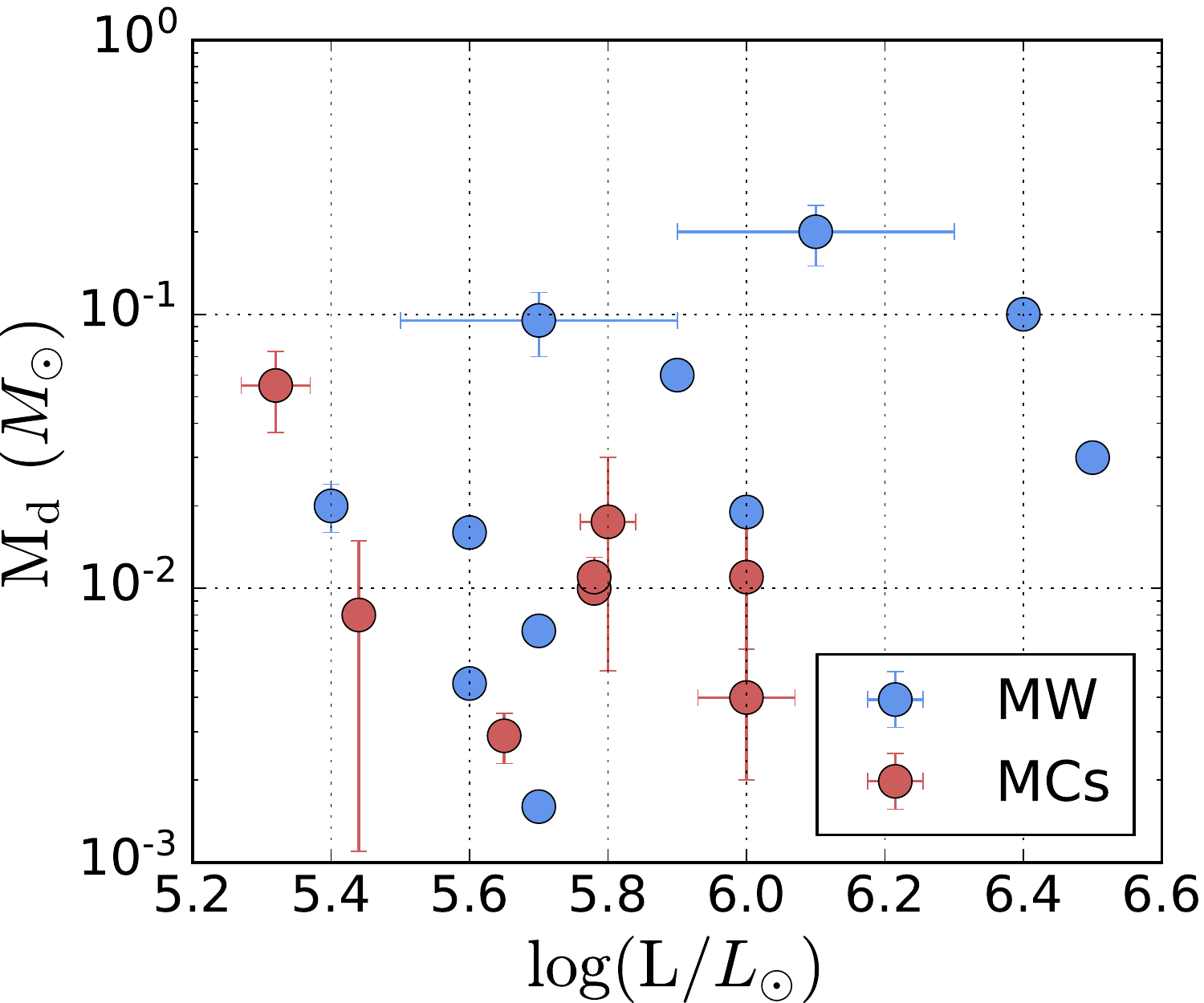}
    \caption{Estimated dust masses for individual stars as a function of the stellar bolometric luminosity. Red circles: LMC LBVs (this work, see Table~\ref{tab:dustproperties}). Blue circles: Galactic LBVs, specifically AG~Car \citep{2015Vamvatira}; G24.73+0.69, G25.5+0.2, HD~168625, Hen~3-519, AFGL~2298 \citep[][and references therein]{2003Clark}; G26.47+0.02 \citep{2012Umana}; G79.49+0.26 \citep{2014Agliozzo};
HR Car \citep{2002Machado,2017Buemi}; Pistol star \citep{2014Lau}; Wra 751 \citep{2013Vamvatira}. Error bars are shown when available from the literature.}
    \label{fig:mdvL}
\end{figure}
The uncertainty on the distance is  only $\sim\!2\%$ for the Magellanic sources (see \S\ref{sec:modelling:grey}), 
and thus their stellar luminosity is well constrained ($\sim\!4\%$ uncertainty due to the distance). This is a great opportunity to investigate any possible dependence of the dust mass on the stellar parameters and on the nebular size. 
Fig.~\ref{fig:mdvL} shows the dust masses estimated in Sect.~\ref{sec:modelling:grey} for 7 Class 1 LBVs in the LMC and RMC\,40 in the SMC versus the stellar bolometric luminosity as in the literature (Table~\ref{tab:stellparam}). In the same plot, dust masses for some well-known nebulae around candidate and confirmed LBVs in the Milky Way are displayed. Dust production in these stars does not seem correlated to the stellar properties. 

For five objects the nebula size is known in the literature and one has been determined in this work. 
Sizes span from 0.3 to 2.2~pc (Table~\ref{tab:stellparam}). We do not observe a systematic trend in the dust mass versus the nebular size.

\subsection{Total dust mass in the LMC}
\begin{table}
\centering
\caption{Summary of estimated LBV dust masses in the LMC. }
\label{tab:dustmasses}
\resizebox{0.475\textwidth}{!}{
\begin{tabular}{lcc}
\hline\hline 
Source & \multicolumn{2}{c}{Grey-body fitting}  \\ 
&($\kappa_{\lambda}$ as SD87)  & $(\kappa_{\lambda}$ as GD14)  \\
\hline
Stack 18& $9^{+33}_{-6}\times 10^{-2}\,M_{\odot}$&$11^{+6}_{-3}\times10^{-2}\,M_{\odot}$ \\
Mean per source& $0.5^{+1.8}_{-0.3}\times10^{-2}\,M_{\odot}$&$0.6^{+0.3}_{-0.2}\times10^{-2}\,M_{\odot}$ \\
\hline
\end{tabular}
}
\tablefoot{Results are given for two different opacity values, as in \citealt[][]{1987Sodroski} (SD87) and \citealt[][]{2014Gordon} (GD14).}
\end{table}
Since the ultimate goal of this work is to compare the LBV dust masses with the entire LMC dust yield 
we also fit the stack data with the simple grey model that \citet{2014Gordon}  used to derive the global dust properties of the ISM in the LMC and SMC. 
In particular, we adopt their absorption coefficient $\kappa_{\lambda}$ (calibrated with ultraviolet and optical gas-phase absorption measurements in the diffuse ISM of the Galaxy, which provide the expected amount of dust and are independent from the far-infrared and submillimetre dust emission). At $160\,\rm \mu m$, Gordon et al. derived $\kappa_{160} = 9.6\pm0.4 \pm 2.5$ where the first error is statistical and the second is the systematic uncertainty due to the assumption that the dust in the MCs is like that in the Galaxy. 
With this $\kappa_{160}$, our stack PACS160 flux and our stack SED fit temperature (top panel in Fig. \ref{fig:sedstot}), we obtain for the 18 LMC (Class~1 and 2) LBVs $M_{\rm dust}=0.11^{+0.06}_{-0.03}\, M
_{\odot}$, and an average of $\langle M_{\rm dust}\rangle=0.006^{+0.003}_{-0.002}\, M
_{\odot}$ per source. 
The uncertainty of about $25\%$ in $\kappa_{\lambda}$ due to the systematic error  is not taken into account. 

Table~\ref{tab:dustmasses} summarises the dust masses derived from grey body fitting of the stack data. We notice that the integrated dust mass from 18 sources is of the same order of magnitude of the dust mass estimated in the Homunculus nebula of $\eta$~Carinae \citep{2017Morris}. 
This difference may be the result of different metallicities. It may also suggest that several stars have still not experienced the required mass loss to form dust (giant eruptions)  or that they do not have colliding winds within a close binary.  
It should be noted that only a handful of sources contribute significantly to the stacked dust mass (compare Tables~\ref{tab:dustproperties}   
and \ref{tab:dustmasses}). 
Additionally, the dust masses estimated here could be a lower limit, as cooler dust may have escaped the observations. Sensitive sub-millimetre observations with ALMA of all Class~1 and 2 objects would be an asset for further studies.

For the LMC, \citet{2014Gordon} derived an integrated dust mass of $(7.3\pm1.7)\times 10^{5}\,M_{\odot}$ (averaging the results from three different models, or $(8.1\pm2.8)\times 10^{5}\,M_{\odot}$ with the simple grey body model only). A more recent analysis based on the same dataset reports estimates about a factor of two lower \citep{2017Chastenet}. 
 In the classical scenario, the LBV phase is estimated to last around $10^{4}-10^{5}\,\rm yr$, which is a small fraction compared to the age of the LMC \citep[about 15~Gyr,][]{2009H&Z}. 
To evaluate the potential contribution of LBVs to the ISM dust yield, we must account for the entire population of LBVs  over the lifetime of the LMC.

This is not an easy task, because stellar evolution models are not able to predict how many BSGs become LBVs (or manifest the LBV phenomenon). 
Ideally, we would like to compare the ratio of LBVs and BSGs of similar luminosity, but finding a list near to completeness is challenging. 

\citet{2017Urbaneja} analysed 90 LMC BSGs to study the gravity-luminosity relationship, but this number is probably a lower limit. In fact, in \citet{2009Bonanos} we count 133 stars earlier than B4 and of luminosity classes Ia and Iab in the LMC, excluding four objects that overlap with the LBV sample. 
However, \citet{2009Bonanos} could not claim completeness, as they targeted O, B supergiant stars with both accurate coordinates and spectral classification. We queried in Simbad for stars with spectral types earlier than B4 and of luminosity class I, within 2.5 deg from the LMC centre, and we find around 230 BSGs \citep[several of these objects appear in][]{2002Massey}. \citet{2010Massey}  estimated $\sim6100$ unevolved massive stars with $M>20\,M_{\odot}$, and $\sim700$ with $M>50\,M_{\odot}$. 

\subsubsection{Simple case: Constant number of LBVs over time}
The simplest estimate we can make of the total dust mass produced by LBVs in the LMC over all time is by assuming that the number of LBVs has on average been the same as today. This simple assumption avoids having to consider the OB to LBV rate, but it does depend directly on the assumed LBV phase duration. For the range of accepted LBV phase durations of $10^4$--$10^5$~yr, the LMC age of 15~Gyr, and the mass produced by the current LBV sample of $0.11^{+0.06}_{-0.03}\,M_\odot$, the range of total dust mass is from $1.2\times10^4\,M_\odot$ to $1.7\times10^5\,M_\odot$, with a nominal value of $6.6^{+10.4}_{-5.4}\times10^4\,M_\odot$ for an assumed LBV phase duration of $25\,{\rm kyr}$.

\subsubsection{Accounting for star formation history and initial mass function}
We assume that the LBV population closely followed the fluctuations in the star formation history (SFH) of the LMC. The latter was computed by \citet{2009H&Z}, who derived star formation rates (SFR) in about 1400 LMC sub-regions in time bins of $\Delta \log(t) = 0.2$. We integrate the SFR over these sub-regions, in order to obtain the total SFH for the LMC as in Fig.~11  of \citet{2009H&Z}. 
For the Initial Mass Function (IMF) we adopt the power law slope derived by \citet{2018Schneider}, who found an excess of massive stars by $30\%$ in the 30~Dor star forming region. 
We also use the Salpeter IMF as a possible lower limit in the LMC.
Finally, we derive the total dust yield by LBVs as,
\begin{multline*} 
\frac{\left(\displaystyle\sum_{i} {\rm SFR}(t_{i})\Delta t_i\right) \times\ f_{\rm LBV, MS}\times \displaystyle\int_{25M_\odot}^{100M_\odot} \xi(m) \,dm}{\displaystyle\int_{0.1M_\odot}^{100M_\odot} m \xi(m) \,dm} \times \langle M_{\rm dust}\rangle, \\
\end{multline*}

where $\xi(m)=\xi_{0}\times m^{\gamma}$ is the IMF, ${\rm SFR}(t_{i})$ is the SFR in the time bin $i$, 
$\langle M_{\rm dust}\rangle$ is the mean dust mass produced per LBV, and $f_{\rm LBV, MS}$ is the fraction of stars in mass range $25$--$100M_\odot$ experiencing the LBV phenomenon. The IMF is  normalised to the total mass of stars with masses between 0.1 and 100$\,M_{\odot}$.
Assuming the slope $\gamma=-1.9$ from \citet{2018Schneider} (Case 1) and $\gamma=-2.35$ from Salpeter (Case 2), we find the following ranges of dust mass:
   $ 17.1^{+14.0}_{-3.7}\times10^{4}\times f_{\rm LBV, MS}  \,M_{\odot}$ (Case 1) and
$  4.4^{+3.6}_{-1.0}\times10^{4}\times f_{\rm LBV, MS} \,M_{\odot}$ (Case 2). The error bars include the uncertainty on the dust mass and on the SFR. A major uncertainty in this relation lies in the fraction of massive main-sequence stars that will become LBV stars ($f_{\rm{LBV, MS}}$).  Obviously, the number cannot exceed unity. Because the conditions leading to the LBV phenomenon are unknown, it is unconstrained by evolutionary models.  

The observed fraction with respect to BSGs, $f_{\rm{LBV, obs}} = n_{\rm LBV, obs}/n_{\rm BSG, obs}$, needs to be corrected for incompleteness and, more importantly, for the ratio of the duration of the BSG and LBV phases, $t_{\rm BSG} / t_{\rm LBV}$. Evolutionary models suggest $t_{\rm BSG} \leq 10^{6}\,\rm yr$ for stars more massive than $25\,M_{\odot}$ in the He-burning phase, and $t_{\rm MS} \leq 10^{7}\,\rm yr$ for stars in the H-burning phase \citep{1989M&M}. While the number of stars in the LMC currently exhibiting the LBV phenomenon should be reasonably correct ($n_{LBV, obs}$), it should be subject to fluctuations, and some current LBVs might be inactive.  The number of BSGs in the LMC that may become LBVs ($n_{\rm BSG, obs} \sim 230$, see above) is also uncertain.  
The number of MS O,B stars (6700) may be more reliable \citep{2002Massey, 2010Massey}. 
From the currently available information, it seems plausible that $f_{\rm LBV,MS}$ ranges between about $30\%$\footnote{Comparing the LBVs with the MS OBs, taking $10^7{\rm yr}$ as the MS lifetime and an upper limit of $10^5{\rm yr}$ for the LBV phase duration gives,\\
$f_{\rm LBV} = (18\,{\rm LBVs}/6700\,{\rm MS\,OBs})\times(10^7{\rm yr}/10^5{\rm yr})=27\%$.}
and $100\%$\footnote{Comparing the LBVs with the BSGs, taking $10^6{\rm yr}$ as the BSG phase duration and  $10^5{\rm yr}$ for the LBV phase duration gives,\\
$(18\,{\rm LBVs}/230\,{\rm BSGs})\times(10^6{\rm yr}/10^5{\rm yr})=78\%$.\\
As the LBV phase may be an order of magnitude shorter than this, it suggests that the number of 230 BSGs may be incomplete.}.

We repeat a similar calculation for the AGB/RSG stars. We use the average dust mass per star determined from the stack of all extreme-AGBs ($9^{+2.2}_{-3.0}\times10^{-8}\,M_{\odot}$) and we integrate over stellar masses between 1 and 25~$M_\odot$ to also include RSGs. We find a total dust mass of $136^{+56}_{-35}\,M_{\odot}$ (Case 1) or $144^{+59}_{-37}\,M_{\odot}$ (Case 2), which are much lower values than LBVs, but maybe consistent with the long dust injection timescales of AGBs  \citep[$\sim2\times10^{9}\,\rm yr^{-1}$][]{2018Micelotta}. 
Moreover, the dust lifetime against processing and destruction by supernova shocks in the ISM is shorter, around $2\times10^{7}\,\rm yr$ \citep{2015Temim}. The inefficiency of existing AGB stars to produce the dust mass observed in the ISM was already pointed out by \citet{2009Matsuura} for the LMC and by \citet{2012Boyer} and \citet{2013Matsuura} for the SMC. Dust injection timescales from LBVs (and subsequent SNe) are much shorter ($\leq10-100\,\rm Myr$). In their study of the total dust input from stars in the SMC \citet{2012Boyer} estimated a dust production rate of $\sim10^{-6}\,M_{\odot}\,\rm yr^{-1}$ from LBVs, matching the input from all cool stars, but concluded that the grains would not survive the subsequent supernova explosion. 

\subsection{Potential dust survival from SN explosions}

Now we evaluate how likely this LBV dust is to survive the SN forward shock (blast wave) which will follow  the LBV or WR evolutionary stage. 
The case of ISM grains processing by the forward shock was analysed by \citet{1978Barlow}, \citet{ 2006Nozawa}, and \citet{2014Bocchio}. The processes of interest are generally sputtering and shattering. There are a few arguments to mention in favour of dust survival in the SN blast wave in LBVNe. 1) LBV eruptions are a source of large grains ($>0.1\,\rm \mu m$) due to the favourable conditions for particle growth \citep{2011Kochanek}. The relatively low $\beta$ values found in this work are also  consistent with large grains. However, they could also be low due to a temperature gradient over the nebula. 2) The material in LBVNe seems often distributed in a clumpy configuration, spread at large distances from the star. 3) Such nebulae can be asymmetric (e.g. RMC~143, LHA$\,$120-S$\,$119), subtending solid angles of less than $4\pi$ so that some of the SN energy will leak out without interacting with the nebula.

 Such properties reduce the importance of the destructive action by the shock, as for example evidenced in studies of the newly formed dust in SN ejecta impacted by the reverse shock. In particular, survival rates vary between 40 and $98\%$ \citep{2008Nath, 2010Silvia, 2012Silvia, 2016Biscaro, 2018Gall} for dust reached by the reverse shock, and around $85\%$ for large grains in the ISM \citep[e.g.][]{2006Nozawa}. More results are summarised in Table~2 of \citet{2018Micelotta}. 
Some studies also show that shocks can induce dust (re)formation and growth \citep[e.g.][]{2011KochanekSN, 2019Matsuura}. Recent 3-D hydrodynamical simulations report that the wind-driven bubbles created during the entire life of a massive star create the conditions for the dust to largely survive \citep{2019MartinezGonzalez}, or that under favourable ambient conditions implantation and trapping of ions in dust grains can counteract grain destruction by sputtering \citep{2020Kirchschlager}. Dust destruction is particularly negligible if the pre-existing shell is massive \citep{2019MartinezGonzalez}, which is the case of LBV nebulae. 

From the observational point of view, a great example is SN1987A, the progenitor of which may have been an LBV \citep[as speculated by][]{2007Smith}. Along with the discovery of dust formation in the SN ejecta a few years after the explosion \citep{1989Danziger}, followed up by far-infrared and sub-millimetre observations about three decades later \citep[which revealed large amounts $\sim 0.5\,M_{\odot}$ of cold dust, ][]{2011Matsuura,2014Indebetouw,2015Wesson}, another dust component distributed in an equatorial ring lost by the progenitor $20-40\,\rm kyr$ ago was found \citep{1989Arnett, 1993McCray, 2006Bouchet}. The fastest part of the forward shock has passed the ring \citep{2015Fransson}, offering unique insights into the resulting dust destructive processing. The silicate grains in the ring are collisionally heated by the SN blast wave \citep{2006Bouchet}. \citet{2016Arendt} proposed that some dust grains were destroyed by sputtering, supporting the theory \citep{2008Dwek,2010Dwek}. Subsequently, \citet{2019Matsuura} reported a $30-70\,\rm \mu m$ excess at day $\sim10\,000$ on top of the earlier modelled hot and warm component (in the ring) and cold component (in the SN ejecta), and a dust mass about a factor of 10 larger than previous estimates at day $\sim8\,000$. They identified two possible mechanisms behind this infrared excess: dust re-formation (dust grains formed from the gas phase) and dust growth (survived grains accreting atoms from the gas phase). Dust re-formation was also suggested in the cool dense shell created between the forward and reverse shocks as the SN ejecta plows into the existing circumstellar material. Several studies of type IIn SNe  (which, in some cases, may have an LBV as   progenitor) are a good example \citep[e.g.][]{2008SmithFoleyFilippenko, 2009Fox, 2014Gall,2016Andrews,2018Chugai}. Grain growth may also be a viable mechanism as explained by \citet{2020Kirchschlager}.

\section{Summary and conclusions}
We have analysed the infrared images of Magellanic LBVs available in the  archives, adding for a few sources new photometry that either was missing or not accurate because of extended objects or crowded fields. 
We have revised the Magellanic LBV sample on the basis of a new infrared SED-based classification scheme and literature information. The sources can be divided in four classes. All strongly active LBVs show
either a cool dust excess peaking in the mid-infrared (Class 1a and 1b) or an excess of near-infrared emission due to free-free emission from the ionised stellar wind (Class 2), or both (Class 1a). These features are also often present in several candidate LBVs, which appear in the literature as ex-/dormant LBVs. Four sources can be removed from our initial sample, as we find that in past studies these stars did not satisfy the LBV requirements and, in addition, they do not present any of the features mentioned above (Class 4). Six other sources are known as sgB[e] (Class 3): their flux density distribution does not resemble those of confirmed LBVs, consistent with a different mechanism of dust formation, as suggested in the literature. For this reason we excluded them from further analysis.

 We have employed a simple grey-body fitting method to model the infrared SED of individual sources.
In the LMC, which has the most numerous list of LBVs, large amounts of dust are observed  ($\sim\!10^{-3}$--$10^{-2}\,M_{\odot}$), similar to Galactic LBVNe. We stacked the infrared images of the LMC LBVs and extracted the photometry of the resulting source, detected up to $160$--$250\,\rm \mu m$. The integrated SED from the stacks resembles that of LBVs with a strong ionised stellar wind and an extended dusty nebula. The SED can be fitted with only two components: a power-law describing the free-free spectrum of ionised stellar winds and the stellar photosphere, and a single-component grey-body for the dust. For the grey-body we adopted two different values of the $\kappa$ parameter, including the value determined by \citet{2014Gordon} to fit the integrated ISM SED of the LMC. A significant contribution to the stack SED comes from a few sources, the most important one is RMC143. This was already identified as a massive nebula \citep{2019Agliozzo}. We obtain an integrated present dust mass of $0.11^{+0.06}_{-0.03}\,M_{\odot}$.  
We have repeated a similar analysis on the sample of AGBs and RSGs by \citet{2012Riebel}. We obtain a detection in the stacked images only when considering the ``extreme''-AGBs. We find that the integrated $160\,\mu{\rm m}$ emission of 1342 extreme-AGBs is of the same order of magnitude as that of 18 LBVs. The integrated dust mass from these sources is $1.2^{+0.3}_{-0.4}\times10^{-4}\,M_{\odot}$. 
We do not find any correlation between the dust masses and the stellar luminosities. This could be due to the fact that such stars have different evolutionary histories or that the dust production mechanism does not depend on the initial mass of the star. Most likely we are also unable to detect the lowest-mass nebulae.  
To estimate the total dust mass produced by LBVs in the LMC during its full lifetime, we consider two cases: constant number of LBVs across time and a case accounting for IMF and SFH. The uncertainty on the duration of LBV phase in the first case, or on the population of LBVs in the second case, add a significant uncertainty in the total mass produced by LBVs. 

The range of total dust masses produced by LBVs over the lifetime of the LMC spans between $10^{4}$ and $10^{5}\,M_{\odot}$, suggesting that LBVs are potentially significant dust factories in the LMC. The LBV dust masses per source of $10^{-3}-10^{-2}\,M_{\odot}$ are at least one order of magnitude smaller than the amount of dust ($0.1-1\,M_{\odot}$) postulated in SN ejecta by \citet{2001TF} and \citet{2003Nozawa}. However, those works adopted ideal assumptions for the seed material condensing into dust grains. 
Mid-infrared observations of CC-SNe suggest smaller dust masses \citep[$10^{-4}-10^{-2}\,M_{\odot}$,][although estimates from single epoch observations of the rapidly evolving SN ejecta may not be representative of the total dust produced or mid-infrared observations may have not captured the emission from colder dust]{2009Kotak,2011Fox}. On the other hand ALMA sub-millimetre observations revealed larger amounts of dust in both SN1987A \citep[$0.2\,M_{\odot}$, ][]{2014Indebetouw}
and LBV RMC143 \citep[$0.055\,M_{\odot}$, ][]{2019Agliozzo}, however these works assume different opacities at $850\,\rm \mu m$, making the comparison difficult. Still the LBV phenomenon has the potential to be the second most important source of dust, before low- and intermediate-mass stars. This was also the conclusion by \citet{2012Boyer} in the SMC. 
It should be noted that if, on the one hand, the LBV dust mass has a large uncertainty after analysing a relatively large sample, in the case of CC-SNe the dust mass is also very uncertain and based on observations of only a handful of objects. Destruction of dust in LBVNe by a possible subsequent SN blast wave remains to be determined. 
A significant fraction of this dust mass may survive thanks to several properties of these nebulae, such as large nebular masses, big grains, clumpiness and asymmetry. Dust grain (re)formation or growth may also occur, like in the case of SN1987A \citep{2019Matsuura}.

%
%
\begin{acknowledgements}  

D.A. acknowledges funding through the European Union’s Horizon 2020 and Innovation programme 
under the Marie Sk{\l}odowska-Curie grant agreement no. 793499 (DUSTDEVILS).
F.K.~is supported by the Ministry of Science and Technology (MoST) of Taiwan, under grant number MOST107-2119-M-001-031-MY3, and also by Academia Sinica, in the form of Investigator Award AS-IA-106-MO3. 
GP is supported by ANID – Millennium Science Initiative – ICN12$\_$009.  
 This paper makes use of the following ALMA data:
ADS/JAO.ALMA\#2017.A.00054.S and ADS/JAO.ALMA\#2018.A.00049.S. ALMA is a partnership of ESO
(representing its member states), NSF (USA) and NINS (Japan), together
with NRC (Canada) and NSC and ASIAA (Taiwan) and KASI (Republic of
Korea), in cooperation with the Republic of Chile. The Joint ALMA
Observatory is operated by ESO, AUI/NRAO and NAOJ. This paper also
includes data collected: at the European Organisation for Astronomical
Research in the Southern Hemisphere under ESO programmes 096.D-0047(A), 097.D-0612(A,B), and 0100.D-0469(A,B). 
This work made use of PyAstronomy. This research has made use of the International Variable Star Index (VSX) database, operated at AAVSO, Cambridge, Massachusetts, USA, and the VizieR catalogue access tool, CDS, Strasbourg, France. 
The original description of the VizieR service was
published in A\&AS 143, 23. This research has made use of the NASA/IPAC Infrared Science Archive, which is funded by the National Aeronautics and Space Administration and operated by the California Institute of Technology. This research made use of Photutils, an Astropy package for
detection and photometry of astronomical sources (Bradley et al.
2021). The Cornell Atlas of Spitzer/IRS Sources (CASSIS) is a product of the Infrared Science Center at Cornell University, supported by NASA and JPL. 
\end{acknowledgements}



\bibliographystyle{aa} 
\bibliography{mybib} 
%
%

%

\begin{appendix} 

\section{Table of stellar and nebular parameters}
\label{appendix:stellarparam}
\begin{sidewaystable*}
\caption{Stellar and nebular parameters.}
\label{tab:stellparam}
\centering
\begin{tabular}{rlcccccccl}
\hline\hline
ID & Name 1  &  T$_{\rm eff}$	& $\sigma_{T_{\rm eff}}$ &  $\log_{\frac{L}{L_{\odot}}}$ &    $\sigma_{\log_{\frac{L}{L_{\odot}}}}$ & E(B-V) & $\sigma_{E(B-V)}$& Nebula diameter & References \\
&	      & [K]	& [K]	&	&	&	&   &  [pc] & \\
\hline
 1&RMC$\,$143	  & 8500  & 300   & 5.32 &  0.05 &  0.42  & 0.02  & 2.2$\times$0.7  & 1\\
 2&RMC$\,$127	  & 28840 & 3000  & 6	 &  ---   &  0.12  & ---  & 1.6$\times$1.4  & 2, 29, 30\\
 3&S$\,$Doradus	  & 34670 & --- & 6.3  &  ---   &  0.2   & ---    & ---  & 2\\
 4&RMC$\,$110	  & 9120  & ---    & 5.36 &  ---   &  0.15  & --- & ---  & 2\\
 5&HD$\,$269582	  & 23700 & 3700  & 5.68 &  0.16 &  0.24  & ---   & ---  & 2, 3, 4 \\
 6&RMC$\,$116	  & 19500 & 500   & 5.92 &  ---   &  0.15  & ---  & ---  & 2\\
 7&HD$\,$269216	  & 12000 & ---    & 5.44 &  ---   &  0.12  & --- & ---  & 2\\
 8&RMC$\,$71	  & 15500 & 500   & 5.78 &  0.02 &  0.18  & ---   & <0.4  & 5, this work\\
 9&RMC$\,$85	  & 12300 & 1600  & 5.67 &  0.1  &  0.21  & 0.04  & ---  & 2, 6 \\
10&RMC$\,$123	 & 34700 & ---	 & 6.5  &  0.1  &  0.32  & 0.08   & ---  & 2, 7 \\
11&RMC$\,$99	 & 32870 & 3830  & 6.26 &  0.47 &  0.39  & 0.06   & ---  & 2, 3, 6, 8, 9 \\
12&Sk$\,-69\,279$	 & 27100 & 3100  & 5.6  &  0.07 &  0.25  & --- &  4.36    & 2, 10, 31	\\
13&LHA$\,$120-S$\,$119 & 27300 & 800   & 5.78 &  0.02 &  0.21  & 0.12  &  1.7$\times$1.6    & 2, 3, 6, 32 \\
14&LHA$\,$120-S$\,$61  & 27860 & 320   & 5.8  &  0.04 &  0.255 & 0.075 &  1.2    & 2, 3, 30, 33 \\
15&RMC$\,$74	 & 12880 & ---	 & 5.4  &  ---	&  0.15  & ---    & ---  & 11  \\
16&RMC$\,$78	 & 25125 & 600   & 5.78 &  0.07 &  0.15  & ---	  & ---  & 2, 12 \\
17&RMC$\,$81	 & 19900 & ---	 & 6	&  0.07 &  0.16  & 0.03   & ---  & 2, 13, 14, 15  \\
18&LHA$\,$120-S$\,$18  & 25100 & ---    & 5.58 &  ---   &  0.4    & --- &   	& 2\\
19&RMC$\,$66	 & 12000 & ---	 & 5.47 &  0.01 &  0.175 & 0.025  & ---  &	16, 17 \\
20&RMC$\,$84	 & 28800 & 3100  & 6.01 &  0.6  &  0.17  & 0.07   & ---  &	2, 3, 6, 8, 18, 26  \\
21&HD$\,$34664	 & 19800 & 4700  & 5.56 &  0.24 &  0.28  & ---	&  <0.2   &	17, 19, 20, this work  \\
22&HD$\,$38489	 & 29400 & 3700  & 6.13 &  0.23 &  0.35  & 0.15 & <0.2 &	2, 17, this work \\
23&RMC$\,$126	 & 22500 & ---  & 6.15 &  --- &  0.25  & --- &  ---      &	17, 27 \\
24&Sk$\,-69\,271$	 & 9204 & ---  & 4.67 &  --- &  ---  & --- & --- & 	28 \\
25&RMC$\,$128	 & 17300 & ---    & 5.76 &  ---  &  0.12  & ---  &  ---  &	2\\
26&RMC$\,$149	 & 19900 & ---    & 5.16 &  ---  &  0.1   & ---  & ---   & 2\\
27&HD2$\,$69604	 & 8100  & ---    & 5.2  &  ---  &  0.1   & ---  & ---   & 2\\
\hline
28&RMC$\,$14	 & 60300 & 2800  & 6.5  &  0.16 &  ---	 & ---	& ---    & 21, 22   \\
29&RMC$\,$40	 & 10610 & 610   & 5.65 &  0.01 &  0.14  & ---	&  ---   & 23, 24   \\
30&RMC$\,$4	 & 20300 & 7700  & 5.29 &  0.23 &  0.11  & 0.06 &  ---   & 2, 17\\
31&LHA$\,$115-S$\,$18  & 25100 & ---    & 5.3  &  ---   &  ---    & --- &  &	25 \\
\hline
\end{tabular}
\tablebib{(1) \citet{2019Agliozzo}; (2) \citet{vanGenderen2001}; (3) \citet{1997Crowther}; (4) \citet{1996vanGHD}; (5)  \citet{2017Mehner};
(6) \citet{1984Stahl}; (7)  \citet{1987S&W}; (8) \citet{2014Hainich}; (9) \citet{1997Pasquali}; (10) \citet{2018Gvaramadze}; (11) \citet{1999vanG}; 
(12) \citet{2004Evans}; (13) \citet{2002Tubb}; (14) \citet{1981Wolf}; (15) \citet{1987StahlR81}; (16) \citet{1983StahlR66}; (17) \citet{1986Zickgraf};
(18) \citet{1991Schmutz}; (19)  \citet{1996ZickgrafBestars}; (20)  \citet{1988Muratorio}; (21) \citet{1998Koenig}; (22) \citet{2014Koenig};
(23) \citet{1993Szeifert}; (24)   \citet{1960Feast}; (25) \citet{2013ClarkS18}; (26) \citet{1987Wolfvarie}; (27) \citet{2006Kastner}; (28) \citet{2012Neugent};
(29) \citet{2017AgliozzoB}; (30) \citet{2003Weis}; (31) \citet{2002Weis}; (32) \citet{2003Weis-s119}; (33) \citet{2012Agliozzo}.
}
\end{sidewaystable*}

We use stellar parameters collected from the literature. Only values corresponding to the quiescent state of the stars were taken into account. To compensate for the lack of error estimates, we take the mean value and the corresponding standard deviation when more than one estimate is available. In this case all the relevant works are mentioned. We also include information on the nebula size when known from the literature and from this work. 

\section{IR images and SEDs}
\label{appendix:imagesSEDs}

Flux density distribution plots of individual sources and selected space telescope images are provided here.
 The photometric data are corrected for interstellar dust extinction, using the extinction curve from \citet{1999PASP..111...63F}, E(B-V) as in Table \ref{tab:stellparam} and $R_{V} = 3.1$. For RMC~143 an $R_{V} = 4.0$ was adopted as previously determined by \citet{2019Agliozzo}. 
The images are useful for assessing the confusion and thus the reliability of the photometry in each band. The field-of-view in each image  is $240''\times240''$. The position of the source is highlighted with a cyan cross-hair, which is rotated accordingly with the sky projection. The colour-scale is squared-root in order to highlight the local environment around each source.

\begin{figure*}
    \centering
    \begin{minipage}{.35\textwidth}
        \centering
        \includegraphics[width=1\linewidth]{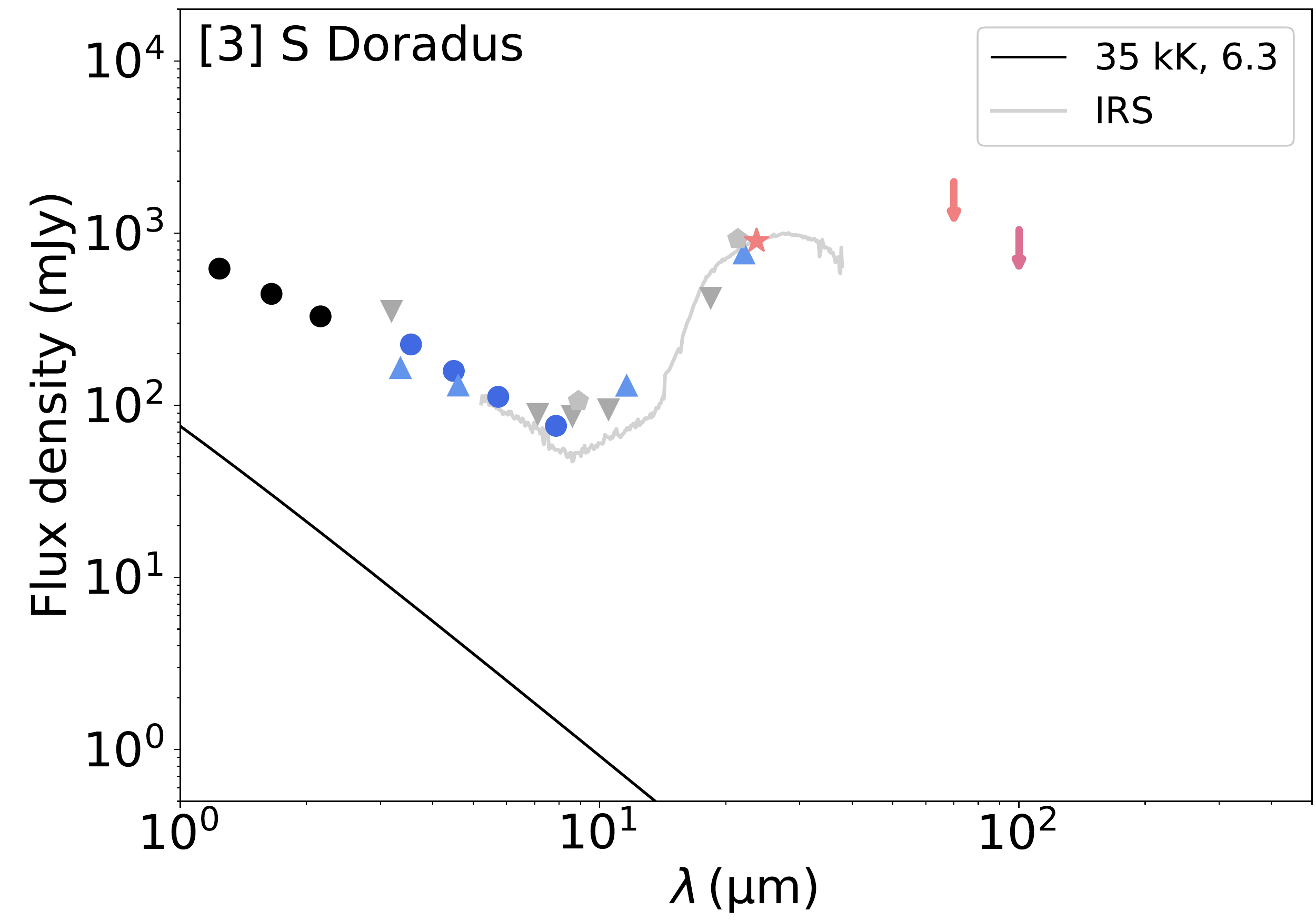}
    \end{minipage}%
    \begin{minipage}{0.65\textwidth}
        \centering
        \includegraphics[width=1\linewidth]{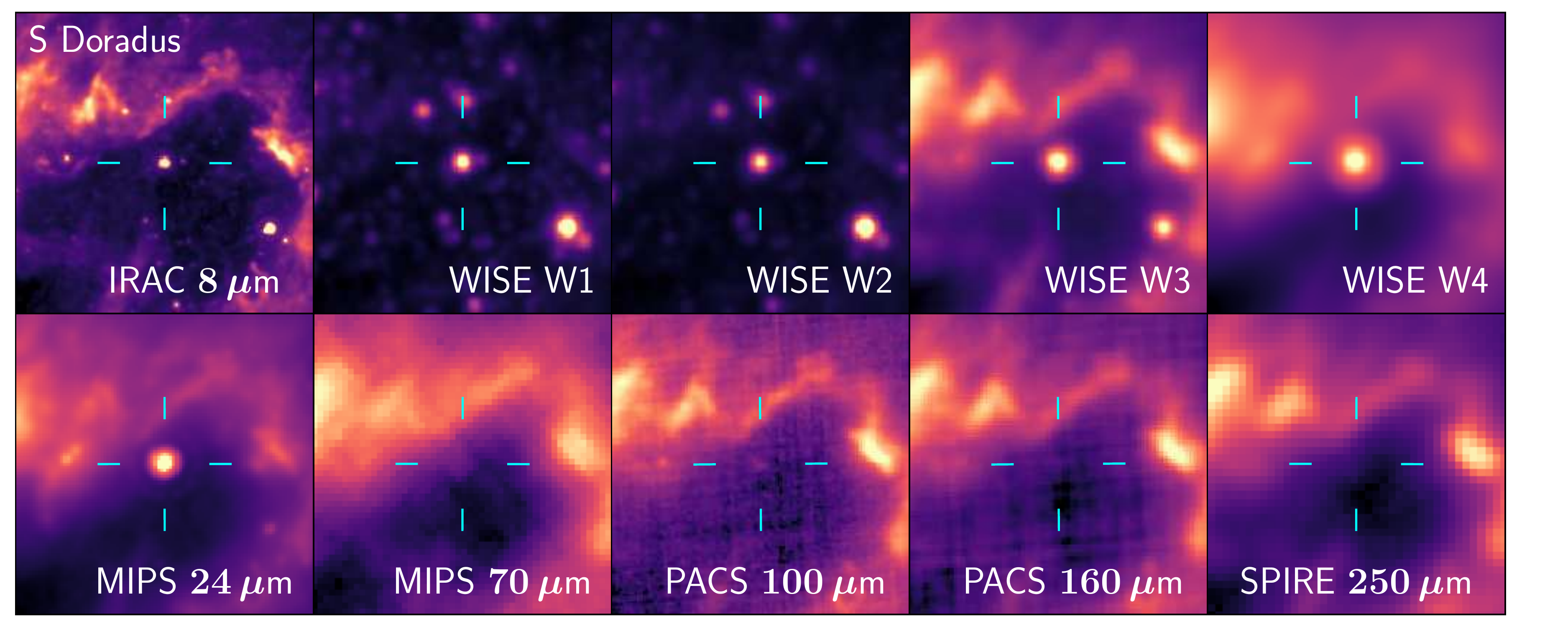}
    \end{minipage}       
    \begin{minipage}{.35\textwidth}
        \centering
        \includegraphics[width=1\linewidth]{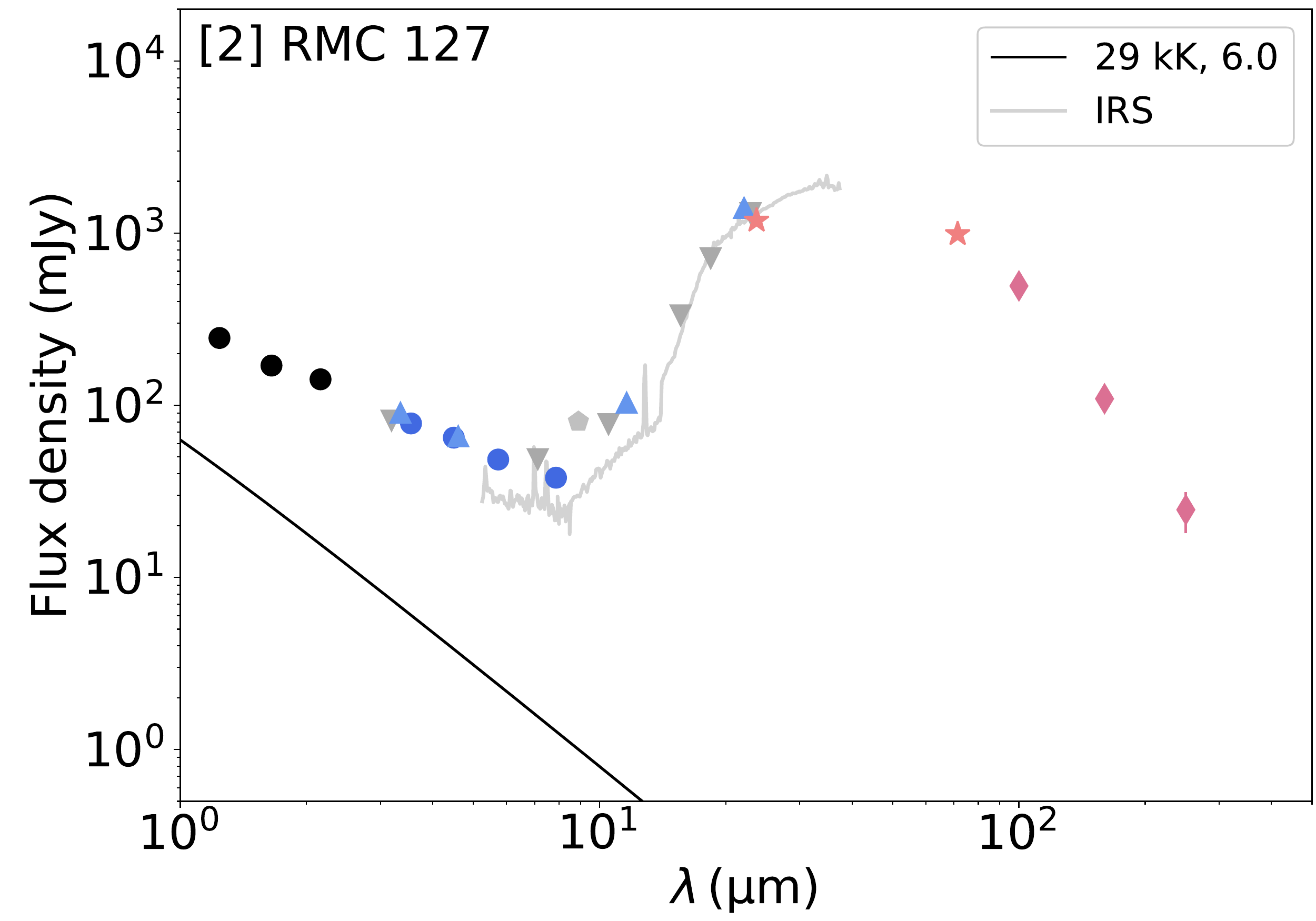}
    \end{minipage}%
    \begin{minipage}{0.65\textwidth}
        \centering
        \includegraphics[width=1\linewidth]{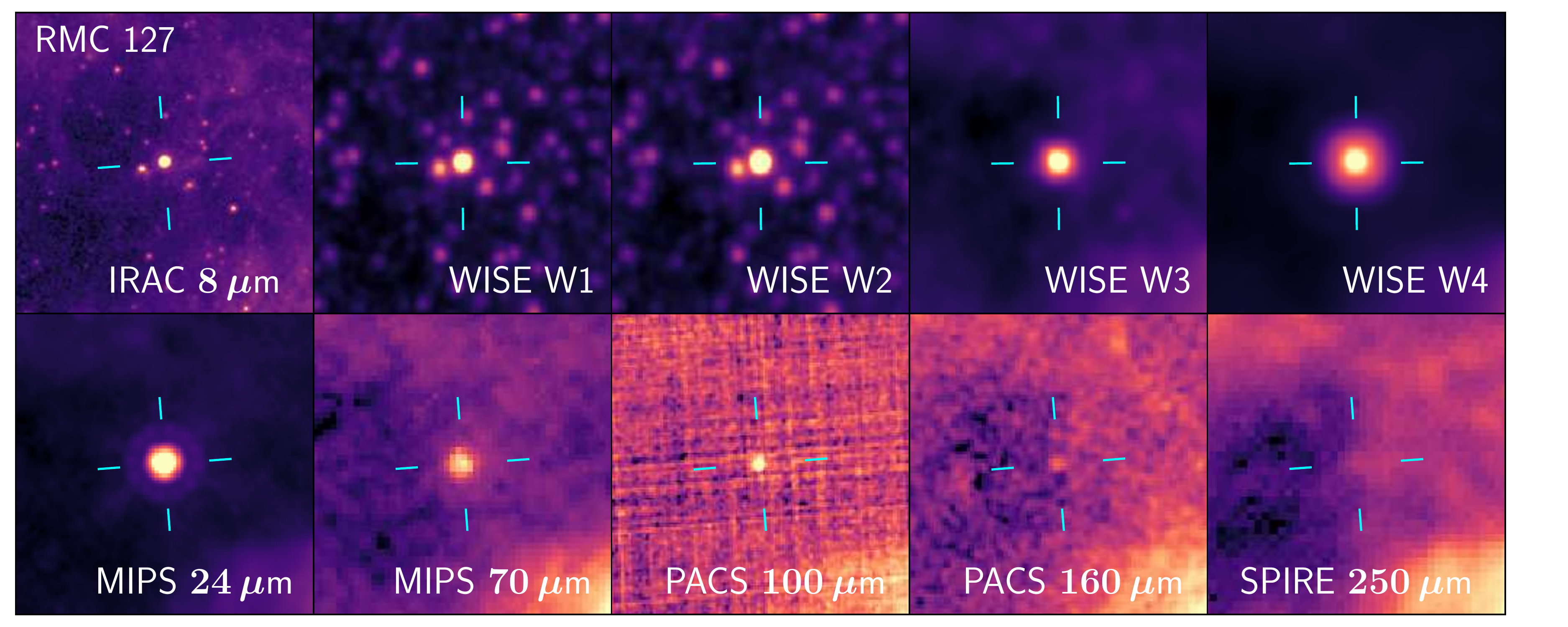}
    \end{minipage}
        \begin{minipage}{.35\textwidth}
        \centering
    \includegraphics[width=1\linewidth]{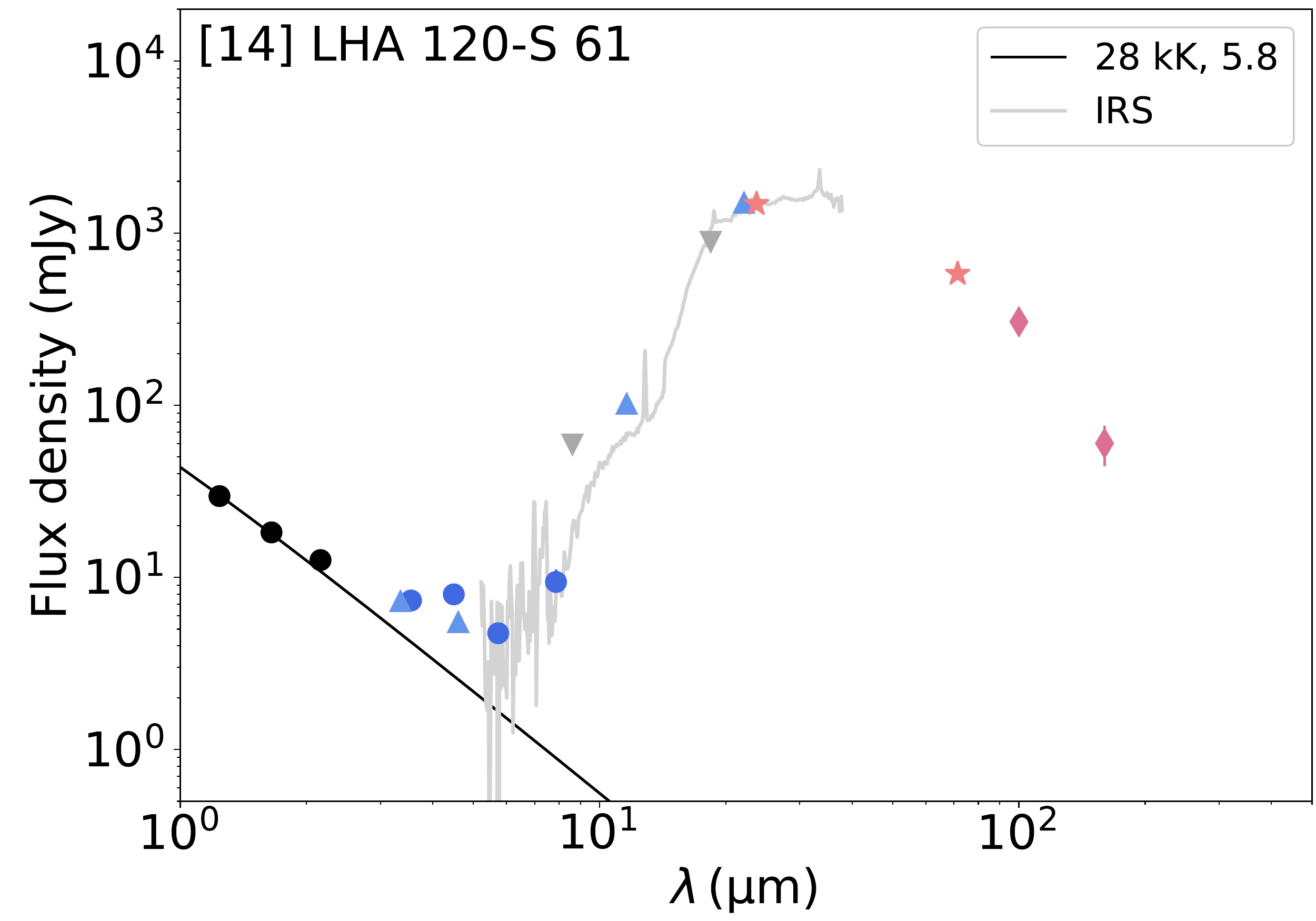}
    \end{minipage}%
    \begin{minipage}{0.65\textwidth}
        \centering
        \includegraphics[width=1\linewidth]{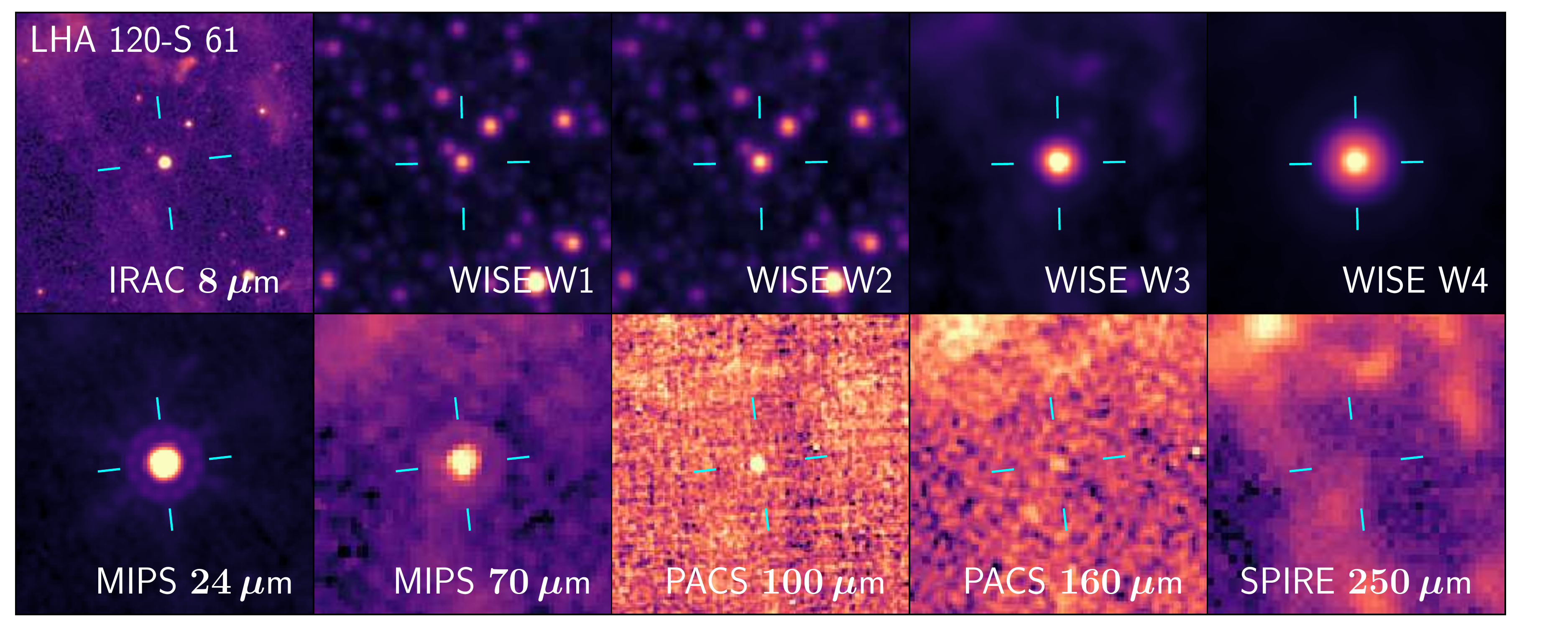}
    \end{minipage}   
    \begin{minipage}{.35\textwidth}
        \centering
        \includegraphics[width=1\linewidth]{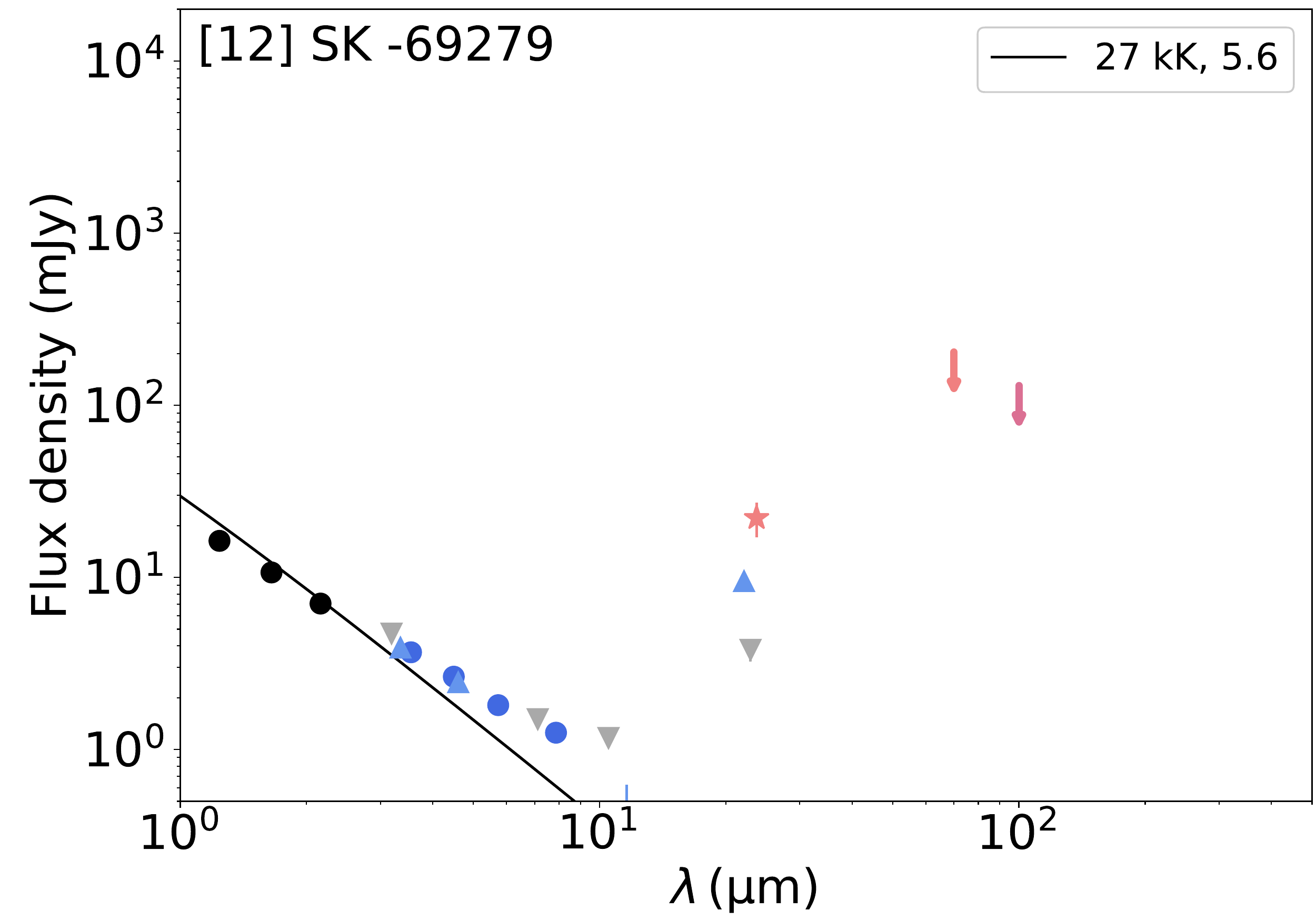}
    \end{minipage}%
    \begin{minipage}{0.65\textwidth}
        \centering
        \includegraphics[width=1\linewidth]{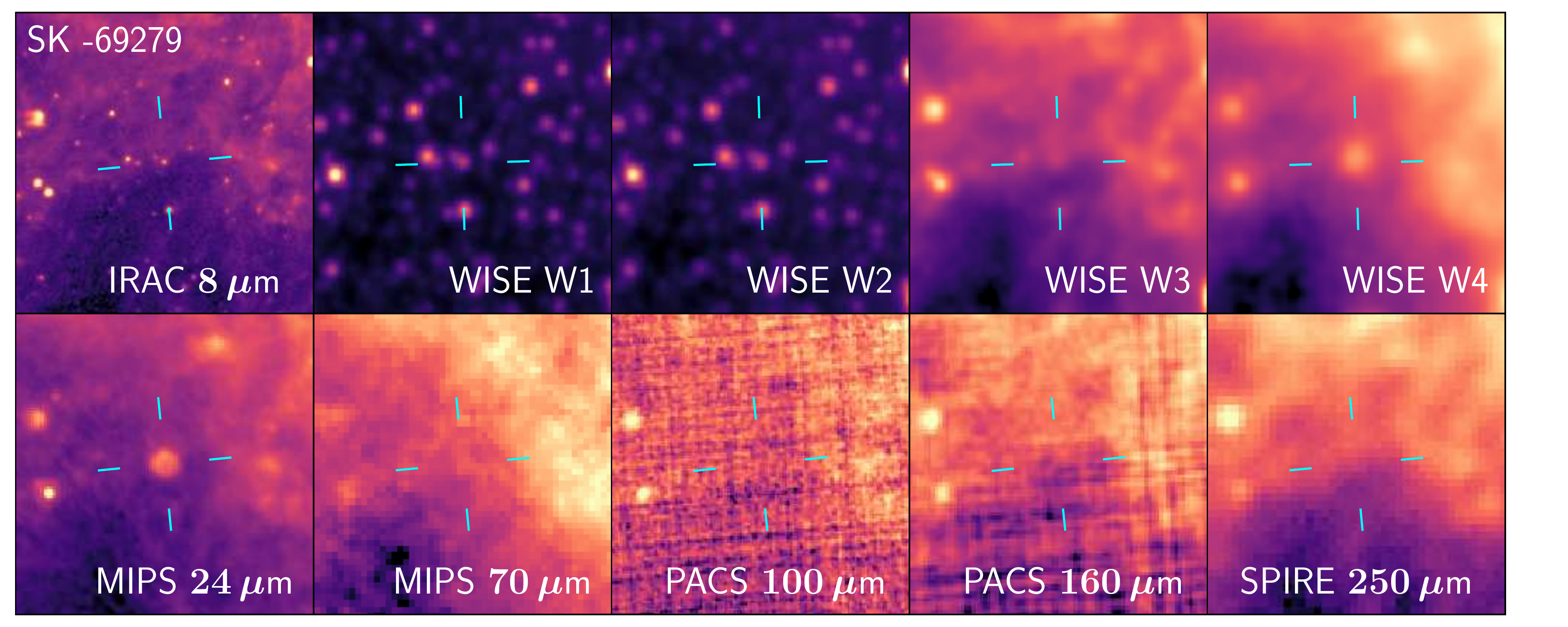}
    \end{minipage}
    \begin{minipage}{.35\textwidth}
        \centering
        \includegraphics[width=1\linewidth]{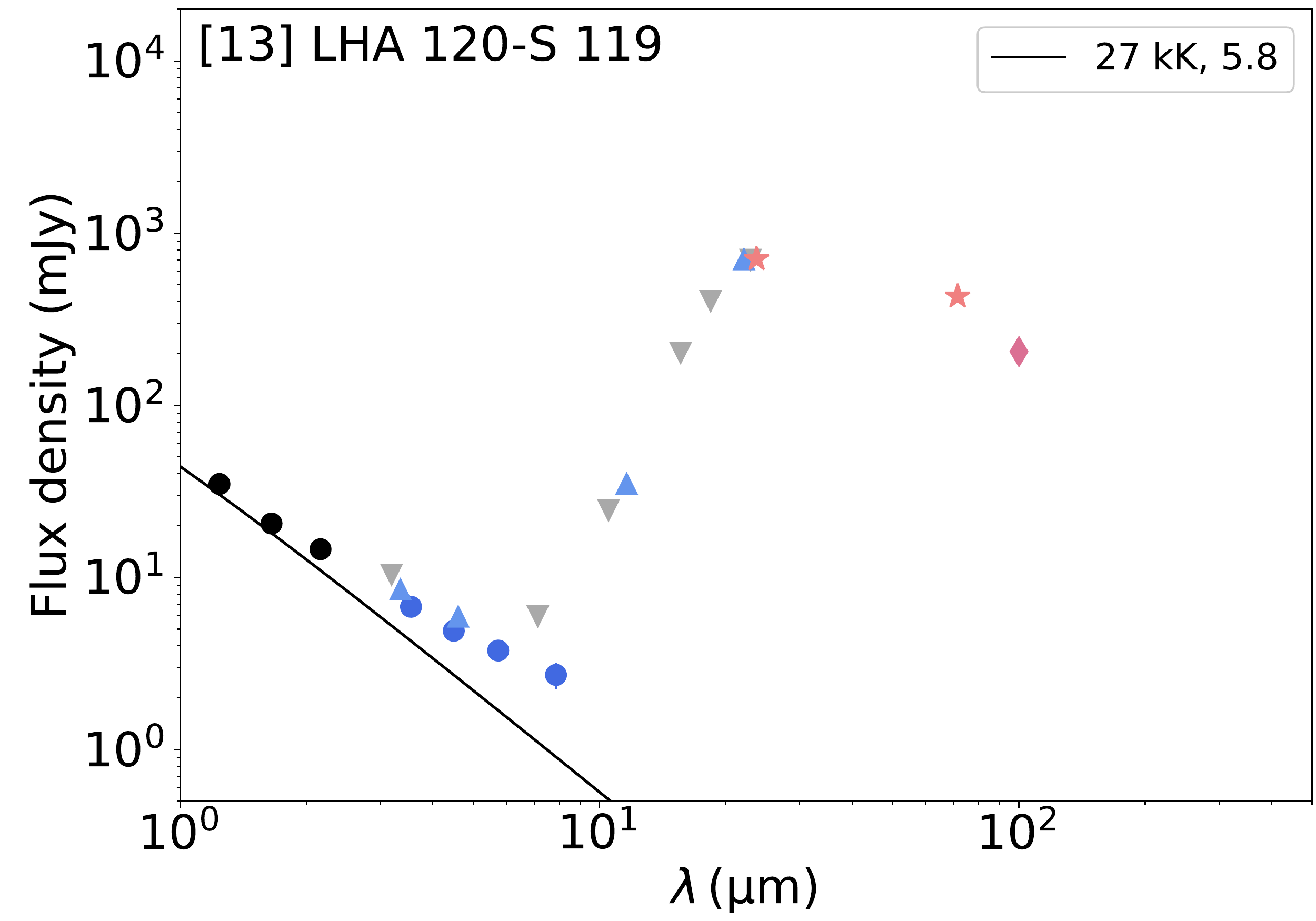}
    \end{minipage}%
    \begin{minipage}{0.65\textwidth}
        \centering
        \includegraphics[width=1\linewidth]{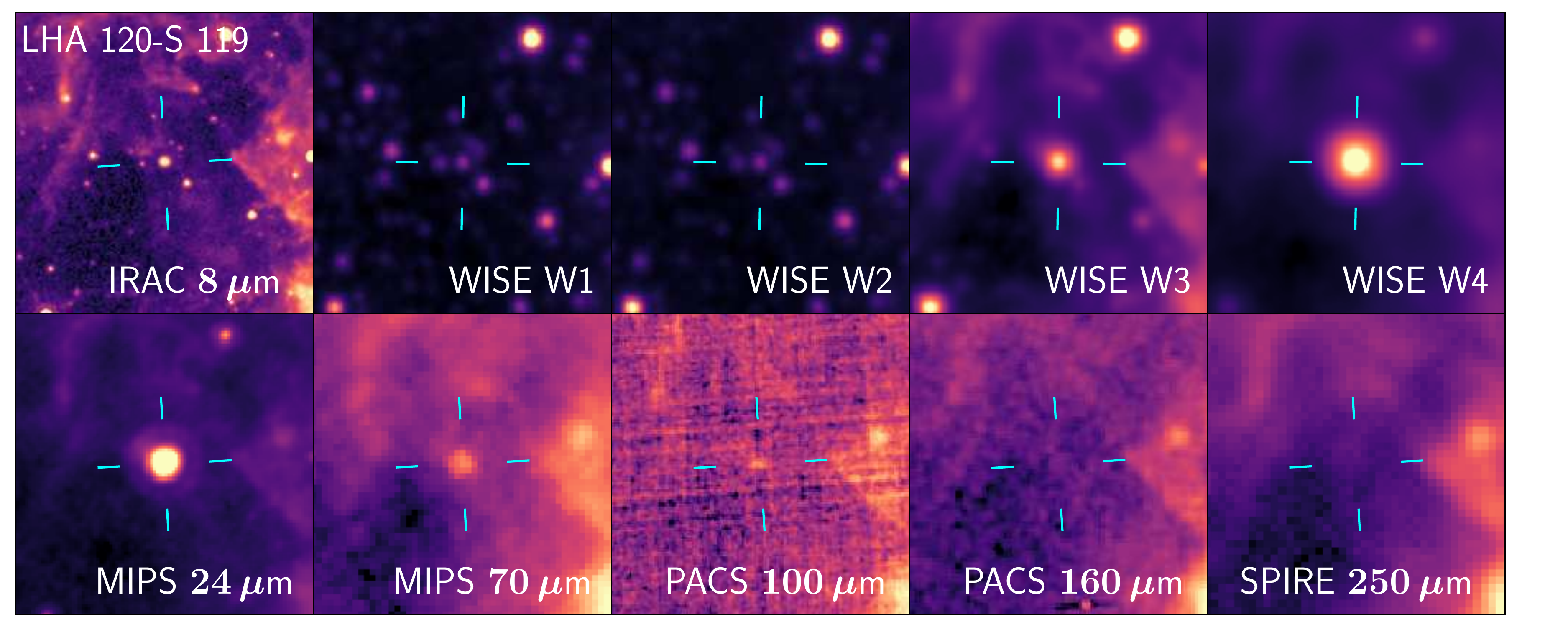}
    \end{minipage}
    
       \caption{Confirmed and candidate LBVs with a SED characterised by a free-free continuum from their stellar wind and by a thermal continuum due to dust in their circumstellar ejecta (see Class~1a sources in Sect.  \ref{sec:sedreview}). The legend for the photometric points can be found in Fig. \ref{fig:cmd-LMC} of Sect. \ref{sec:review}.}
     \label{fig:group1}
\end{figure*}
\begin{figure*}
    \centering
    \captionsetup{labelformat=empty}
    \begin{minipage}{.35\textwidth}
        \centering
        \includegraphics[width=1\linewidth]{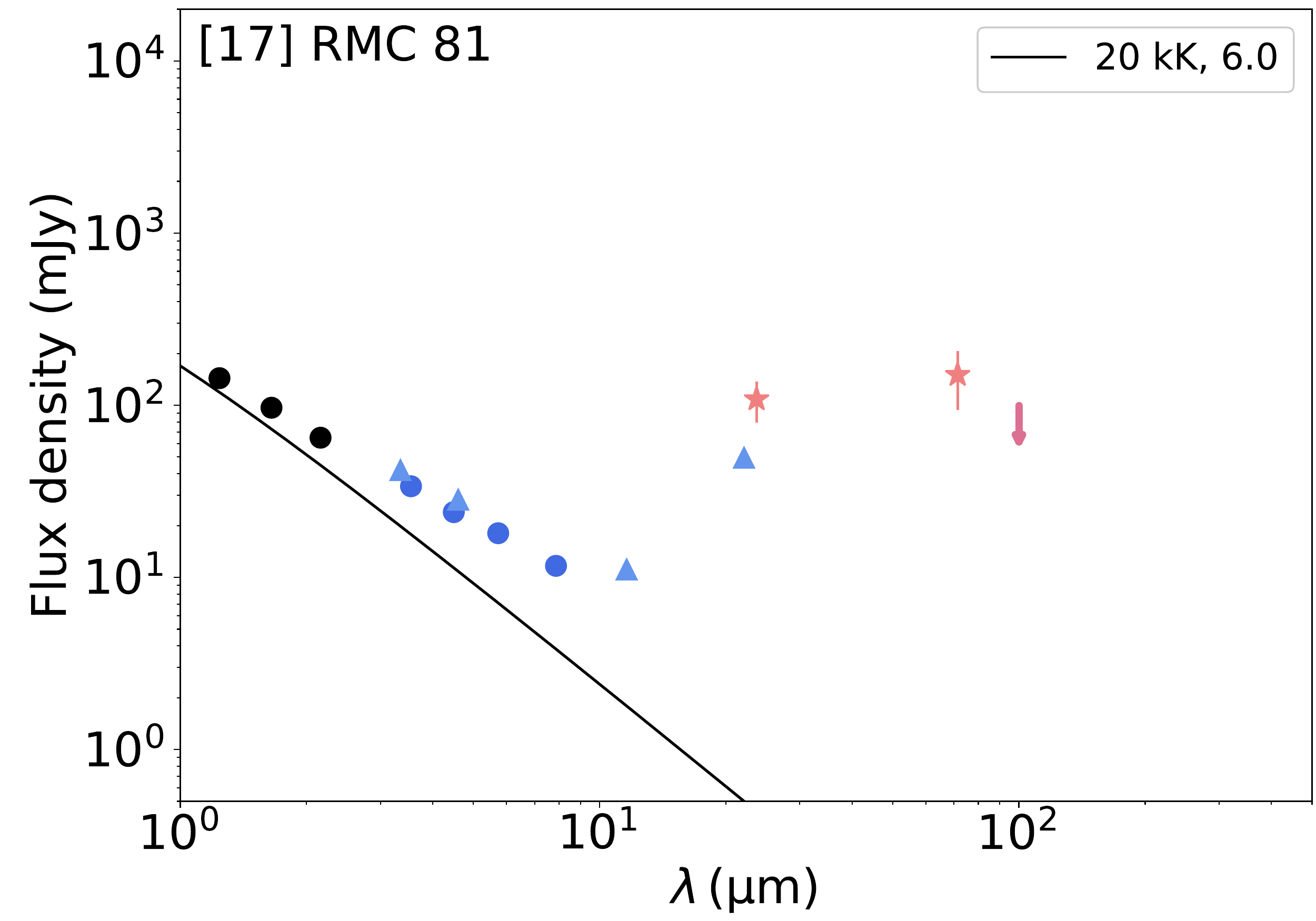}
    \end{minipage}%
    \begin{minipage}{0.65\textwidth}
        \centering
        \includegraphics[width=1\linewidth]{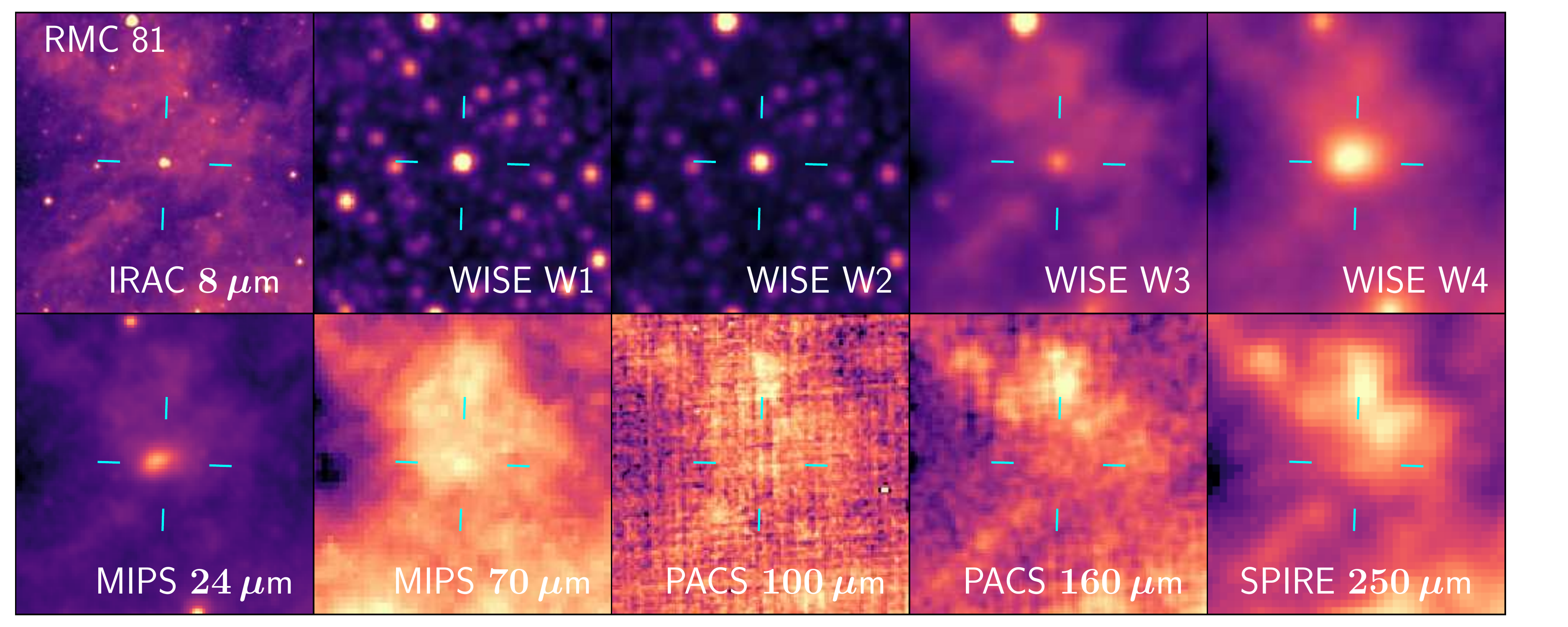}
    \end{minipage}  
    \begin{minipage}{.35\textwidth}
        \centering
        \includegraphics[width=1\linewidth]{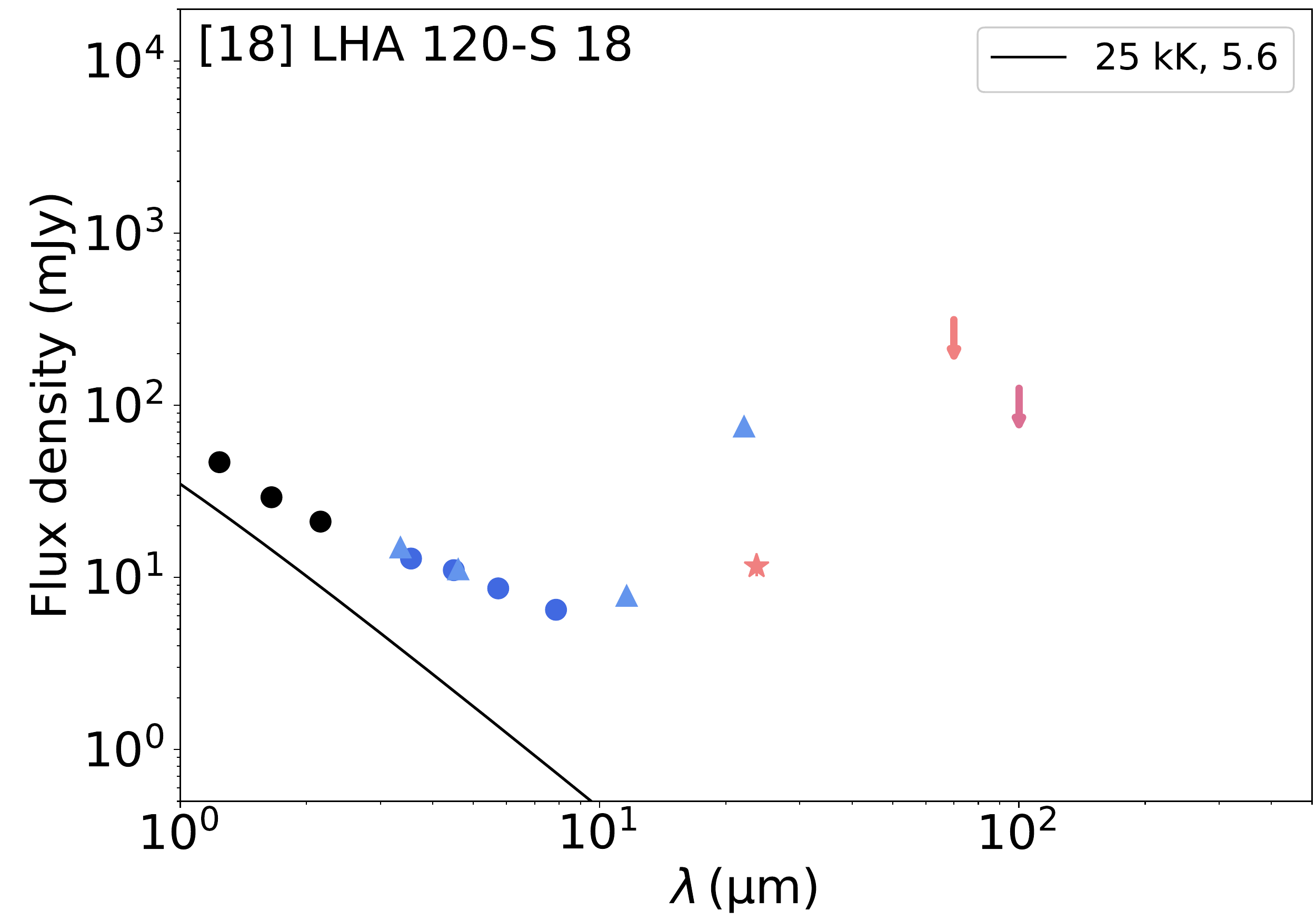}
    \end{minipage}%
    \begin{minipage}{0.65\textwidth}
        \centering
        \includegraphics[width=1\linewidth]{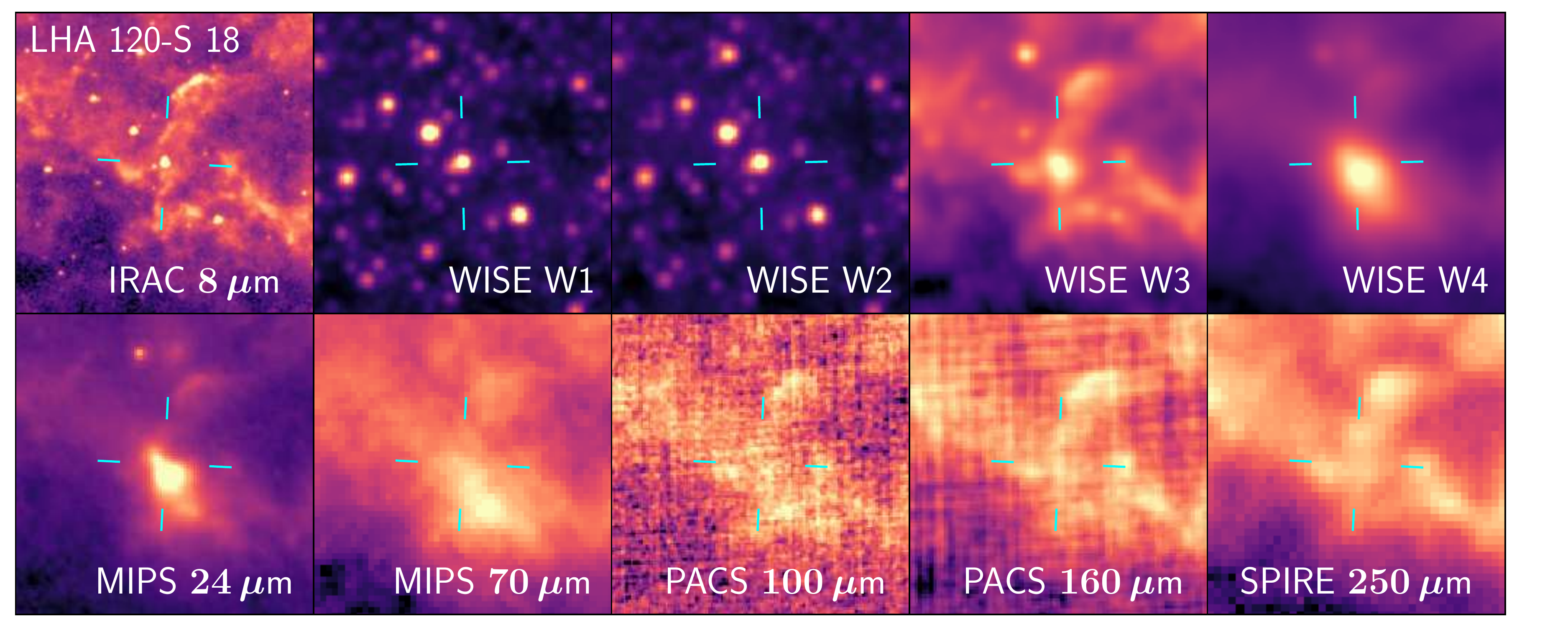}
    \end{minipage}    
\caption*{Fig.~\ref{fig:group1} continued.}

\end{figure*}

\begin{figure*}
    \centering
    \begin{minipage}{.35\textwidth}
        \centering
        \includegraphics[width=1\linewidth]{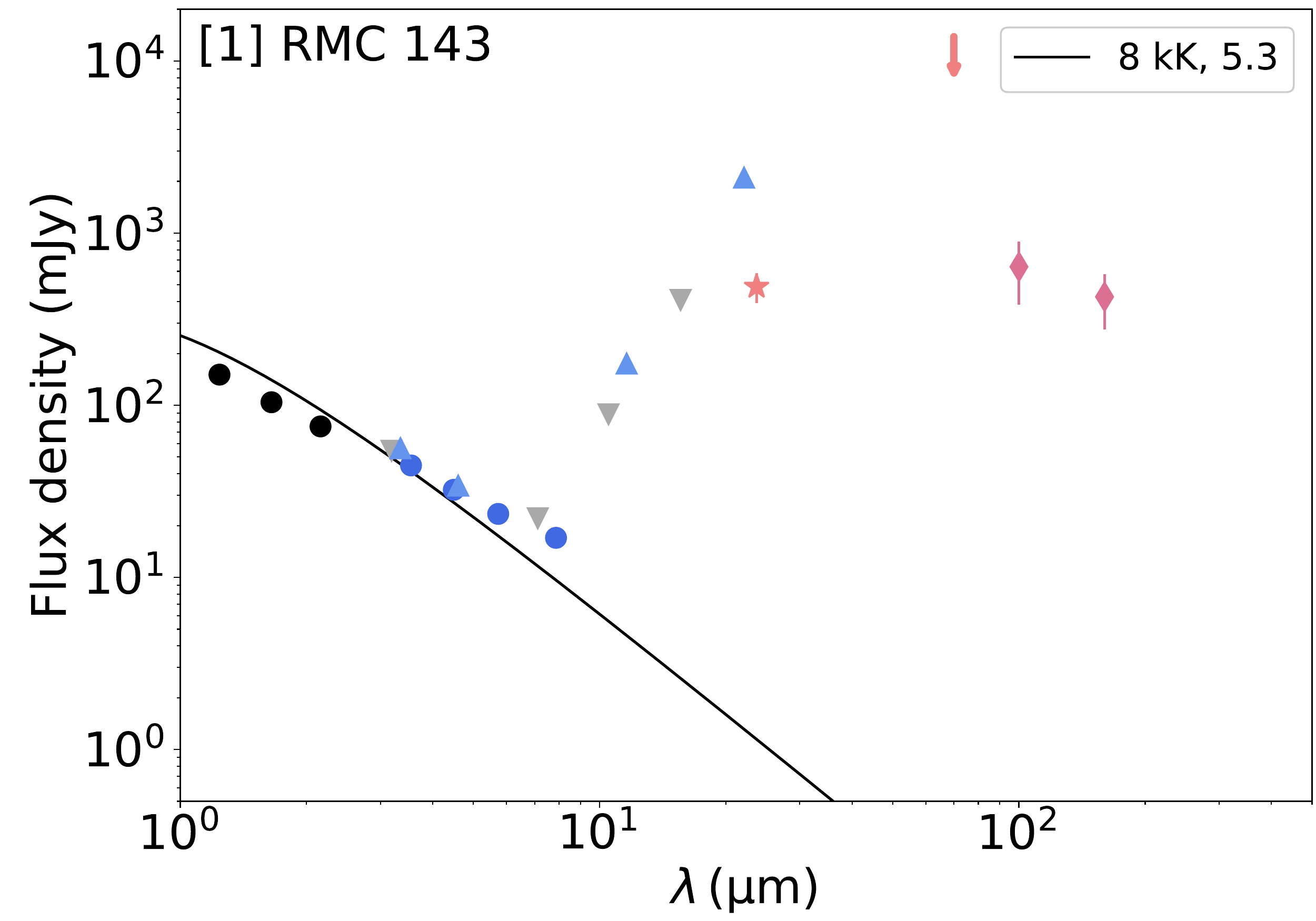}
    \end{minipage}%
    \begin{minipage}{0.65\textwidth}
        \centering
        \includegraphics[width=1\linewidth]{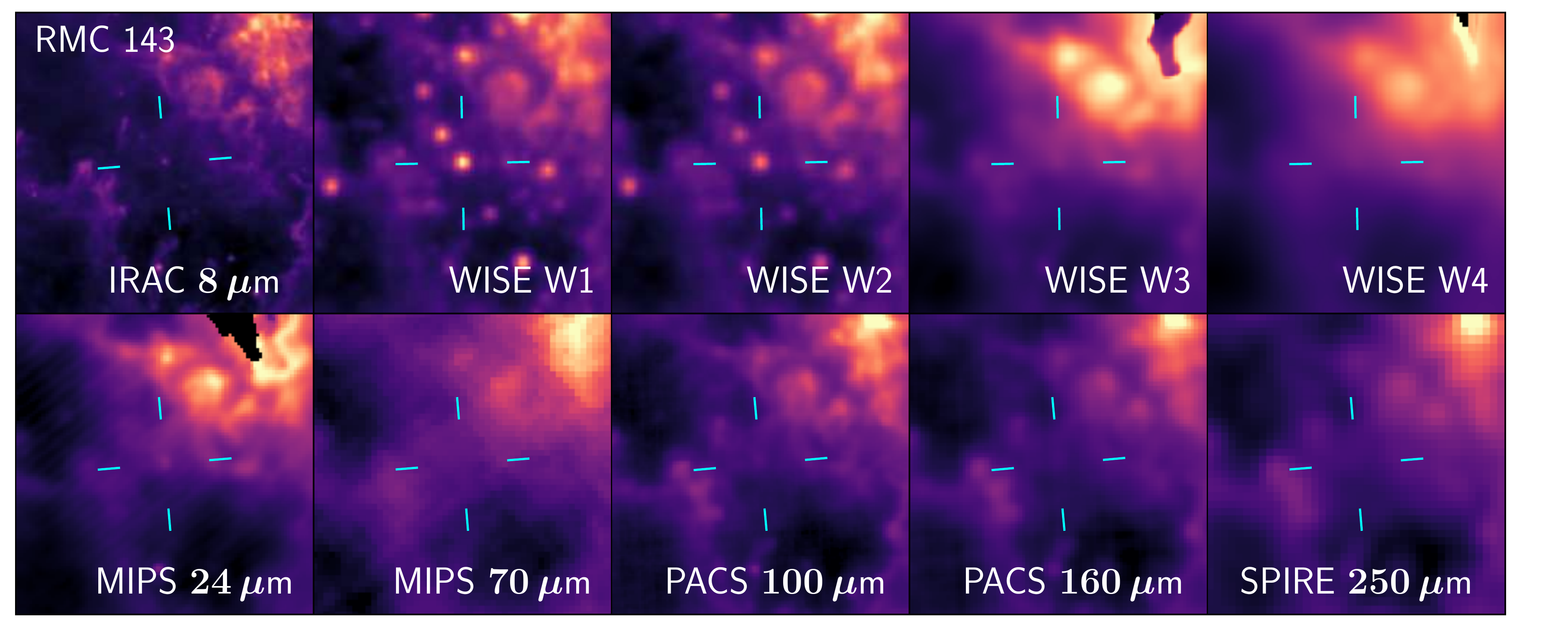}
    \end{minipage}
    \begin{minipage}{.35\textwidth}
        \centering
        \includegraphics[width=1\linewidth]{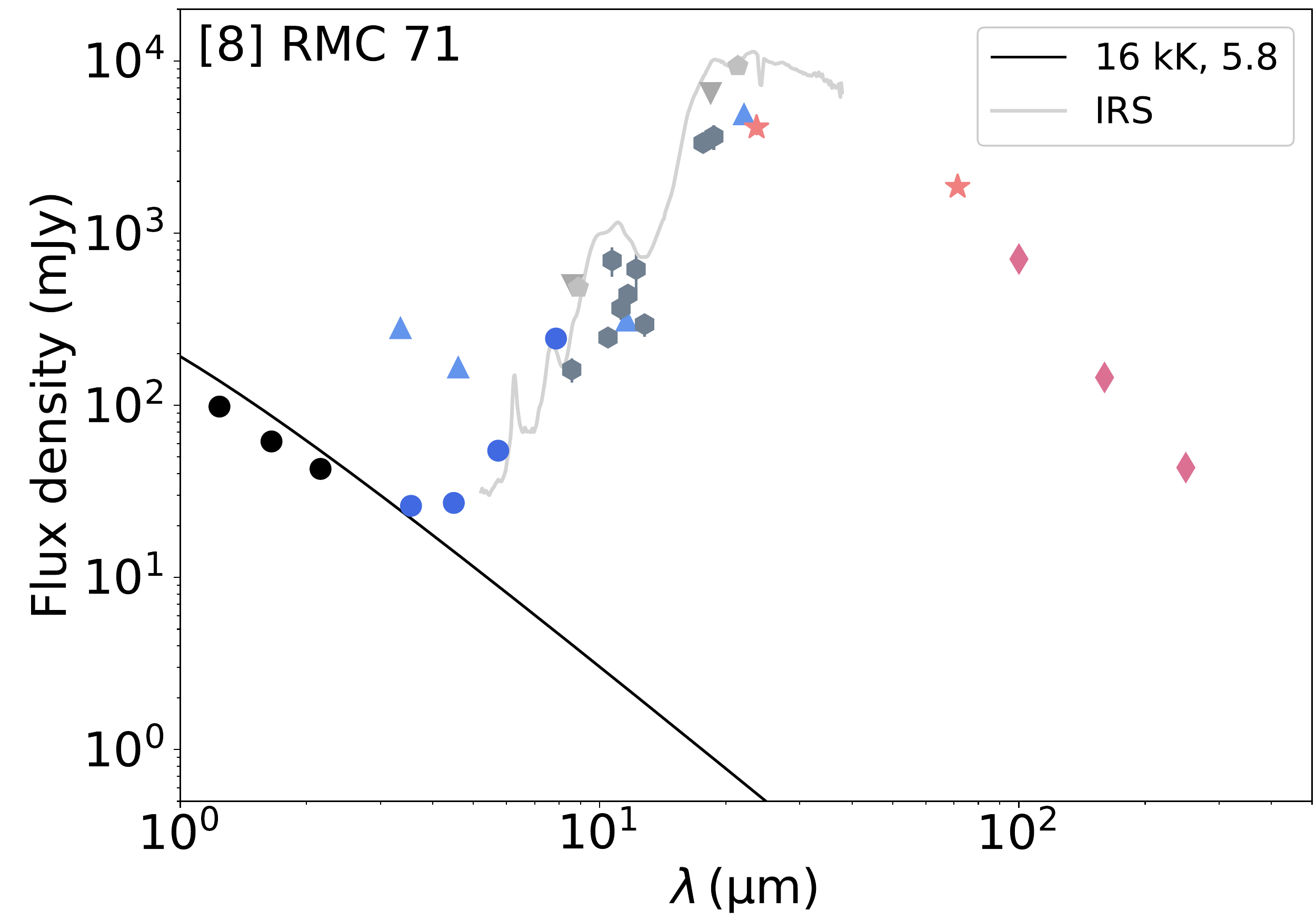}
    \end{minipage}%
    \begin{minipage}{0.65\textwidth}
        \centering
        \includegraphics[width=1\linewidth]{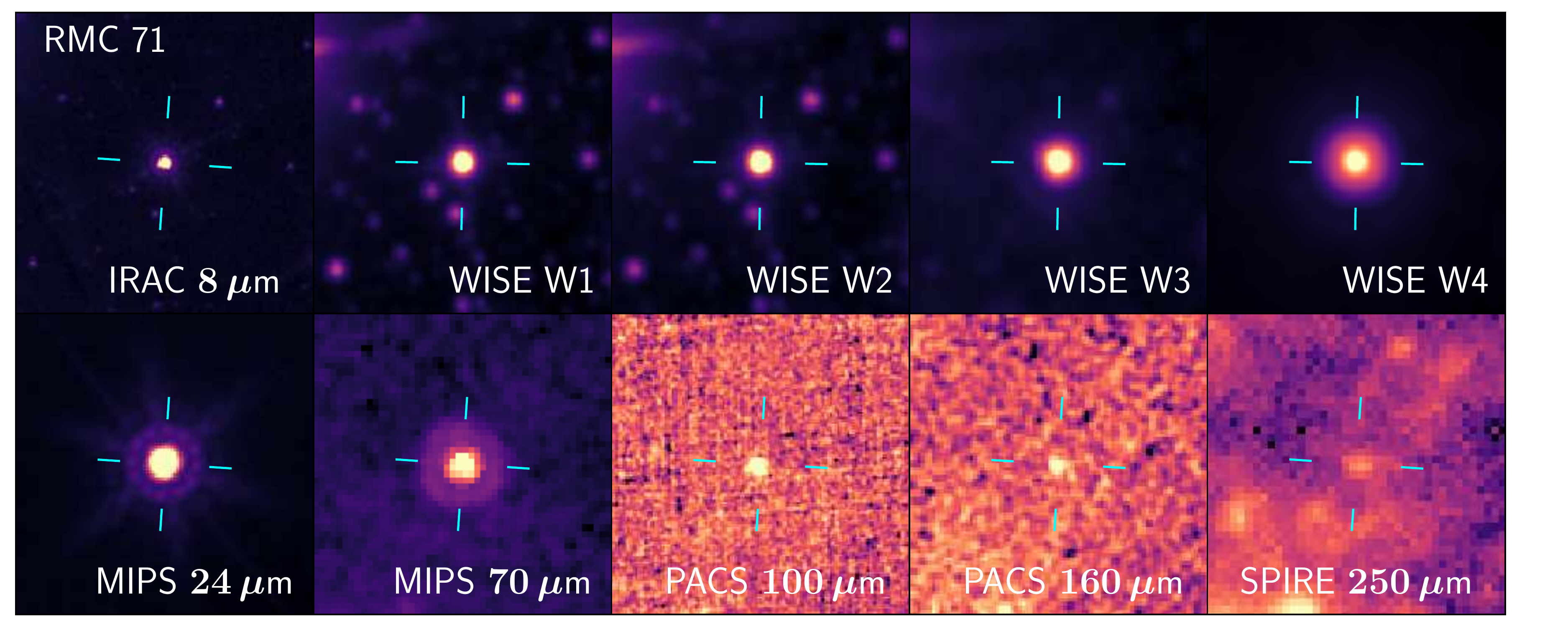}
    \end{minipage}    
    \begin{minipage}{.35\textwidth}
        \centering
        \includegraphics[width=1\linewidth]{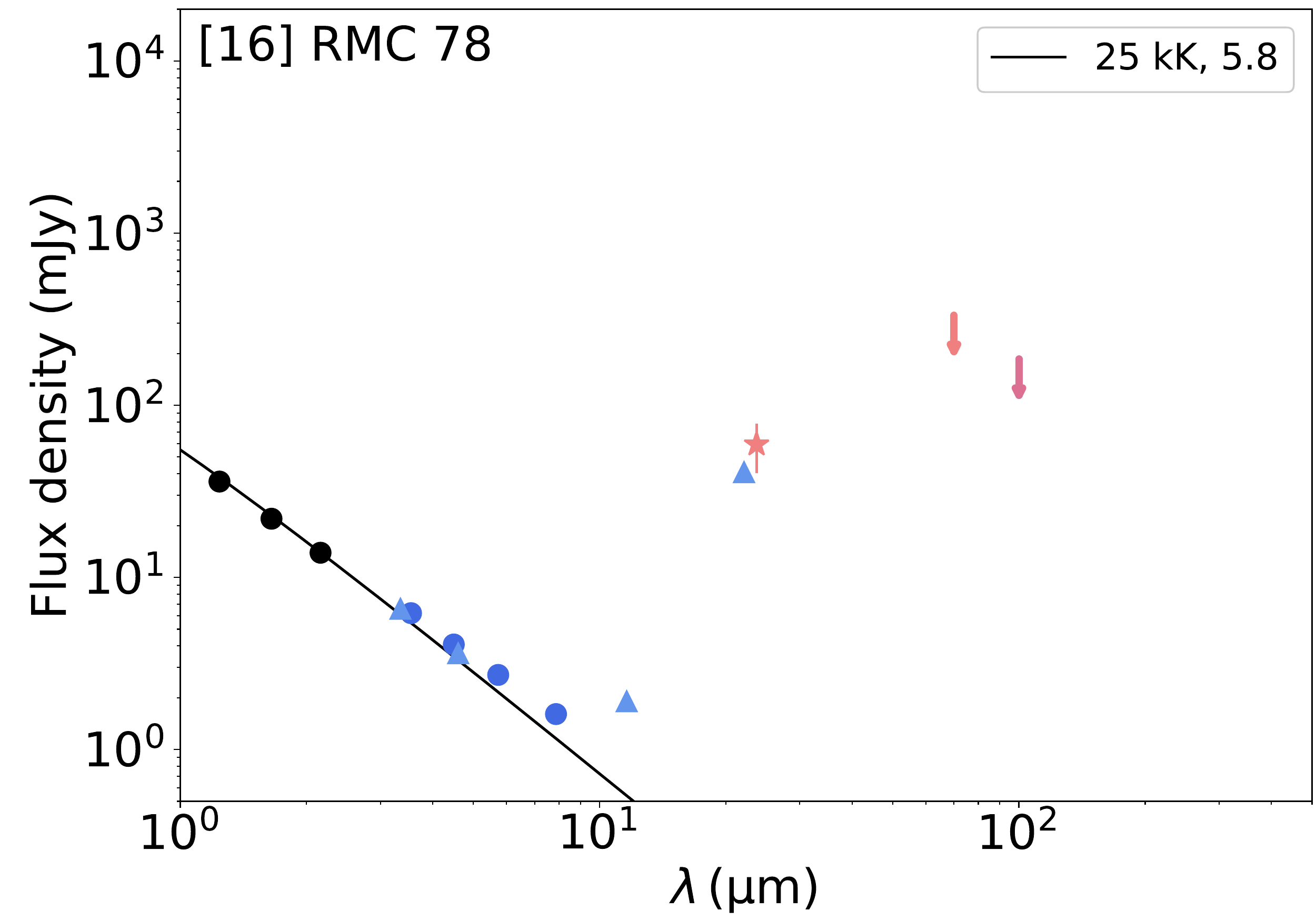}
    \end{minipage}%
    \begin{minipage}{0.65\textwidth}
        \centering
        \includegraphics[width=1\linewidth]{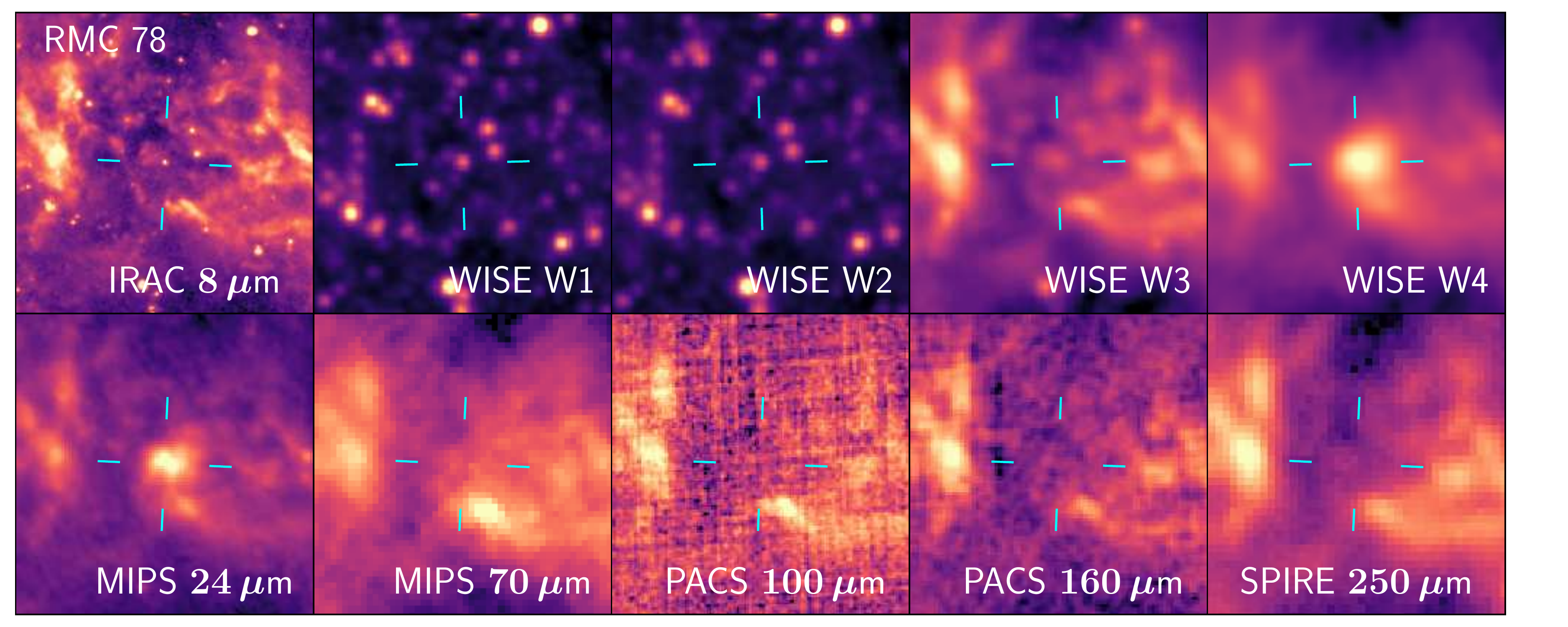}
    \end{minipage}   
    \begin{minipage}{.35\textwidth}
        \centering
        \includegraphics[width=1\linewidth]{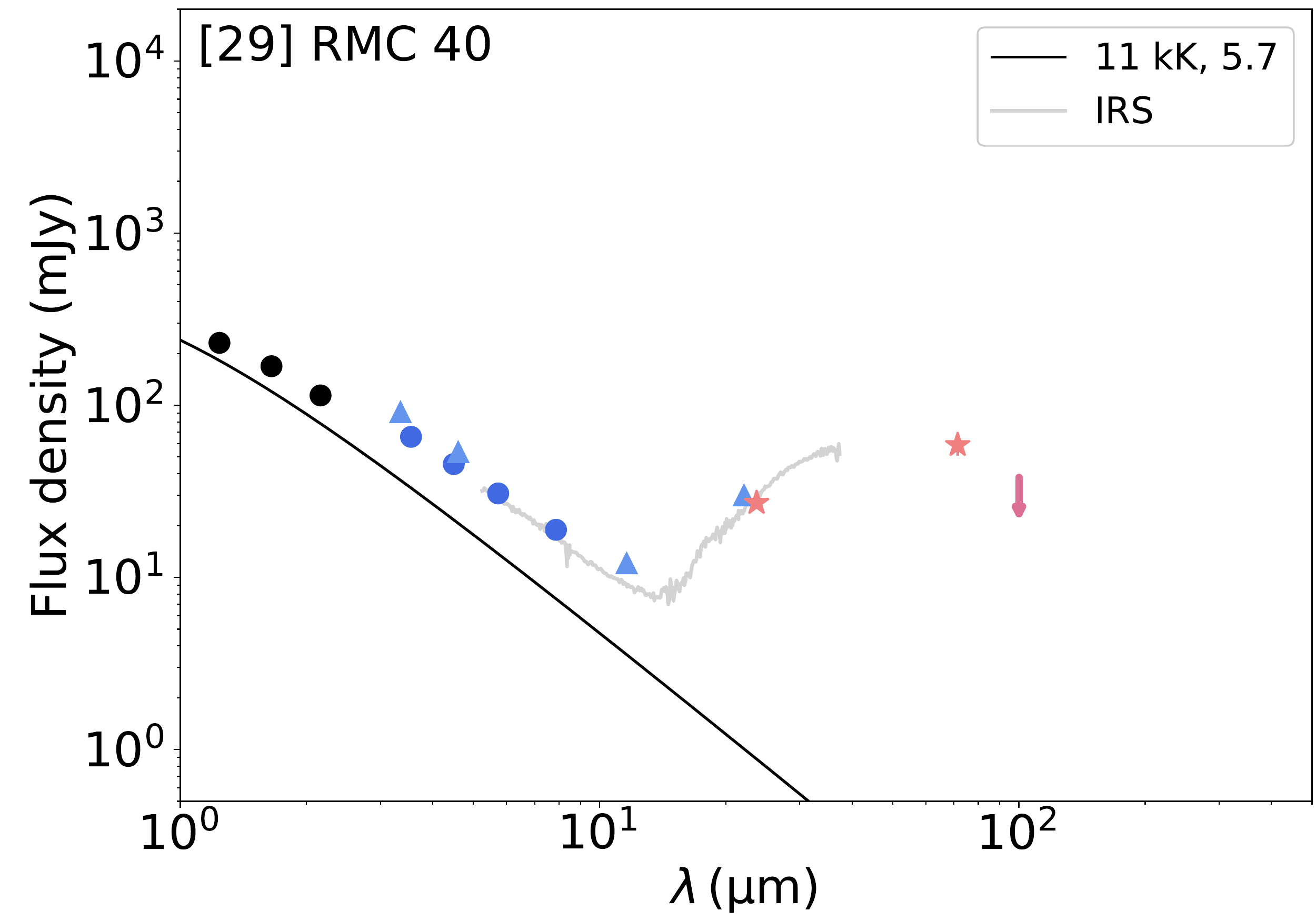}
    \end{minipage}%
    \begin{minipage}{0.65\textwidth}
        \centering
        \includegraphics[width=1\linewidth]{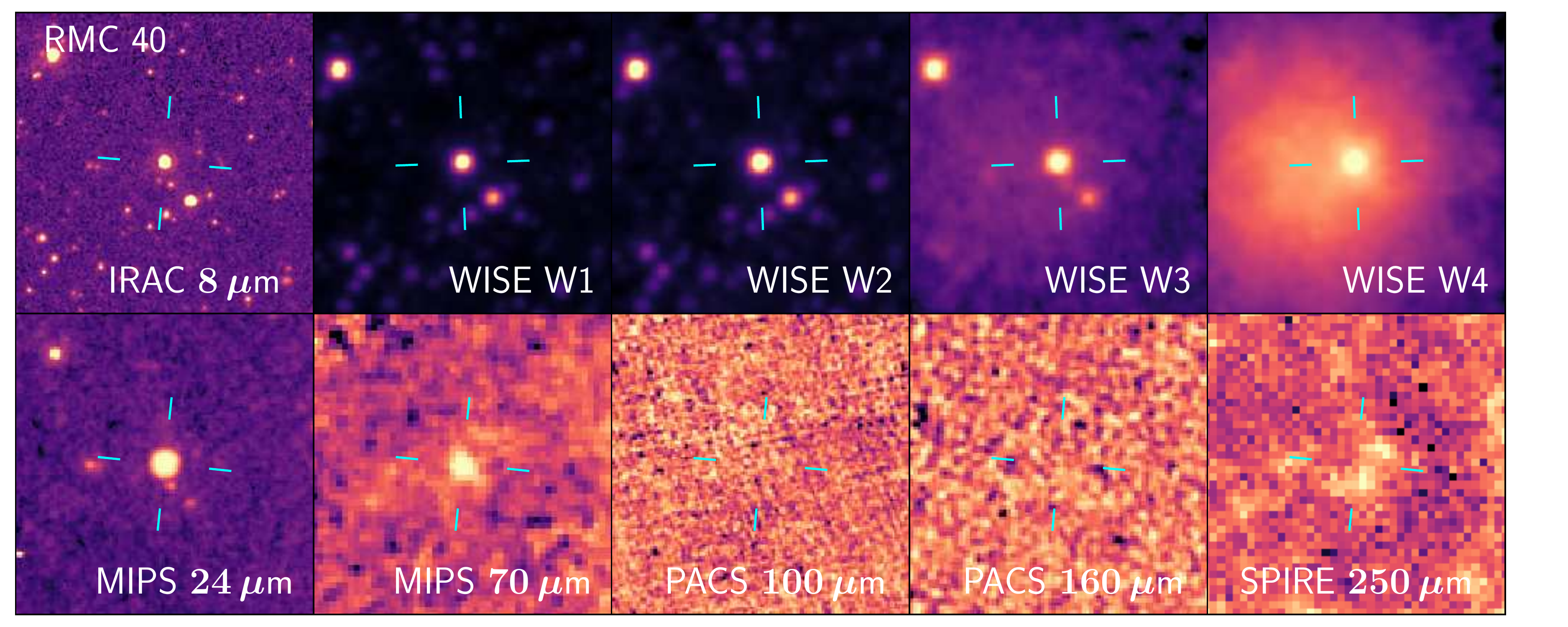}
    \end{minipage}

    \caption{Confirmed and candidate LBVs with a SED characterised by a thermal continuum due to dust in their circumstellar ejecta and relatively weak or absent  free-free continuum from their stellar wind (see Class~1b sources in Sect. \ref{sec:sedreview}). The legend for the photometric points can be found in Fig. \ref{fig:cmd-LMC} of Sect. \ref{sec:review}.}
    \label{fig:group1b}
\end{figure*}

\begin{figure*}
    \centering
    \begin{minipage}{.35\textwidth}
        \centering
        \includegraphics[width=1\linewidth]{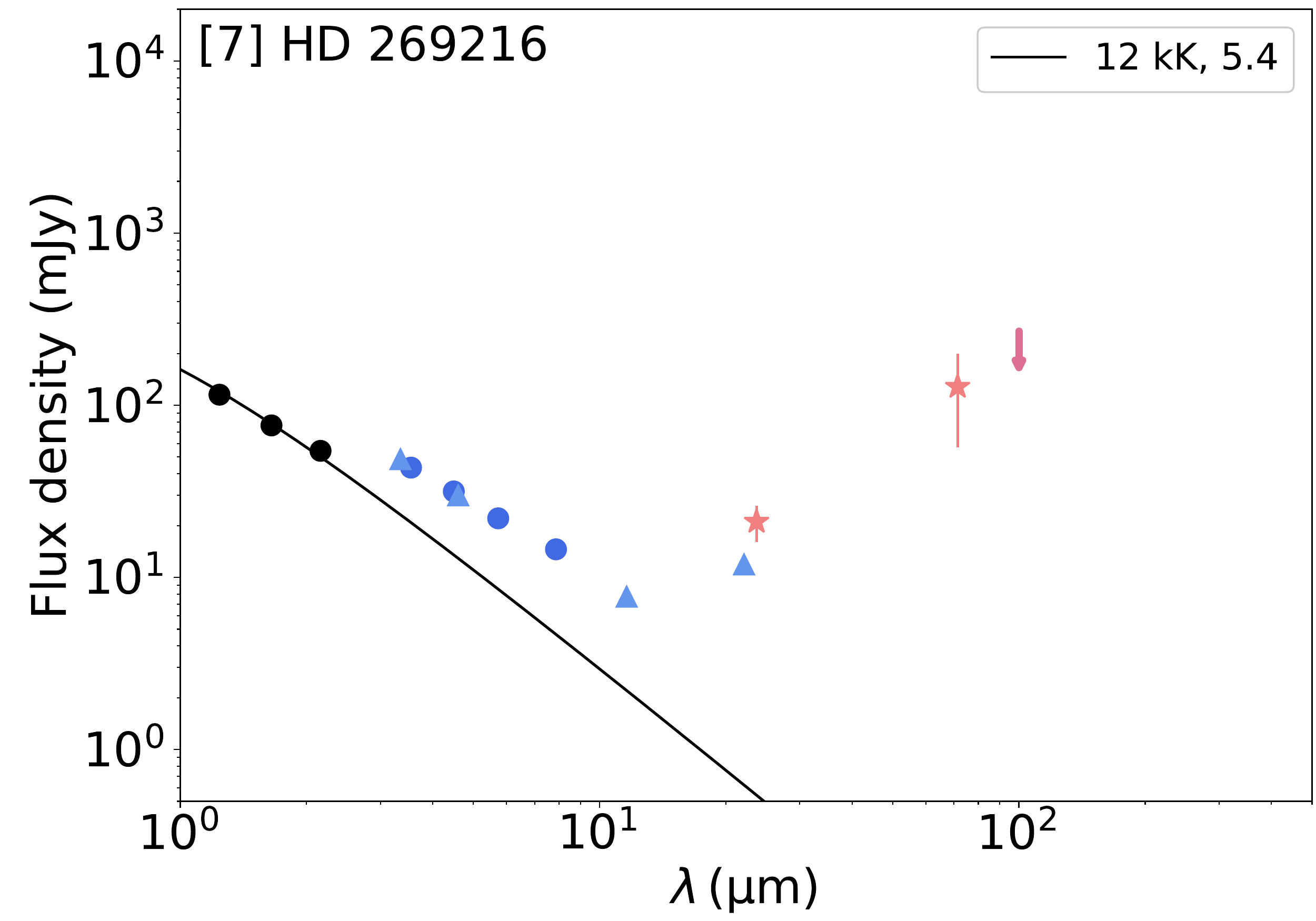}
    \end{minipage}%
    \begin{minipage}{0.65\textwidth}
        \centering
        \includegraphics[width=1\linewidth]{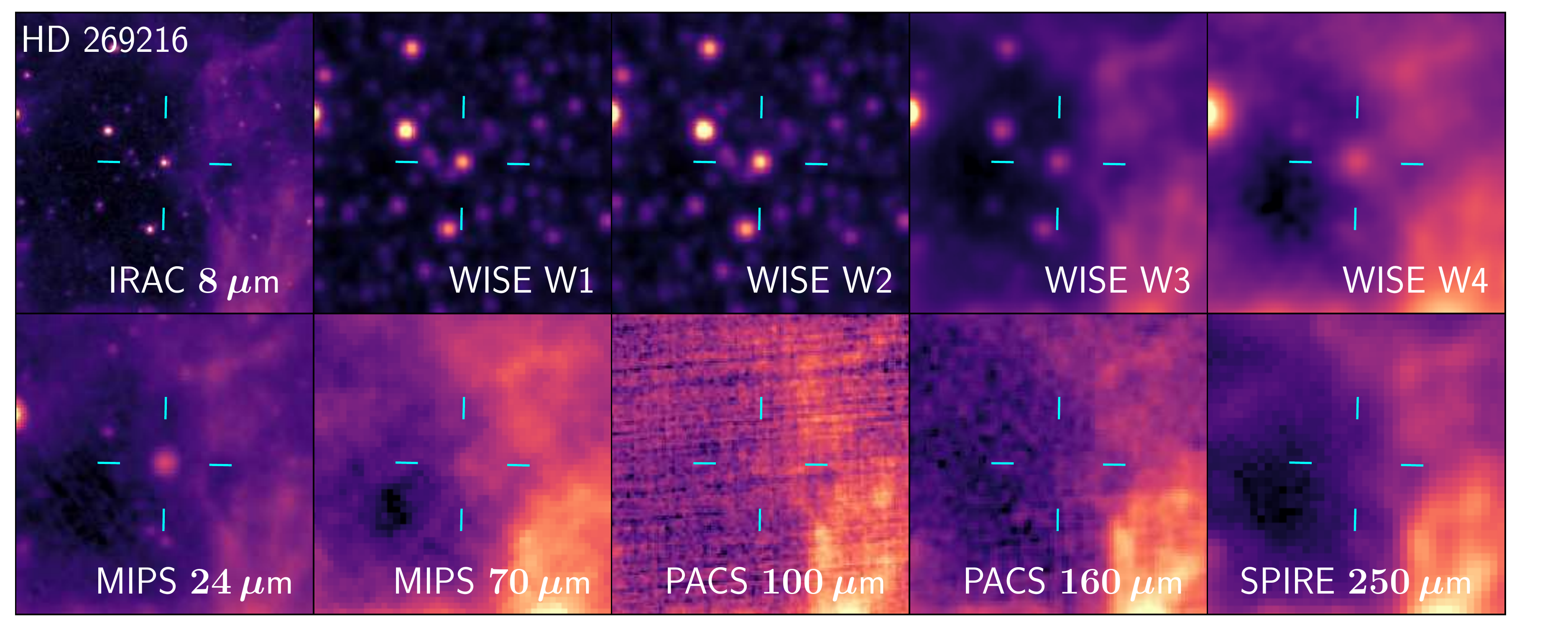}
    \end{minipage}
    \begin{minipage}{.35\textwidth}
        \centering
        \includegraphics[width=1\linewidth]{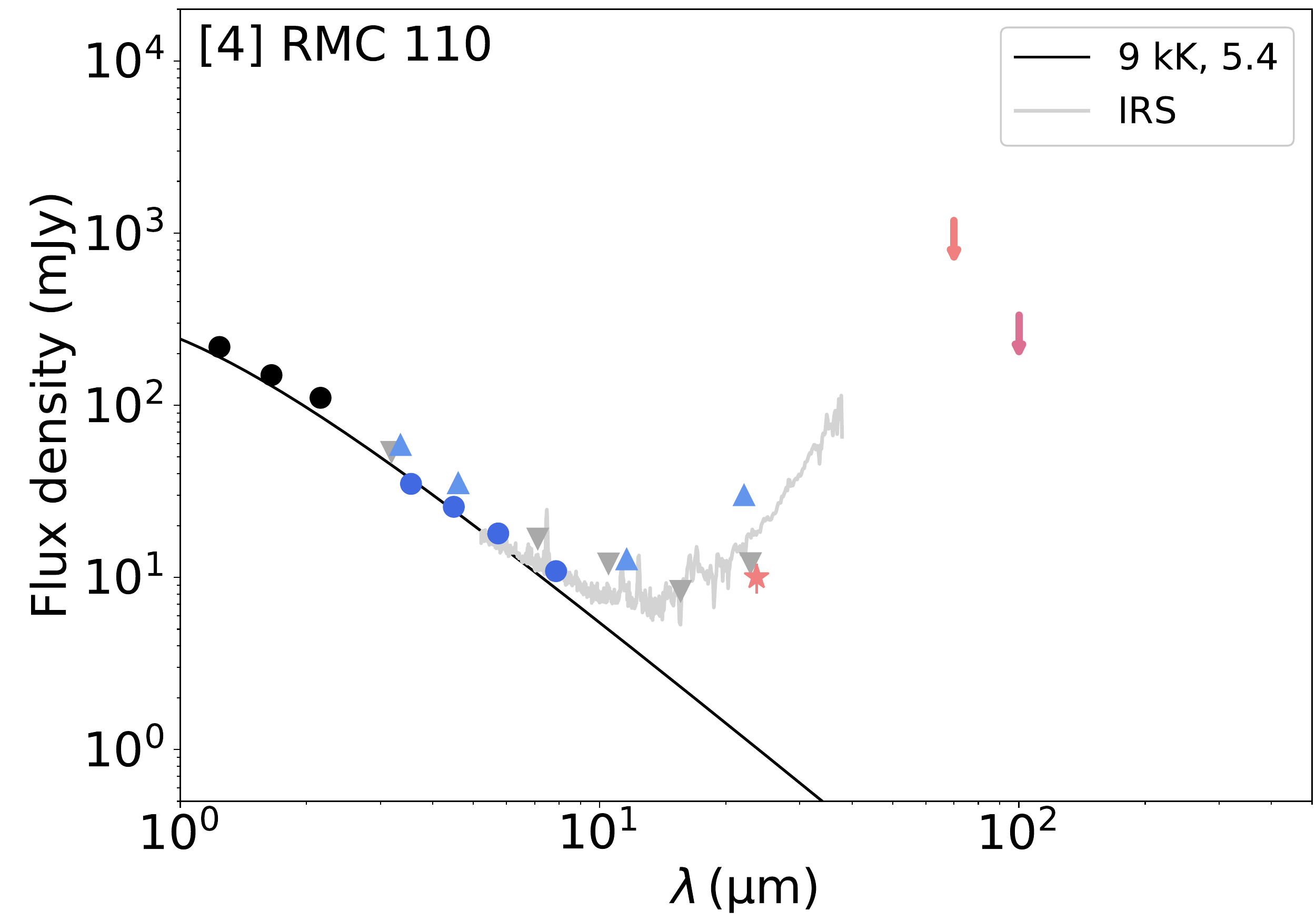}
    \end{minipage}%
    \begin{minipage}{0.65\textwidth}
        \centering
        \includegraphics[width=1\linewidth]{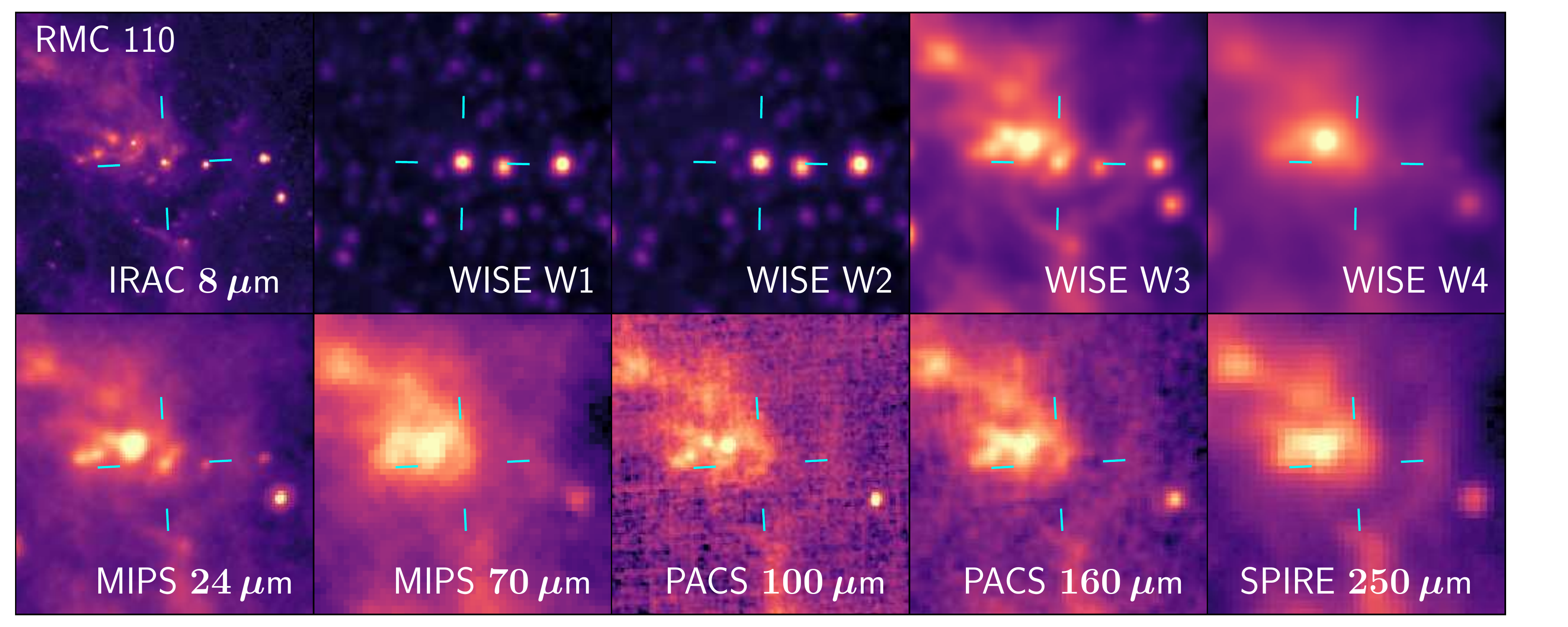}
    \end{minipage}
\caption*{Fig.~\ref{fig:group1b} continued}
\end{figure*}

\begin{figure*}
    \centering
    \begin{minipage}{.35\textwidth}
        \centering
        \includegraphics[width=1\linewidth]{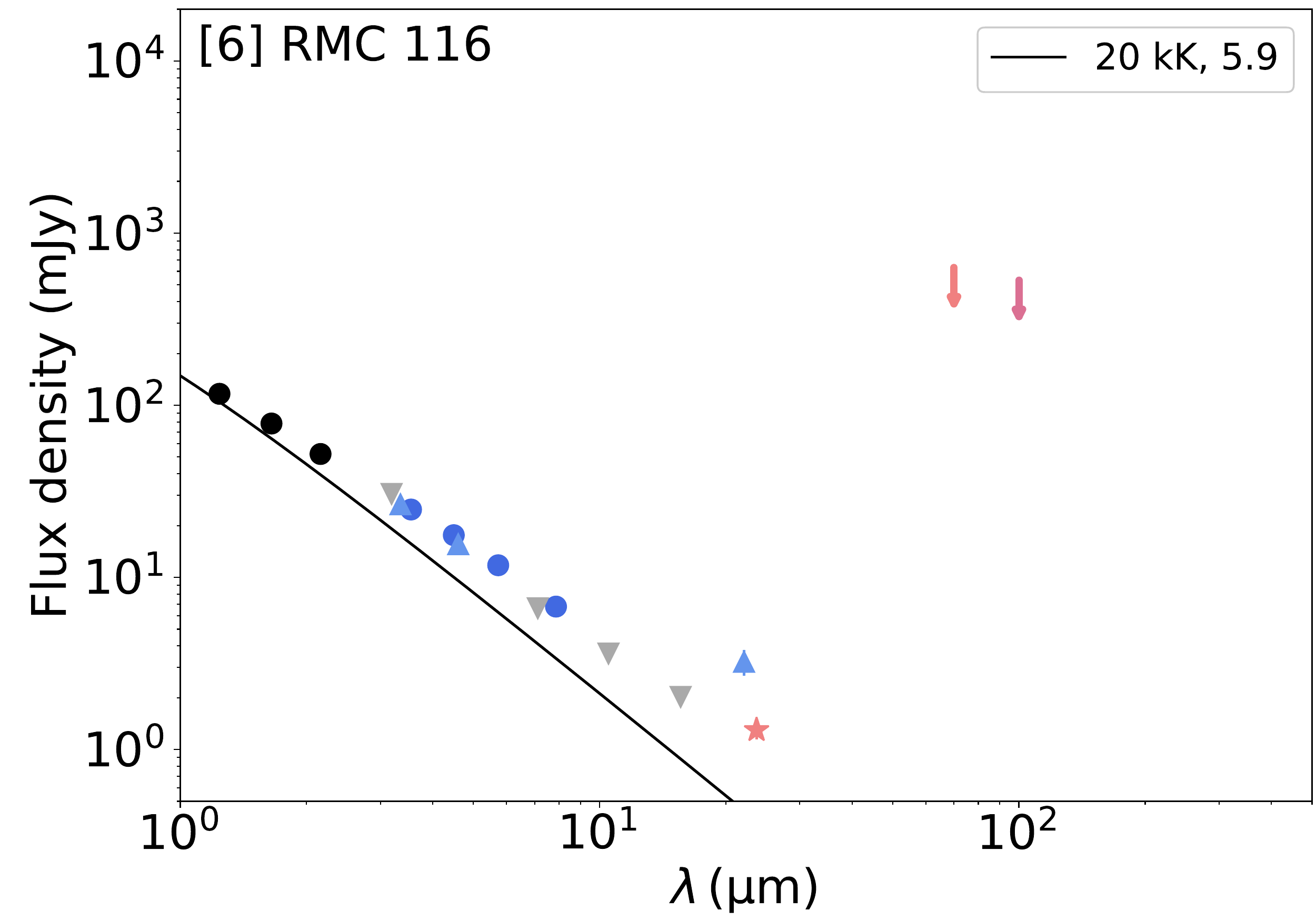}
    \end{minipage}%
    \begin{minipage}{0.65\textwidth}
        \centering
        \includegraphics[width=1\linewidth]{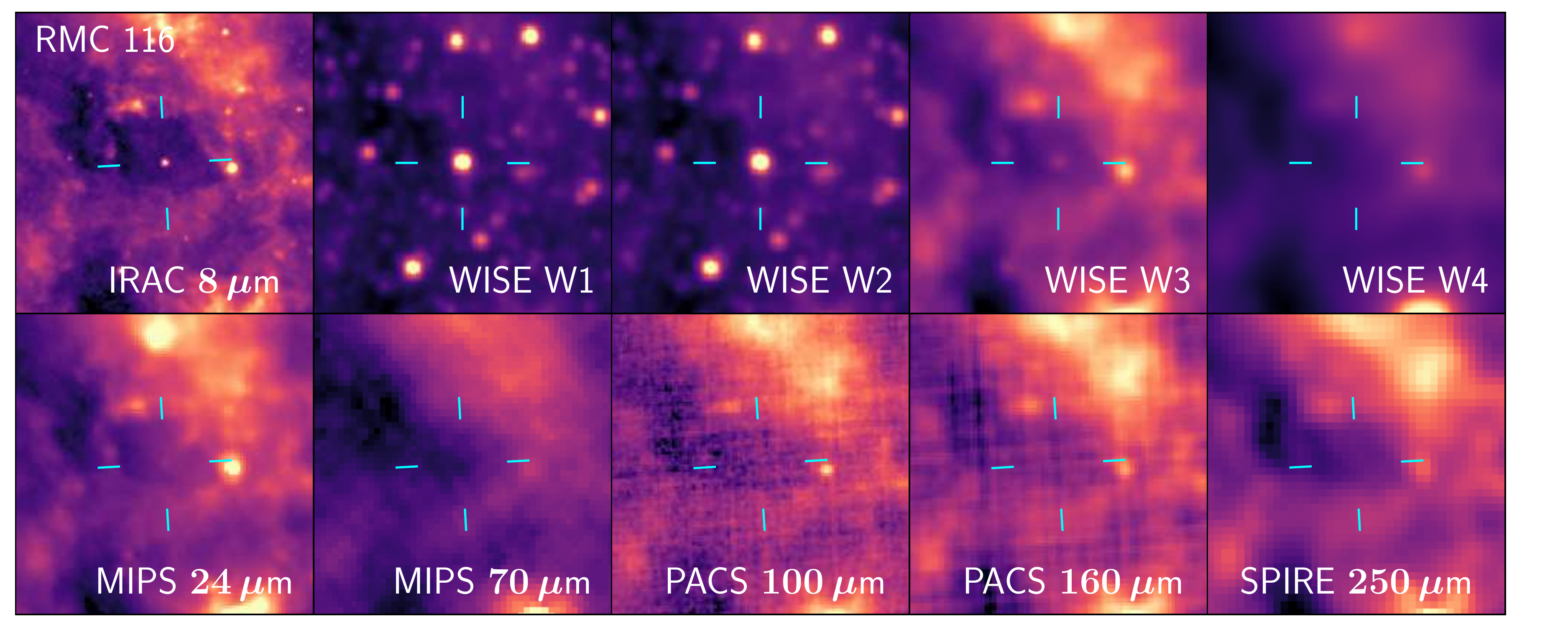}
    \end{minipage}  
    \begin{minipage}{.35\textwidth}
        \centering
        \includegraphics[width=1\linewidth]{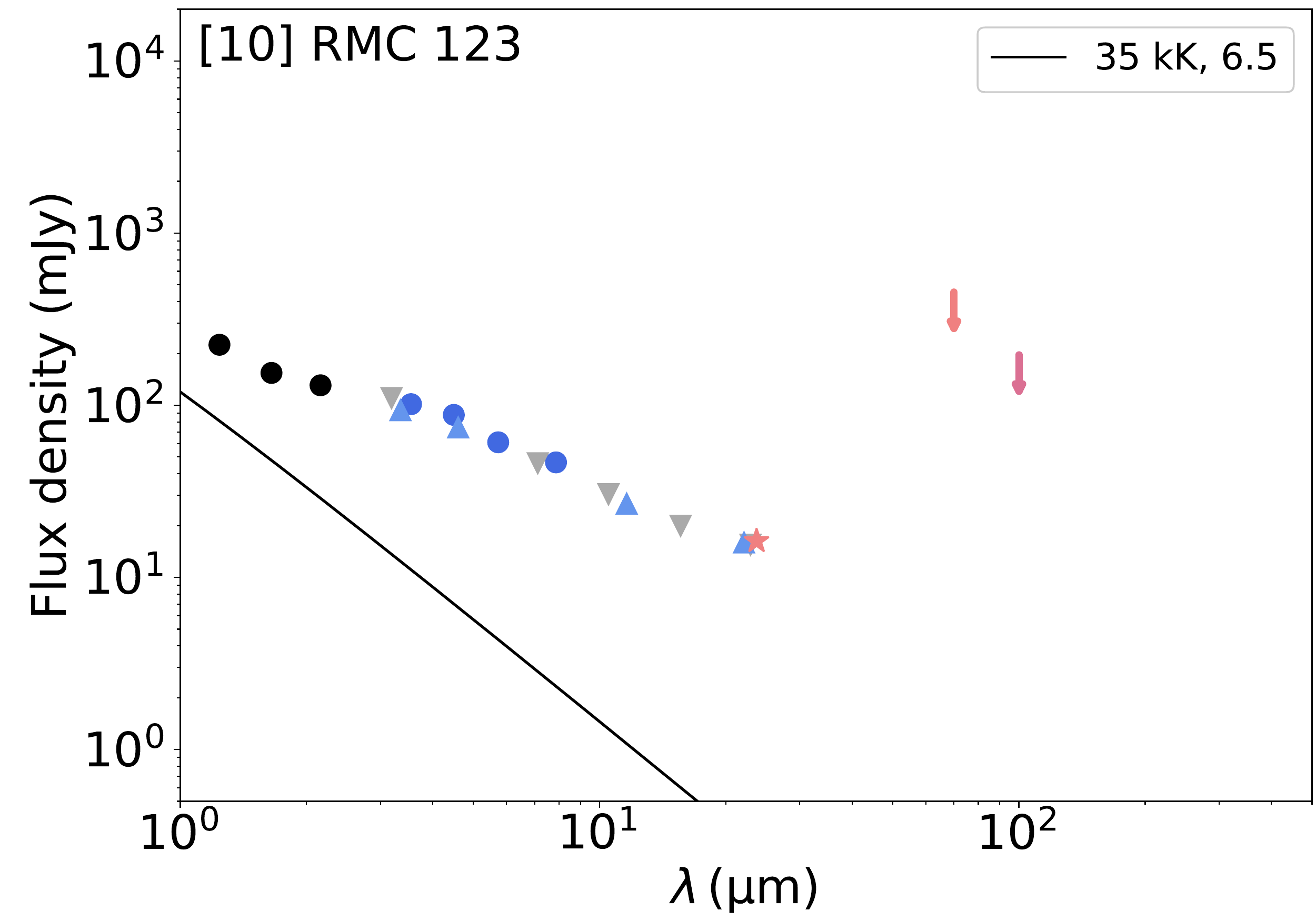}
    \end{minipage}%
    \begin{minipage}{0.65\textwidth}
        \centering
        \includegraphics[width=1\linewidth]{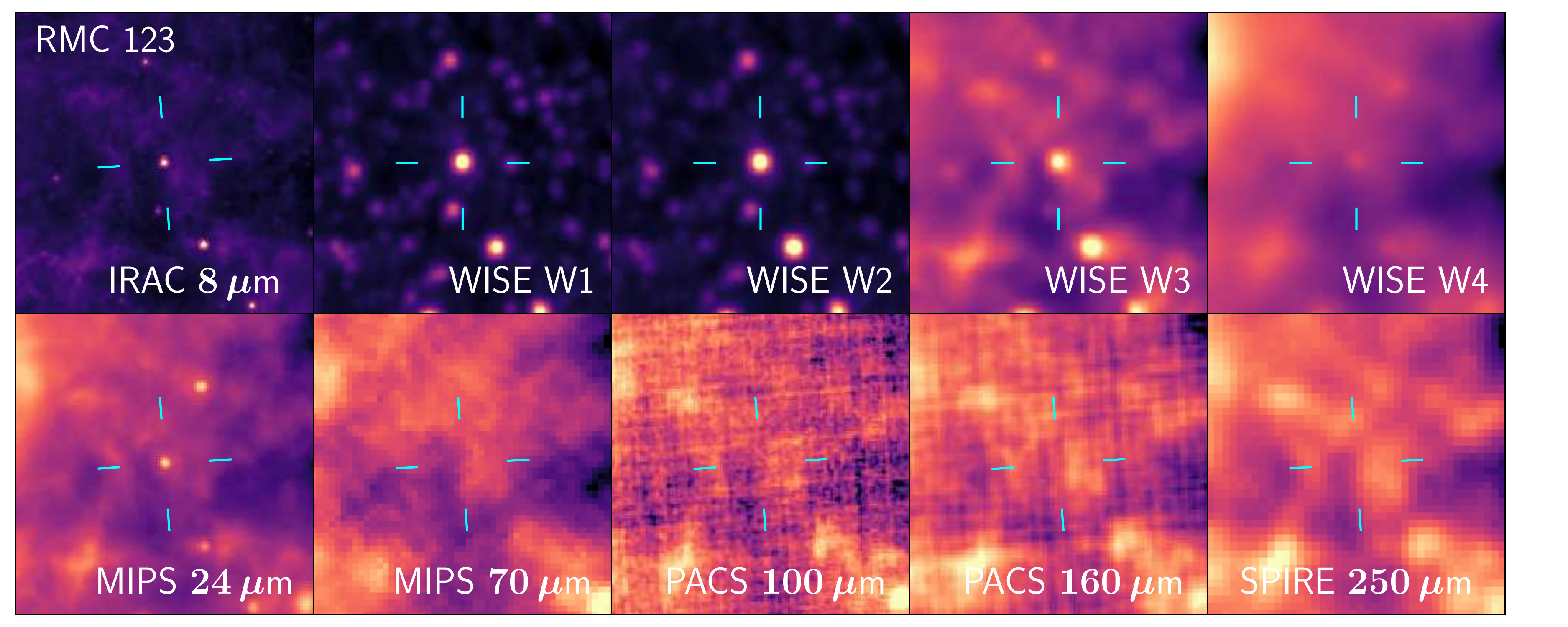}
    \end{minipage}      
    \begin{minipage}{.35\textwidth}
        \centering
        \includegraphics[width=1\linewidth]{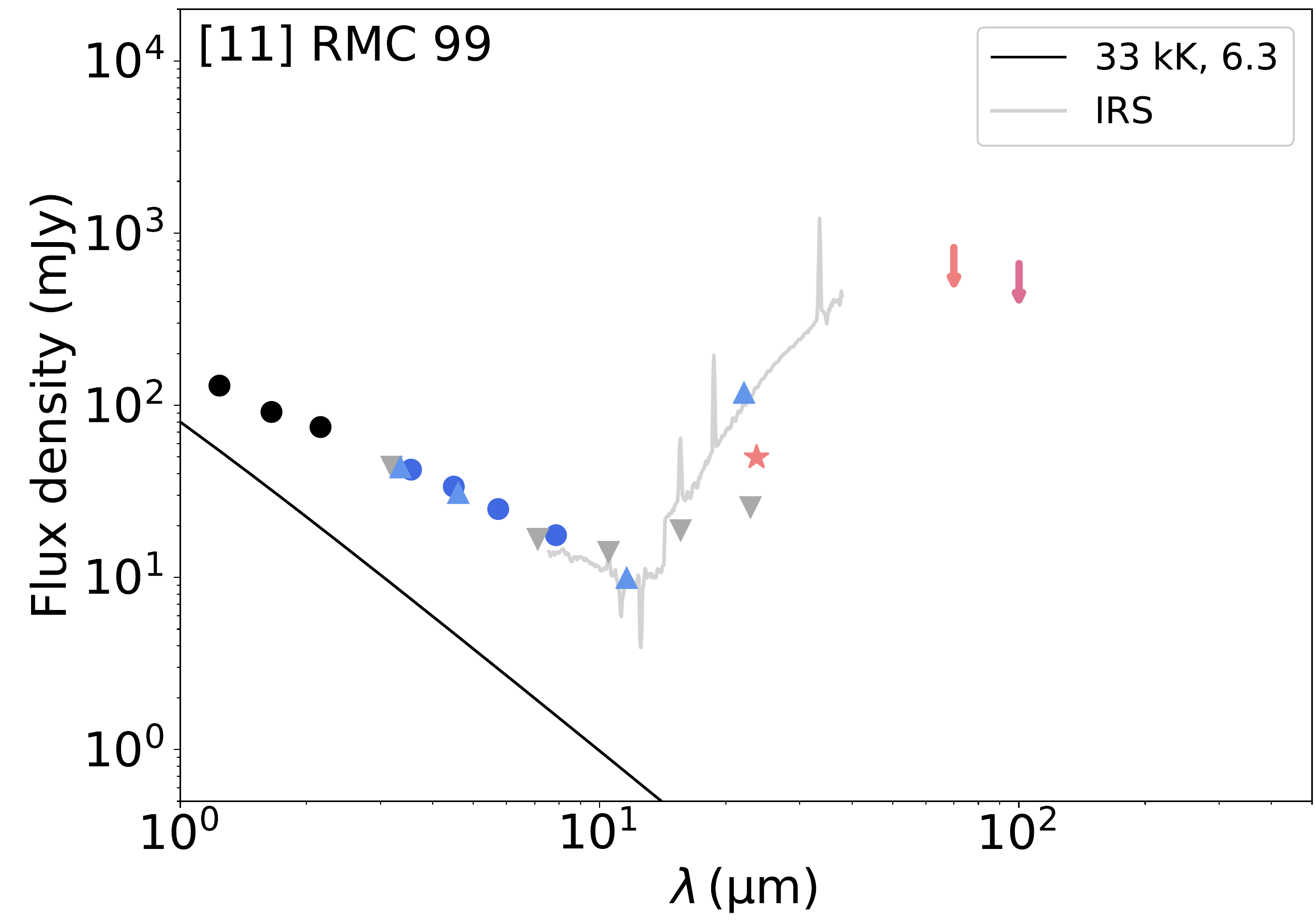}
    \end{minipage}%
    \begin{minipage}{0.65\textwidth}
        \centering
        \includegraphics[width=1\linewidth]{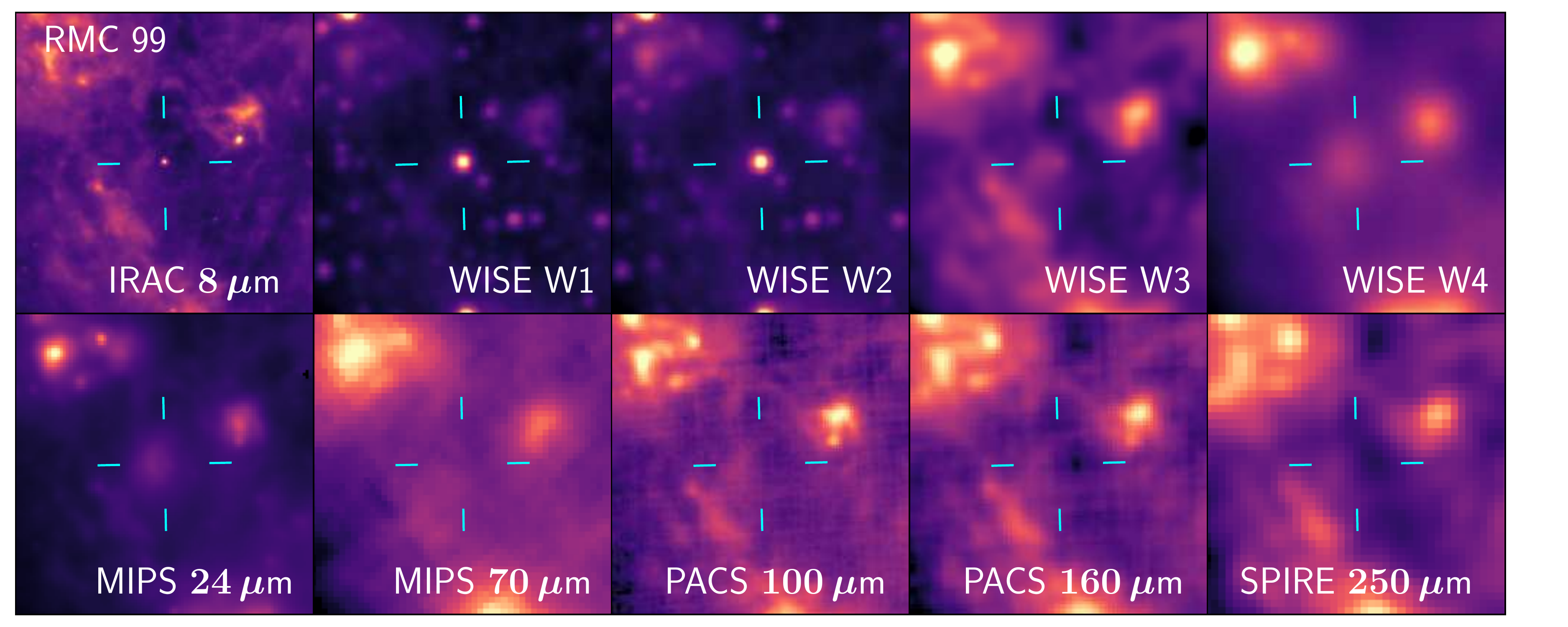}
    \end{minipage}   
    \begin{minipage}{.35\textwidth}
        \centering
        \includegraphics[width=1\linewidth]{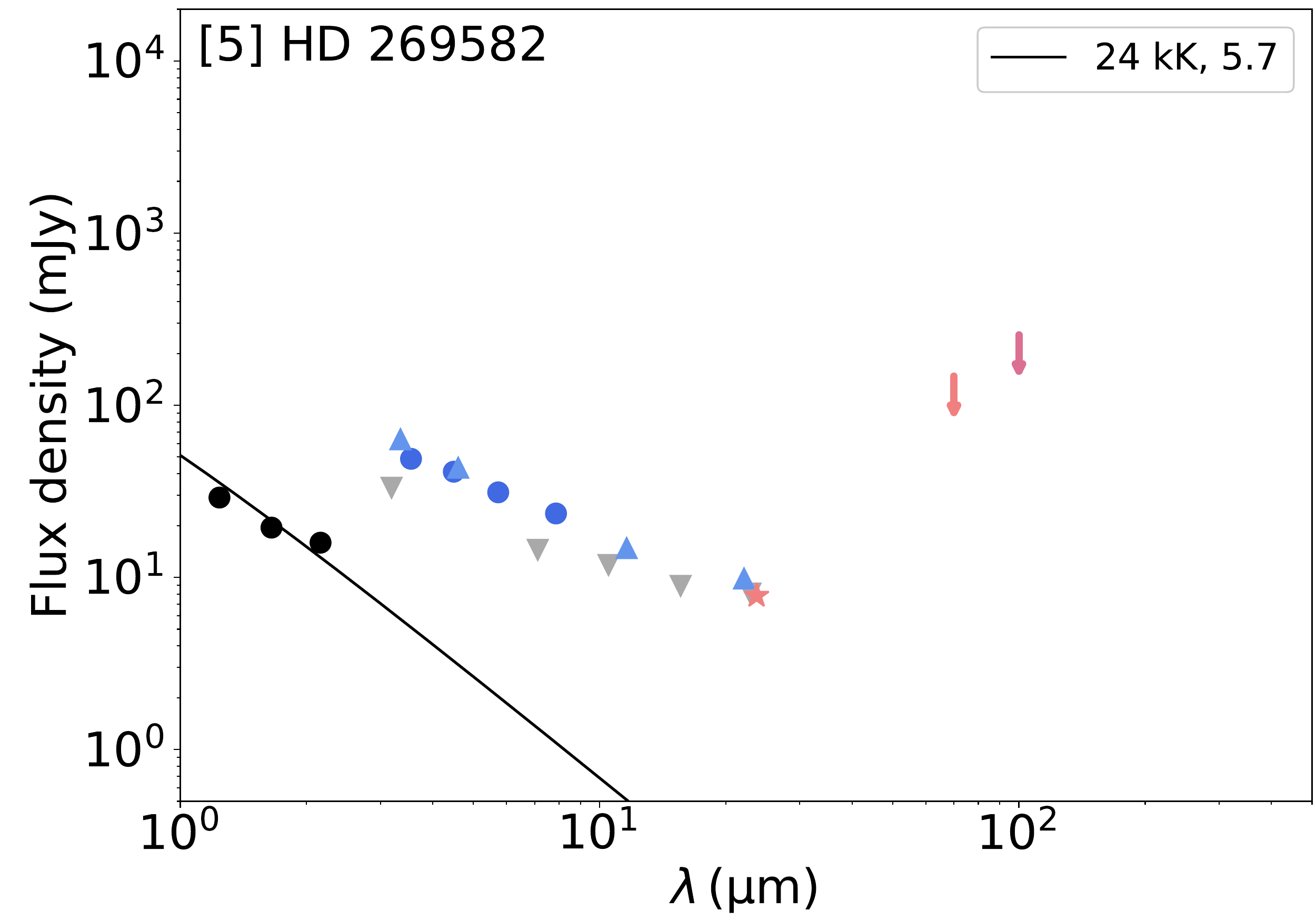}
    \end{minipage}%
    \begin{minipage}{0.65\textwidth}
        \centering
        \includegraphics[width=1\linewidth]{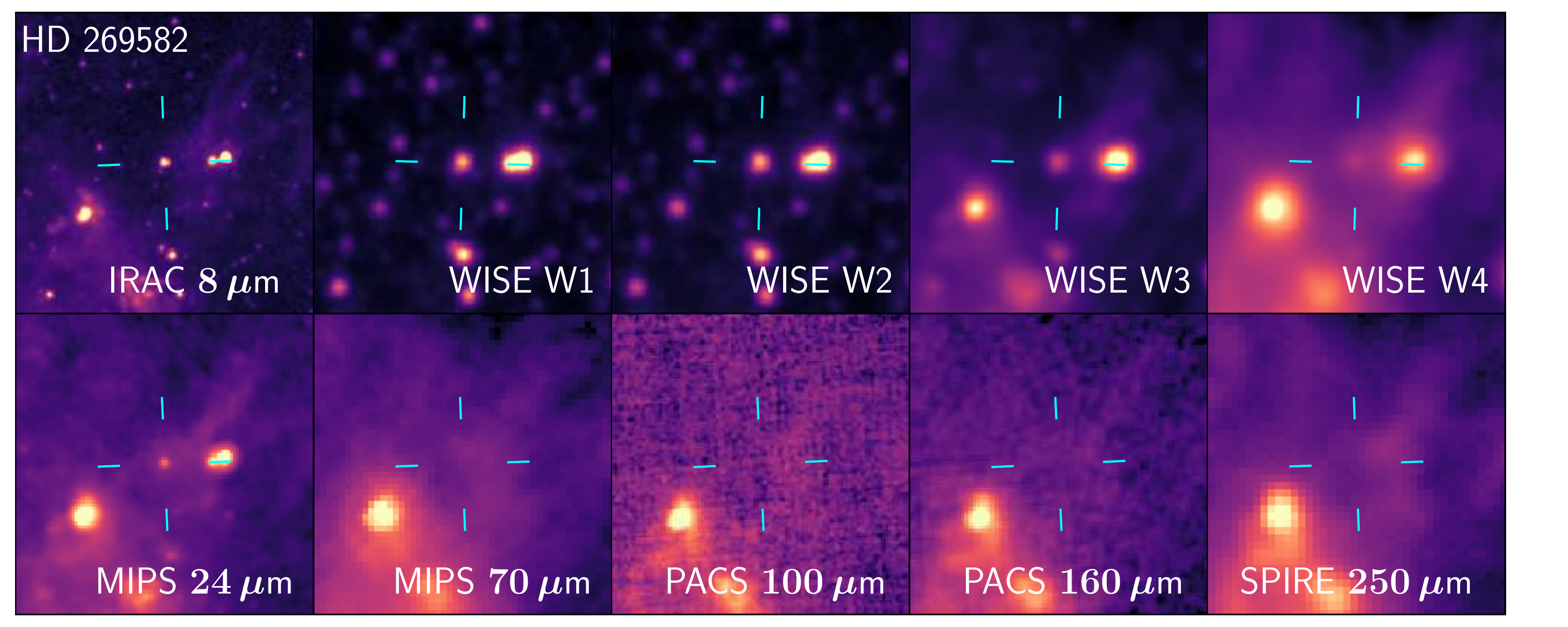}
    \end{minipage}

    \caption{Confirmed and candidate LBVs with a SED dominated by free-free continuum up to $\rm 24\,\mu$m and by confusion at longer wavelengths (Class ~2 sources in Sect.~\ref{sec:sedreview}). For HD$\,$269582 the jump between 2MASS and other photometry is due to real source variability. The legend for the photometric points can be found in Fig. \ref{fig:cmd-LMC} of Sect. \ref{sec:review}.
    }
           \label{fig:group2}
\end{figure*}
\begin{figure*}[!htb]
    \centering
    \begin{minipage}{.35\textwidth}
        \centering
        \includegraphics[width=1\linewidth]{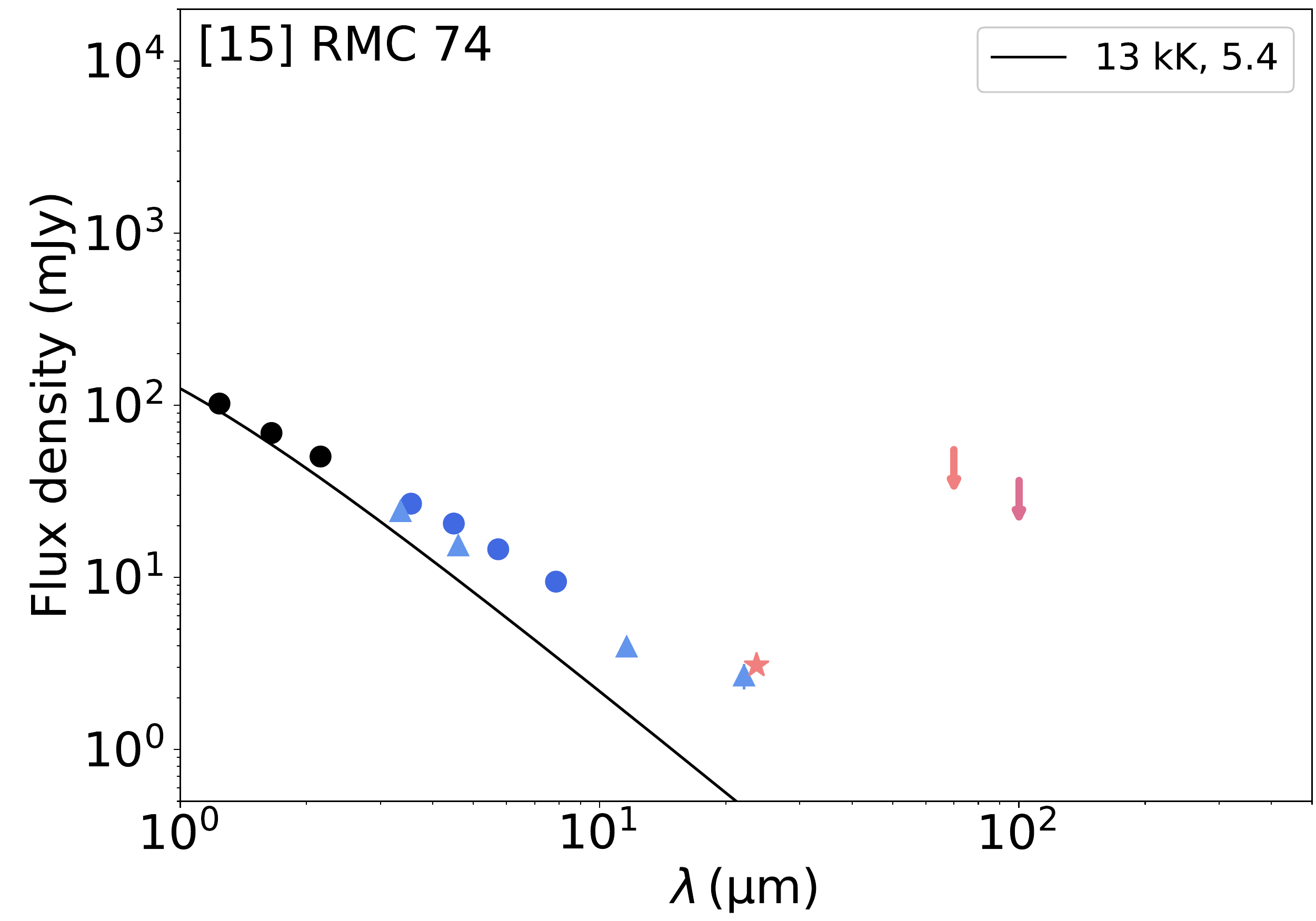}
    \end{minipage}%
    \begin{minipage}{0.65\textwidth}
        \centering
        \includegraphics[width=1\linewidth]{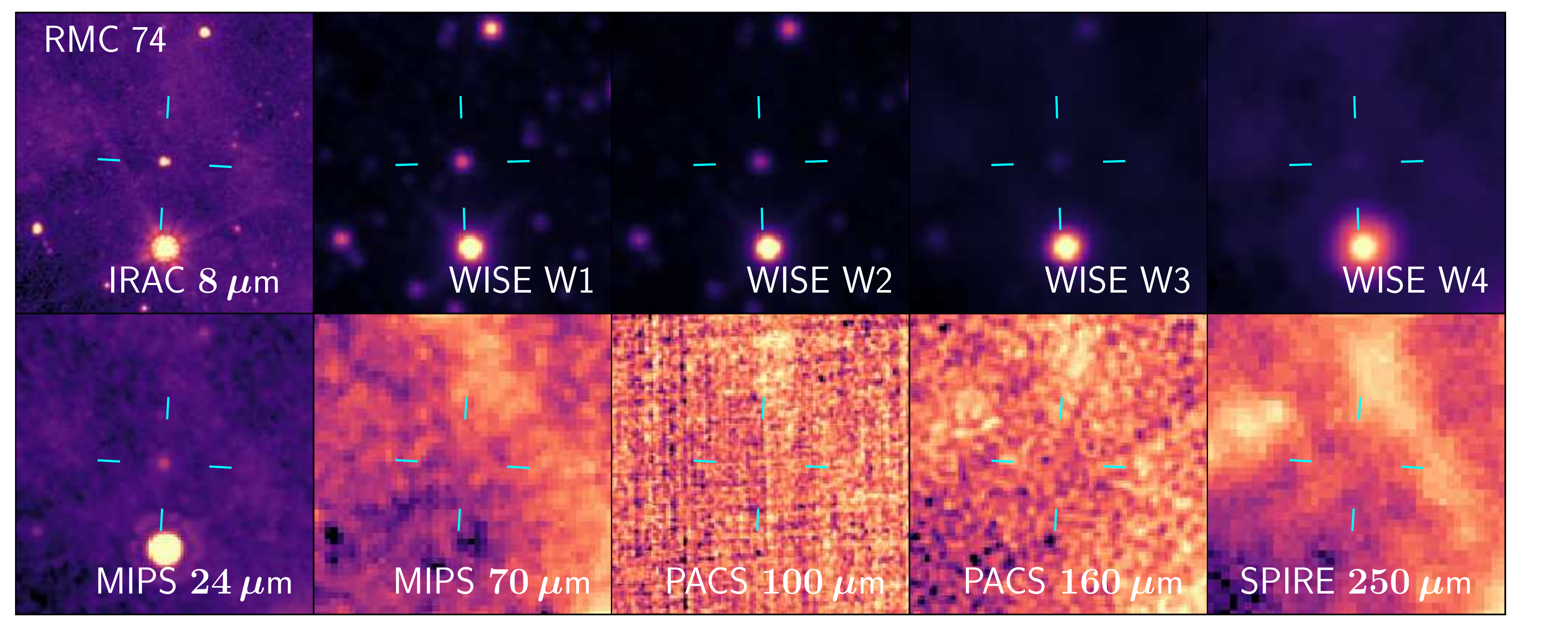}
    \end{minipage}
    \begin{minipage}{.35\textwidth}
        \centering
        \includegraphics[width=1\linewidth]{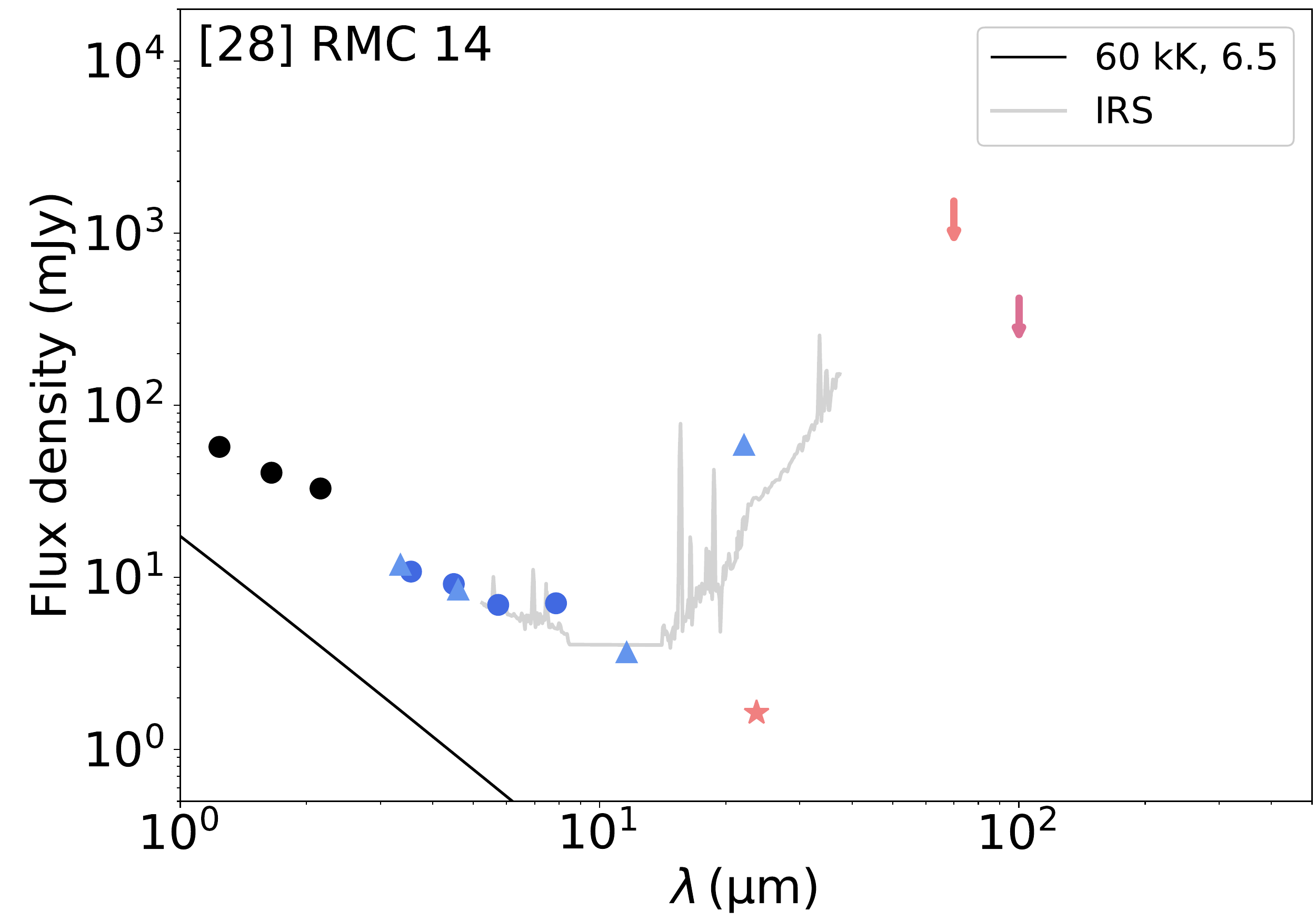}
    \end{minipage}%
    \begin{minipage}{0.65\textwidth}
        \centering
        \includegraphics[width=1\linewidth]{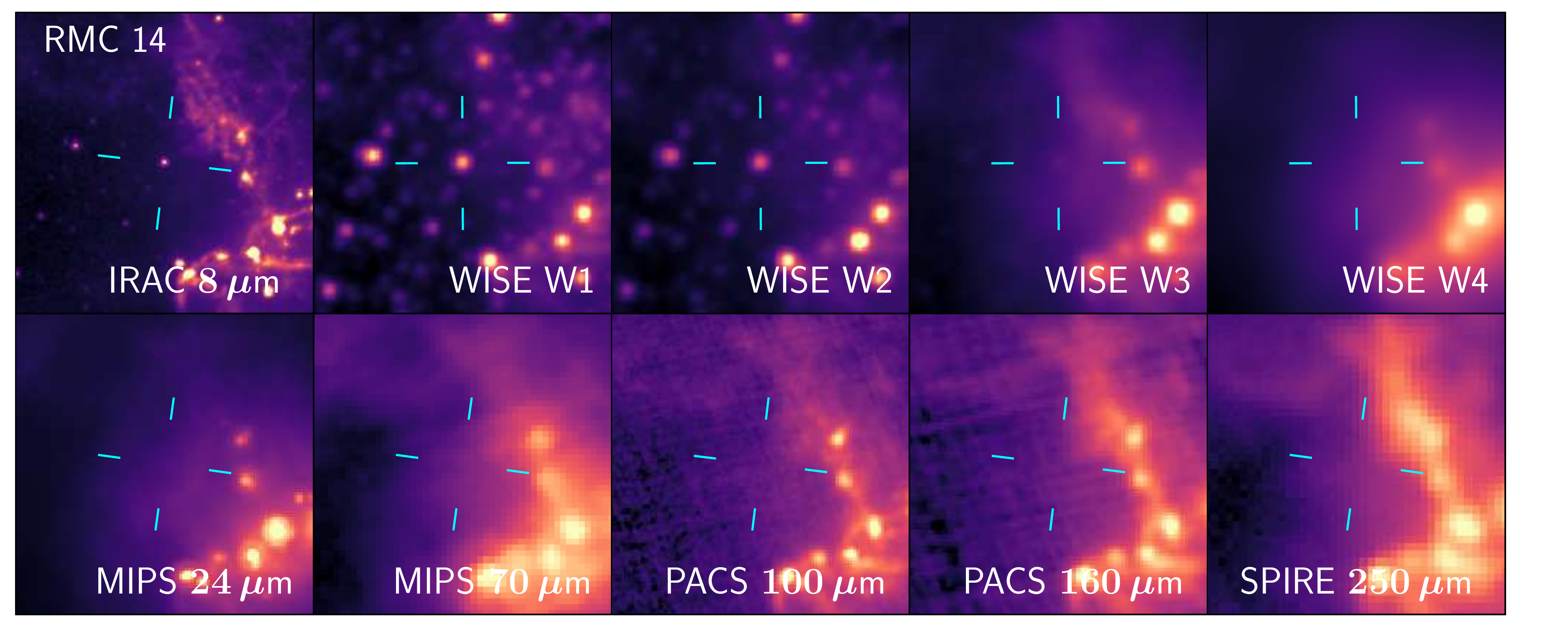}
    \end{minipage}
\caption*{Fig.~\ref{fig:group2} continued. We note that photometry for RMC$\,14$ has not been extinction corrected due to no $A_{\rm V}$ value in the literature.}
\end{figure*}

\begin{figure*}
    \centering
    \begin{minipage}{.35\textwidth}
        \centering
        \includegraphics[width=1\linewidth]{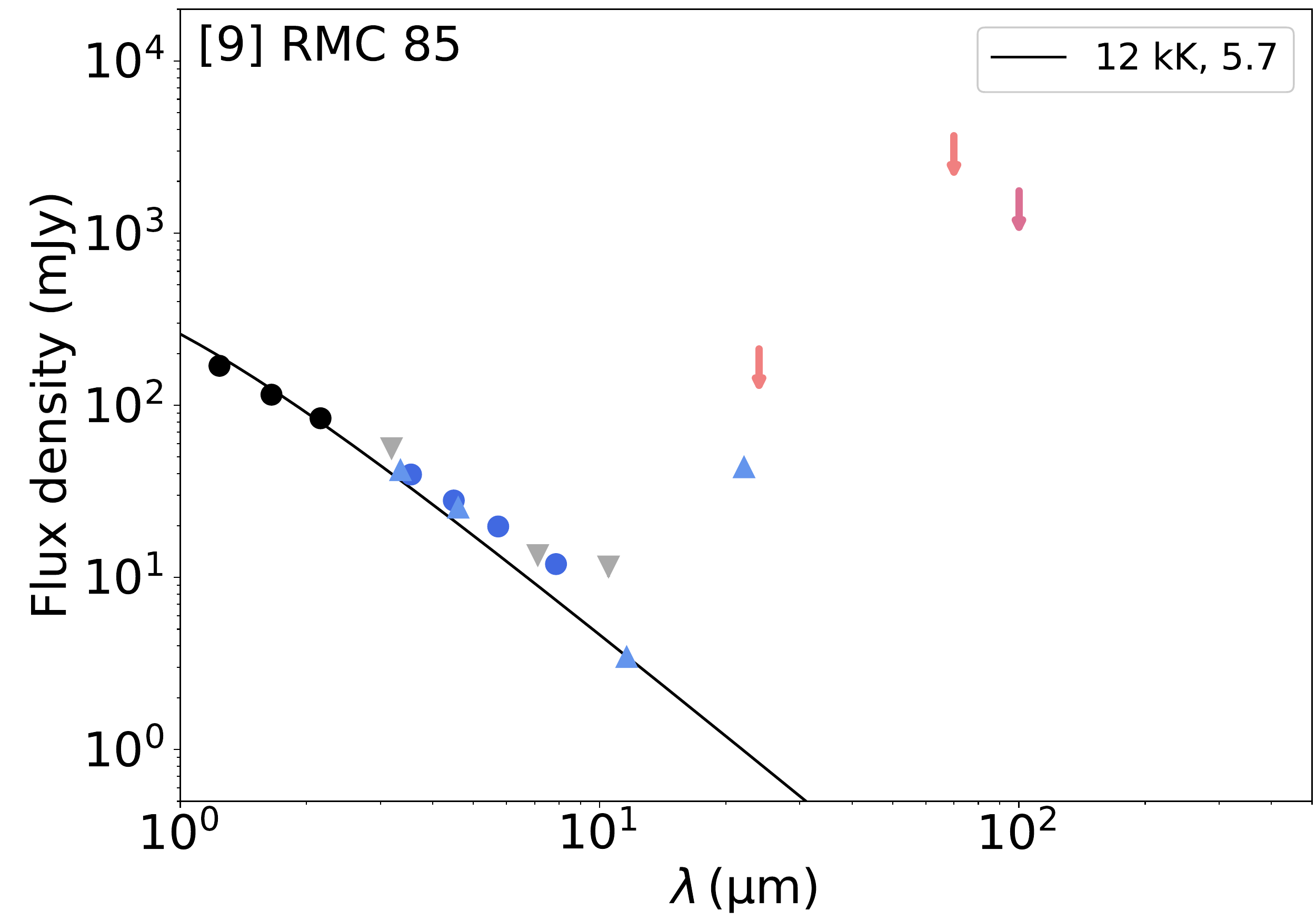}
    \end{minipage}%
    \begin{minipage}{0.65\textwidth}
        \centering
        \includegraphics[width=1\linewidth]{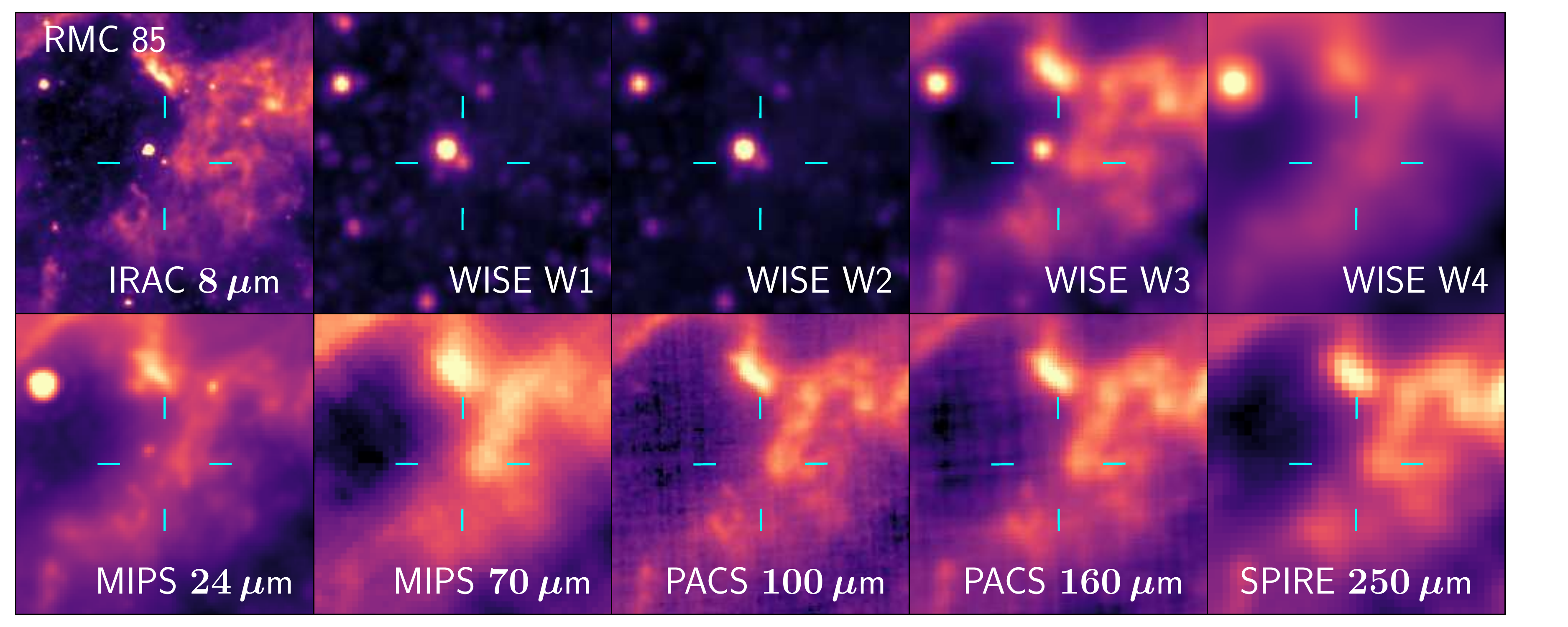}
    \end{minipage}
\caption{Unclassified LBV RMC$\,$85. No reliable photometry could be obtained longward of $20\,\mu\rm m$ due to confusion with bright ISM dust. The legend for the photometric points can be found in Fig. \ref{fig:cmd-LMC} of Sect. \ref{sec:review}.}

\label{fig:ungrouped}
\end{figure*}
\begin{figure*}
    \centering
    \begin{minipage}{.35\textwidth}
        \centering
        \includegraphics[width=1\linewidth]{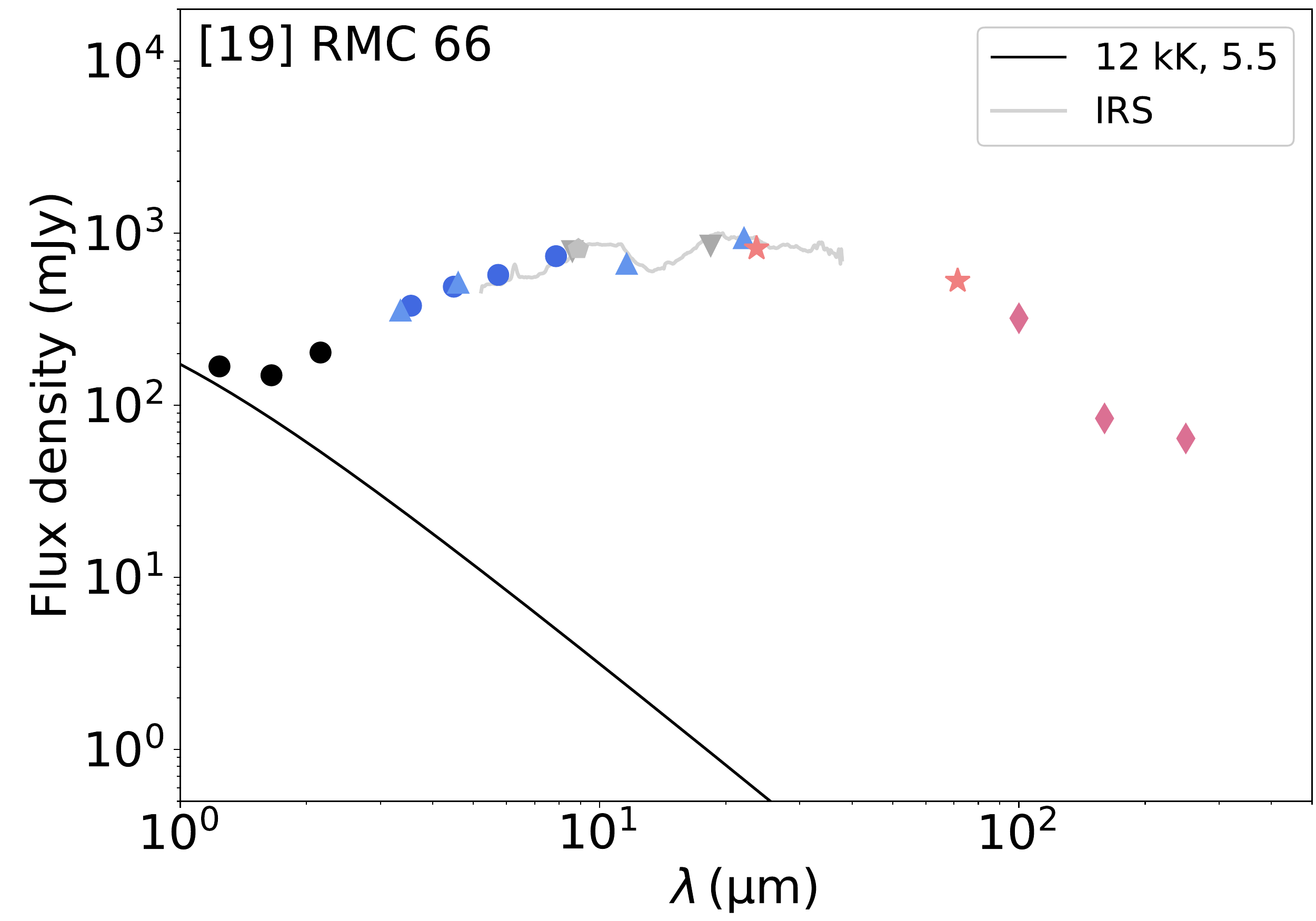}
    \end{minipage}%
    \begin{minipage}{0.65\textwidth}
        \centering
        \includegraphics[width=1\linewidth]{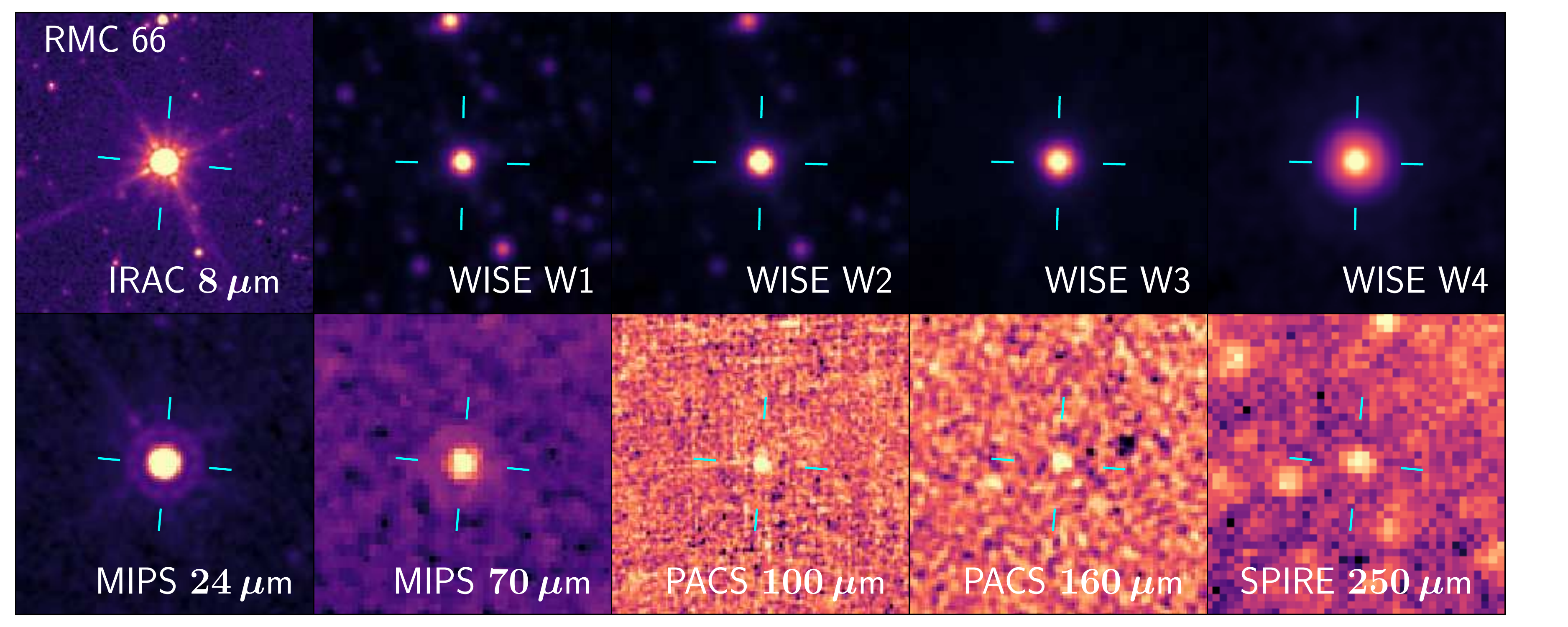}
    \end{minipage} 
    \begin{minipage}{.35\textwidth}
        \centering
        \includegraphics[width=1\linewidth]{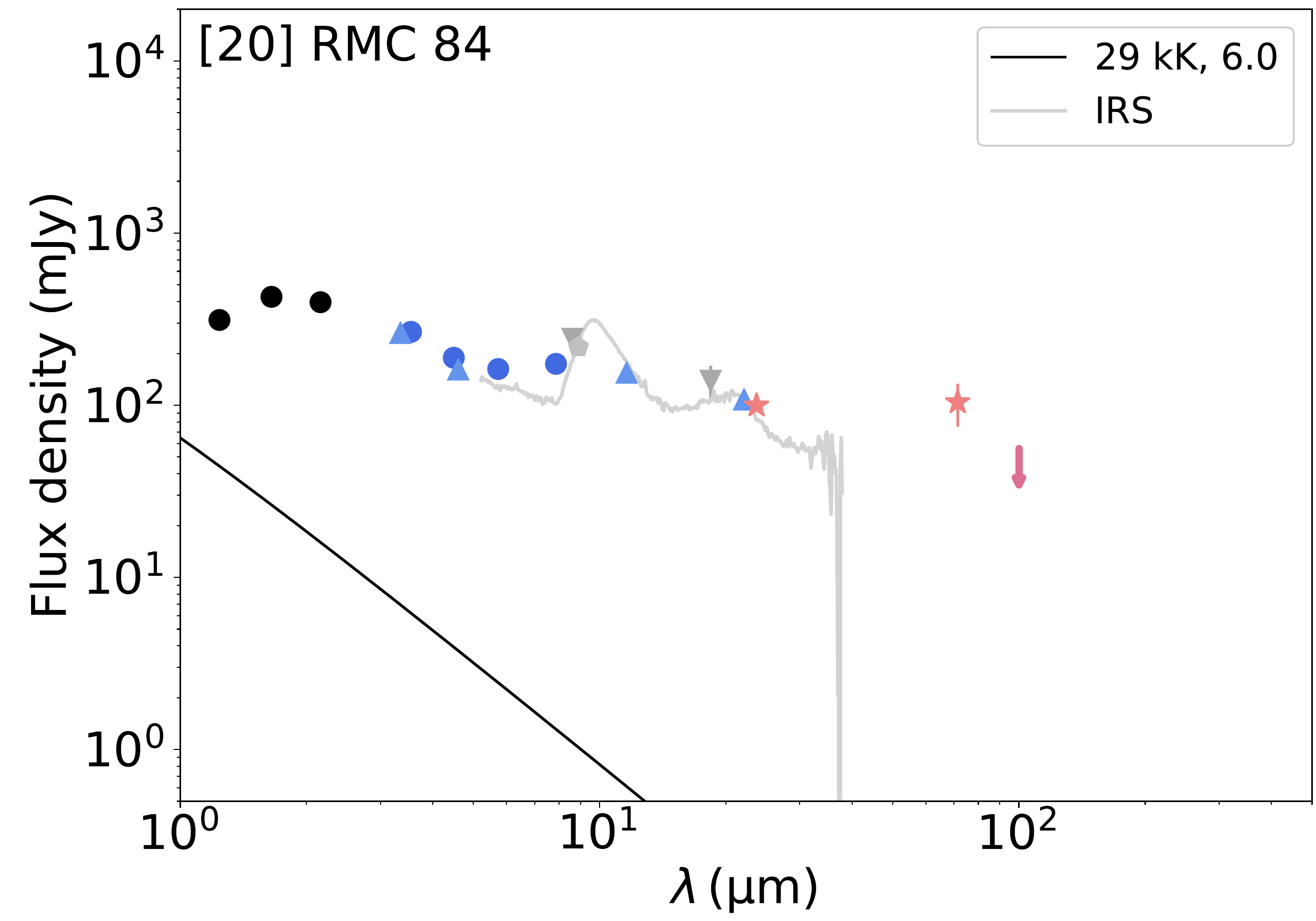}
    \end{minipage}%
    \begin{minipage}{0.65\textwidth}
        \centering
        \includegraphics[width=1\linewidth]{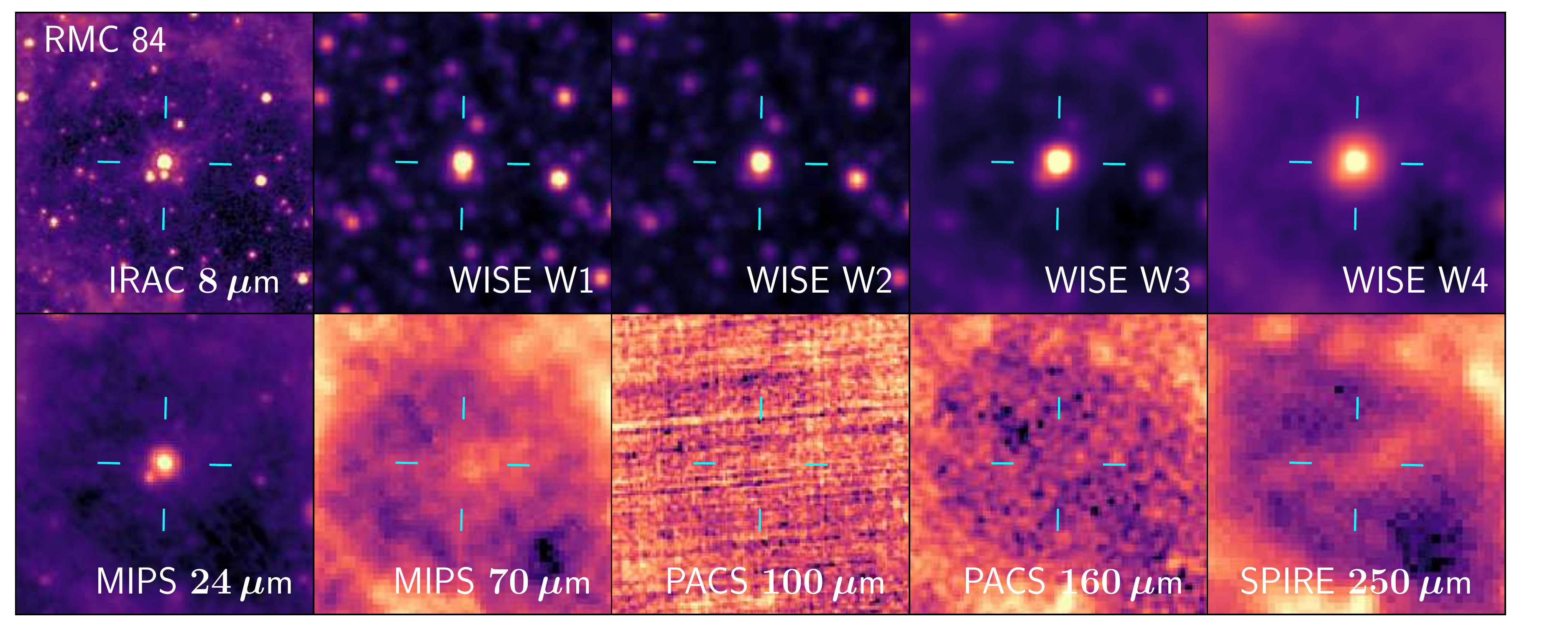}
    \end{minipage}     
    \begin{minipage}{.35\textwidth}
        \centering
        \includegraphics[width=1\linewidth]{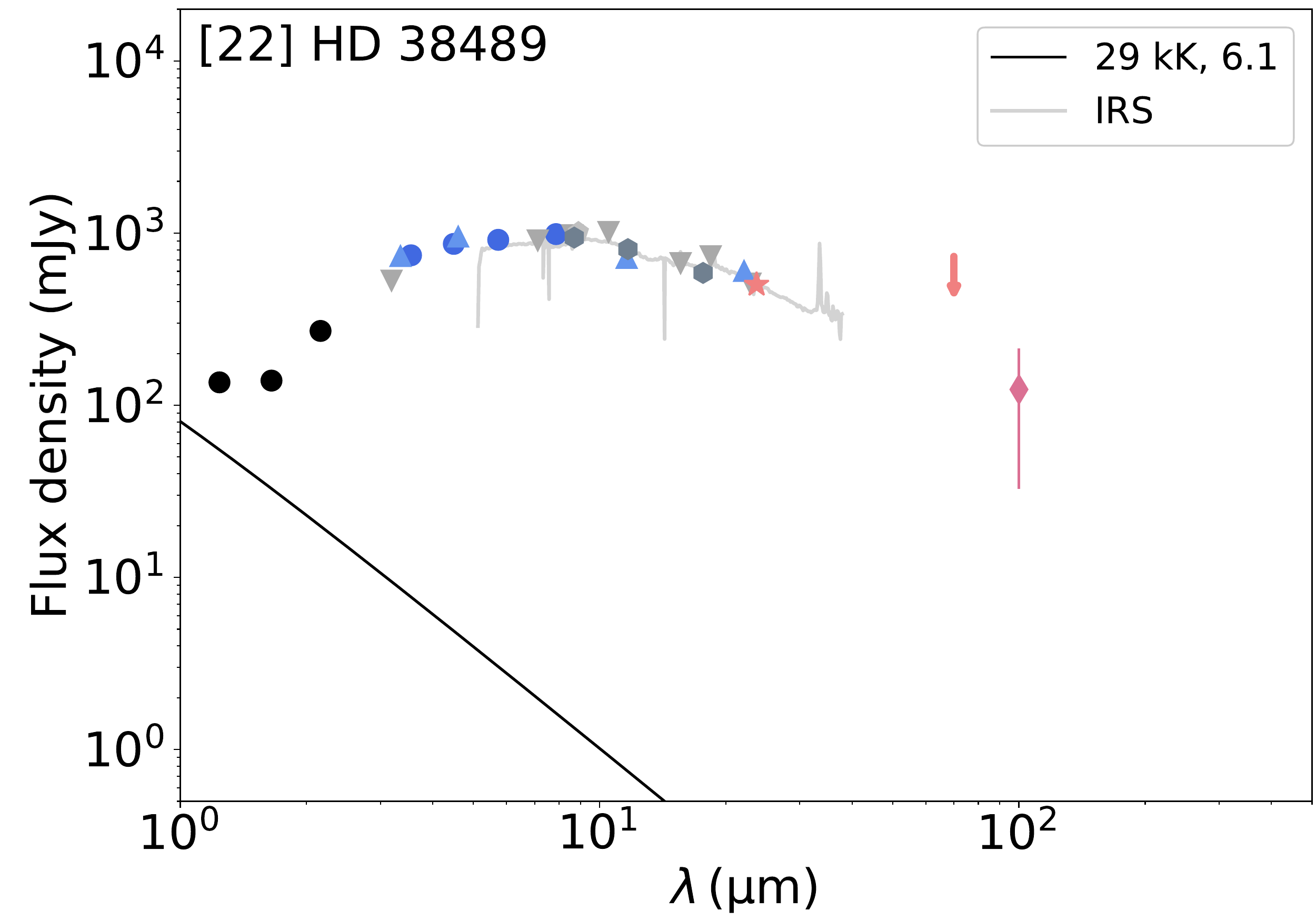}
   \end{minipage}%
    \begin{minipage}{0.65\textwidth}
        \centering
        \includegraphics[width=1\linewidth]{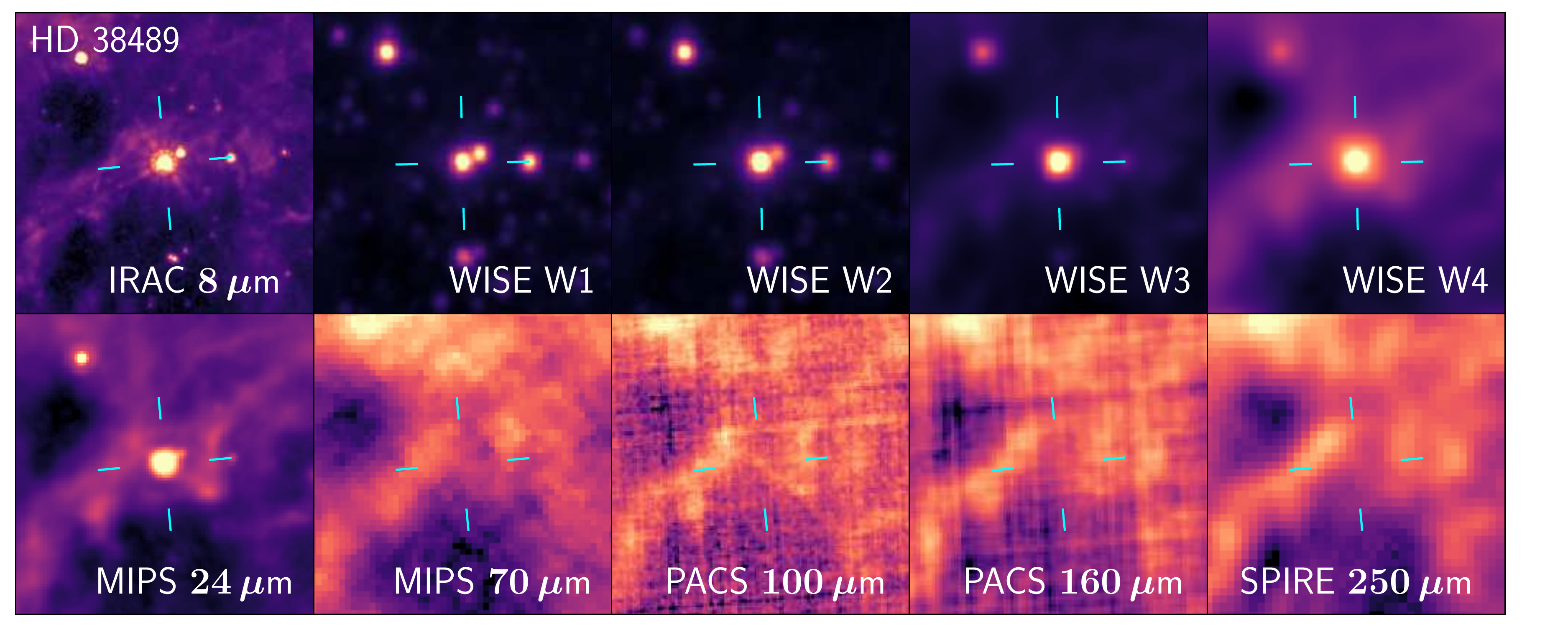}
    \end{minipage}   
   \begin{minipage}{.35\textwidth}
        \centering
        \includegraphics[width=1\linewidth]{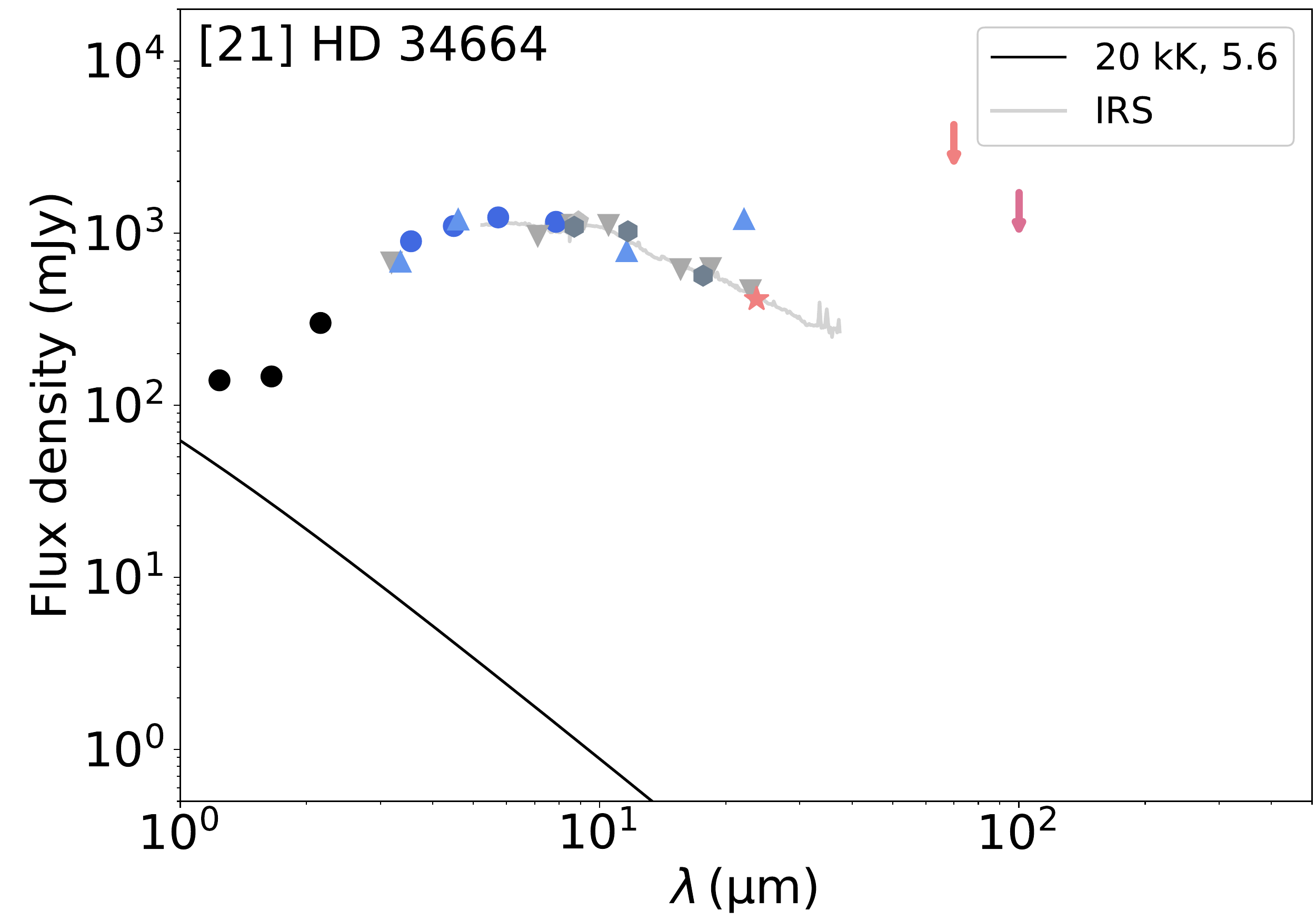}
    \end{minipage}%
    \begin{minipage}{0.65\textwidth}
        \centering
        \includegraphics[width=1\linewidth]{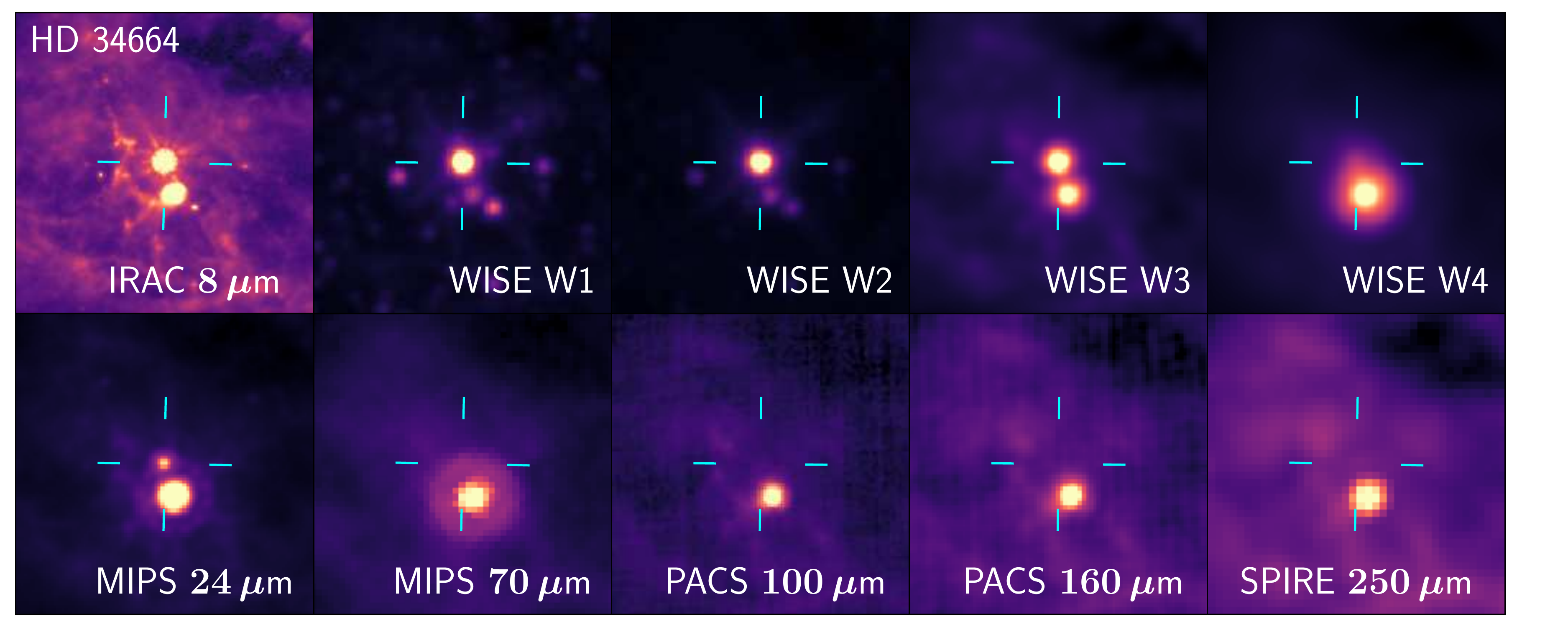}
    \end{minipage}   
    \begin{minipage}{.35\textwidth}
        \centering
        \includegraphics[width=1\linewidth]{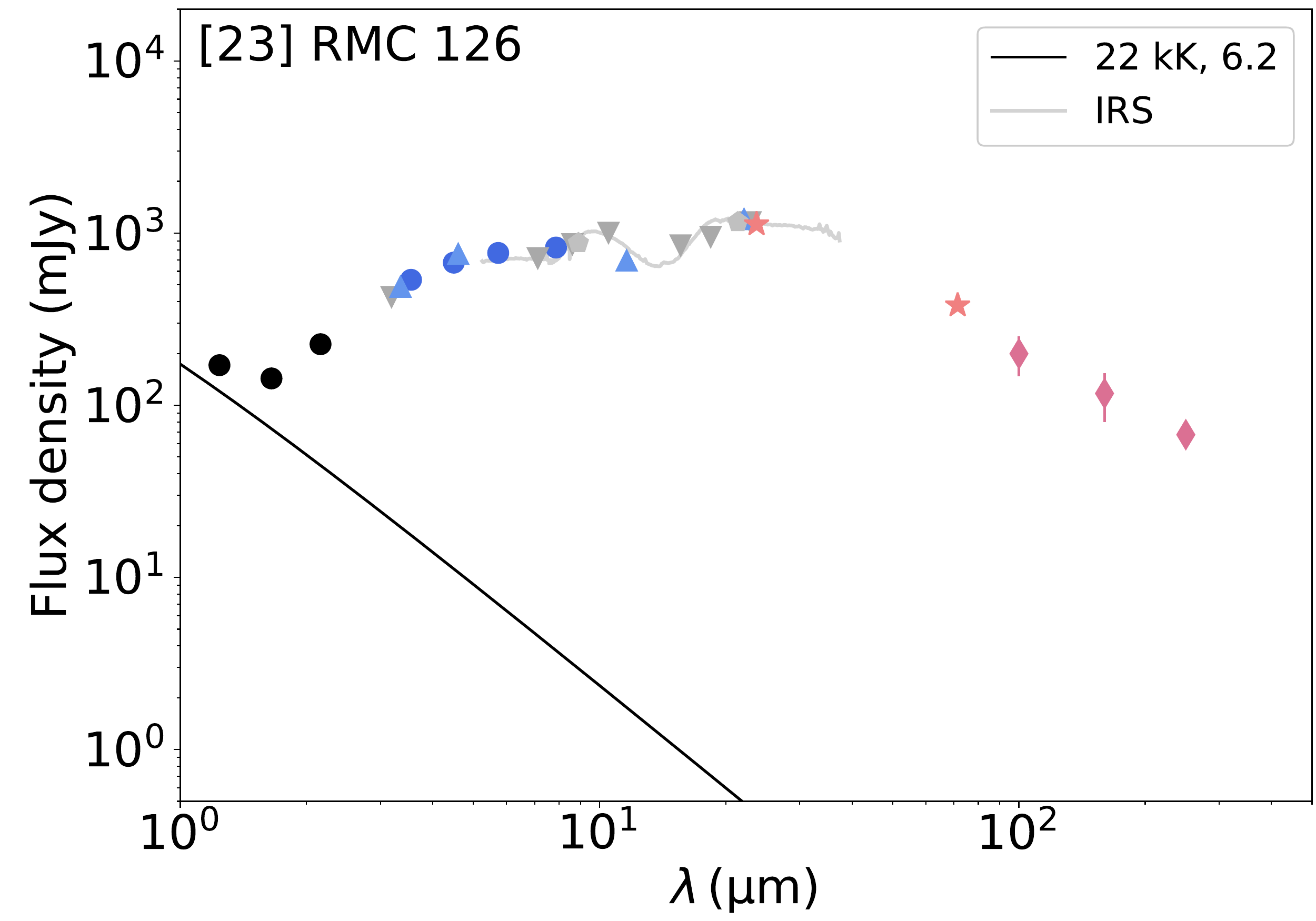}

    \end{minipage}%
    \begin{minipage}{0.65\textwidth}
        \centering
        \includegraphics[width=1\linewidth]{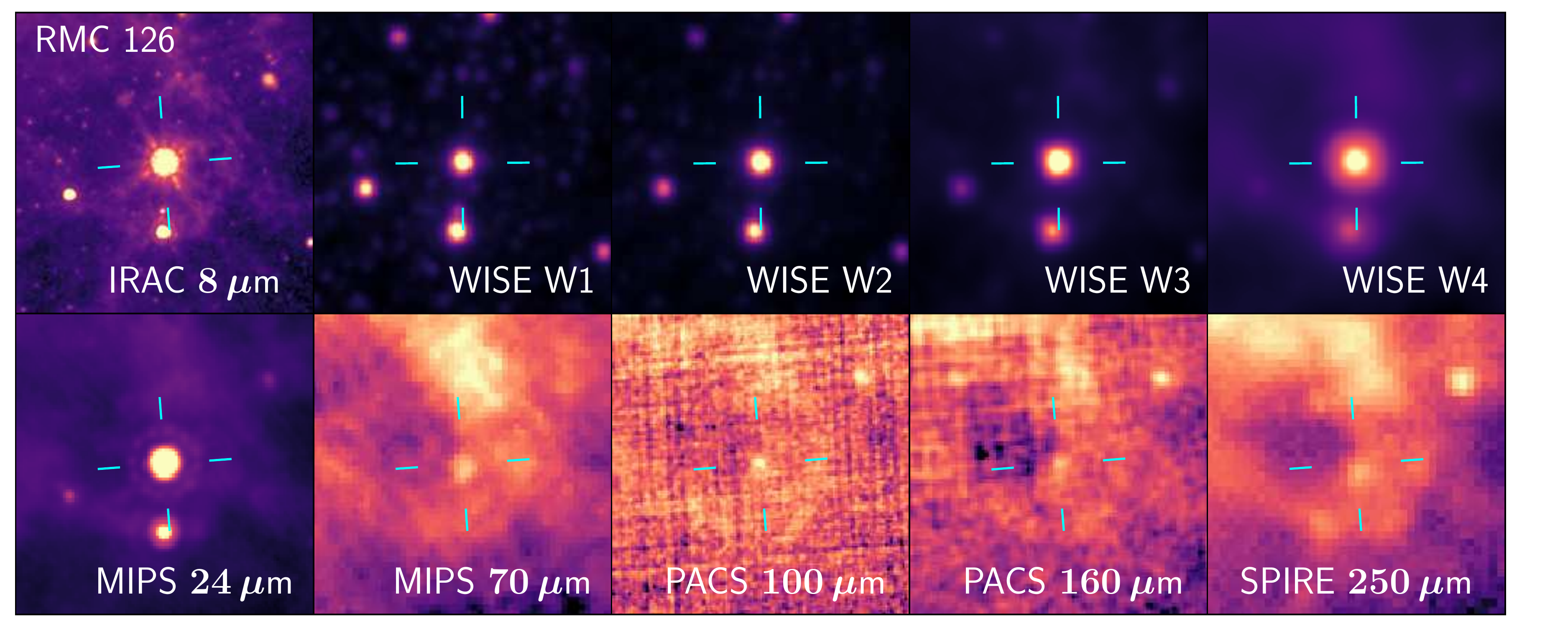}
    \end{minipage}    
   \caption{sgB[e] stars as in literature (see Class~3 sources in  Sect.~\ref{sec:sedreview}). The legend for the photometric points can be found in Fig. \ref{fig:cmd-LMC} of Sect. \ref{sec:review}.}
\label{fig:group3}
\end{figure*}

\begin{figure*}

    \centering
    \begin{minipage}{.35\textwidth}
        \centering
        \includegraphics[width=1\linewidth]{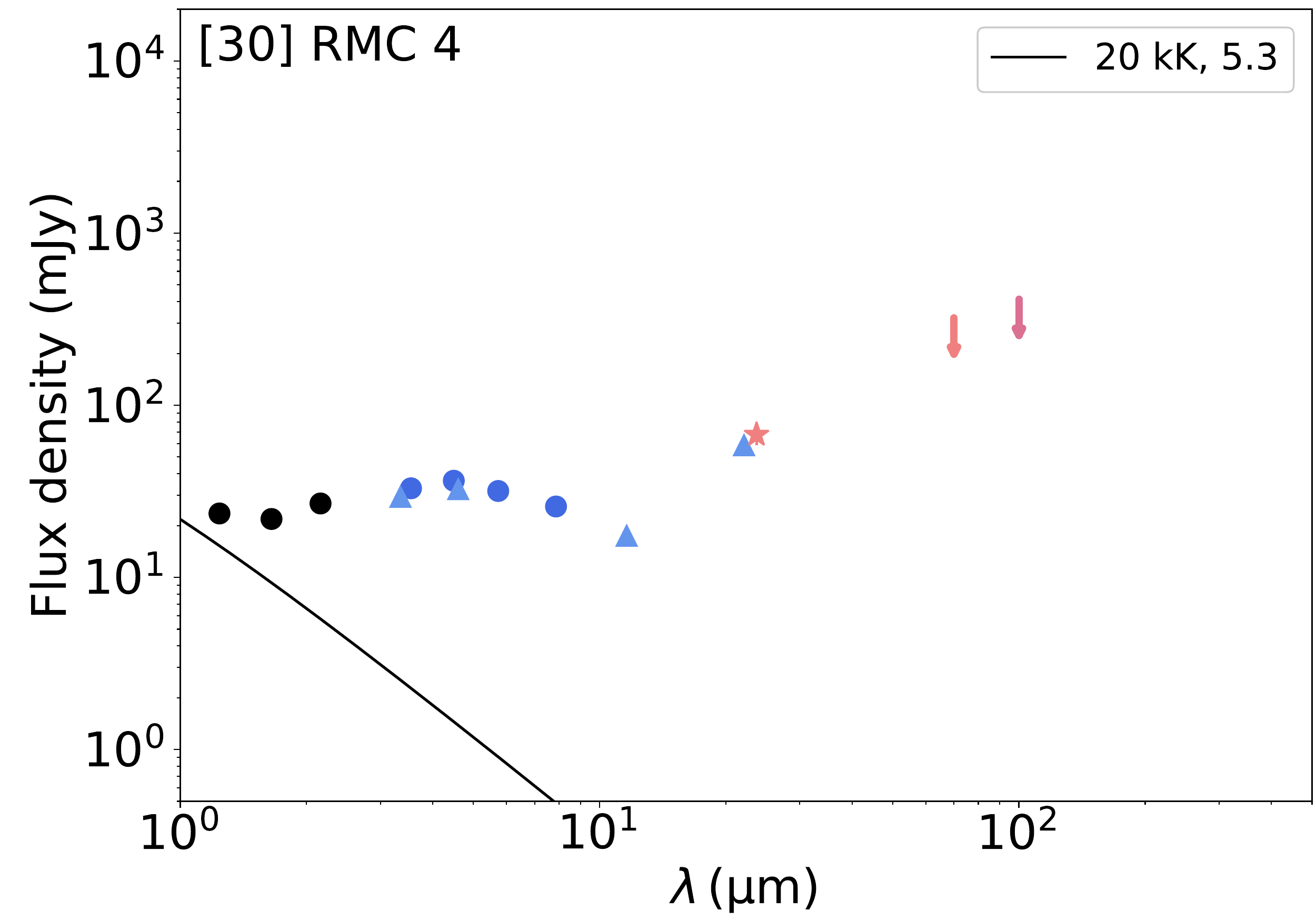}

    \end{minipage}%
    \begin{minipage}{0.65\textwidth}
        \centering
        \includegraphics[width=1\linewidth]{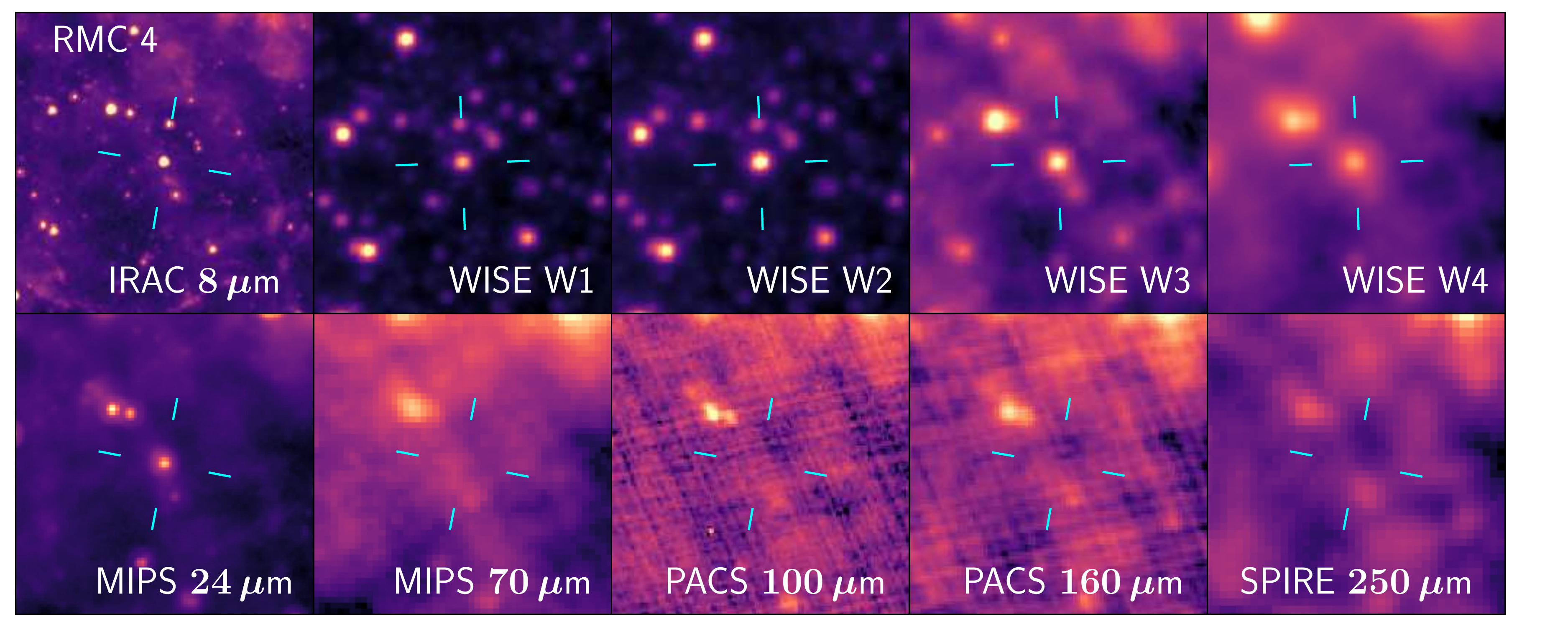}
    \end{minipage}        
    \begin{minipage}{.35\textwidth}
        \centering
        \includegraphics[width=1\linewidth]{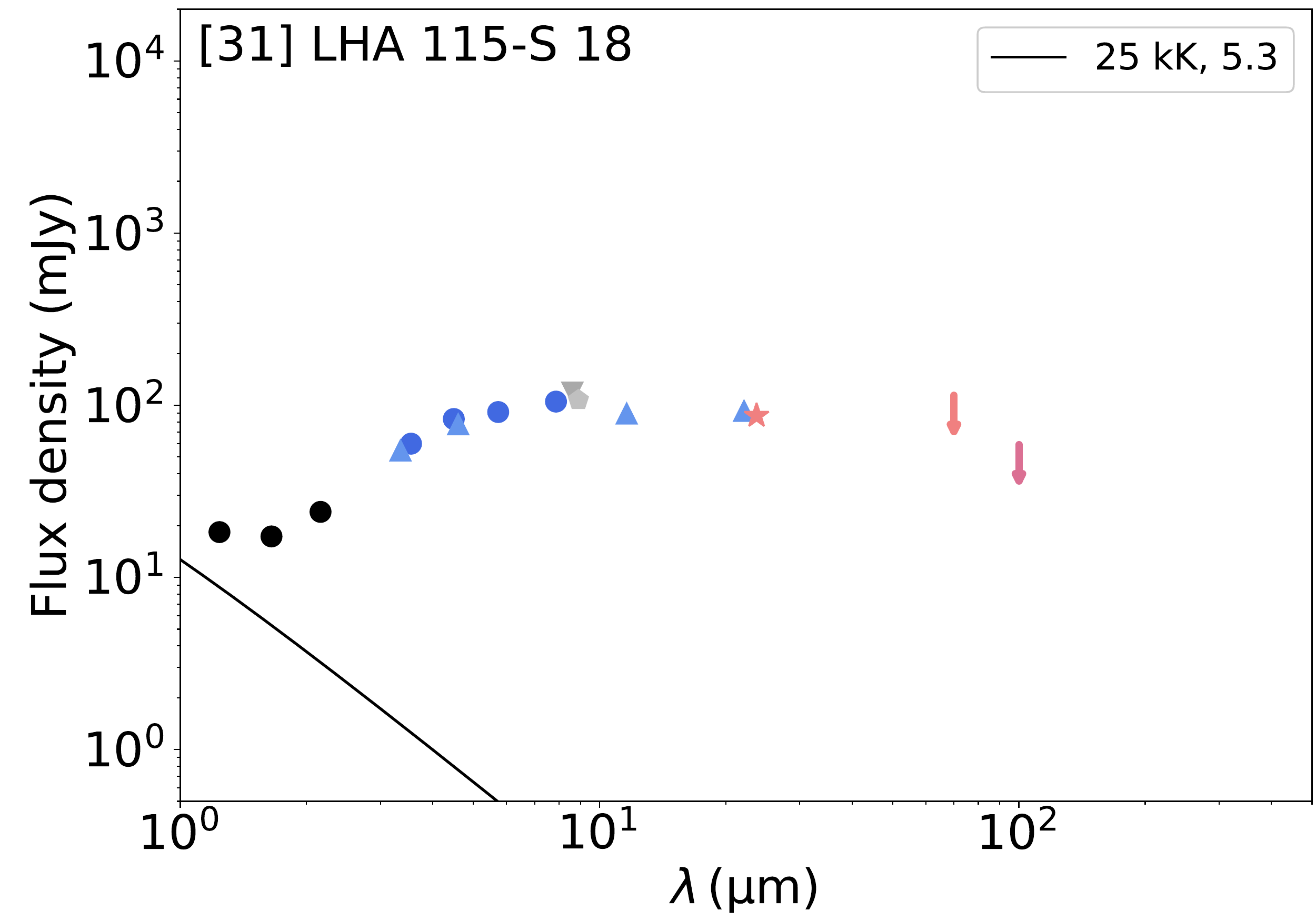}

    \end{minipage}%
    \begin{minipage}{0.65\textwidth}
        \centering
        \includegraphics[width=1\linewidth]{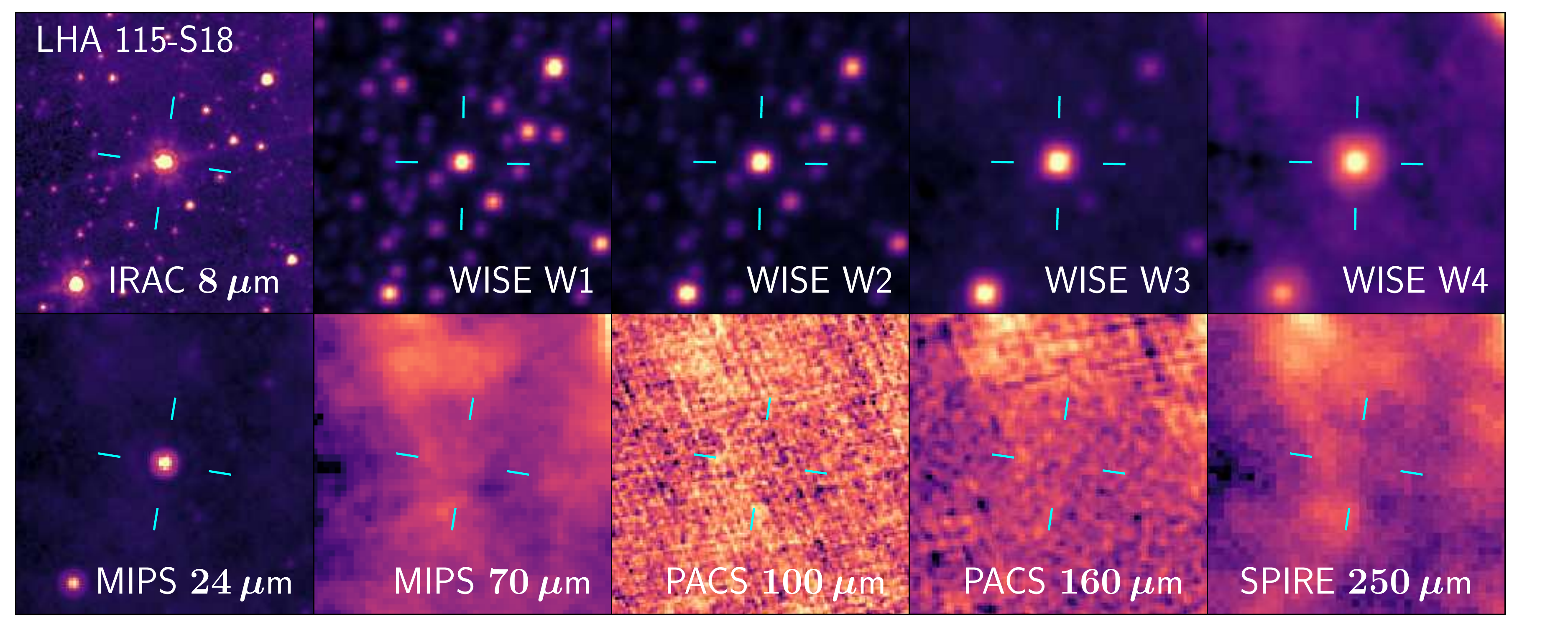}
    \end{minipage}   
       \caption*{Fig.~\ref{fig:group3} continued. We note that photometry for LHA$\,$115-S$\,$18 has not been extinction corrected due to no $A_{\rm V}$ value in the literature.}

\end{figure*}

\begin{figure*}

    \centering
    \begin{minipage}{.35\textwidth}
        \centering
        \includegraphics[width=1\linewidth]{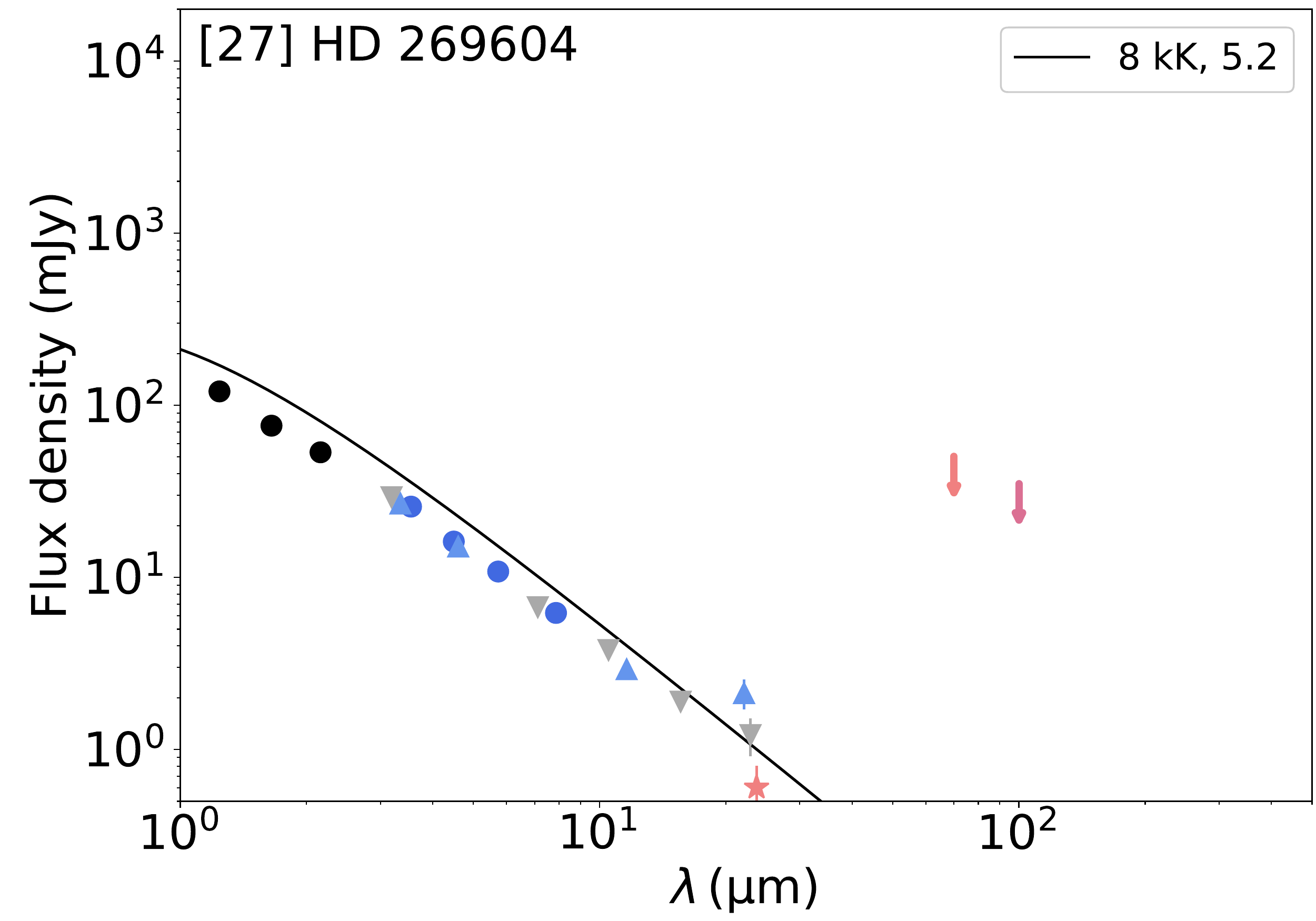}
    \end{minipage}%
    \begin{minipage}{0.65\textwidth}
        \centering
        \includegraphics[width=1\linewidth]{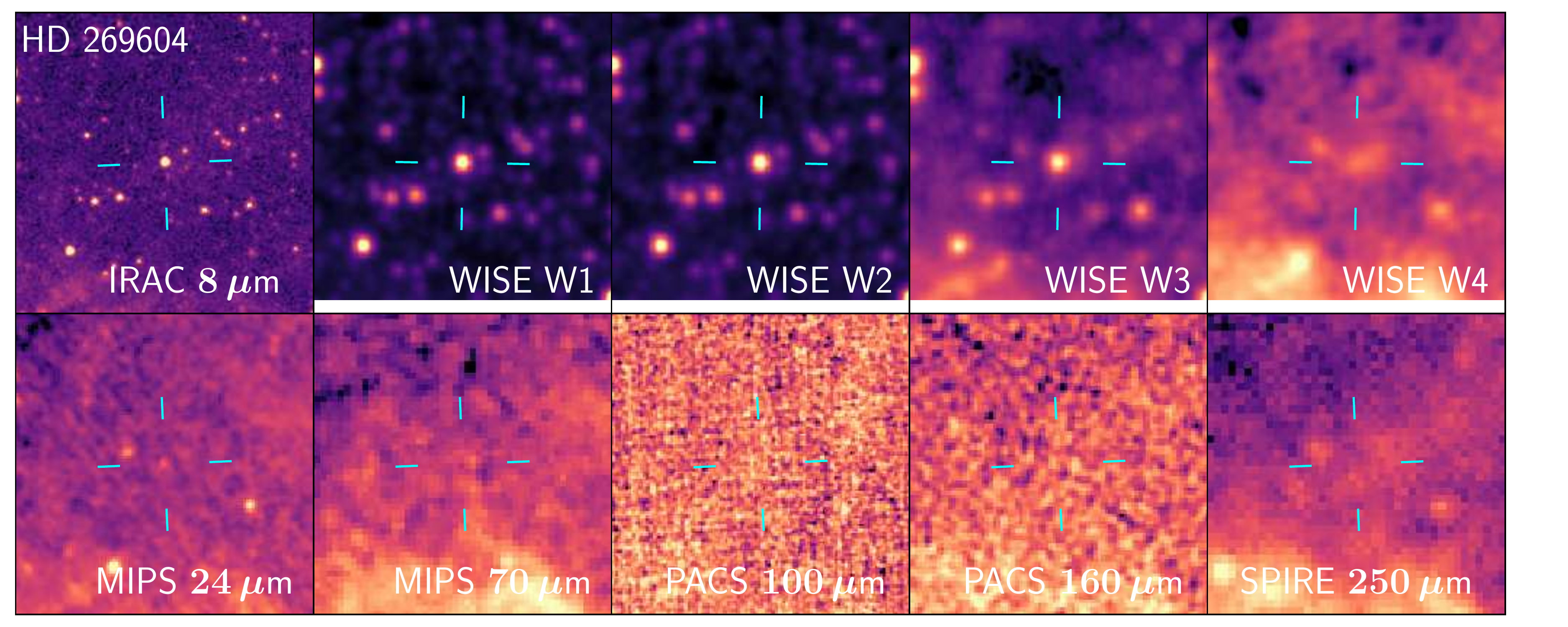}
    \end{minipage}

    \begin{minipage}{.35\textwidth}
        \centering
        \includegraphics[width=1\linewidth]{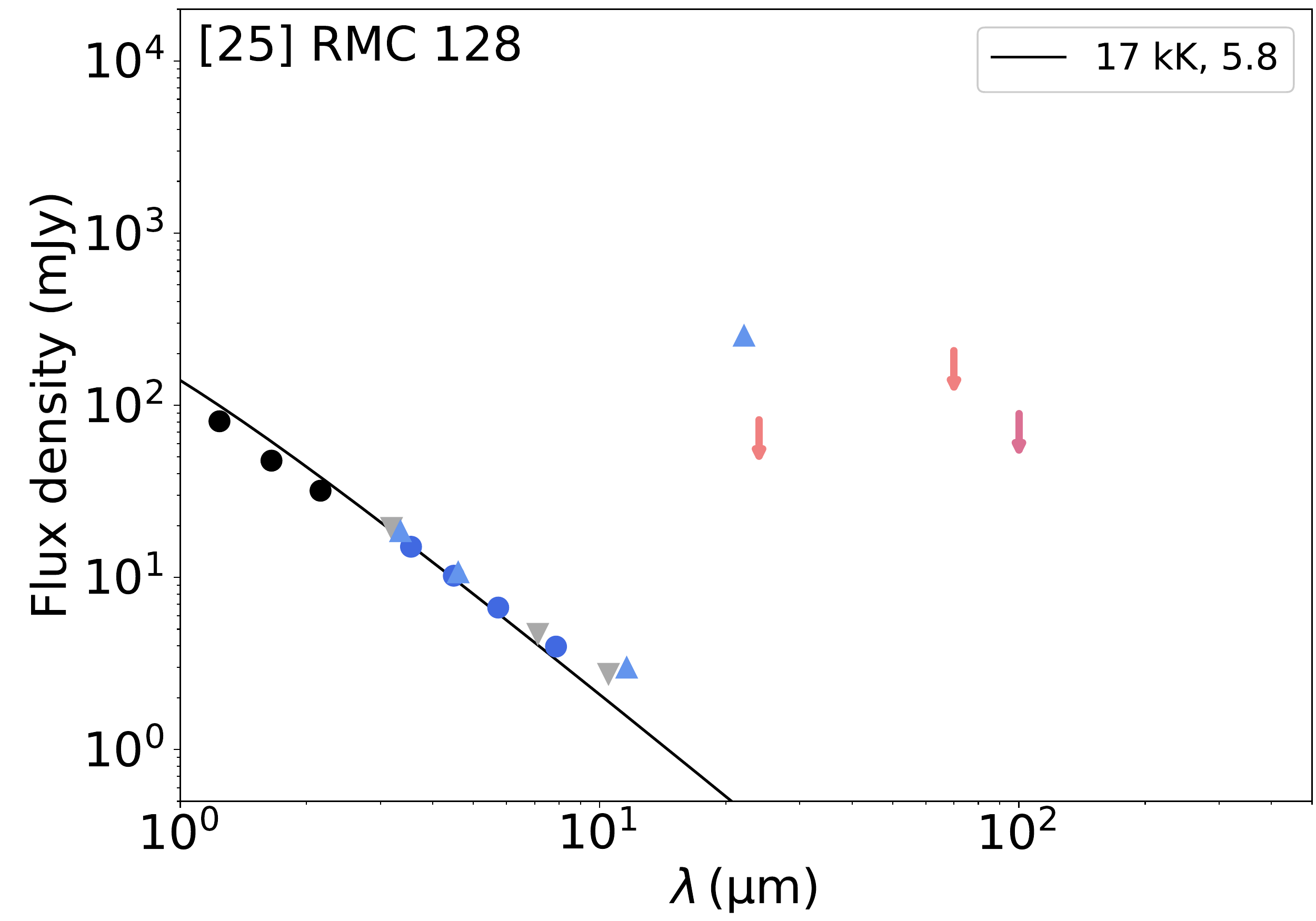}

    \end{minipage}%
    \begin{minipage}{0.65\textwidth}
        \centering
        \includegraphics[width=1\linewidth]{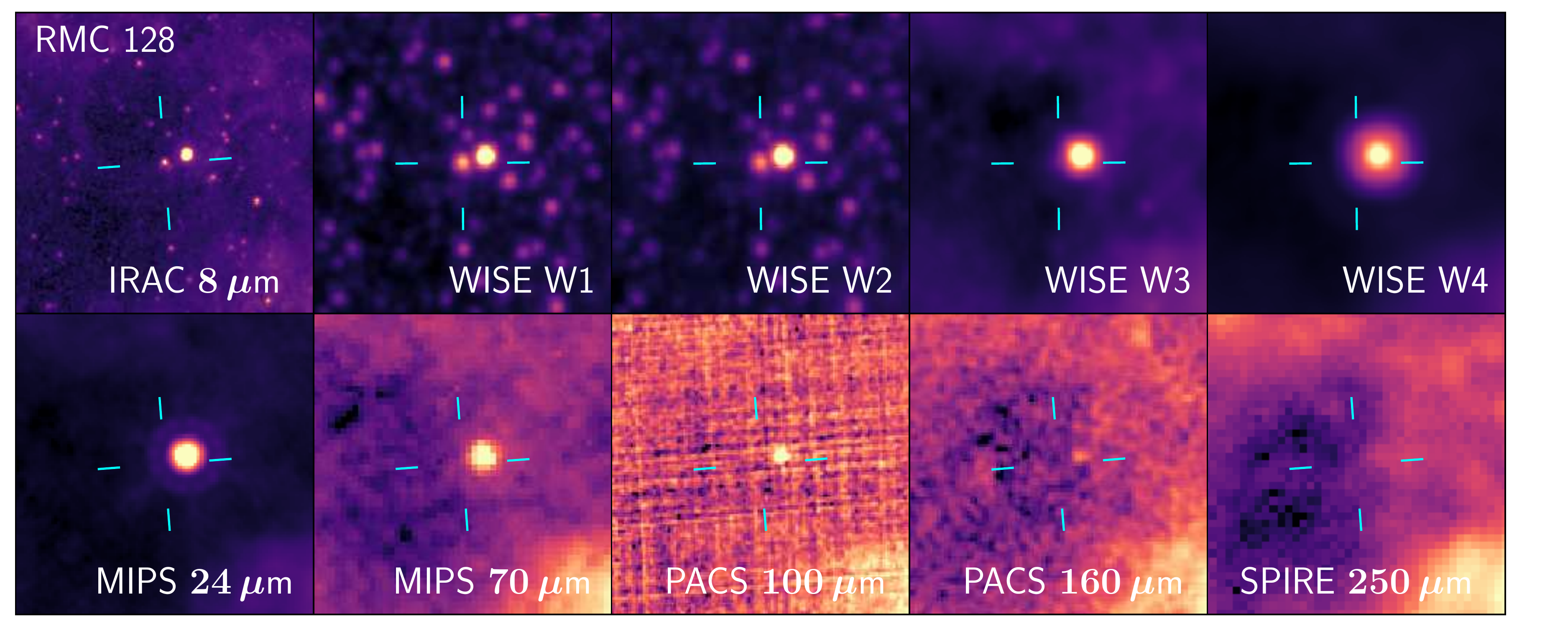}
    \end{minipage}    
    
    \begin{minipage}{.35\textwidth}
        \centering
        \includegraphics[width=1\linewidth]{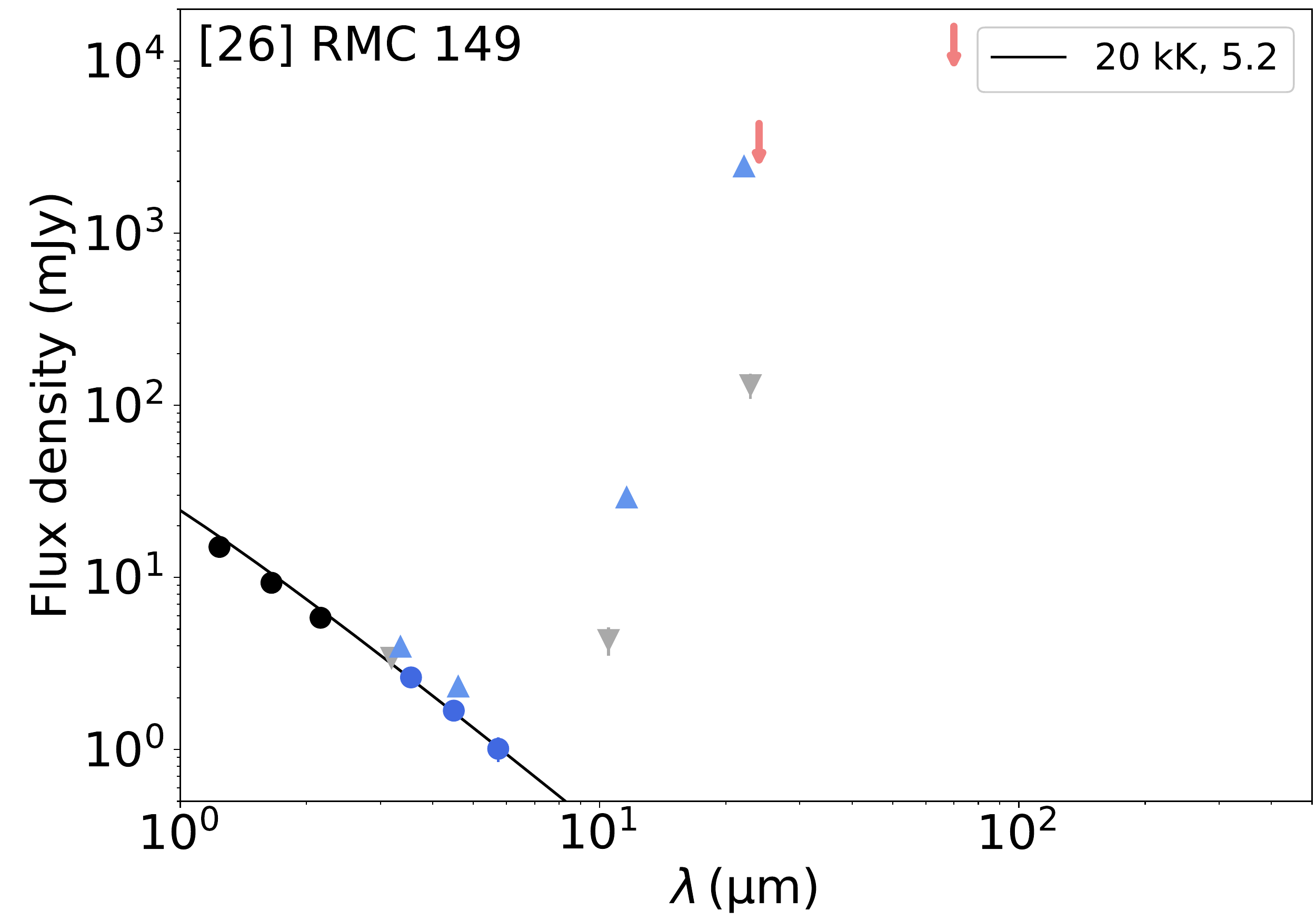}

    \end{minipage}%
    \begin{minipage}{0.65\textwidth}
        \centering
        \includegraphics[width=1\linewidth]{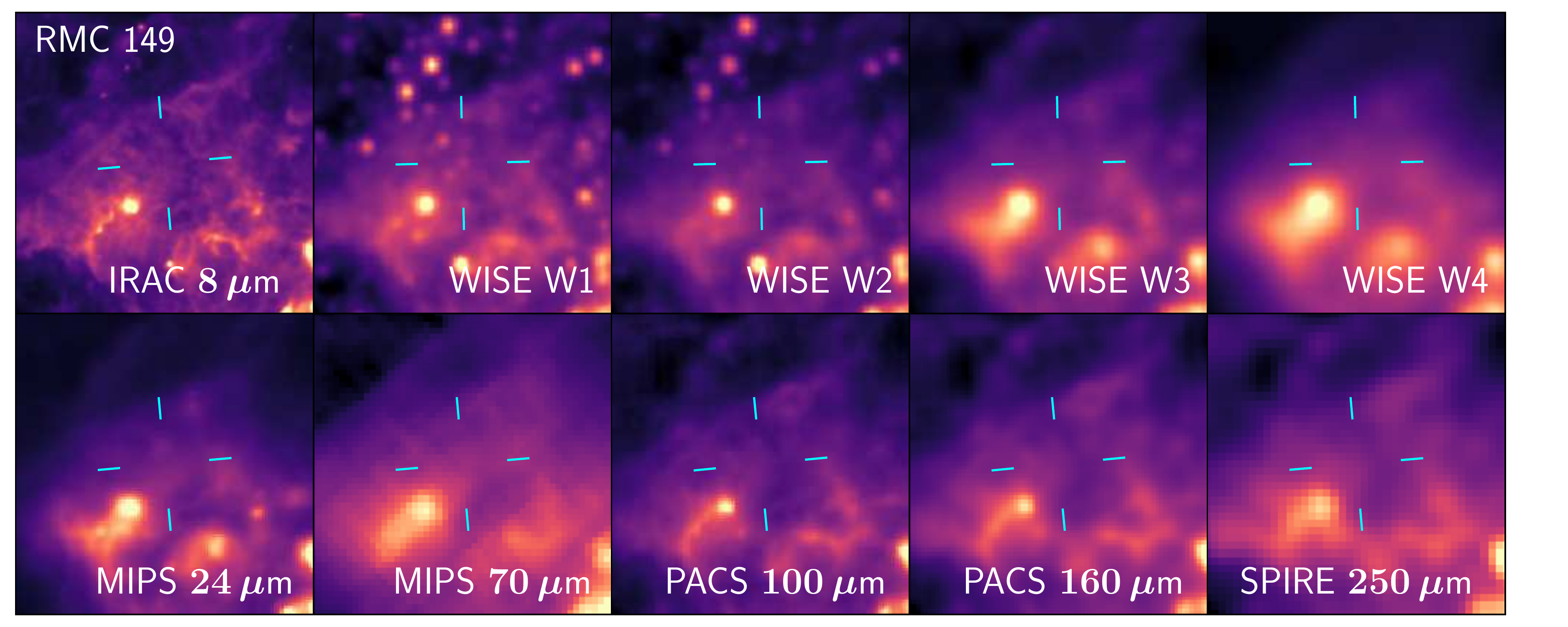}
    \end{minipage}

        \begin{minipage}{.35\textwidth}
        \centering
        \includegraphics[width=1\linewidth]{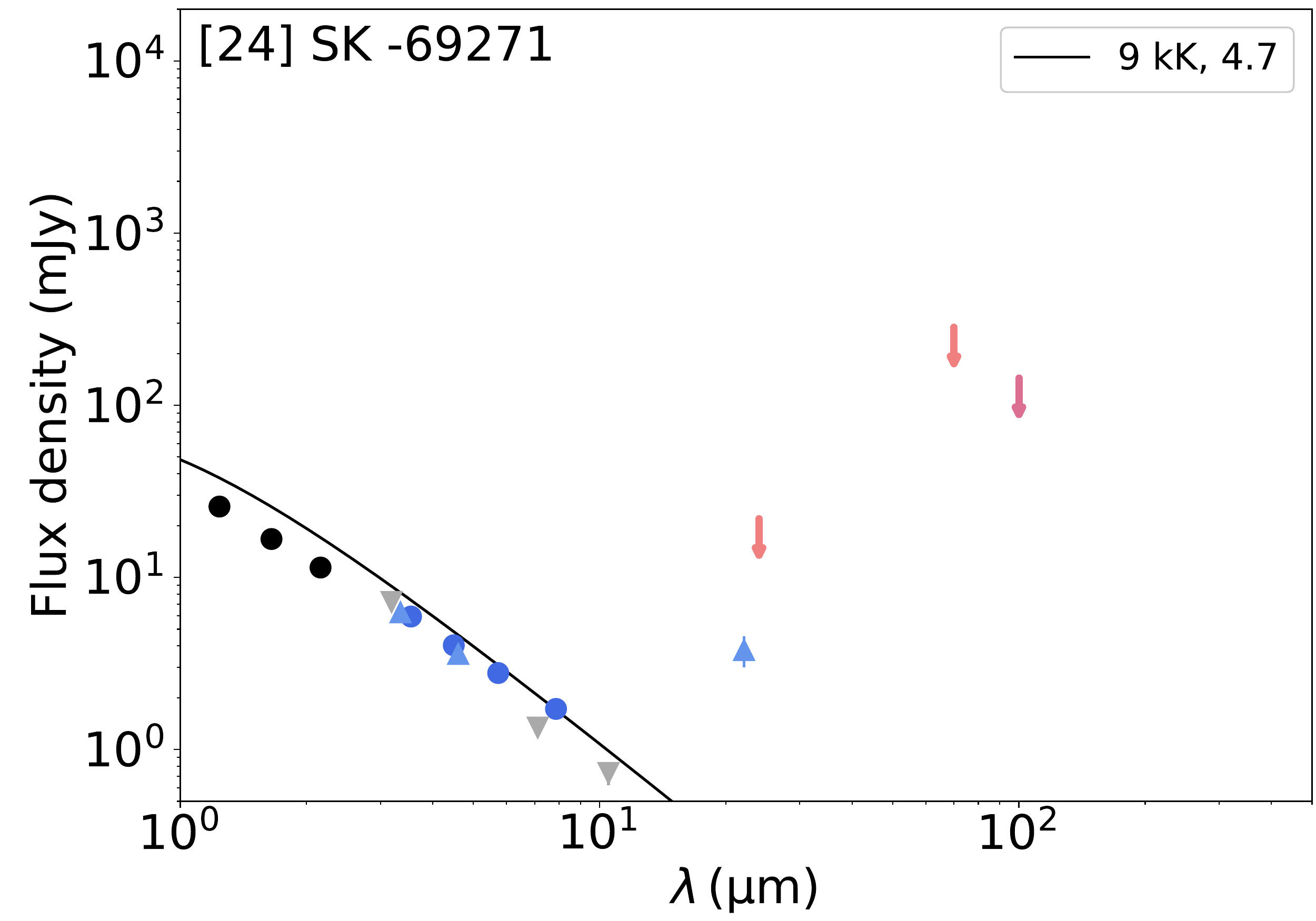}

    \end{minipage}%
    \begin{minipage}{0.65\textwidth}
        \centering
        \includegraphics[width=1\linewidth]{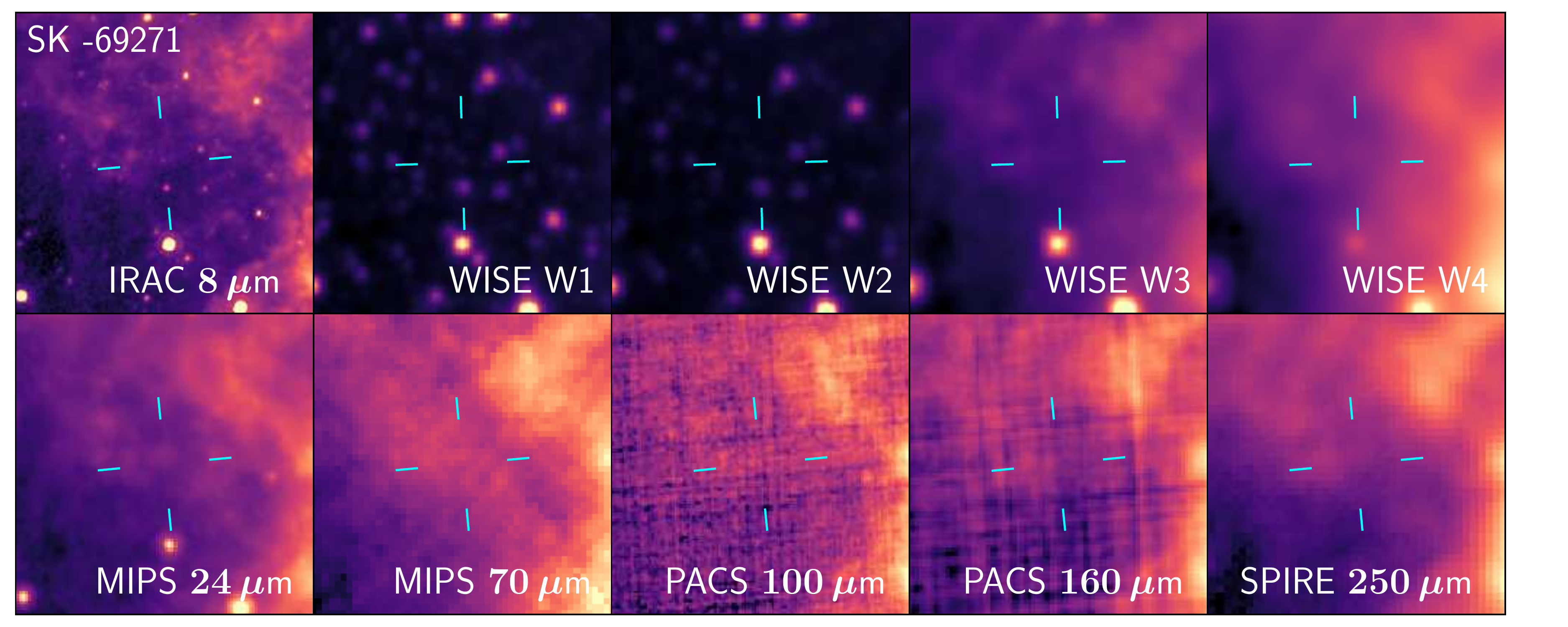}
    \end{minipage}
     \caption{Featureless sources with no recognised LBV membership (see Class~4 sources in  Sect.~\ref{sec:sedreview}). We note that photometry for Sk$\,-69\,271$ has not been extinction corrected due to no $A_{\rm V}$ value in the literature. The legend for the photometric points can be found in Fig. \ref{fig:cmd-LMC} of Sect. \ref{sec:review}.}
     \label{fig:group4}
\end{figure*}

\clearpage

\section{Tables of infrared photometry}
\label{appendix:allphoto}
In this appendix the adopted photometry for all sources from the near-infrared to far-infrared are provided. In the following tables all photometry is provided as frequency flux density in units of mJy. Where catalogue values were provided as magnitudes, these have been converted to flux densities using zero point (zero magnitude) fluxes which are indicated in the table header. 

All 2MASS photometry comes from the 2MASS Point Source Catalogue \citep{2003Cutri}, and were converted to flux densities using the zero point fluxes from \S6.4a of the corresponding Explanatory Supplement\footnote{\url{https://old.ipac.caltech.edu/2mass/releases/allsky/doc/sec6_4a.html}}. AKARI N3, S7, S11, L15 and L24 photometry come from the AKARI-LMC Point-source catalog \citep{2012Kato}, and were converted to flux densities using the zero point fluxes from \citet{2012Kato} (which, although of low precision, are the logical choice to use, as they explain that the photometry was first calibrated into units of Jy and then converted to the magnitudes in the catalogue using these zero point fluxes). AKARI S9W and L18W photometry come from the AKARI/IRC mid-infrared all-sky Survey \citep{2010aIshihara}, originally in Jy units. WISE photometry almost exclusively comes from The AllWISE data Release (updated version, 16-Feb-2021) \citep{2014Cutri}. WISE W4 values for RMC\,116 (ID 6) and Sk$\,-69\,271$ (ID 24) come from The WISE All-Sky data Release \citep{2012Cutri} as they are not detected in ALLWISE (both are faint and hard to distinguish from the background emission). WISE magnitudes were converted to flux densities using zero point fluxes for a $F_\nu\propto\nu^{-2}$ spectrum from \S IV.4.i.1 of the Explanatory Supplement to the WISE All-Sky Data Release Products\footnote{\url{https://wise2.ipac.caltech.edu/docs/release/allsky/expsup/sec4_4h.html}}. All MSX photometry comes from the MSX Point Source Catalog Version 2.3 \citep{2003Egan}, originally in Jy units.

The \emph{Spitzer} and \emph{Herschel} photometry are discussed in the main text, with some values newly extracted in this work.

For completeness, catalogue values that we have identified as incorrect are included, with their value suffixed by a flag character. A `c' denotes that the value is most likely affected by confusion with nearby bright emission, and a `r' denotes that the value is likely affected by the source being resolved. These were assessed with the aid of the flux density distributions and images in Appendix~\ref{appendix:imagesSEDs}.

\begin{sidewaystable*}
\caption{Photometry at infrared wavelengths shorter than $5\,\mu{\rm m}$. }
\label{tab:PhotometryNearIR}
\centering
\begin{tabular}{rr@{}c@{}l@{ }c@{ }r@{}c@{}l@{}lr@{}c@{}l@{ }c@{ }r@{}c@{}l@{}lr@{}c@{}l@{ }c@{ }r@{}c@{}l@{}lr@{}c@{}l@{ }c@{ }r@{}c@{}l@{}lr@{}c@{}l@{ }c@{ }r@{}c@{}l@{}lr@{}c@{}l@{ }c@{ }r@{}c@{}l@{}lr@{}c@{}l@{ }c@{ }r@{}c@{}l@{}lr@{}c@{}l@{ }c@{ }r@{}c@{}l@{}l}
\hline\hline
ID & \multicolumn{8}{c}{2MASS $J$} & \multicolumn{8}{c}{2MASS $H$} & \multicolumn{8}{c}{2MASS $K_{\rm s}$} & \multicolumn{8}{c}{AKARI N3} & \multicolumn{8}{c}{WISE W1} & \multicolumn{8}{c}{IRAC [3.6]} & \multicolumn{8}{c}{IRAC [4.5]} & \multicolumn{8}{c}{WISE W2} \\
   & \multicolumn{8}{c}{$\lambda_{\rm eff}=1.235\,\mu{\rm m}$} & \multicolumn{8}{c}{$\lambda_{\rm eff}=1.662\,\mu{\rm m}$} & \multicolumn{8}{c}{$\lambda_{\rm eff}=2.159\,\mu{\rm m}$} & \multicolumn{8}{c}{$\lambda_{\rm eff}=3.130\,\mu{\rm m}$} & \multicolumn{8}{c}{$\lambda_{\rm eff}=3.353\,\mu{\rm m}$} & \multicolumn{8}{c}{$\lambda_{\rm eff}=3.550\,\mu{\rm m}$} & \multicolumn{8}{c}{$\lambda_{\rm eff}=4.493\,\mu{\rm m}$} & \multicolumn{8}{c}{$\lambda_{\rm eff}=4.603\,\mu{\rm m}$} \\
   & \multicolumn{8}{c}{${\rm ZP}=1594\,{\rm Jy}$} & \multicolumn{8}{c}{${\rm ZP}=1024\,{\rm Jy}$} & \multicolumn{8}{c}{${\rm ZP}=666.7\,{\rm Jy}$} & \multicolumn{8}{c}{${\rm ZP}=343\,{\rm Jy}$} & \multicolumn{8}{c}{${\rm ZP}=306.682\,{\rm Jy}$} & \multicolumn{8}{c}{ } & \multicolumn{8}{c}{ } & \multicolumn{8}{c}{${\rm ZP}=170.663\,{\rm Jy}$} \\
\hline
1&$100$&$.$&$76$&$\pm$&$2$&$.$&$13$&&$81$&$.$&$04$&$\pm$&$1$&$.$&$79$&&$63$&$.$&$49$&$\pm$&$1$&$.$&$52$&&$48$&$.$&$94$&$\pm$&$2$&$.$&$03$&&$50$&$.$&$90$&$\pm$&$1$&$.$&$08$&&$40$&$.$&$7$&$\pm$&$1$&$.$&$6$&&$30$&&&$\pm$&$2$&&&&$31$&$.$&$78$&$\pm$&$0$&$.$&$61$&\\
2&$224$&$.$&$95$&$\pm$&$5$&$.$&$39$&&$160$&$.$&$66$&$\pm$&$4$&$.$&$00$&&$136$&$.$&$12$&$\pm$&$2$&$.$&$88$&&$79$&$.$&$60$&$\pm$&$2$&$.$&$13$&&$88$&$.$&$45$&$\pm$&$1$&$.$&$79$&&$76$&$.$&$8$&$\pm$&$2$&$.$&$3$&&$63$&$.$&$8$&$\pm$&$2$&$.$&$1$&&$65$&$.$&$00$&$\pm$&$1$&$.$&$20$&\\
3&$536$&$.$&$15$&$\pm$&$9$&$.$&$38$&&$403$&$.$&$92$&$\pm$&$18$&$.$&$98$&&$308$&$.$&$13$&$\pm$&$7$&$.$&$10$&&$339$&$.$&$86$&$\pm$&$19$&$.$&$73$&&$158$&$.$&$59$&$\pm$&$3$&$.$&$21$&&$218$&&&$\pm$&$5$&&&&$154$&$.$&$1$&$\pm$&$3$&$.$&$5$&&$126$&$.$&$98$&$\pm$&$2$&$.$&$22$&\\
4&$195$&$.$&$02$&$\pm$&$4$&$.$&$13$&&$139$&$.$&$41$&$\pm$&$3$&$.$&$21$&&$105$&$.$&$28$&$\pm$&$2$&$.$&$42$&&$52$&$.$&$01$&$\pm$&$1$&$.$&$68$&&$57$&$.$&$21$&$\pm$&$1$&$.$&$21$&&$34$&$.$&$04$&$\pm$&$0$&$.$&$76$&&$25$&$.$&$19$&$\pm$&$0$&$.$&$72$&&$34$&$.$&$59$&$\pm$&$0$&$.$&$64$&\\
5&$24$&$.$&$33$&$\pm$&$0$&$.$&$58$&&$17$&$.$&$37$&$\pm$&$0$&$.$&$51$&&$14$&$.$&$73$&$\pm$&$0$&$.$&$35$&&$31$&$.$&$66$&$\pm$&$1$&$.$&$02$&&$60$&$.$&$69$&$\pm$&$1$&$.$&$29$&&$46$&$.$&$86$&$\pm$&$2$&$.$&$10$&&$39$&$.$&$83$&$\pm$&$1$&$.$&$31$&&$42$&$.$&$05$&$\pm$&$0$&$.$&$77$&\\
6&$104$&$.$&$25$&$\pm$&$2$&$.$&$30$&&$72$&$.$&$96$&$\pm$&$1$&$.$&$41$&&$49$&$.$&$65$&$\pm$&$1$&$.$&$05$&&$29$&$.$&$68$&$\pm$&$1$&$.$&$09$&&$25$&$.$&$82$&$\pm$&$0$&$.$&$55$&&$24$&$.$&$15$&$\pm$&$0$&$.$&$76$&&$17$&$.$&$24$&$\pm$&$0$&$.$&$37$&&$15$&$.$&$28$&$\pm$&$0$&$.$&$30$&\\
7&$105$&$.$&$31$&$\pm$&$2$&$.$&$13$&&$72$&$.$&$16$&$\pm$&$1$&$.$&$66$&&$52$&$.$&$28$&$\pm$&$1$&$.$&$16$&&\multicolumn{7}{c}{---}&&$47$&$.$&$50$&$\pm$&$0$&$.$&$96$&&$42$&$.$&$52$&$\pm$&$1$&$.$&$19$&&$31$&$.$&$07$&$\pm$&$0$&$.$&$90$&&$29$&$.$&$49$&$\pm$&$0$&$.$&$54$&\\
8&$85$&$.$&$92$&$\pm$&$1$&$.$&$82$&&$56$&$.$&$58$&$\pm$&$1$&$.$&$41$&&$40$&$.$&$28$&$\pm$&$0$&$.$&$96$&&\multicolumn{7}{c}{---}&&$271$&$.$&$57$&$\pm$&$7$&$.$&$25$&&$25$&$.$&$24$&$\pm$&$0$&$.$&$74$&&$26$&$.$&$44$&$\pm$&$0$&$.$&$69$&&$161$&$.$&$93$&$\pm$&$2$&$.$&$98$&\\
9&$144$&$.$&$97$&$\pm$&$3$&$.$&$20$&&$104$&$.$&$30$&$\pm$&$2$&$.$&$69$&&$78$&$.$&$55$&$\pm$&$1$&$.$&$81$&&$54$&$.$&$01$&$\pm$&$3$&$.$&$04$&&$40$&$.$&$39$&$\pm$&$0$&$.$&$86$&&$38$&$.$&$19$&$\pm$&$0$&$.$&$93$&&$27$&$.$&$2$&$\pm$&$0$&$.$&$6$&&$24$&$.$&$74$&$\pm$&$0$&$.$&$48$&\\
10&$177$&$.$&$05$&$\pm$&$3$&$.$&$59$&&$132$&$.$&$53$&$\pm$&$3$&$.$&$05$&&$117$&$.$&$90$&$\pm$&$2$&$.$&$17$&&$103$&$.$&$49$&$\pm$&$2$&$.$&$38$&&$88$&$.$&$20$&$\pm$&$1$&$.$&$87$&&$96$&$.$&$05$&$\pm$&$3$&$.$&$76$&&$84$&$.$&$27$&$\pm$&$2$&$.$&$42$&&$71$&$.$&$28$&$\pm$&$1$&$.$&$31$&\\
11&$97$&$.$&$29$&$\pm$&$2$&$.$&$06$&&$76$&$.$&$05$&$\pm$&$1$&$.$&$75$&&$65$&$.$&$94$&$\pm$&$1$&$.$&$52$&&$40$&$.$&$78$&$\pm$&$1$&$.$&$54$&&$40$&$.$&$54$&$\pm$&$0$&$.$&$82$&&$39$&$.$&$51$&$\pm$&$1$&$.$&$49$&&$31$&$.$&$99$&$\pm$&$0$&$.$&$79$&&$29$&$.$&$39$&$\pm$&$0$&$.$&$51$&\\
12&$13$&$.$&$53$&$\pm$&$0$&$.$&$36$&&$9$&$.$&$49$&$\pm$&$0$&$.$&$32$&&$6$&$.$&$50$&$\pm$&$0$&$.$&$23$&&$4$&$.$&$49$&$\pm$&$0$&$.$&$17$&&$3$&$.$&$73$&$\pm$&$0$&$.$&$09$&&$3$&$.$&$52$&$\pm$&$0$&$.$&$12$&&$2$&$.$&$56$&$\pm$&$0$&$.$&$10$&&$2$&$.$&$38$&$\pm$&$0$&$.$&$05$&\\
13&$29$&$.$&$82$&$\pm$&$0$&$.$&$63$&&$18$&$.$&$60$&$\pm$&$0$&$.$&$46$&&$13$&$.$&$62$&$\pm$&$0$&$.$&$31$&&$9$&$.$&$94$&$\pm$&$0$&$.$&$30$&&$8$&$.$&$18$&$\pm$&$0$&$.$&$17$&&$6$&$.$&$50$&$\pm$&$0$&$.$&$32$&&$4$&$.$&$76$&$\pm$&$0$&$.$&$36$&&$5$&$.$&$75$&$\pm$&$0$&$.$&$11$&\\
14&$24$&$.$&$53$&$\pm$&$0$&$.$&$59$&&$16$&$.$&$18$&$\pm$&$0$&$.$&$40$&&$11$&$.$&$61$&$\pm$&$0$&$.$&$27$&&\multicolumn{7}{c}{---}&&$6$&$.$&$95$&$\pm$&$0$&$.$&$15$&&$7$&$.$&$03$&$\pm$&$0$&$.$&$33$&&$7$&$.$&$70$&$\pm$&$0$&$.$&$31$&&$5$&$.$&$32$&$\pm$&$0$&$.$&$10$&\\
15&$91$&$.$&$47$&$\pm$&$1$&$.$&$77$&&$64$&$.$&$25$&$\pm$&$1$&$.$&$48$&&$47$&$.$&$99$&$\pm$&$0$&$.$&$84$&&\multicolumn{7}{c}{---}&&$23$&$.$&$63$&$\pm$&$0$&$.$&$50$&&$26$&$.$&$12$&$\pm$&$0$&$.$&$99$&&$20$&$.$&$14$&$\pm$&$0$&$.$&$53$&&$15$&$.$&$04$&$\pm$&$0$&$.$&$28$&\\
16&$32$&$.$&$22$&$\pm$&$0$&$.$&$62$&&$20$&$.$&$41$&$\pm$&$0$&$.$&$47$&&$13$&$.$&$25$&$\pm$&$0$&$.$&$32$&&\multicolumn{7}{c}{---}&&$6$&$.$&$39$&$\pm$&$0$&$.$&$14$&&$6$&$.$&$04$&$\pm$&$0$&$.$&$24$&&$3$&$.$&$99$&$\pm$&$0$&$.$&$14$&&$3$&$.$&$55$&$\pm$&$0$&$.$&$07$&\\
17&$127$&$.$&$55$&$\pm$&$2$&$.$&$58$&&$89$&$.$&$68$&$\pm$&$1$&$.$&$82$&&$61$&$.$&$54$&$\pm$&$1$&$.$&$19$&&\multicolumn{7}{c}{---}&&$40$&$.$&$73$&$\pm$&$0$&$.$&$83$&&$32$&$.$&$91$&$\pm$&$0$&$.$&$98$&&$23$&$.$&$43$&$\pm$&$0$&$.$&$60$&&$27$&$.$&$70$&$\pm$&$0$&$.$&$54$&\\
18&$34$&$.$&$68$&$\pm$&$0$&$.$&$70$&&$24$&$.$&$18$&$\pm$&$0$&$.$&$58$&&$18$&$.$&$55$&$\pm$&$0$&$.$&$32$&&\multicolumn{7}{c}{---}&&$13$&$.$&$85$&$\pm$&$0$&$.$&$29$&&$12$&$.$&$01$&$\pm$&$0$&$.$&$38$&&$10$&$.$&$45$&$\pm$&$0$&$.$&$29$&&$10$&$.$&$57$&$\pm$&$0$&$.$&$19$&\\
19&$147$&$.$&$67$&$\pm$&$2$&$.$&$99$&&$137$&$.$&$50$&$\pm$&$3$&$.$&$17$&&$191$&$.$&$39$&$\pm$&$3$&$.$&$88$&&\multicolumn{7}{c}{---}&&$343$&$.$&$79$&$\pm$&$10$&$.$&$77$&&$367$&$.$&$40$&$\pm$&$17$&$.$&$76$&&$477$&$.$&$40$&$\pm$&$18$&$.$&$67$&&$503$&$.$&$66$&$\pm$&$11$&$.$&$60$&\\
20&$275$&$.$&$73$&$\pm$&$5$&$.$&$59$&&$393$&$.$&$64$&$\pm$&$7$&$.$&$61$&&$375$&$.$&$95$&$\pm$&$7$&$.$&$27$&&\multicolumn{7}{c}{---}&&$254$&$.$&$85$&$\pm$&$7$&$.$&$04$&&$259$&$.$&$30$&$\pm$&$13$&$.$&$09$&&$184$&$.$&$60$&$\pm$&$5$&$.$&$93$&&$157$&$.$&$81$&$\pm$&$2$&$.$&$62$&\\
21&$113$&$.$&$37$&$\pm$&$2$&$.$&$30$&&$128$&$.$&$80$&$\pm$&$2$&$.$&$97$&&$274$&$.$&$87$&$\pm$&$5$&$.$&$32$&&$639$&$.$&$28$&$\pm$&$25$&$.$&$91$&&$649$&$.$&$07$&$\pm$&$41$&$.$&$28$&&$855$&&&$\pm$&$22$&&&&$1060$&&&$\pm$&$30$&&&&$1158$&$.$&$08$&$\pm$&$53$&$.$&$35$&\\
22&$104$&$.$&$73$&$\pm$&$2$&$.$&$03$&&$117$&$.$&$90$&$\pm$&$2$&$.$&$28$&&$241$&$.$&$62$&$\pm$&$4$&$.$&$67$&&$495$&$.$&$33$&$\pm$&$21$&$.$&$45$&&$690$&$.$&$38$&$\pm$&$40$&$.$&$72$&&$700$&$.$&$20$&$\pm$&$35$&$.$&$43$&&$827$&$.$&$10$&$\pm$&$26$&$.$&$82$&&$912$&$.$&$30$&$\pm$&$43$&$.$&$71$&\\
23&$141$&$.$&$93$&$\pm$&$2$&$.$&$75$&&$127$&$.$&$26$&$\pm$&$3$&$.$&$05$&&$208$&$.$&$90$&$\pm$&$4$&$.$&$04$&&$405$&$.$&$97$&$\pm$&$14$&$.$&$96$&&$465$&$.$&$90$&$\pm$&$18$&$.$&$89$&&$513$&&&$\pm$&$13$&&&&$652$&&&$\pm$&$17$&&&&$733$&$.$&$40$&$\pm$&$29$&$.$&$05$&\\
24&$25$&$.$&$78$&$\pm$&$0$&$.$&$62$&&$16$&$.$&$70$&$\pm$&$0$&$.$&$52$&&$11$&$.$&$40$&$\pm$&$0$&$.$&$28$&&$7$&$.$&$20$&$\pm$&$0$&$.$&$17$&&$6$&$.$&$26$&$\pm$&$0$&$.$&$13$&&$5$&$.$&$92$&$\pm$&$0$&$.$&$17$&&$4$&$.$&$03$&$\pm$&$0$&$.$&$10$&&$3$&$.$&$59$&$\pm$&$0$&$.$&$07$&\\
25&$73$&$.$&$81$&$\pm$&$1$&$.$&$84$&&$45$&$.$&$03$&$\pm$&$1$&$.$&$29$&&$30$&$.$&$67$&$\pm$&$0$&$.$&$82$&&$18$&$.$&$92$&$\pm$&$0$&$.$&$78$&&$18$&$.$&$13$&$\pm$&$0$&$.$&$38$&&$14$&$.$&$75$&$\pm$&$0$&$.$&$67$&&$10$&$.$&$06$&$\pm$&$0$&$.$&$24$&&$10$&$.$&$53$&$\pm$&$0$&$.$&$19$&\\
26&$13$&$.$&$95$&$\pm$&$0$&$.$&$36$&&$8$&$.$&$87$&$\pm$&$0$&$.$&$28$&&$5$&$.$&$64$&$\pm$&$0$&$.$&$15$&&$3$&$.$&$34$&$\pm$&$0$&$.$&$20$&&$3$&$.$&$88$&$\pm$&$0$&$.$&$29$&&$2$&$.$&$58$&$\pm$&$0$&$.$&$14$&&$1$&$.$&$66$&$\pm$&$0$&$.$&$09$&&$2$&$.$&$29$&$\pm$&$0$&$.$&$24$&\\
27&$111$&$.$&$71$&$\pm$&$2$&$.$&$26$&&$72$&$.$&$56$&$\pm$&$1$&$.$&$60$&&$51$&$.$&$61$&$\pm$&$1$&$.$&$00$&&$28$&$.$&$66$&$\pm$&$0$&$.$&$84$&&$26$&$.$&$34$&$\pm$&$0$&$.$&$56$&&$25$&$.$&$32$&$\pm$&$0$&$.$&$67$&&$15$&$.$&$93$&$\pm$&$0$&$.$&$43$&&$14$&$.$&$91$&$\pm$&$0$&$.$&$27$&\\
28&$57$&$.$&$29$&$\pm$&$1$&$.$&$37$&&$40$&$.$&$54$&$\pm$&$0$&$.$&$93$&&$32$&$.$&$83$&$\pm$&$0$&$.$&$70$&&\multicolumn{7}{c}{---}&&$11$&$.$&$85$&$\pm$&$0$&$.$&$25$&&$10$&$.$&$8$&$\pm$&$0$&$.$&$4$&&$9$&$.$&$14$&$\pm$&$0$&$.$&$21$&&$8$&$.$&$49$&$\pm$&$0$&$.$&$16$&\\
29&$207$&$.$&$63$&$\pm$&$5$&$.$&$16$&&$157$&$.$&$72$&$\pm$&$3$&$.$&$49$&&$109$&$.$&$03$&$\pm$&$2$&$.$&$31$&&\multicolumn{7}{c}{---}&&$88$&$.$&$94$&$\pm$&$1$&$.$&$88$&&$64$&$.$&$0$&$\pm$&$2$&$.$&$5$&&$44$&$.$&$7$&$\pm$&$1$&$.$&$5$&&$52$&$.$&$50$&$\pm$&$0$&$.$&$97$&\\
30&$21$&$.$&$66$&$\pm$&$0$&$.$&$48$&&$20$&$.$&$73$&$\pm$&$0$&$.$&$46$&&$25$&$.$&$94$&$\pm$&$0$&$.$&$50$&&\multicolumn{7}{c}{---}&&$28$&$.$&$78$&$\pm$&$0$&$.$&$61$&&$32$&$.$&$3$&$\pm$&$0$&$.$&$5$&&$35$&$.$&$9$&$\pm$&$0$&$.$&$3$&&$32$&$.$&$04$&$\pm$&$0$&$.$&$59$&\\
31&$18$&$.$&$32$&$\pm$&$0$&$.$&$56$&&$17$&$.$&$29$&$\pm$&$0$&$.$&$62$&&$24$&$.$&$01$&$\pm$&$0$&$.$&$57$&&\multicolumn{7}{c}{---}&&$54$&$.$&$54$&$\pm$&$1$&$.$&$16$&&$59$&$.$&$8$&$\pm$&$2$&$.$&$3$&&$83$&$.$&$2$&$\pm$&$3$&$.$&$1$&&$77$&$.$&$44$&$\pm$&$1$&$.$&$43$&\\
\hline
\end{tabular}
\tablefoot{Values are frequency flux density in units of mJy.}
\end{sidewaystable*}

\begin{sidewaystable*}
\caption{Photometry at wavelengths $5$--$20\,\mu{\rm m}$.  }
\label{tab:PhotometryMidIR}
\centering

\begin{footnotesize}
\begin{tabular}{rr@{}c@{}l@{ }c@{ }r@{}c@{}l@{}lr@{}c@{}l@{ }c@{ }r@{}c@{}l@{}lr@{}c@{}l@{ }c@{ }r@{}c@{}l@{}lr@{}c@{}l@{ }c@{ }r@{}c@{}l@{}lr@{}c@{}l@{ }c@{ }r@{}c@{}l@{}lr@{}c@{}l@{ }c@{ }r@{}c@{}l@{}lr@{}c@{}l@{ }c@{ }r@{}c@{}l@{}lr@{}c@{}l@{ }c@{ }r@{}c@{}l@{}lr@{}c@{}l@{ }c@{ }r@{}c@{}l@{}l}
\hline\hline
ID & \multicolumn{8}{c}{IRAC [5.8]} & \multicolumn{8}{c}{AKARI S7} & \multicolumn{8}{c}{IRAC [8]} & \multicolumn{8}{c}{MSX A} & \multicolumn{8}{c}{AKARI S9W} & \multicolumn{8}{c}{AKARI S11} & \multicolumn{8}{c}{WISE W3} & \multicolumn{8}{c}{AKARI L15} & \multicolumn{8}{c}{AKARI L18W} \\
   & \multicolumn{8}{c}{$\lambda_{\rm eff}=5.731\,\mu{\rm m}$} & \multicolumn{8}{c}{$\lambda_{\rm eff}=6.954\,\mu{\rm m}$} & \multicolumn{8}{c}{$\lambda_{\rm eff}=7.872\,\mu{\rm m}$} & \multicolumn{8}{c}{$\lambda_{\rm eff}=7.951\,\mu{\rm m}$} & \multicolumn{8}{c}{$\lambda_{\rm eff}=8.228\,\mu{\rm m}$} & \multicolumn{8}{c}{$\lambda_{\rm eff}=10.193\,\mu{\rm m}$} & \multicolumn{8}{c}{$\lambda_{\rm eff}=11.560\,\mu{\rm m}$} & \multicolumn{8}{c}{$\lambda_{\rm eff}=15.231\,\mu{\rm m}$} & \multicolumn{8}{c}{$\lambda_{\rm eff}=17.609\,\mu{\rm m}$} \\
   & \multicolumn{8}{c}{ } & \multicolumn{8}{c}{${\rm ZP}=75.0\,{\rm Jy}$} & \multicolumn{8}{c}{ } & \multicolumn{8}{c}{ } & \multicolumn{8}{c}{${\rm ZP}=38.3\,{\rm Jy}$} & \multicolumn{8}{c}{ } & \multicolumn{8}{c}{${\rm ZP}=29.045\,{\rm Jy}$} & \multicolumn{8}{c}{${\rm ZP}=16.0\,{\rm Jy}$} & \multicolumn{8}{c}{ } \\
\hline
1&$22$&$.$&$1$&$\pm$&$0$&$.$&$8$&&$21$&$.$&$08$&$\pm$&$1$&$.$&$07$&&$16$&$.$&$3$&$\pm$&$1$&$.$&$1$&&\multicolumn{7}{c}{---}&&\multicolumn{7}{c}{---}&&$85$&$.$&$98$&$\pm$&$9$&$.$&$52$&c&$170$&$.$&$09$&$\pm$&$3$&$.$&$76$&c&$400$&$.$&$79$&$\pm$&$41$&$.$&$79$&c&\multicolumn{7}{c}{---}&\\
2&$47$&$.$&$8$&$\pm$&$1$&$.$&$1$&&$48$&$.$&$16$&$\pm$&$0$&$.$&$93$&&$37$&$.$&$6$&$\pm$&$1$&$.$&$1$&&$79$&$.$&$69$&$\pm$&$6$&$.$&$14$&&\multicolumn{7}{c}{---}&&$77$&$.$&$13$&$\pm$&$1$&$.$&$28$&&$102$&$.$&$87$&$\pm$&$1$&$.$&$33$&&$331$&$.$&$53$&$\pm$&$12$&$.$&$83$&&$713$&$.$&$1$&$\pm$&$8$&$.$&$5$&\\
3&$109$&$.$&$70$&$\pm$&$2$&$.$&$10$&&$87$&$.$&$63$&$\pm$&$1$&$.$&$53$&&$74$&$.$&$82$&$\pm$&$1$&$.$&$11$&&$104$&$.$&$5$&$\pm$&$6$&$.$&$5$&&$85$&$.$&$33$&$\pm$&$4$&$.$&$48$&&$93$&$.$&$58$&$\pm$&$1$&$.$&$55$&&$128$&$.$&$67$&$\pm$&$1$&$.$&$66$&&\multicolumn{7}{c}{---}&&$417$&$.$&$8$&$\pm$&$5$&$.$&$3$&\\
4&$17$&$.$&$72$&$\pm$&$0$&$.$&$58$&&$16$&$.$&$58$&$\pm$&$0$&$.$&$76$&&$10$&$.$&$75$&$\pm$&$0$&$.$&$54$&&\multicolumn{7}{c}{---}&&\multicolumn{7}{c}{---}&&$11$&$.$&$95$&$\pm$&$0$&$.$&$54$&&$12$&$.$&$64$&$\pm$&$0$&$.$&$23$&&$8$&$.$&$29$&$\pm$&$0$&$.$&$71$&&\multicolumn{7}{c}{---}&\\
5&$30$&$.$&$44$&$\pm$&$0$&$.$&$93$&&$14$&$.$&$17$&$\pm$&$0$&$.$&$52$&&$23$&$.$&$1$&$\pm$&$0$&$.$&$6$&&\multicolumn{7}{c}{---}&&\multicolumn{7}{c}{---}&&$11$&$.$&$65$&$\pm$&$0$&$.$&$40$&&$14$&$.$&$62$&$\pm$&$0$&$.$&$30$&&$8$&$.$&$85$&$\pm$&$0$&$.$&$47$&&\multicolumn{7}{c}{---}&\\
6&$11$&$.$&$58$&$\pm$&$0$&$.$&$33$&&$6$&$.$&$54$&$\pm$&$0$&$.$&$45$&&$6$&$.$&$69$&$\pm$&$0$&$.$&$22$&&\multicolumn{7}{c}{---}&&\multicolumn{7}{c}{---}&&$3$&$.$&$58$&$\pm$&$0$&$.$&$27$&&\multicolumn{7}{c}{---}&&$2$&$.$&$01$&$\pm$&$0$&$.$&$23$&&\multicolumn{7}{c}{---}&\\
7&$21$&$.$&$76$&$\pm$&$0$&$.$&$73$&&\multicolumn{7}{c}{---}&&$14$&$.$&$42$&$\pm$&$0$&$.$&$50$&&\multicolumn{7}{c}{---}&&\multicolumn{7}{c}{---}&&\multicolumn{7}{c}{---}&&$7$&$.$&$68$&$\pm$&$0$&$.$&$21$&&\multicolumn{7}{c}{---}&&\multicolumn{7}{c}{---}&\\
8&$53$&$.$&$48$&$\pm$&$1$&$.$&$29$&&\multicolumn{7}{c}{---}&&$240$&$.$&$90$&$\pm$&$7$&$.$&$12$&&$477$&$.$&$2$&$\pm$&$20$&$.$&$5$&&$493$&$.$&$1$&$\pm$&$7$&$.$&$8$&&\multicolumn{7}{c}{---}&&$307$&$.$&$80$&$\pm$&$4$&$.$&$25$&&\multicolumn{7}{c}{---}&&$6468$&&&$\pm$&$88$&&&\\
9&$19$&$.$&$34$&$\pm$&$0$&$.$&$52$&&$13$&$.$&$23$&$\pm$&$1$&$.$&$66$&&$11$&$.$&$76$&$\pm$&$0$&$.$&$54$&&\multicolumn{7}{c}{---}&&\multicolumn{7}{c}{---}&&$11$&$.$&$42$&$\pm$&$1$&$.$&$49$&&$3$&$.$&$41$&$\pm$&$0$&$.$&$13$&c&\multicolumn{7}{c}{---}&&\multicolumn{7}{c}{---}&\\
10&$58$&$.$&$97$&$\pm$&$1$&$.$&$65$&&$44$&$.$&$90$&$\pm$&$0$&$.$&$91$&&$45$&$.$&$49$&$\pm$&$0$&$.$&$99$&&\multicolumn{7}{c}{---}&&\multicolumn{7}{c}{---}&&$29$&$.$&$90$&$\pm$&$0$&$.$&$69$&&$26$&$.$&$37$&$\pm$&$0$&$.$&$53$&&$19$&$.$&$72$&$\pm$&$1$&$.$&$35$&&\multicolumn{7}{c}{---}&\\
11&$23$&$.$&$98$&$\pm$&$0$&$.$&$69$&&$16$&$.$&$21$&$\pm$&$0$&$.$&$72$&&$17$&$.$&$09$&$\pm$&$0$&$.$&$49$&&\multicolumn{7}{c}{---}&&\multicolumn{7}{c}{---}&&$13$&$.$&$75$&$\pm$&$0$&$.$&$56$&&$9$&$.$&$72$&$\pm$&$0$&$.$&$43$&&$18$&$.$&$54$&$\pm$&$1$&$.$&$98$&c&\multicolumn{7}{c}{---}&\\
12&$1$&$.$&$76$&$\pm$&$0$&$.$&$09$&&$1$&$.$&$47$&$\pm$&$0$&$.$&$10$&&$1$&$.$&$23$&$\pm$&$0$&$.$&$07$&&\multicolumn{7}{c}{---}&&\multicolumn{7}{c}{---}&&$1$&$.$&$15$&$\pm$&$0$&$.$&$14$&&$0$&$.$&$44$&$\pm$&$0$&$.$&$18$&cr&\multicolumn{7}{c}{---}&&\multicolumn{7}{c}{---}&\\
13&$3$&$.$&$68$&$\pm$&$0$&$.$&$27$&&$5$&$.$&$85$&$\pm$&$0$&$.$&$27$&&$2$&$.$&$67$&$\pm$&$0$&$.$&$48$&&\multicolumn{7}{c}{---}&&\multicolumn{7}{c}{---}&&$24$&$.$&$25$&$\pm$&$0$&$.$&$56$&&$34$&$.$&$63$&$\pm$&$0$&$.$&$51$&&$200$&$.$&$13$&$\pm$&$13$&$.$&$84$&&$400$&$.$&$5$&$\pm$&$16$&$.$&$4$&\\
14&$4$&$.$&$62$&$\pm$&$0$&$.$&$34$&&\multicolumn{7}{c}{---}&&$9$&$.$&$23$&$\pm$&$1$&$.$&$72$&&\multicolumn{7}{c}{---}&&$58$&$.$&$07$&$\pm$&$0$&$.$&$45$&&\multicolumn{7}{c}{---}&&$100$&$.$&$90$&$\pm$&$1$&$.$&$12$&&\multicolumn{7}{c}{---}&&$883$&$.$&$6$&$\pm$&$6$&$.$&$8$&\\
15&$14$&$.$&$34$&$\pm$&$0$&$.$&$52$&&\multicolumn{7}{c}{---}&&$9$&$.$&$34$&$\pm$&$0$&$.$&$46$&&\multicolumn{7}{c}{---}&&\multicolumn{7}{c}{---}&&\multicolumn{7}{c}{---}&&$3$&$.$&$93$&$\pm$&$0$&$.$&$10$&&\multicolumn{7}{c}{---}&&\multicolumn{7}{c}{---}&\\
16&$2$&$.$&$67$&$\pm$&$0$&$.$&$12$&&\multicolumn{7}{c}{---}&&$1$&$.$&$59$&$\pm$&$0$&$.$&$09$&&\multicolumn{7}{c}{---}&&\multicolumn{7}{c}{---}&&\multicolumn{7}{c}{---}&&$1$&$.$&$89$&$\pm$&$0$&$.$&$11$&&\multicolumn{7}{c}{---}&&\multicolumn{7}{c}{---}&\\
17&$17$&$.$&$76$&$\pm$&$0$&$.$&$49$&&\multicolumn{7}{c}{---}&&$11$&$.$&$53$&$\pm$&$0$&$.$&$47$&&\multicolumn{7}{c}{---}&&\multicolumn{7}{c}{---}&&\multicolumn{7}{c}{---}&&$11$&$.$&$03$&$\pm$&$0$&$.$&$30$&&\multicolumn{7}{c}{---}&&\multicolumn{7}{c}{---}&\\
18&$8$&$.$&$29$&$\pm$&$0$&$.$&$31$&&\multicolumn{7}{c}{---}&&$6$&$.$&$30$&$\pm$&$0$&$.$&$25$&&\multicolumn{7}{c}{---}&&\multicolumn{7}{c}{---}&&\multicolumn{7}{c}{---}&&$7$&$.$&$66$&$\pm$&$0$&$.$&$22$&&\multicolumn{7}{c}{---}&&\multicolumn{7}{c}{---}&\\
19&$562$&&&$\pm$&$18$&&&&\multicolumn{7}{c}{---}&&$725$&$.$&$40$&$\pm$&$24$&$.$&$50$&&$799$&$.$&$4$&$\pm$&$33$&$.$&$6$&&$776$&$.$&$4$&$\pm$&$6$&$.$&$0$&&\multicolumn{7}{c}{---}&&$657$&$.$&$46$&$\pm$&$8$&$.$&$48$&&\multicolumn{7}{c}{---}&&$841$&$.$&$2$&$\pm$&$12$&$.$&$9$&\\
20&$159$&$.$&$60$&$\pm$&$3$&$.$&$60$&&\multicolumn{7}{c}{---}&&$171$&$.$&$90$&$\pm$&$5$&$.$&$47$&&$215$&$.$&$5$&$\pm$&$10$&$.$&$1$&&$240$&$.$&$7$&$\pm$&$5$&$.$&$4$&&\multicolumn{7}{c}{---}&&$153$&$.$&$28$&$\pm$&$1$&$.$&$41$&&\multicolumn{7}{c}{---}&&$138$&$.$&$1$&$\pm$&$30$&$.$&$2$&\\
21&$1200$&&&$\pm$&$30$&&&&$942$&$.$&$46$&$\pm$&$13$&$.$&$89$&&$1140$&&&$\pm$&$30$&&&&$1149$&&&$\pm$&$47$&&&&$1097$&&&$\pm$&$6$&&&&$1099$&$.$&$51$&$\pm$&$16$&$.$&$20$&&$776$&$.$&$73$&$\pm$&$10$&$.$&$73$&&$611$&$.$&$67$&$\pm$&$28$&$.$&$74$&&$621$&$.$&$4$&$\pm$&$34$&$.$&$8$&\\
22&$882$&$.$&$40$&$\pm$&$27$&$.$&$40$&&$882$&$.$&$80$&$\pm$&$13$&$.$&$01$&&$962$&$.$&$40$&$\pm$&$30$&$.$&$20$&&$992$&$.$&$3$&$\pm$&$40$&$.$&$7$&&$951$&$.$&$0$&$\pm$&$4$&$.$&$5$&&$996$&$.$&$32$&$\pm$&$14$&$.$&$68$&&$705$&$.$&$78$&$\pm$&$9$&$.$&$10$&&$660$&$.$&$88$&$\pm$&$21$&$.$&$31$&&$725$&$.$&$7$&$\pm$&$14$&$.$&$9$&\\
23&$748$&&&$\pm$&$15$&&&&$701$&$.$&$88$&$\pm$&$10$&$.$&$34$&&$810$&&&$\pm$&$19$&&&&$862$&$.$&$2$&$\pm$&$35$&$.$&$4$&&$848$&$.$&$9$&$\pm$&$6$&$.$&$2$&&$991$&$.$&$74$&$\pm$&$14$&$.$&$62$&&$684$&$.$&$65$&$\pm$&$9$&$.$&$46$&&$842$&$.$&$02$&$\pm$&$40$&$.$&$34$&&$944$&$.$&$9$&$\pm$&$8$&$.$&$4$&\\
24&$2$&$.$&$78$&$\pm$&$0$&$.$&$08$&&$1$&$.$&$34$&$\pm$&$0$&$.$&$14$&&$1$&$.$&$72$&$\pm$&$0$&$.$&$06$&&\multicolumn{7}{c}{---}&&\multicolumn{7}{c}{---}&&$0$&$.$&$73$&$\pm$&$0$&$.$&$11$&&\multicolumn{7}{c}{---}&&\multicolumn{7}{c}{---}&&\multicolumn{7}{c}{---}&\\
25&$6$&$.$&$59$&$\pm$&$0$&$.$&$24$&&$4$&$.$&$65$&$\pm$&$0$&$.$&$24$&&$3$&$.$&$92$&$\pm$&$0$&$.$&$18$&&\multicolumn{7}{c}{---}&&\multicolumn{7}{c}{---}&&$2$&$.$&$73$&$\pm$&$0$&$.$&$12$&&$2$&$.$&$97$&$\pm$&$0$&$.$&$12$&&\multicolumn{7}{c}{---}&&\multicolumn{7}{c}{---}&\\
26&$1$&$.$&$00$&$\pm$&$0$&$.$&$17$&&\multicolumn{7}{c}{---}&&\multicolumn{7}{c}{---}&&\multicolumn{7}{c}{---}&&\multicolumn{7}{c}{---}&&$4$&$.$&$30$&$\pm$&$0$&$.$&$81$&c&$28$&$.$&$99$&$\pm$&$0$&$.$&$69$&c&\multicolumn{7}{c}{---}&&\multicolumn{7}{c}{---}&\\
27&$10$&$.$&$7$&$\pm$&$0$&$.$&$4$&&$6$&$.$&$66$&$\pm$&$0$&$.$&$28$&&$6$&$.$&$17$&$\pm$&$0$&$.$&$26$&&\multicolumn{7}{c}{---}&&\multicolumn{7}{c}{---}&&$3$&$.$&$76$&$\pm$&$0$&$.$&$16$&&$2$&$.$&$91$&$\pm$&$0$&$.$&$09$&&$1$&$.$&$89$&$\pm$&$0$&$.$&$18$&&\multicolumn{7}{c}{---}&\\
28&$6$&$.$&$92$&$\pm$&$0$&$.$&$23$&&\multicolumn{7}{c}{---}&&$7$&$.$&$07$&$\pm$&$0$&$.$&$27$&&\multicolumn{7}{c}{---}&&\multicolumn{7}{c}{---}&&\multicolumn{7}{c}{---}&&$3$&$.$&$67$&$\pm$&$0$&$.$&$18$&&\multicolumn{7}{c}{---}&&\multicolumn{7}{c}{---}&\\
29&$30$&$.$&$3$&$\pm$&$1$&$.$&$1$&&\multicolumn{7}{c}{---}&&$18$&$.$&$7$&$\pm$&$0$&$.$&$5$&&\multicolumn{7}{c}{---}&&\multicolumn{7}{c}{---}&&\multicolumn{7}{c}{---}&&$12$&$.$&$00$&$\pm$&$0$&$.$&$27$&&\multicolumn{7}{c}{---}&&\multicolumn{7}{c}{---}&\\
30&$31$&$.$&$4$&$\pm$&$0$&$.$&$5$&&\multicolumn{7}{c}{---}&&$25$&$.$&$6$&$\pm$&$0$&$.$&$2$&&\multicolumn{7}{c}{---}&&\multicolumn{7}{c}{---}&&\multicolumn{7}{c}{---}&&$17$&$.$&$36$&$\pm$&$0$&$.$&$34$&&\multicolumn{7}{c}{---}&&\multicolumn{7}{c}{---}&\\
31&$91$&$.$&$3$&$\pm$&$2$&$.$&$4$&&\multicolumn{7}{c}{---}&&$105$&&&$\pm$&$2$&&&&$107$&$.$&$3$&$\pm$&$6$&$.$&$5$&&$118$&$.$&$4$&$\pm$&$9$&$.$&$8$&&\multicolumn{7}{c}{---}&&$89$&$.$&$43$&$\pm$&$1$&$.$&$32$&&\multicolumn{7}{c}{---}&&\multicolumn{7}{c}{---}&\\
\hline
\end{tabular}
\tablefoot{Values are frequency flux density in units of mJy. Suffixes to values of `c' and `r' indicate that the value is affected by confusion or the source being significantly resolved, respectively.}
\end{footnotesize}
\end{sidewaystable*}

\begin{sidewaystable*}
\caption{Photometry at wavelengths longer than $20\,\mu{\rm m}$. }
\label{tab:PhotometryFarIR}
\centering
\begin{tabular}{rr@{}c@{}l@{ }c@{ }r@{}c@{}l@{}lr@{}c@{}l@{ }c@{ }r@{}c@{}l@{}lr@{}c@{}l@{ }c@{ }r@{}c@{}l@{}lr@{}c@{}l@{ }c@{ }r@{}c@{}l@{}lr@{}c@{}l@{ }c@{ }r@{}c@{}l@{}lr@{}c@{}l@{ }c@{ }r@{}c@{}l@{}lr@{}c@{}l@{ }c@{ }r@{}c@{}l@{}lr@{}c@{}l@{ }c@{ }r@{}c@{}l@{}l}
\hline\hline
ID & \multicolumn{8}{c}{MSX E} & \multicolumn{8}{c}{AKARI L24} & \multicolumn{8}{c}{WISE W4} & \multicolumn{8}{c}{MIPS [24]} & \multicolumn{8}{c}{MIPS [70]} & \multicolumn{8}{c}{PACS [100]} & \multicolumn{8}{c}{PACS [160]} & \multicolumn{8}{c}{SPIRE [250]} \\
   & \multicolumn{8}{c}{$\lambda_{\rm eff}=21.018\,\mu{\rm m}$} & \multicolumn{8}{c}{$\lambda_{\rm eff}=22.753\,\mu{\rm m}$} & \multicolumn{8}{c}{$\lambda_{\rm eff}=22.800\,\mu{\rm m}$} & \multicolumn{8}{c}{$\lambda_{\rm eff}=23.680\,\mu{\rm m}$} & \multicolumn{8}{c}{$\lambda_{\rm eff}=71.42\,\mu{\rm m}$} & \multicolumn{8}{c}{$\lambda_{\rm eff}=100\,\mu{\rm m}$} & \multicolumn{8}{c}{$\lambda_{\rm eff}=160\,\mu{\rm m}$} & \multicolumn{8}{c}{$\lambda_{\rm eff}=250\,\mu{\rm m}$} \\
   & \multicolumn{8}{c}{ } & \multicolumn{8}{c}{${\rm ZP}=8.0\,{\rm Jy}$} & \multicolumn{8}{c}{${\rm ZP}=8.284\,{\rm Jy}$} & \multicolumn{8}{c}{ } & \multicolumn{8}{c}{ } & \multicolumn{8}{c}{ } & \multicolumn{8}{c}{ } & \multicolumn{8}{c}{ } \\
\hline
1&\multicolumn{7}{c}{---}&&\multicolumn{7}{c}{---}&&$2071$&$.$&$29$&$\pm$&$95$&$.$&$42$&c&$482$&$.$&$2$&$\pm$&$96$&$.$&$4$&&\multicolumn{7}{c}{---}&&$636$&&&$\pm$&$254$&&&&$426$&&&$\pm$&$151$&&&&\multicolumn{7}{c}{---}&\\
2&\multicolumn{7}{c}{---}&&$1302$&$.$&$24$&$\pm$&$46$&$.$&$79$&&$1396$&$.$&$50$&$\pm$&$12$&$.$&$86$&&$1180$&&&$\pm$&$60$&&&&$987$&&&$\pm$&$27$&&&&$493$&&&$\pm$&$43$&&&&$109$&&&$\pm$&$17$&&&&$24$&$.$&$7$&$\pm$&$6$&$.$&$7$&\\
3&$918$&$.$&$6$&$\pm$&$77$&$.$&$2$&&\multicolumn{7}{c}{---}&&$758$&$.$&$30$&$\pm$&$9$&$.$&$78$&&$898$&&&$\pm$&$22$&&&&\multicolumn{7}{c}{---}&&\multicolumn{7}{c}{---}&&\multicolumn{7}{c}{---}&&\multicolumn{7}{c}{---}&\\
4&\multicolumn{7}{c}{---}&&$12$&$.$&$03$&$\pm$&$1$&$.$&$35$&&$29$&$.$&$91$&$\pm$&$1$&$.$&$13$&c&$10$&&&$\pm$&$2$&&&&\multicolumn{7}{c}{---}&&\multicolumn{7}{c}{---}&&\multicolumn{7}{c}{---}&&\multicolumn{7}{c}{---}&\\
5&\multicolumn{7}{c}{---}&&$8$&$.$&$01$&$\pm$&$0$&$.$&$34$&&$9$&$.$&$78$&$\pm$&$0$&$.$&$69$&&$7$&$.$&$783$&$\pm$&$0$&$.$&$225$&&\multicolumn{7}{c}{---}&&\multicolumn{7}{c}{---}&&\multicolumn{7}{c}{---}&&\multicolumn{7}{c}{---}&\\
6&\multicolumn{7}{c}{---}&&\multicolumn{7}{c}{---}&&$3$&$.$&$22$&$\pm$&$0$&$.$&$55$&c&$1$&$.$&$293$&$\pm$&$0$&$.$&$155$&&\multicolumn{7}{c}{---}&&\multicolumn{7}{c}{---}&&\multicolumn{7}{c}{---}&&\multicolumn{7}{c}{---}&\\
7&\multicolumn{7}{c}{---}&&\multicolumn{7}{c}{---}&&$11$&$.$&$85$&$\pm$&$0$&$.$&$75$&c&$21$&&&$\pm$&$5$&&&&$128$&&&$\pm$&$71$&&&&\multicolumn{7}{c}{---}&&\multicolumn{7}{c}{---}&&\multicolumn{7}{c}{---}&\\
8&$9316$&&&$\pm$&$559$&&&&\multicolumn{7}{c}{---}&&$4882$&$.$&$48$&$\pm$&$26$&$.$&$98$&&$>$$4100$&&&$\pm$&$410$&&&&$1858$&&&$\pm$&$24$&&&&$706$&&&$\pm$&$48$&&&&$145$&&&$\pm$&$16$&&&&$43$&$.$&$4$&$\pm$&$4$&$.$&$1$&\\
9&\multicolumn{7}{c}{---}&&\multicolumn{7}{c}{---}&&$43$&$.$&$40$&$\pm$&$4$&$.$&$32$&c&\multicolumn{7}{c}{---}&&\multicolumn{7}{c}{---}&&\multicolumn{7}{c}{---}&&\multicolumn{7}{c}{---}&&\multicolumn{7}{c}{---}&\\
10&\multicolumn{7}{c}{---}&&$15$&$.$&$36$&$\pm$&$0$&$.$&$58$&&$15$&$.$&$80$&$\pm$&$1$&$.$&$22$&&$16$&$.$&$13$&$\pm$&$0$&$.$&$28$&&\multicolumn{7}{c}{---}&&\multicolumn{7}{c}{---}&&\multicolumn{7}{c}{---}&&\multicolumn{7}{c}{---}&\\
11&\multicolumn{7}{c}{---}&&$25$&$.$&$32$&$\pm$&$3$&$.$&$11$&c&$116$&$.$&$80$&$\pm$&$5$&$.$&$81$&c&$49$&$.$&$4$&$\pm$&$0$&$.$&$5$&c&\multicolumn{7}{c}{---}&&\multicolumn{7}{c}{---}&&\multicolumn{7}{c}{---}&&\multicolumn{7}{c}{---}&\\
12&\multicolumn{7}{c}{---}&&$3$&$.$&$77$&$\pm$&$0$&$.$&$56$&r&$9$&$.$&$45$&$\pm$&$1$&$.$&$14$&r&$22$&&&$\pm$&$5$&&&&\multicolumn{7}{c}{---}&&\multicolumn{7}{c}{---}&&\multicolumn{7}{c}{---}&&\multicolumn{7}{c}{---}&\\
13&\multicolumn{7}{c}{---}&&$696$&$.$&$77$&$\pm$&$52$&$.$&$67$&&$700$&$.$&$55$&$\pm$&$7$&$.$&$10$&&$703$&&&$\pm$&$23$&&&&$428$&$.$&$3$&$\pm$&$15$&$.$&$1$&&$204$&$.$&$7$&$\pm$&$33$&$.$&$6$&&\multicolumn{7}{c}{---}&&\multicolumn{7}{c}{---}&\\
14&\multicolumn{7}{c}{---}&&\multicolumn{7}{c}{---}&&$1493$&$.$&$62$&$\pm$&$12$&$.$&$38$&&$1470$&&&$\pm$&$74$&&&&$579$&$.$&$3$&$\pm$&$8$&$.$&$8$&&$305$&&&$\pm$&$31$&&&&$60$&&&$\pm$&$16$&&&&\multicolumn{7}{c}{---}&\\
15&\multicolumn{7}{c}{---}&&\multicolumn{7}{c}{---}&&$2$&$.$&$68$&$\pm$&$0$&$.$&$45$&&$3$&$.$&$074$&$\pm$&$0$&$.$&$145$&&\multicolumn{7}{c}{---}&&\multicolumn{7}{c}{---}&&\multicolumn{7}{c}{---}&&\multicolumn{7}{c}{---}&\\
16&\multicolumn{7}{c}{---}&&\multicolumn{7}{c}{---}&&$40$&$.$&$69$&$\pm$&$1$&$.$&$16$&&$59$&&&$\pm$&$19$&&&&\multicolumn{7}{c}{---}&&\multicolumn{7}{c}{---}&&\multicolumn{7}{c}{---}&&\multicolumn{7}{c}{---}&\\
17&\multicolumn{7}{c}{---}&&\multicolumn{7}{c}{---}&&$49$&$.$&$32$&$\pm$&$1$&$.$&$41$&cr&$108$&&&$\pm$&$29$&&&cr&$150$&&&$\pm$&$56$&&&c&\multicolumn{7}{c}{---}&&\multicolumn{7}{c}{---}&&\multicolumn{7}{c}{---}&\\
18&\multicolumn{7}{c}{---}&&\multicolumn{7}{c}{---}&&$74$&$.$&$38$&$\pm$&$1$&$.$&$92$&c&$11$&$.$&$5$&$\pm$&$1$&$.$&$5$&&\multicolumn{7}{c}{---}&&\multicolumn{7}{c}{---}&&\multicolumn{7}{c}{---}&&\multicolumn{7}{c}{---}&\\
19&\multicolumn{7}{c}{---}&&\multicolumn{7}{c}{---}&&$931$&$.$&$19$&$\pm$&$16$&$.$&$30$&&$813$&&&$\pm$&$6$&&&&$528$&&&$\pm$&$9$&&&&$320$&&&$\pm$&$26$&&&&$83$&$.$&$8$&$\pm$&$14$&$.$&$5$&&$64$&$.$&$1$&$\pm$&$5$&$.$&$4$&\\
20&\multicolumn{7}{c}{---}&&\multicolumn{7}{c}{---}&&$107$&$.$&$41$&$\pm$&$2$&$.$&$28$&&$99$&$.$&$3$&$\pm$&$0$&$.$&$9$&&$104$&&&$\pm$&$29$&&&c&\multicolumn{7}{c}{---}&&\multicolumn{7}{c}{---}&&\multicolumn{7}{c}{---}&\\
21&\multicolumn{7}{c}{---}&&$462$&$.$&$90$&$\pm$&$11$&$.$&$09$&&$1197$&$.$&$40$&$\pm$&$19$&$.$&$85$&c&$411$&&&$\pm$&$3$&&&&\multicolumn{7}{c}{---}&&\multicolumn{7}{c}{---}&&\multicolumn{7}{c}{---}&&\multicolumn{7}{c}{---}&\\
22&\multicolumn{7}{c}{---}&&$505$&$.$&$70$&$\pm$&$15$&$.$&$37$&&$597$&$.$&$37$&$\pm$&$13$&$.$&$21$&&$498$&&&$\pm$&$4$&&&&\multicolumn{7}{c}{---}&&$123$&$.$&$3$&$\pm$&$90$&$.$&$8$&&\multicolumn{7}{c}{---}&&\multicolumn{7}{c}{---}&\\
23&$1152$&&&$\pm$&$85$&&&&$1151$&$.$&$04$&$\pm$&$22$&$.$&$26$&&$1197$&$.$&$40$&$\pm$&$17$&$.$&$65$&&$1120$&&&$\pm$&$10$&&&&$380$&&&$\pm$&$24$&&&&$199$&&&$\pm$&$52$&&&&$117$&&&$\pm$&$37$&&&&$67$&$.$&$4$&$\pm$&$8$&$.$&$9$&\\
24&\multicolumn{7}{c}{---}&&\multicolumn{7}{c}{---}&&$3$&$.$&$78$&$\pm$&$0$&$.$&$77$&c&\multicolumn{7}{c}{---}&&\multicolumn{7}{c}{---}&&\multicolumn{7}{c}{---}&&\multicolumn{7}{c}{---}&&\multicolumn{7}{c}{---}&\\
25&\multicolumn{7}{c}{---}&&\multicolumn{7}{c}{---}&&$252$&$.$&$72$&$\pm$&$5$&$.$&$82$&c&\multicolumn{7}{c}{---}&&\multicolumn{7}{c}{---}&&\multicolumn{7}{c}{---}&&\multicolumn{7}{c}{---}&&\multicolumn{7}{c}{---}&\\
26&\multicolumn{7}{c}{---}&&$130$&$.$&$46$&$\pm$&$21$&$.$&$85$&c&$2433$&$.$&$55$&$\pm$&$40$&$.$&$35$&c&\multicolumn{7}{c}{---}&&\multicolumn{7}{c}{---}&&\multicolumn{7}{c}{---}&&\multicolumn{7}{c}{---}&&\multicolumn{7}{c}{---}&\\
27&\multicolumn{7}{c}{---}&&$1$&$.$&$21$&$\pm$&$0$&$.$&$30$&&$2$&$.$&$12$&$\pm$&$0$&$.$&$42$&&$0$&$.$&$6$&$\pm$&$0$&$.$&$2$&&\multicolumn{7}{c}{---}&&\multicolumn{7}{c}{---}&&\multicolumn{7}{c}{---}&&\multicolumn{7}{c}{---}&\\
28&\multicolumn{7}{c}{---}&&\multicolumn{7}{c}{---}&&$58$&$.$&$81$&$\pm$&$2$&$.$&$76$&c&$1$&$.$&$634$&$\pm$&$0$&$.$&$131$&&\multicolumn{7}{c}{---}&&\multicolumn{7}{c}{---}&&\multicolumn{7}{c}{---}&&\multicolumn{7}{c}{---}&\\
29&\multicolumn{7}{c}{---}&&\multicolumn{7}{c}{---}&&$29$&$.$&$83$&$\pm$&$1$&$.$&$35$&&$27$&$.$&$0$&$\pm$&$0$&$.$&$4$&&$58$&$.$&$7$&$\pm$&$8$&$.$&$0$&&\multicolumn{7}{c}{---}&&\multicolumn{7}{c}{---}&&\multicolumn{7}{c}{---}&\\
30&\multicolumn{7}{c}{---}&&\multicolumn{7}{c}{---}&&$58$&$.$&$43$&$\pm$&$1$&$.$&$67$&&$67$&$.$&$3$&$\pm$&$8$&$.$&$8$&&\multicolumn{7}{c}{---}&&\multicolumn{7}{c}{---}&&\multicolumn{7}{c}{---}&&\multicolumn{7}{c}{---}&\\
31&\multicolumn{7}{c}{---}&&\multicolumn{7}{c}{---}&&$92$&$.$&$61$&$\pm$&$2$&$.$&$47$&&$87$&$.$&$0$&$\pm$&$0$&$.$&$6$&&\multicolumn{7}{c}{---}&&\multicolumn{7}{c}{---}&&\multicolumn{7}{c}{---}&&\multicolumn{7}{c}{---}&\\
\hline
\end{tabular}
\tablefoot{Values are frequency flux density in units of mJy. Suffixes to values of `c' and `r' indicate that the value is affected by confusion or the source being significantly resolved, respectively.}
\end{sidewaystable*}

\end{appendix}
\end{document}